\documentclass[fleqn,11pt]{article}
\usepackage[figuresright]{rotating}
\usepackage{a4wide,epsfig}
\usepackage{epsfig}
\usepackage{xspace}
\usepackage{graphicx}
\usepackage{multirow}
\usepackage{rotating,amsmath,amssymb}
\usepackage{eurosym}

\def\Journal#1#2#3#4{{#1} {\bf #2}, #3 (#4)}

\def\NIM{\em Nucl. Instrum. Methods}
\def\NIMA{{\em Nucl. Inst. Meth.} A}
\def\NPB{{\em Nucl. Phys.} B}

\def\CPC{{\em Comp. Phys. Com.}}

\newcommand{\numu}{\mbox{$\nu_{\mu}~$}}
\newcommand{\numubar}{\mbox{$\overline{\nu}_{\mu}~$}}
\newcommand{\nue}{\mbox{$\nu_{e}~$}}
\newcommand{\nuebar}{\mbox{$\overline{\nu}_{e}~$}}

\newcommand{\nova}{\ensuremath{\mbox{NO} \nu \mbox{A}}\xspace}

\newcommand{\GeV}{\ensuremath{\mbox{GeV}}\xspace}
\newcommand{\MeV}{\ensuremath{\mbox{MeV}}\xspace}
\newcommand{\GeVc}{\ensuremath{\mbox{GeV}/c}\xspace}
\newcommand{\MeVc}{\ensuremath{\mbox{MeV}/c}\xspace}

\newcommand{\Km}{\ensuremath{\mbox{km}}\xspace}
\newcommand{\cm}{\ensuremath{\mbox{cm}}\xspace}
\newcommand{\mm}{\ensuremath{\mbox{mm}}\xspace}

\newcommand{\ns}{\ensuremath{\mbox{ns}}\xspace}
\newcommand{\m}{\ensuremath{\mbox{m}}\xspace}

\newcommand{\Tesla}{\ensuremath{\mbox{Tesla}}\xspace}

\newcommand{\Kton}{\ensuremath{\mbox{kton}}\xspace}

\def\enu{E_\nu}
\def\ehad{E_{h}}
\def\evis{E_{vis}}
\def\dcp{\delta_{CP}}
\def\pmu{P_{\mu}}
\def\qt{Q_t}

\begin{document}

\begin{titlepage}

\vspace*{-3.0cm}
\begin{center}
\begin{figure}[htbp]
\epsfig{figure=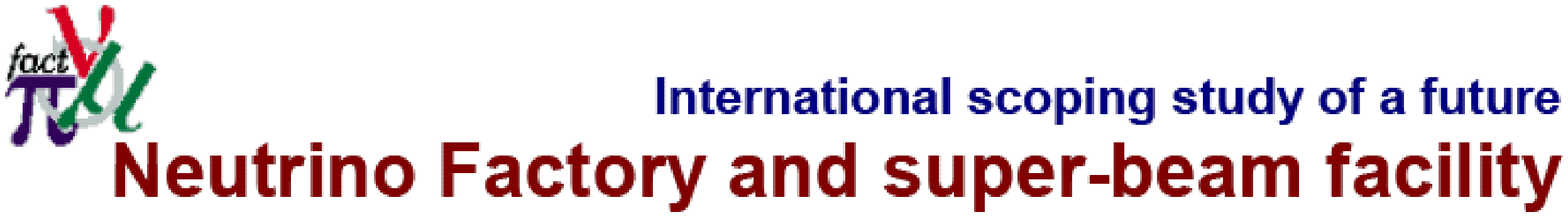, height=2.5cm}
\end{figure}
\end{center}
\vspace*{-1.1cm}
{\bf RAL-TR-2007-24 }
\vspace*{-0.2cm}

\begin{center}
{\bf \LARGE Detectors and flux instrumentation for future \\}
\vspace*{0.2cm}
{\bf \LARGE neutrino facilities}
\end{center}

\def\tokyo{\mathrm{a}}  
\def\ral{\mathrm{b}}    
\def\eth{\mathrm{c}}    
\def\infnmilano{\mathrm{d}}  
\def\geneva{\mathrm{e}}    
\def\saclay{\mathrm{f}}    
\def\cern{\mathrm{g}}      
\def\fnal{\mathrm{h}}      
\def\granada{\mathrm{i}}   
\def\lal{\mathrm{j}}       
\def\frascati{\mathrm{k}}  
\def\ific{\mathrm{l}}      
\def\napoli{\mathrm{m}}      
\def\bern{\mathrm{n}}      
\def\gransasso{\mathrm{o}}      
\def\toho{\mathrm{p}}    
\def\sofia{\mathrm{q}}    
\def\hamamtsu{\mathrm{r}}    
\def\nagoya{\mathrm{s}}    
\def\sgooulu{\mathrm{t}}    
\def\inr{\mathrm{u}}    
\def\padova{\mathrm{v}}    
\def\suny{\mathrm{w}}    
\def\york{\mathrm{x}}    
\def\mumbai{\mathrm{y}}    
\def\pavia{\mathrm{z}}    
\def\kek{{\mathrm{aa}}}    
\def\virginia{{\mathrm{ab}}}    
\def\cuppoulu{{\mathrm{ac}}}    
\def\milano{{\mathrm{ad}}}    
\def\ifae{{\mathrm{ae}}}    
\def\bologna{{\mathrm{af}}}    
\def\glasgow{{\mathrm{ag}}}    
\def\imperial{{\mathrm{ah}}}    
\def\oxford{{\mathrm{ai}}}    
\def\wroclaw{{\mathrm{aj}}}    
\def\louisiana{{\mathrm{ak}}}    

\def\AFtokyo{     ICRR, University of Tokyo, Tokyo, Japan}  
\def\AFral{       CCLRC, Rutherford Appleton Laboratory, UK}
\def\AFeth{       ETH, Zurich, Switzerland}
\def\AFinfnmilano{INFN Sezione di Milano, Milano, Italy}
\def\AFgeneva{    University of Geneva, Geneva, Switzerland}
\def\AFsaclay{    CEA/DAPNIA-Saclay \& APC Paris}
\def\AFcern{      CERN, Geneva, Switzerland}
\def\AFfnal{      FNAL, USA}
\def\AFgranada{   University of Granada, Granada, Spain}
\def\AFlal{       LAL, Univ Paris-Sud, IN2P3/CNRS, Orsay, France}
\def\AFfrascati{  Laboratori Nazionali di Frascati dell'INFN, Frascati (Roma), Italy}
\def\AFific{      IFIC, University of Valencia and CSIC, Valencia, Spain}
\def\AFnapoli{    Universit\`a di Napoli Federico II and INFN, Napoli, Italy}
\def\AFbern{      University of Bern, Bern, Switzerland}
\def\AFgransasso{ Laboratori Nazionali del Gran Sasso dell'INFN, Assergi (L'Aquila), Italy}
\def\AFtoho{      Toho University, Funabashi 274-8510, Japan}
\def\AFsofia{     University of Sofia, "St Kliment Ohridski"}
\def\AFhamamtsu{  Hamamatsu Photonics, Japan}
\def\AFnagoya{    Nagoya University, 464-01 Nagoya, Japan}
\def\AFsgooulu{   Sodankylä Geophysical Observatory, University of Oulu, Oulu, Finland}
\def\AFinr{       INR, Moscow, Russia}
\def\AFpadova{    Universit\`a di Padova and INFN, Padova, Italy}
\def\AFsuny{      SUNY Stony Brook, NY, USA} 
\def\AFyork{      York University, Toronto, Ontario, Canada}
\def\AFmumbai{    Tata Institute, Mumbai, India}
\def\AFpavia{     Universit\`a di Pavia and INFN, Pavia, Italy}
\def\AFkek{       KEK, Tsukuba, Japan}
\def\AFvirginia{  College of William \& Mary, Williamsburg, Virginia, USA}
\def\AFcuppoulu{  Centre for Underground Physics Pyh\"asalmi, Univ of Oulu, Oulu, Finland}
\def\AFmilano{    Universit\`a di Milano and INFN, Milano, Italy}
\def\AFifae{      IFAE, Barcelona, Spain}
\def\AFbologna{   Universit\`a di Bologna and INFN, Italy}
\def\AFglasgow{   University of Glasgow, Glasgow, Scotland, UK}
\def\AFimperial{  Imperial College, London, UK}
\def\AFoxford{    University of Oxford, Oxford, UK}
\def\AFwroclaw{  Inst. Th. Physics,  University of Wroclaw, Poland}
\def\AFlouisiana{  Louisiana State University, Louisiana, USA}

\def\back{\!\!}

\begin{center}

T.~Abe                                                  \back    $^\tokyo$,   
H.~Aihara                                               \back    $^\tokyo$, 
C.~Andreopoulos                                         \back    $^\ral$, 
A. Ankowski       	              \back    $^{\wroclaw}$, 
A.~Badertscher                                          \back    $^\eth$, 
G.~Battistoni                                           \back    $^\infnmilano$, 
A.~Blondel                                              \back    $^\geneva$, 
J.~Bouchez                                              \back    $^\saclay$,    
%
%
A.~Bross                                                \back    $^\fnal$, 
A.~Bueno                                                \back    $^\granada$, 
L.~Camilleri                                            \back    $^\cern$, 
J.E.~Campagne                                           \back    $^\lal$, 
A.~Cazes                                                \back    $^\frascati$, 
A.~Cervera-Villanueva                                   \back    $^\ific$,  
%
%
G.~De Lellis                                            \back    $^\napoli$,  
F.~Di Capua                                             \back    $^\napoli$,  
M.~Ellis                                                \back    $^\fnal$,  
A.~Ereditato                                            \back    $^\bern$,  
L.S.~Esposito$^\gransasso$,  
C.~Fukushima                                            \back    $^\toho$,
E.~Gschwendtner                                         \back    $^\cern$,  
J.J.~Gomez-Cadenas                                      \back    $^\ific$,  
M.~Iwasaki                                              \back    $^\tokyo$,  
%
%
K.~Kaneyuki                                             \back    $^\tokyo$,  
Y.~Karadzhov                                           \back    $^\sofia$,
V.~Kashikhin                                            \back    $^\fnal$,  
Y.~Kawai                                                \back    $^\hamamtsu$, 
M.~Komatsu                                              \back    $^\nagoya$,   
E.~Kozlovskaya                                          \back    $^\sgooulu$,
Y.~Kudenko                                              \back    $^\inr$,
A.~Kusaka                                               \back    $^\tokyo$,
H.~Kyushima                                             \back    $^\hamamtsu$, 
A.~Longhin                                              \back    $^\padova$, 
A.~Marchionni                                            \back    $^\eth$,  
A.~Marotta                                              \back    $^\napoli$, 
C.~McGrew                                               \back    $^\suny$,  
S.~Menary                                               \back    $^{\fnal,\york}$,  
A.~Meregaglia                                           \back    $^\eth$,  
M.~Mezzeto                                              \back    $^\padova$,  
P.~Migliozzi                                            \back    $^\napoli$,  
N.K.~Mondal                                             \back    $^\mumbai$,  
C.~Montanari                                            \back    $^\pavia$,  
T.~Nakadaira                                            \back    $^\kek$,
M.~Nakamura                                             \back    $^\nagoya$,  
H.~Nakumo                                               \back    $^\tokyo$,  
H.~Nakayama                                             \back    $^\tokyo$,  
J.~Nelson                                               \back    $^\virginia$,  
J.~Nowak                \back    $^\louisiana$, 
S.~Ogawa                                                \back    $^\toho$,  
J.~Peltoniemi                                           \back    $^\cuppoulu$,  
A.~Pla-Dalmau                                          \back    $^\fnal$,  
S.~Ragazzi                                              \back    $^\milano$, 
A.~Rubbia                                               \back    $^\eth$,  
F.~Sanchez                                              \back    $^\ifae$, 
J.~Sarkamo                                              \back    $^\cuppoulu$, 
O.~Sato                                                 \back    $^\nagoya$,
%
%
M.~Selvi                                                \back    $^\bologna$, 
H.~Shibuya                                              \back    $^\toho$,
M.~Shozawa                                              \back    $^\tokyo$,   
J.~Sobczyk                \back    $^\wroclaw$,
F.J.P.~Soler                                            \back    $^\glasgow$,  
P.~Strolin                                              \back    $^\napoli$,
M.~Suyama                                               \back    $^\hamamtsu$,
M.~Tanak                                                \back    $^\kek$,
F.~Terranova                                            \back    $^\frascati$,
R.~Tsenov                                               \back    $^\sofia$,  
Y.~Uchida$^\imperial$,  
A.~Weber$^{\oxford,\ral}$,  
%
%
A.~Zlobin                                               \back    $^\fnal$                                                                                         

\vspace*{0.5cm}

{\it \scriptsize

{$^\tokyo$              \AFtokyo}    \\                
{$^\ral$                \AFral}    \\                
{$^\eth$                \AFeth}    \\                
{$^\infnmilano$          \AFinfnmilano}    \\         
{$^\geneva$             \AFgeneva}    \\             
{$^\saclay$             \AFsaclay}    \\             
{$^\cern$               \AFcern}    \\               
{$^\fnal$               \AFfnal}    \\               
{$^\granada$            \AFgranada}    \\            
{$^\lal$                \AFlal}    \\                
{$^\frascati$           \AFfrascati}    \\           
{$^\ific$               \AFific}    \\               
{$^\napoli$             \AFnapoli}    \\             
{$^\bern$               \AFbern}    \\               
{$^\gransasso$           \AFgransasso}    \\          
{$^\toho$               \AFtoho}    \\               
{$^\sofia$              \AFsofia}    \\              
{$^\hamamtsu$           \AFhamamtsu}    \\           
{$^\nagoya$             \AFnagoya}    \\             
{$^\sgooulu$            \AFsgooulu}    \\            
{$^\inr$                \AFinr}    \\                
{$^\padova$             \AFpadova}    \\             
{$^\suny$               \AFsuny}    \\               
{$^\york$               \AFyork}    \\               
{$^\mumbai$             \AFmumbai}    \\             
{$^\pavia$              \AFpavia}    \\              
{$^\kek$                \AFkek}    \\                
{$^\virginia$           \AFvirginia}    \\           
{$^\cuppoulu$           \AFcuppoulu}    \\           
{$^\milano$             \AFmilano}    \\         
{$^\ifae$               \AFifae}    \\               
{$^\bologna$            \AFbologna}    \\            
{$^\glasgow$            \AFglasgow}    \\            
{$^\imperial$           \AFimperial}    \\           
{$^\oxford$             \AFoxford}   \\
{$^\wroclaw$             \AFwroclaw}  \\
{$^\louisiana$             \AFlouisiana} 
                 
}

\vspace{0.5cm}
{\small The ISS Detector Working Group}
\end{center}

\begin{center}
{\bf \Large Abstract}\\
\end{center}
This report summarises the conclusions from the detector group of the International Scoping Study of a future Neutrino Factory and Super-Beam neutrino facility. The baseline detector options for each possible neutrino beam are defined as follows:                                                                                                                           
\begin{enumerate}
\item 
A very massive (Megaton) water Cherenkov detector is the baseline option for a sub-GeV Beta Beam and Super Beam facility.  
\item    
There are a number of possibilities for either a Beta Beam or Super Beam (SB) medium energy facility between 1-5 GeV. These include a totally active scintillating detector (TASD),
a liquid argon TPC or a water Cherenkov detector.

\item   
A 100 kton magnetized iron neutrino detector (MIND) is the baseline to detect the wrong sign muon final states (golden channel) at a high energy (20-50 GeV) neutrino factory from muon decay.
A 10 kton hybrid neutrino magnetic emulsion cloud chamber detector for wrong sign tau 
detection (silver channel) is a possible complement to MIND, if one needs to resolve 
degeneracies that appear in the $\delta$-$\theta_{13}$ parameter space.
\end{enumerate}

\tableofcontents

\end{titlepage}




\section{Introduction}

The International Scoping Study (ISS) for a future accelerator neutrino complex was carried out by
the international community between NuFact05, Frascati, 21-26 June 2005, and NuFact06, Irvine,
24-30 August 2006. The physics case for the facility was evaluated and options for the accelerator
complex and neutrino detection systems were studied. One of the novel characteristics of the
ISS with respect to previous studies was the systematic investigation of detector options for
future long base line neutrino experiments, as a necessary step towards optimising the performance
of the whole facility. In addition to the study of far detectors it was felt
necessary to add a study of the near detectors and instrumentation for the primary beam line. 
These are crucial to understand the performance of the facilities from the point of view of
systematic errors.  This applies to the Beta-beam or Neutrino Factory storage ring, or to the
Superbeam decay tunnel. Two additional topics of critical relevance for the choice of facility
were added to the discussion: matter effect uncertainties and systematic errors due to
uncertainties in the cross-sections and efficiencies of low energy neutrino interactions.

\subsection{Organization}

Following the initial guidelines given at NUFACT05~\cite{nufact05}, 
the working groups have largely built on existing studies 
to delineate the main avenues where further investigations would be most beneficial, 
and initiated the required simulation work. The work was carried out in five working groups: 

\begin{itemize}
\item Segmented magnetic detectors; 
\item Large Water Cherenkov detectors (WC); 
\item Large Liquid Argon TPCs (LAr TPC);
\item Emulsion-based detectors: Emulsion Cloud Chamber (ECC) and Magnetized ECC (MECC);
\item Near detector and beam instrumentation. 
\end{itemize}

The important issue of novel detector techniques of common interest 
(such as Silicon Photo Multipliers and large area photo-detectors) 
was treated in common dedicated sessions of the working group. 
Finally, the need of large magnetic volumes required for the neutrino factory detector was considered.

The mandate of the study was to establish a set of baseline detectors to be carried forward for further study. 
It is clear that accomplishing such a goal would require an extremely tight collaboration between the physics 
performance group and the detector design group. However a number of choices could be 
made
from known feasibility/cost considerations.  

\subsection{Main beam and far detector options}

The main far detector options are listed below:

\begin{enumerate}
\item	{\bf Single flavour sub-GeV neutrino beams: low energy superbeam and beta-beam.} This 
is the scenario advocated for instance for the off-axis beam from J-PARC, the SPL superbeam and 
$^6He$ or $^{18}Ne$ beta-beams at CERN. In this energy range detectors need not be magnetized, quasi-elastic 
reactions dominate and pion production is small. 
A very massive water Cherenkov (WC) detector is the baseline option. 
The small and poorly understood  cross-sections, and the low $Q^2$ of the interactions pose considerable 
systematic problems which make the design of the near detectors very critical. The possibility to use 
very large LArTPCs has been envisaged, but the relative merit
would need to be better 
justified, and indications are that this is not the case. 
\item	{\bf Few GeV beams: off-axis and wide band beam and high energy beta-beam}. This is what one would obtain 
with an off-axis NUMI beam or equivalent, wide-band pion/kaon decay beam (WBB) from a 20-50~\GeV proton beam, or from a high energy 
beta-beam, either from high $\gamma$  $^6He$ or $^{18}Ne$ or from accelerating higher 
Q (e.g. $^8B$ or $^8Li$ ) isotopes. 
 Here the situation is more complex since multi-pion production becomes common and event identification 
requires more sophistication. This is not an easy energy domain to work at, and there is not a clear winner 
in this domain between the WC, the totally active scintillating detector (TASD) (\`{a} la \nova), a 
LArTPC or even an iron-scintillator sandwich.   
\item	{\bf High energy beams from muon decay (Neutrino Factory)}.  Magnetic detectors are compulsory since two 
leptonic charges of neutrinos are present at the same time. The baseline detector here is the magnetized 
iron neutrino detector (MIND) for the wrong sign muon final states, but the full exploitation of the richness of 
possible oscillation channels strongly motivates the study of other types of detectors: magnetized low Z fine 
grain detector (scintillator or LAr) for wrong-sign electron final states, emulsion detector (ECC) for 
wrong-sign tau detection and magnetized emulsion (MECC) for all the above. 
\end{enumerate}  

In all three scenarios appropriate near detectors and beam instrumentation are essential. 
Indeed, the precision era poses new challenges for the flux and 
cross-section monitoring systems. Appearance measurements require that the product of cross-section times 
acceptance be measured for the appearance channel in relation to that of initial neutrino flavour. This is a major difficulty for the conventional pion decay superbeam, since little 
instrumentation can be installed to monitor the secondary flux of mesons in a high intensity environment; 
there is a clear need for specific hadro-production experiments backed up with fine grained near detectors, to measure precisely \numu, \numubar, \nue~  and ~\nuebar, topological  cross-sections. 
The issue is much easier for the beta-beam or the neutrino factory, 
where the stored parent beam can be monitored 
precisely and the known decay provides a potentially well known flux. In addition, purely leptonic reactions 
can be used as absolute candles. A new domain of precision cross-section measurements at the $10^{-3}$ 
level opens up. 
Of course a detailed simulation and study of the near detector station and of the associated near detectors and 
beam instrumentation is required to firm up these claims. 

More details and the presentations can be found on the detector study web site~\cite{det_web}. 
The physics performance\footnote{Including signal and background efficiencies in some cases} 
and the sensitivity to the oscillation parameters of the different 
far detectors (and combinations) can be found in the ISS Physics Report~\cite{iss_physics_report}.

\subsection{Main achievements and open issues}

Given that this study is not the first one, it is worthwhile emphasizing in this introduction what is the new 
information content, and what are the issues which remain open after its completion. 

The main achievements or new information gathered through this study are as follows. 
\begin{itemize}
\item A Magnetized Iron Neutrino Detector (MIND) of 100~\Kton should be feasible for a hardware 
  cost of $\sim$200~M\euro. 
\item The threshold for muon detection in an optimised   
  MIND can be lowered down to 1-3~GeV/c for a dominant background of wrong charge assignment 
  of $\mathcal{O}(10^{-3})$. The efficiency above 5~\GeV can be set to 70\%. 
  \item A large air-core coil can be envisaged to host 20-30~\Kton of fully active 
  fine grained detector (scintillator, LAr or emulsion) for a reasonable cost ($\mathcal{O}$(100 M\euro) ). 
\item The muon detection threshold can be further lowered down to $\sim$0.4~\GeVc  
  using a Totally Active Scintillating Detector (TASD). This detector should be able to measure 
  the charge of the muon with a negligible mis-identification rate ($\mathcal{O}(10^{-5})$) for muons 
  above $\sim$0.4~\GeVc.  

\item A MECC of 10~\Kton can be designed, which, 
  thanks to the exquisite space and angle resolution of the emulsion, can measure electron and 
  muon charge and momentum up to $\sim$10~\GeV. 
\item The first studies of very large underground excavations have been pursued and cost 
  estimates for a megaton WC detector have been given.
\item A revolution in photo-detection has been brought forward in the last few 
  years with the appearance of new type of avalanche-photodiode-arrays (SiPMs of MPPCs).
\item In the context of the LArTPC-Glacier project  
  the operation of a (small) LAr TPC in a magnetic field was achieved and a comercial company 
  has produced a feasibility study of a LNG tanker for 100~\Kton LAr. 
  There is a very active R\&D program to study i) a two phase detector with very long drift paths, 
  ii) novel charge readout and HV supply and iii) drift properties at high pressure.  
\item A large LAr TPC (15 to 50~\Kton) is being considered in the US as the detector for a long-baseline 
  $\nu_\mu \to \nu_e$ appearance experiment. The efficiency for detecting $\nu_e$'s in such a detector 
  is $\sim$80--90\% with a negligible neutral-current $\pi^0$ event background. An ambitious R\&D 
  program was approved in 2005 and is underway. 
\item Matter effects can be calculated rather precisely down to a matter density uncertainty of 
  about 2\% or better, but a dedicated geological study has to be foreseen once the site has been chosen. 
  A few particular baselines encountering very irregular terrain should be avoided. 
\item A first estimate was performed of the interplay of final state lepton mass, nuclear effects, 
  and non-isoscalar target (water) with the conclusion that at a few 100~\MeV they impact measurements 
  of CP asymmetries by several percent. This effect decreases with energy for the quasi-elastic reaction, and at higher energy may affect also the pion production channel. Detector effects have not been studied yet. 
\item The detectors can take alternative trains of neutrinos produced by stored positive and negative muons 
  as long as the time distance between trains is above 100~\ns.  
\end{itemize}

Nevertheless many issues remain open for further study and R\&D. A few outstanding points are listed below.

{\bf For what concerns the neutrino factory detectors:} 

\begin{itemize}
\item Priority should be given to a solid study of performance, cost estimate and infrastructure 
  requirements of the baseline detector for the neutrino factory (MIND) and of its variants (such as the Indian Neutrino Observatory, INO).  
\item The performance of the TASD detector against hadronic backgrounds should be computed. Pion decay and 
  pion/muon mis-identification could be important given the low detector density. 
\item The study of the large coils and associated infrastructure for the above has only 
  started and this is clearly a field that motivates further studies. The super-conducting 
  transfer line (STL) is probably the most promising option for large magnetic 
  volumes at reasonable cost.  A full engineering design would still need to be done.
  \item The comparative performance study of 'platinum detectors' should be pushed to a conclusion. 
  Efficiency vs charge confusion background for the electron channel for different setups 
  (MECC, TASD or LAr) needs to be understood and compared
\item The monitoring of the muon beam angular divergence in the storage ring is for the 
  moment a very challenging concept (a He Cherenkov with extremely thin windows) that needs 
  to be turned into a demonstrably feasible object. It is not clear that a permanent device 
  can be devised or if a different system needs to be invented.  
  \item The near detector concepts and the near detector area for the neutrino factory needs to be defined, 
  including in a coherent way the necessary shielding and of the purely leptonic detector and DIS-charm 
  detector. 
\item Once a site is considered a study of the matter content of the beam line will be mandatory. 
\end{itemize}
 
{\bf  For what concerns the low energy beta-beam and superbeam detectors:}

\begin{itemize}
\item The priority is rightly given to understanding the feasibility and cost of the Mton-class water Cherenkov detector, in order to exploit the synergy with proton decay and supernovae neutrino detection.
\item How shallow can a LArTPC be operated? This was recently studied \cite{Bueno:2007um} 
for shallow depths ($\sim$ 200 m depth) but it would be good to understand the status for surface operation.
\item Whether a giant LArTPC can usefully compete in this energy range should be ascertained 
  more quantitatively, while the cost and infrastructure/safety implications of it is largely uncertain. 
\item The design and even the concept of the near detector station -- and the problems related to the relative 
  normalization of the beta-beam and superbeam when used in combination -- have not really been addressed and 
  constitute one of the major pending issues in addressing the physics capabilities of this option. There are also 
  fundamental issues associated with doing physics with low energy events: the effects of lepton mass, 
  nuclear effects, Fermi motion and binding energy are some, but the different topologies and their effect 
  on relative acceptance for \numu vs \nue events remains largely untouched. At this point in time any 
  claim of normalization errors (even relative) below 5\% remains 
  unestablished.
\end{itemize}

\section{Beam instrumentation }
\label{Beam_Instrumentation}

\subsection{Flux Control and Resulting Constraints on the Decay Ring
Design for the Neutrino Factory}

One of the most significant qualities of the Neutrino Factory, and more
generally of a system where one stores a beam of decaying particles (such
as the beta beam) is the potential for excellent neutrino flux control.  
The main parameters that govern the systematic uncertainties on the
neutrino fluxes are as follows.

\begin{itemize}
\item 
The monitoring of the total number of muons circulating in the ring,
\item
Theoretical knowledge of the neutrino fluxes from muon decay,
including higher-order radiative effects,
\item
Knowledge of the muon beam polarisation,
\item 
Knowledge of the muon beam energy and energy spread,
\item 
The muon beam angle and angular divergence.
\end{itemize}

Beam shape parameters are crucial for the measurement of oscillation
length, while the absolute normalisation is essential for the measurement
of the mixing angle. The relative normalisation of the two muon charges
plays a crucial role in the measurement of CP asymmetries.

\subsubsection{Neutrino fluxes from muon decay}

The neutrino energy spectra from negative muon decay at rest follow the following
distributions:
\begin{equation}
\label{eq:nufact_fluxes_1}
\frac{d^2 N_{\nu_\mu}}{dx d\Omega}  \propto 
\frac{2 x^2}{4\pi}[(3-2x)+(1-2x)P_\mu \cos\theta]
\end{equation}
\begin{equation}
\label{eq:nufact_fluxes_2}
\frac{d^2 N_{\bar{\nu}_e}}{dx d\Omega}  \propto  \frac{12
  x^2}{4\pi}[(1-x)+(1-x)P_\mu \cos\theta]
\end{equation}
where $x \equiv 2E_\nu/m_\mu$, $P_\mu$ is the muon polarisation, and
$\theta$ is the angle between the muon polarisation vector and the
neutrino direction.  In a long baseline experiment the detector is located 
on the same axis as the Lorentz boost and its size is negligible relative to the 
baseline. In this case the neutrino energy spectrum in the laboratory frame is given by 
the same formula as above but with $x=E_\nu/E_\mu$.


\subsubsection{Absolute flux monitoring}

Monitoring the total number of muons in the ring can be done in a
number of ways. The total beam current can be estimated using a Beam
Current Transformer (BCT), the total number of decay electrons can be
estimated using an electron spectrometer, the product of the flux and
cross section can be inferred from a near-by detector and, finally, the
absolute normalisation can be obtained from semi-leptonic neutrino
interactions in a nearby detector.

The operation of a BCT in the decay ring could provide fast-response
monitoring of the muons in the ring. There are, however, a few potential
difficulties that could limit the precision of such a device, which could
normally reach the $10^{-3}$ level.  The first one is the
presence of decay electrons in the ring, along with the muons. Since all
muons decay, the number of accompanying electrons could potentially be
much larger than the remaining muons after a few muon lifetimes.  A study
of such decay electrons has been made~\cite{keillosses}, with the
conclusion that for 50~GeV muon momentum, the decay electrons are lost in
the beam elements (or the collimators placed to protect them)  after less
than half a turn, either because they are momentum-mismatched or because
they lose energy in the arcs by synchrotron radiation.  Consequently their
number should be always less than about $2 \times 10^{-3}$ of the 
remaining
muons.  In addition, most of the losses arise in the straight sections or
in the early part of the arcs, so that a BCT situated just at the
beginning of a straight section would see an even smaller fraction of
them.  Another worry could be the existence of a moving electron cloud
created by beam-induced multipacting, or by ionization of the residual
gas or of the chamber walls.  This has been studied 
in~\cite{zimmermannlosses}, with the conclusion that the electron cloud
will be several orders of magnitude less than the muon flux itself. In the
absence of a significant parasitic current, it can be concluded that the
BCT readings should be precise to the level of a few $10^{-3}$, or better.
This seems the most practical way to compare the flux induced from $\mu^+$
and $\mu^-$ decays.

\begin{figure}[tbhp]
\begin{center}
\includegraphics[width=0.9\textwidth]{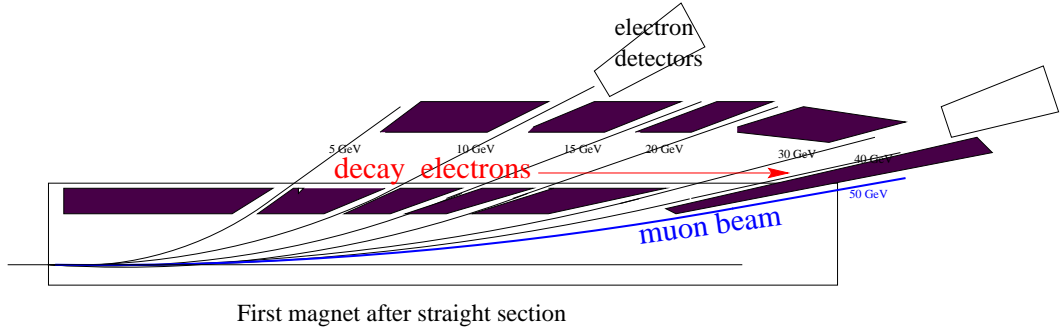}
\caption[muon polarimeter]
{{\sl
A possible muon polarimeter design. The momenta of the decay electrons 
accumulated in a short straight section are analysed in a bending magnet 
in the muon decay ring. Slits in the shielding define the
acceptance of a number of momentum bins.
}}
\label{polarimeter}
\end{center}
\end{figure}

The decay electrons will be used to measure the polarisation of the beam
with a spectrometer as described below, and in Fig.~\ref{polarimeter}. The
same device could in principle be used to monitor the number of muon
decays in an absolute way, especially if one selects the momentum bite
where the electron spectrum is insensitive to the muon polarisation.  
Certainly this will be a useful tool, as a cross-check or for monitoring,
but a very detailed study of the dependence of the acceptance of this
device on the beam parameters must be performed before a conclusion can be
reached.

Knowledge of the flux does not provide knowledge of the cross sections 
folded with the detector acceptance. This task is traditionally delegated to a
near detector.  The high flux should make things very easy. 
Given the high importance of
precision measurements in the Neutrino Factory, it is likely that a near
detector will be an important tool for beam normalisation. Unlike the
situation with conventional pion decay beams, the near detector will in
fact be able to measure absolute cross sections for a large number of
exclusive and inclusive processes.

It is worthwhile mentioning, finally, the possibility offered by the
measurement of purely leptonic interaction processes, which have been
discussed in~\cite{DISreport}. Of practical interest for normalisation is
the measurement of $\nu_{\mu} + e^- \rightarrow \mu^- + \nu_e$, which
appears as a low-angle forward-going muon with no recoil.  Using the
standard electroweak theory, this purely leptonic charged-current process
can be calculated with high precision, and could be measured with a
dedicated detector aimed at measuring also the $\nu_{e} + e^- \rightarrow
e^- + \nu_e$ and $\bar{ \nu_{e}} + e^- \rightarrow e^- + \bar{\nu_e}$
processes.  The weakness of this method is that it only applies to the
$\mu^-$ decay beam, but it could be seen as an overall absolute
normalisation process for the muon flux.

To conclude, there are many tools to monitor and control the absolute flux
normalisation in a neutrino factory, so that the near detectors should be
able to provide very accurate measurements of inclusive and exclusive
cross sections, within the detector acceptance. A flux normalisation at
the level of a few $10^{-3}$ seems an achievable goal. The relative
normalisation of the $\mu^-$ and $\mu^+$ decay beams should be known with
similar precision.

\subsubsection{Theoretical knowledge of the neutrino fluxes from muon
decay}

\begin{figure}[hbt]
\begin{center}
\includegraphics[width=0.49\textwidth]{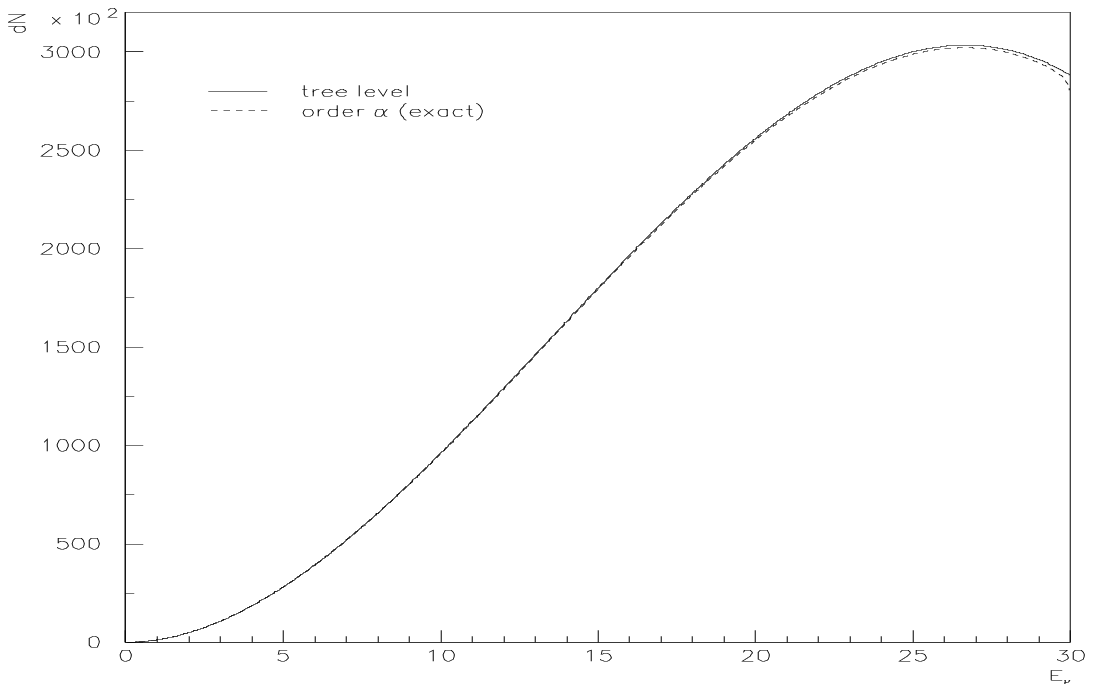}
\includegraphics[width=0.49\textwidth]{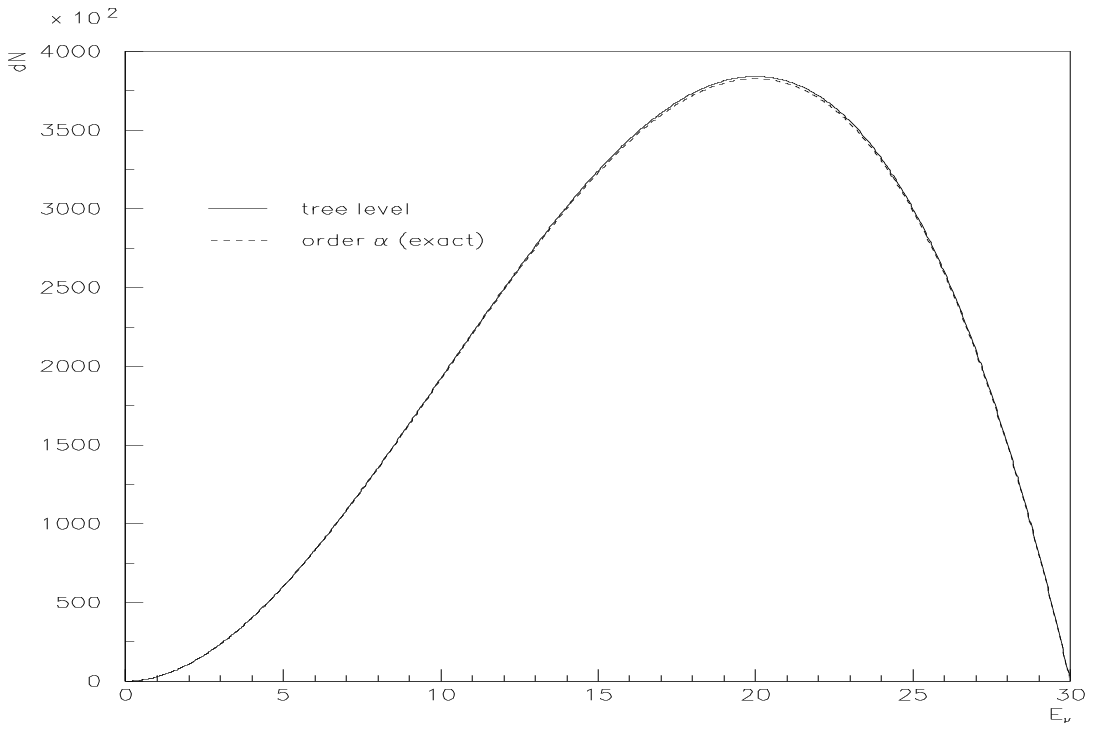}  \\
\includegraphics[width=0.49\textwidth]{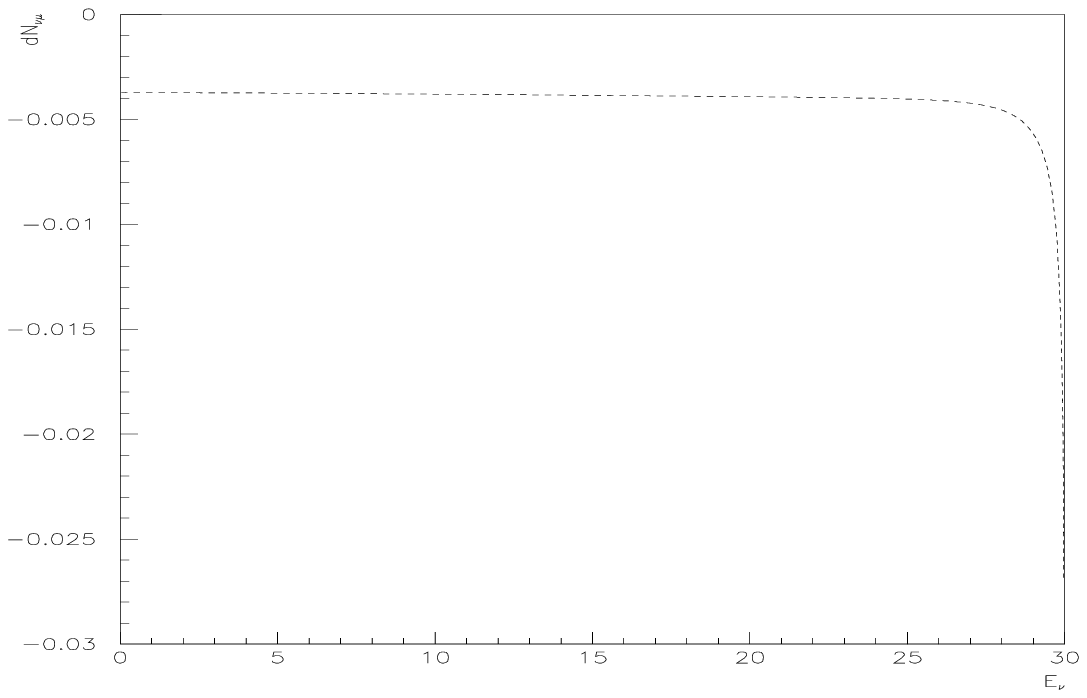}
\includegraphics[width=0.49\textwidth]{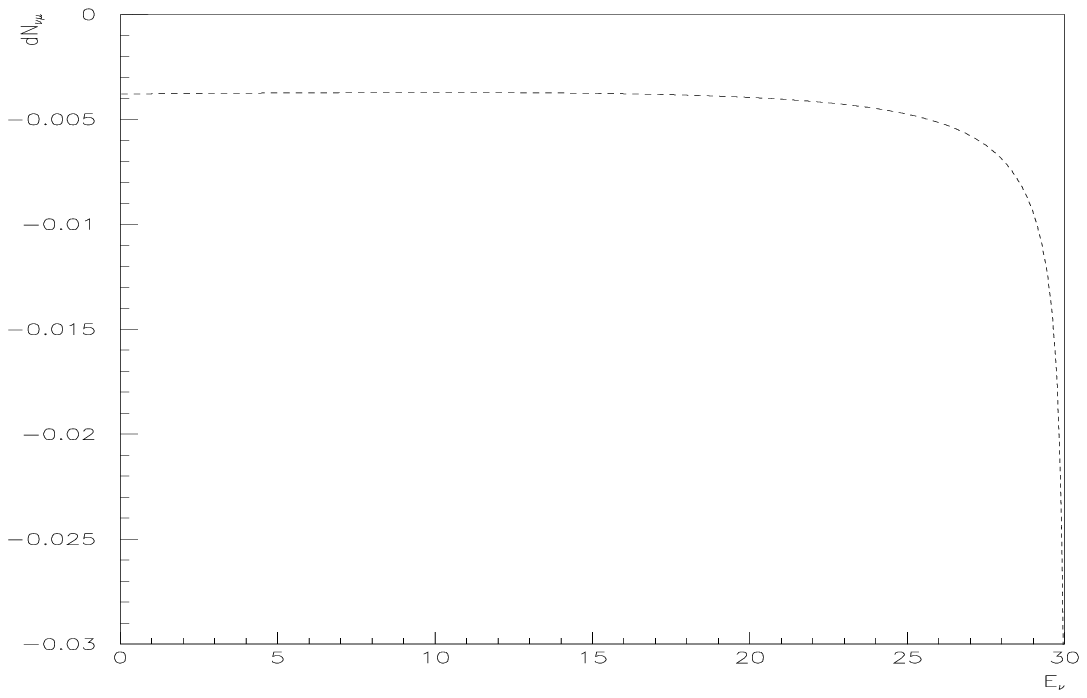}
\caption
{\sl
Radiative corrections to the muon neutrino (left panels) and electron anti-neutrino 
(right panels) fluxes in $\mu^-$ decay. 
Top panels: the resulting energy distribution at zero angle. Bottom  
panels: the relative change due to the ${\cal O}(\alpha)$ correction. 
The overall reduction of flux is due to the additional energy taken away by 
photons, which slightly widens the angular distribution of the neutrinos. 
In order to avoid infinities at the end point, the quantity plotted is 
$\frac{\Phi({\cal O}(\alpha)) -\Phi_0}{\Phi({\cal O}(\alpha)) +\Phi_0}$. 
\label{numufluxcorrection}}
\end{center}
\end{figure}

The expressions given above for the neutrino flux in muon decay,
(equations \ref{eq:nufact_fluxes_1} and \ref{eq:nufact_fluxes_2}), 
do not include QED radiative corrections, which
have been calculated in~\cite{Broncano-Mena} (see Fig.~\ref{numufluxcorrection}). 
The dominant source of corrections is, as can be expected, related to 
photon emission from the decay electron. 
For the electron energy distribution, the corrections 
are of the order of 1\% due to terms proportionanl to 
${\frac{\alpha}{\pi}}\ln {({\frac{m_\mu}{m_e}})}$. It turns out that
the neutrino spectrum is insensitive to the electron mass, i.e., the
integration over the system of electron and photons cancels mass
singularities.  It can be seen that, in the forward direction, an overall
decrease of the neutrino flux of about $4 \times 10^{-3}$ is observed, 
with a
larger decrease near the end point. The global decrease can be understood
by the overall softening and angular widening of the neutrino decay
spectrum due to photon emission.
Since the overall size of the corrections is small, one can certainly
trust the calculated spectrum to a precision much better than $10^{-3}$.
 
\subsubsection{Muon polarisation}

Muons are naturally polarised in pion decay. In the
$\pi^+\rightarrow\mu^+\nu_{\mu}$ rest frame, both the $\nu_{\mu}$ and
$\mu^+$ have negative helicity. In the laboratory frame, the resulting
average helicity of the muon, or longitudinal polarisation, is reduced
from -100\% for a pion at rest to $<h>= -18\%$ for pions above 200-300
MeV momentum~\cite{FGF}. For a pion of given momentum, muon polarisation
is correlated with muon momentum.  It has been argued in~\cite{Palmer}
that monochromatisation of the pions followed by i)~a drift space to 
separate
muons of different momenta, and ii)~collection in successive RF buckets,
should allow separation in different bunches of muons of different
polarisations. This does not change the {\sl average} polarisation, but
creates bunches of different polarisation (up to 50\%), that can be of use
for physics, as long as the times of neutrino interactions are recorded
with a precision of a few nanoseconds.

The muon spin precesses in electric and magnetic fields that
are present during cooling and acceleration, but the muon
spin tune $\nu$ 
-- the number of additional spin precessions happening when the muon
makes a complete turn -- is very low:
$$
\nu = a_\mu \gamma = \frac{g_\mu-2}{2} \frac{\rm E_{ beam}}{m_\mu}
= \frac{\rm E_{beam} (GeV)}{90.6223(6)}\,\,.
$$
It has been evaluated~\cite{FGF} that 80 to 90\% of the original 
polarisation 
will survive all muon handling up to the injection into the storage ring. 
Its orientation will depend on the number of turns that the muons encounter
along the accelerator chain, and can be arranged to be longitudinal 
by an appropriate choice of geometry and of 
the energies in the recirculating linacs~\cite{bdlpol}. 
As we will see, this is not necessarily important. 

What will happen to the muon polarisation in the decay ring depends in the
first instance on whether its geometry is a ring (race track or triangle)
in which the muons undergo one rotation per turn, or a bow-tie, in which
the muon undergoes zero net rotation at each turn.

{\it In the case of a ring}, the polarisation will precess. The
orientation of the polarisation vector will be rotated with respect to the
muon direction by an angle which increases each turn by $2 \pi \nu$.
Unless the energy is chosen very carefully, it will not be aligned, and
reduced on average by a factor 2. At a muon energy of precisely $E =
45.311$~GeV, the spin tune is 0.5 and the polarisation flips during 
each turn.
This would allow the most powerful use of the polarisation for physics
purposes, but absolutely requires that the orientation is correctly chosen
at injection, a condition which is otherwise unnecessary in a ring
geometry.  If no special measure is taken, however, depolarisation will
occur, since particles of different energies will have their spins precess
with different spin-tunes.

The muon polarisation can be monitored by momentum analysis of the decay
electrons, as discussed in~\cite{CERN9902:bdl}, in a polarimeter that
could look like that sketched in Fig.~~\ref{polarimeter}. One can expect
that this measurement will be difficult: the relative normalisation of
electron rates in the different energy bins will depend on various muon
beam parameters such as its exact angle and divergence, and on a precise
modelling of the beam-line geometry. In a ring geometry, the device will
be exposed to a succession of negative and positive helicity muon bunches,
so it will have to perform relative measurements.  These should be
sensitive to small effects, with a {\em relative} precision of a few
percent. 

The spin precession in a storage ring provides a means of high precision
($10^{-6}$ or better) for energy calibration~\cite{Raja}. As shown
in~\cite{CERN9902:bdl}, the measurement of the depolarisation can be used
to measure the energy spread with high precision. In this case, the 
combined effect of precession and depolarisation ensure
that the muon polarisation integrated over a fill averages out to zero with
an excellent precision: simulations show that any residual polarisation is
less than $4 \times 10^{-4}$.

\begin{figure}[ht]
\begin{center}
\includegraphics[width=0.45\textwidth]{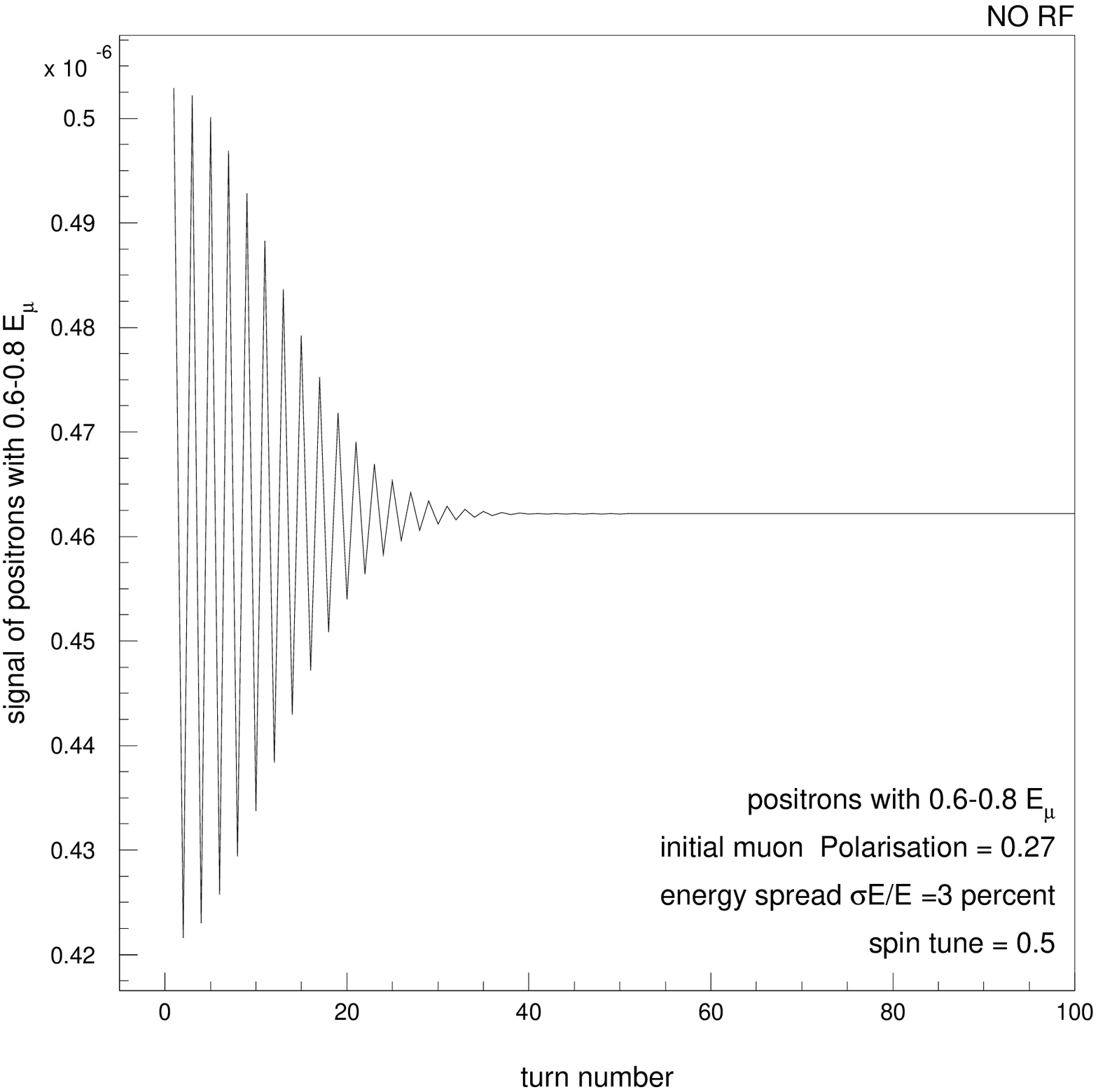}
\includegraphics[width=0.45\textwidth]{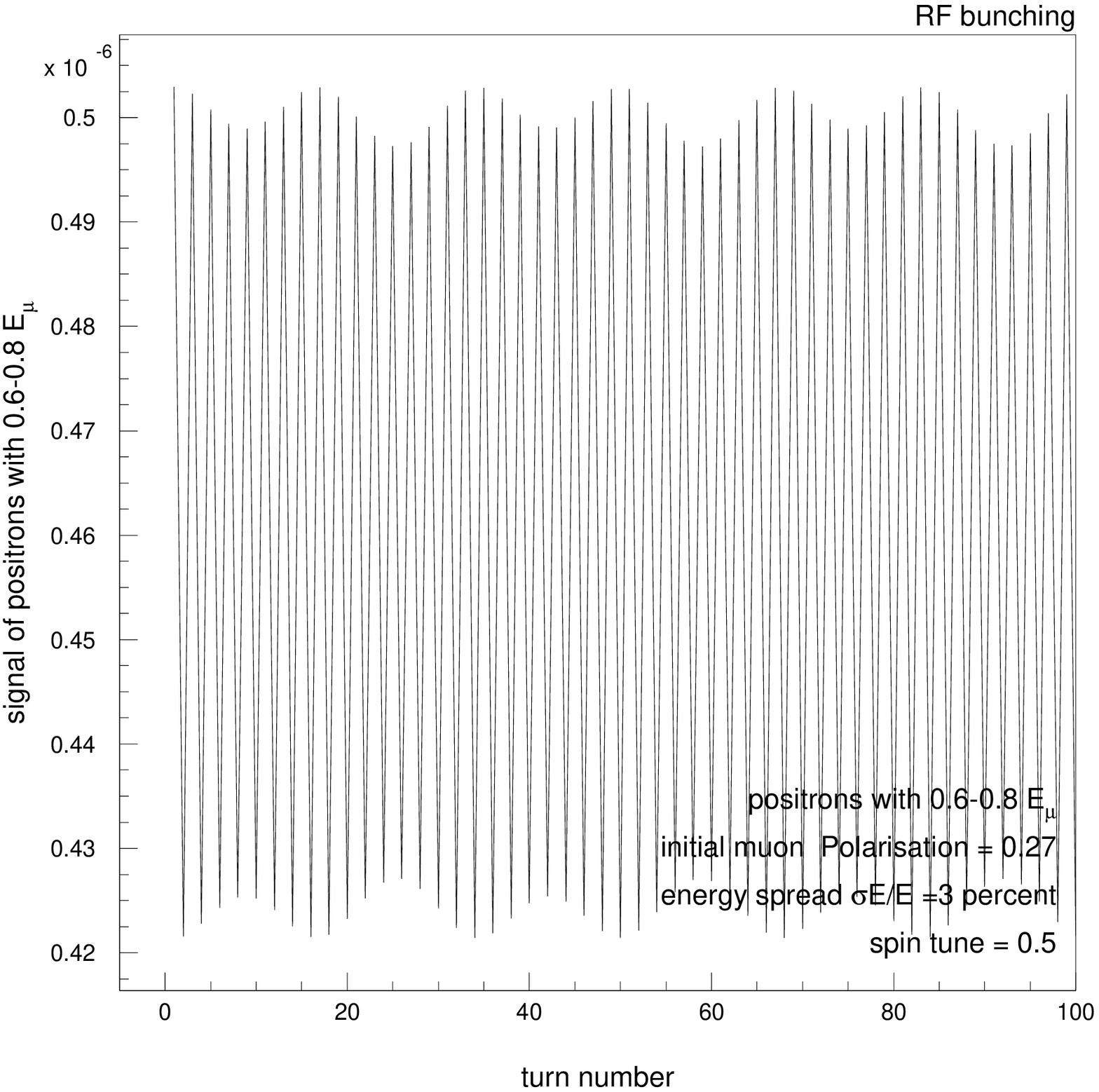}
\caption
{\sl Oscillation with turn number in a fill of the number of electrons
in the energy range 0.6-0.8 $E_{\mu}$, normalised to the 
total number of muon decays during the given turn.
The oscillation amplitude is a measure of the beam polarisation, 
its frequency a measure of the beam energy, 
and, if there is no RF bunching,  
its decrease with time is a measure of energy spread.
The muon lifetime corresponds here to 300 turns.
The beam energy is $E_{\mu}=45.311$~GeV and the energy spread is 
$3\times 
10^{-2}$.
On the left, there is no bunching  RF in the muon storage ring, 
on the right there is RF bunching with $Q_s=0.03$.}
\label{precession} 
\end{center}
\end{figure}

Depolarisation can be avoided, if the storage ring is equipped with an RF
system that ensures that the muons undergo synchrotron
oscillations~\cite{bdlpol}. By doing this, one loses the possibility to
measure the energy spread from the depolarisation, but one can maintain
the muon polarisation.  The average is still essentially zero, but by
recording the exact time of neutrino events, one can infer their bunch
number and turn number, and deduce the polarisation of the decay muons. In
a ring geometry either mode of operation is left open, if one can run with
the required RF system on or off.

{\it In the case of a bow-tie}, the muons will not depolarise: 
spin precession is zero no matter what the muon energy is. This configuration
is not as convenient as the ring for several reasons.

\begin{itemize}
\item{In a bow-tie geometry, there will be no spin precession,
so the energy and energy spread of the muon beam will not be calibrated.} 
\item{The polarisation will not average to zero and 
one will have to measure it based on the measured
electron spectrum. A few percent absolute accuracy seems to be
very challenging in this case, which means that the flux determination 
will be affected  by a sizeable uncertainty, due to the beam polarisation error.}
\item{It will be difficult to change the sign of the muon beam 
polarisation.}
\item{Unless the geometry is very carefully chosen, the beam 
polarisation will be different for the two long straight sections.}
\end{itemize}
For these reasons, and despite the fact that in principle
the useful beam polarisation is higher in the bow-tie geometry, 
{\it the ring geometry is preferred} from the point of view of beam 
control.  

\subsubsection{Neutrino fluxes and muon polarisation}

Neutrino spectra with different beam polarisations are given by 
equations~\ref{eq:nufact_fluxes_1} and \ref{eq:nufact_fluxes_2}. 
In a long-baseline experiment, one is at extremely small angles, so that
$\cos\theta=1$. In this case, the $\overline{\nu}_{e}$ component of the beam is
completely extinct for $ {\mathcal P}=-1$. This is due to spin
conservation in the decay: a left-handed muon cannot decay at zero angle
into a right-handed $\overline{\nu}_{e}$.

\begin{figure}[tbh]
\begin{center}
\includegraphics[width=0.45\textwidth]{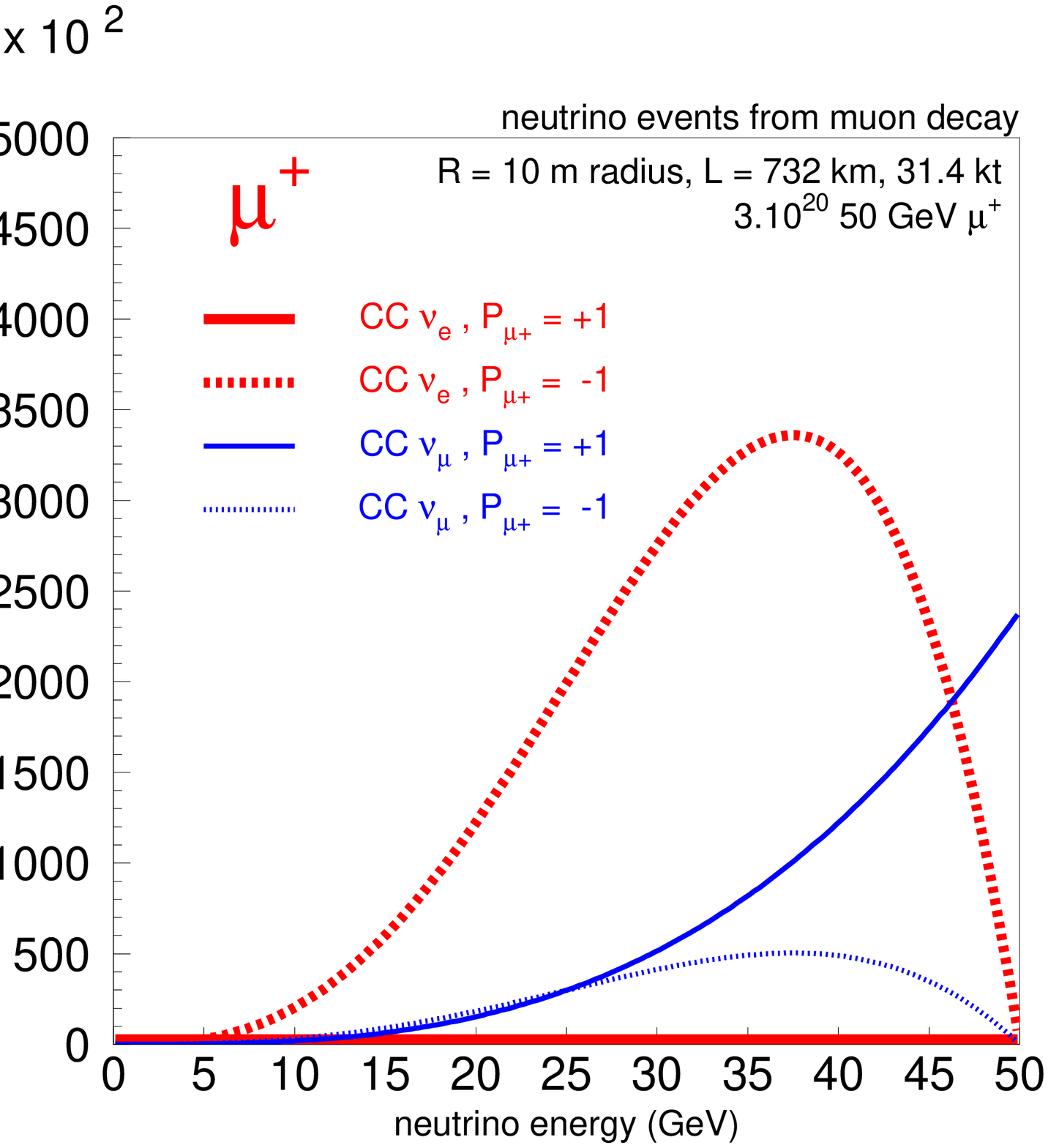}
\includegraphics[width=0.45\textwidth]{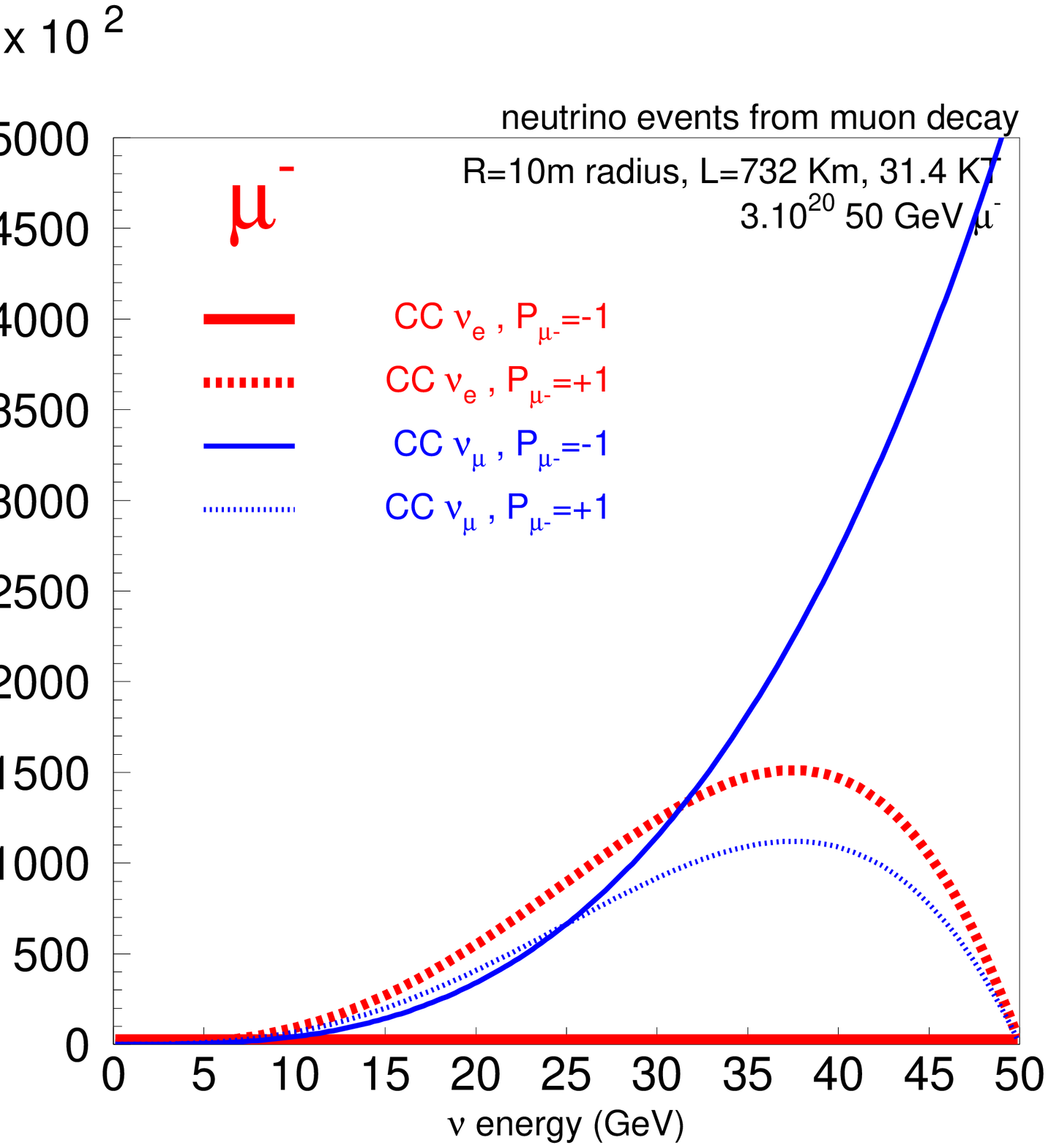}
\\
\caption
{\sl Event numbers for a detector of density 5 with 10~m radius that is
20~m long, situated 732~km away from the muon storage ring,
for $\mu^+  \rightarrow e^+ \nu_e \bar{\nu_{\mu}}$ (left) and  
$\mu^- \rightarrow e^- \bar{\nu_e} \nu_{\mu}$ (right) beams of 50 GeV.
Full lines show the spectra for the `natural' helicity
$\mathcal P =+1$ for $\mu^+$, and dashed ones for the opposite case. 
The CC $\nu_e$ for $\mu^+$ with  $\mathcal P =+1$ 
and  CC $\nu_e$ for $\mu^-$ with  $\mathcal P =-1$ are not visible, 
because the fluxes are almost exactly zero. The
vertical axis gives event numbers per bin of 250~MeV. 
This plot assumes no muon beam 
angular divergence and no beam energy spread. 
}
\label{events} 
\end{center}
\end{figure}

Event numbers can readily be obtained by multiplying by the cross section.
They are shown in Fig.~\ref{events} for a 10~m radius detector 20~m long
situated 730~km away.  Since the neutrino and anti-neutrino cross sections
are in the ratio 1/0.45, negative muons provide enrichment in $\nu_{\mu}$
and positive ones in $\nu_{e}$.

It is clear from Fig.~\ref{events} that the combination of muon sign and
polarisation allows large variations in the composition of the beam, in a
controlled way. Since detector studies show that the muon sign can easily
be determined in a charged-current (CC) (anti)neutrino event, but that the 
electron sign is
much more difficult, we have tried to use the variation of electron
neutrino flux with muon polarisation to infer a signal of $\nu_{\mu}
\rightarrow \nu_e$ oscillations to be compared (for a T-violation test)  
with the $\nu_{e} \rightarrow \nu_{\mu}$ oscillation measured with the
wrong-sign muons. Unfortunately, even for 40\% beam polarisation, the
improvement in the sensitivity to CP/T violation is no more than the
equivalent of a factor of 1.5 to 2 in statistics.  Certainly, it appears
that polarisation is more useful as a tool to measure the beam properties
than as a physics tool. Nevertheless, these statements might be 
parameter-dependent, and should be revisited once the oscillation 
parameters are
better known.

\subsubsection{Effect of beam divergence}

\begin{figure}[ht]
\begin{center}
\includegraphics[width=0.4\textwidth]{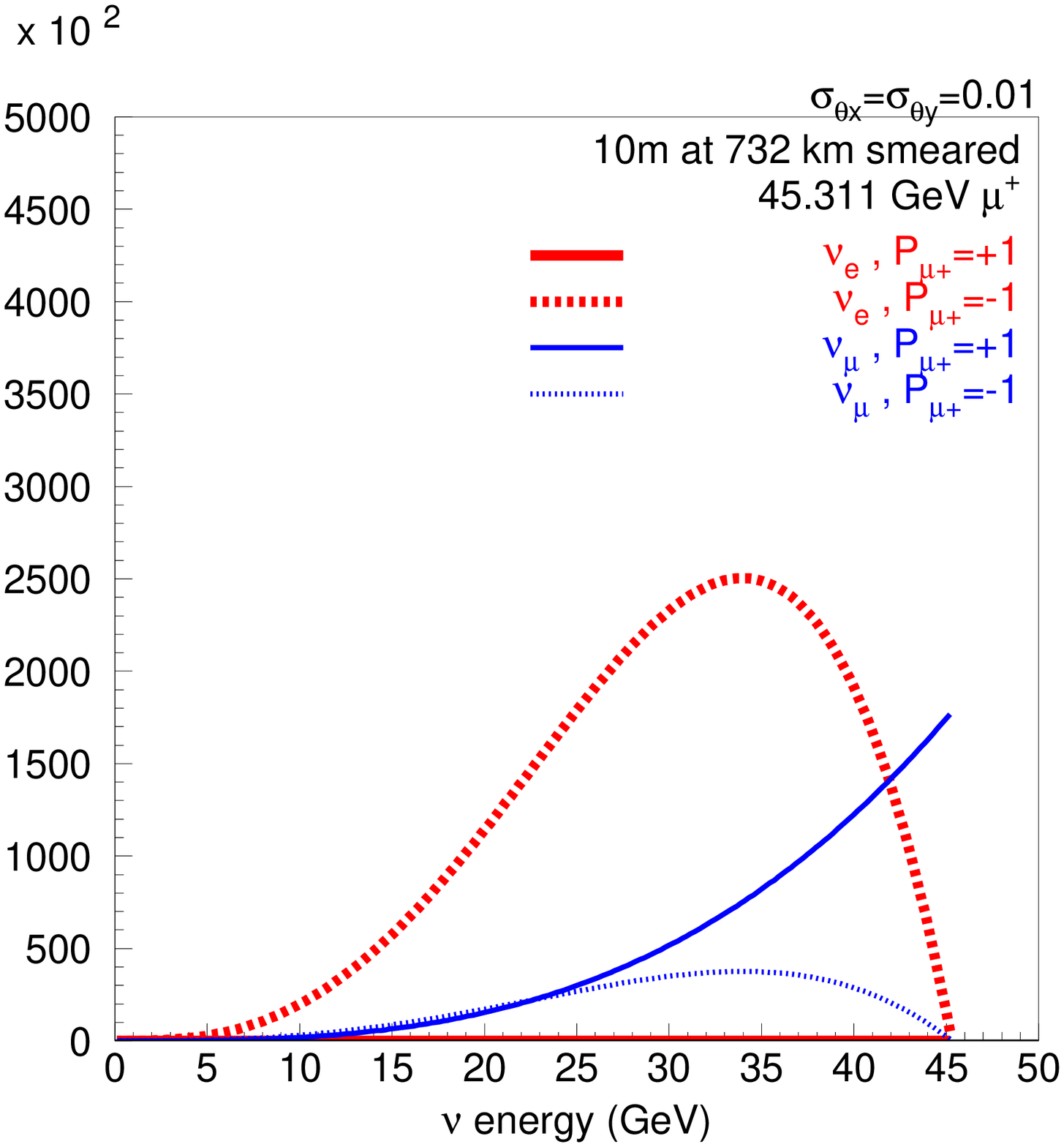}
\includegraphics[width=0.4\textwidth]{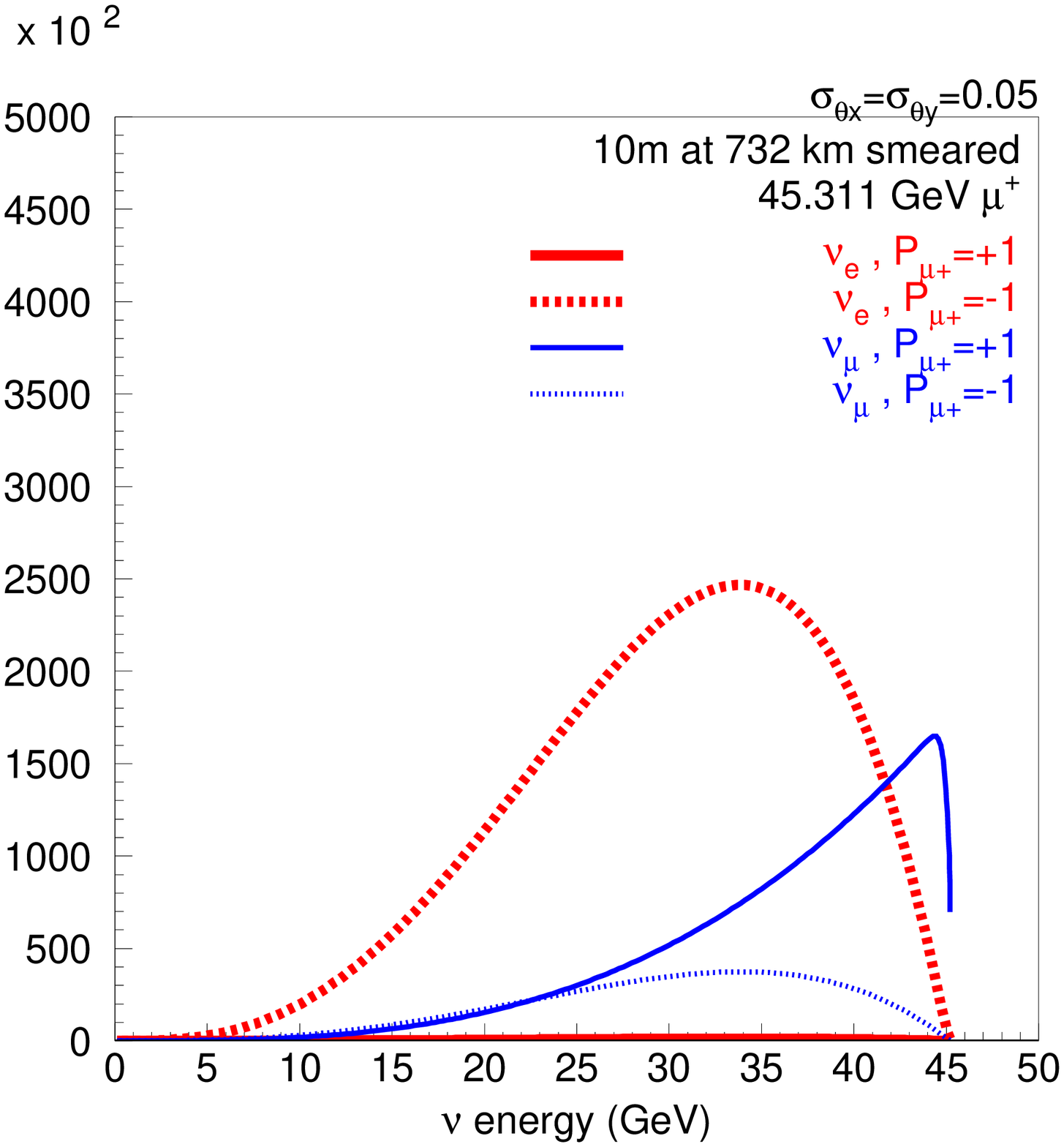}
\\
\includegraphics[width=0.4\textwidth]{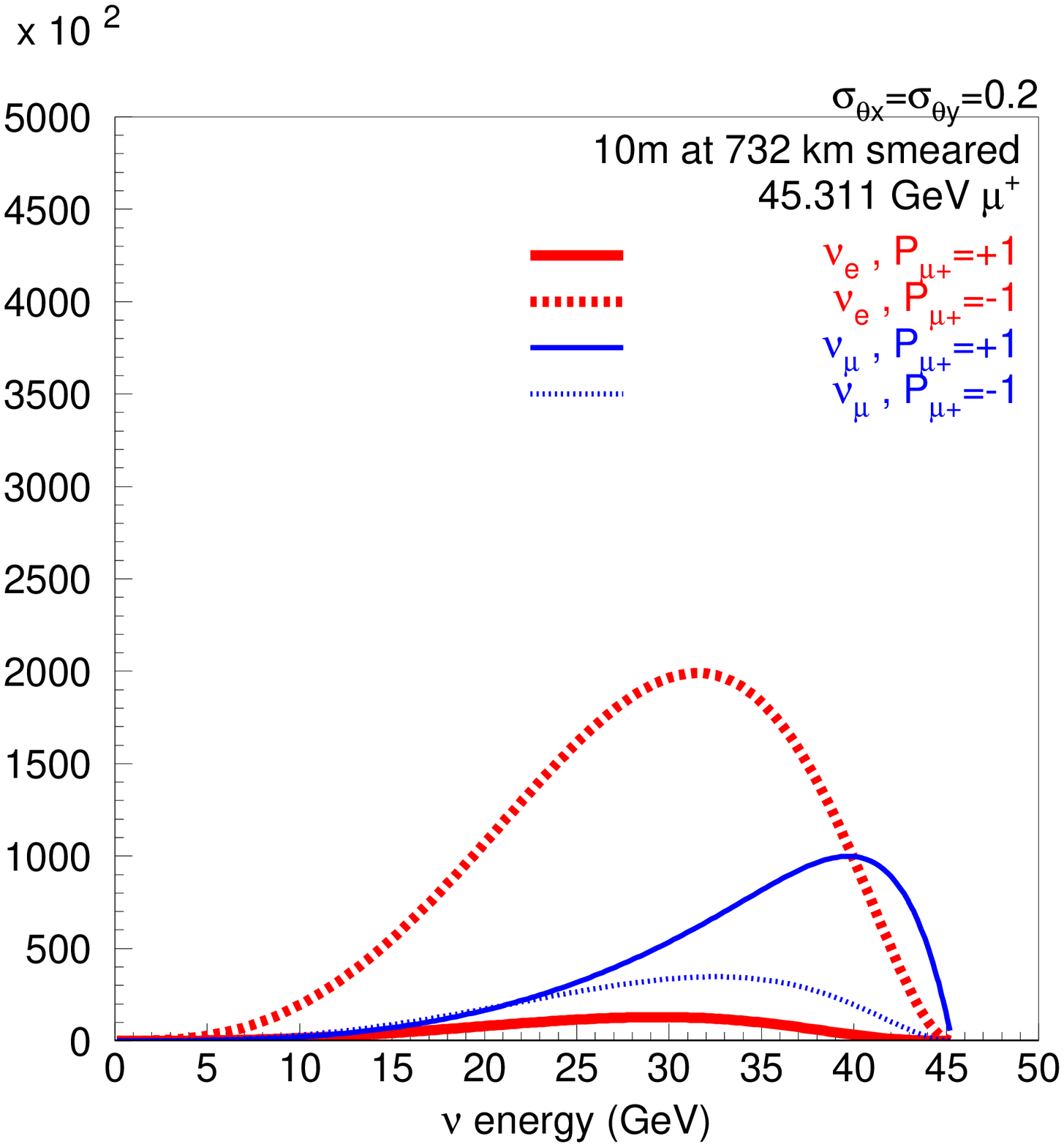}
\includegraphics[width=0.4\textwidth]{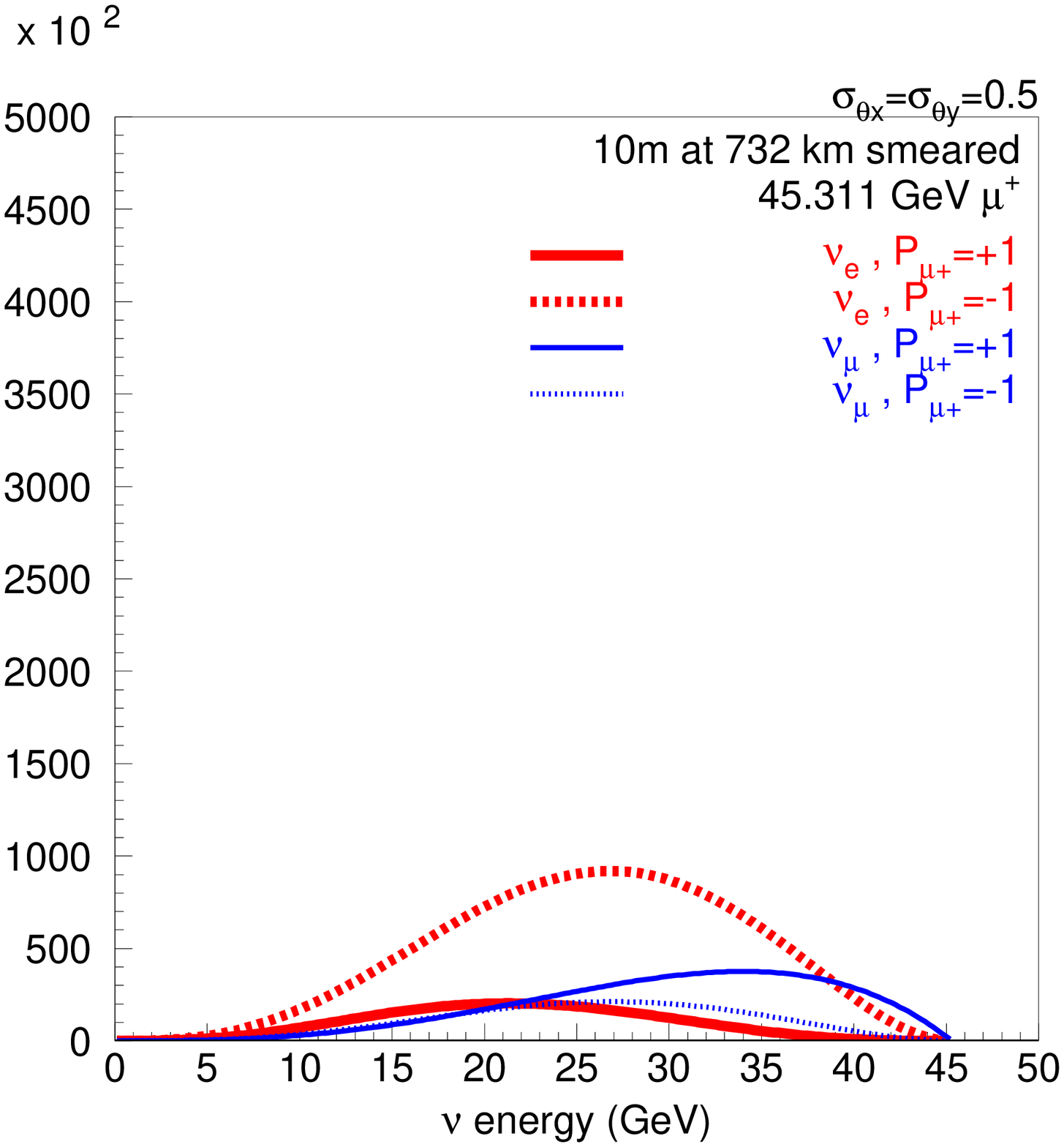}
\\
\caption
{\sl Neutrino event spectra for different beam divergences;
Upper left:  
$\sigma \theta_x = \sigma \theta_y = 0.01~{m_\mu}/{E_\mu}$;
upper right: 
$\sigma \theta_x = \sigma \theta_y = 0.05~{m_\mu}/{E_\mu}$;
lower left:  
$\sigma \theta_x = \sigma \theta_y = 0.2~{m_\mu}/{E_\mu}$ ;
lower right:  
$\sigma \theta_x = \sigma \theta_y = 0.5~{m_\mu}/{E_\mu}$.
It is clear that beam divergence results in a loss of events, and 
in a sizeable distortion of the spectra and of their 
muon polarisation dependence.}
\label{smearedevents} 
\end{center}
\end{figure}

The opening angle of the neutrino beam is typically $1/\gamma$, where
$\gamma = E_{\mu}/m_{\mu}$. As soon as the beam divergence is comparable
with this natural opening angle, a large fraction of the flux will be
lost. This is shown for 45.311~GeV muons in Fig.~\ref{smearedevents}.  It
is clear that beam divergence results in a loss of events, and in a
sizeable distortion of the spectra and of their muon polarisation
dependence. A beam divergence not larger than $\sigma \theta_x= \sigma
\theta_y=0.2{m_\mu}/{E_\mu}$ seems to be desirable, if one wants to avoid a
large sensitivity of physics results upon the experimental determination
of the muon beam parameters.

\begin{figure}[ht]
\begin{center}
\includegraphics[width=0.9\textwidth]{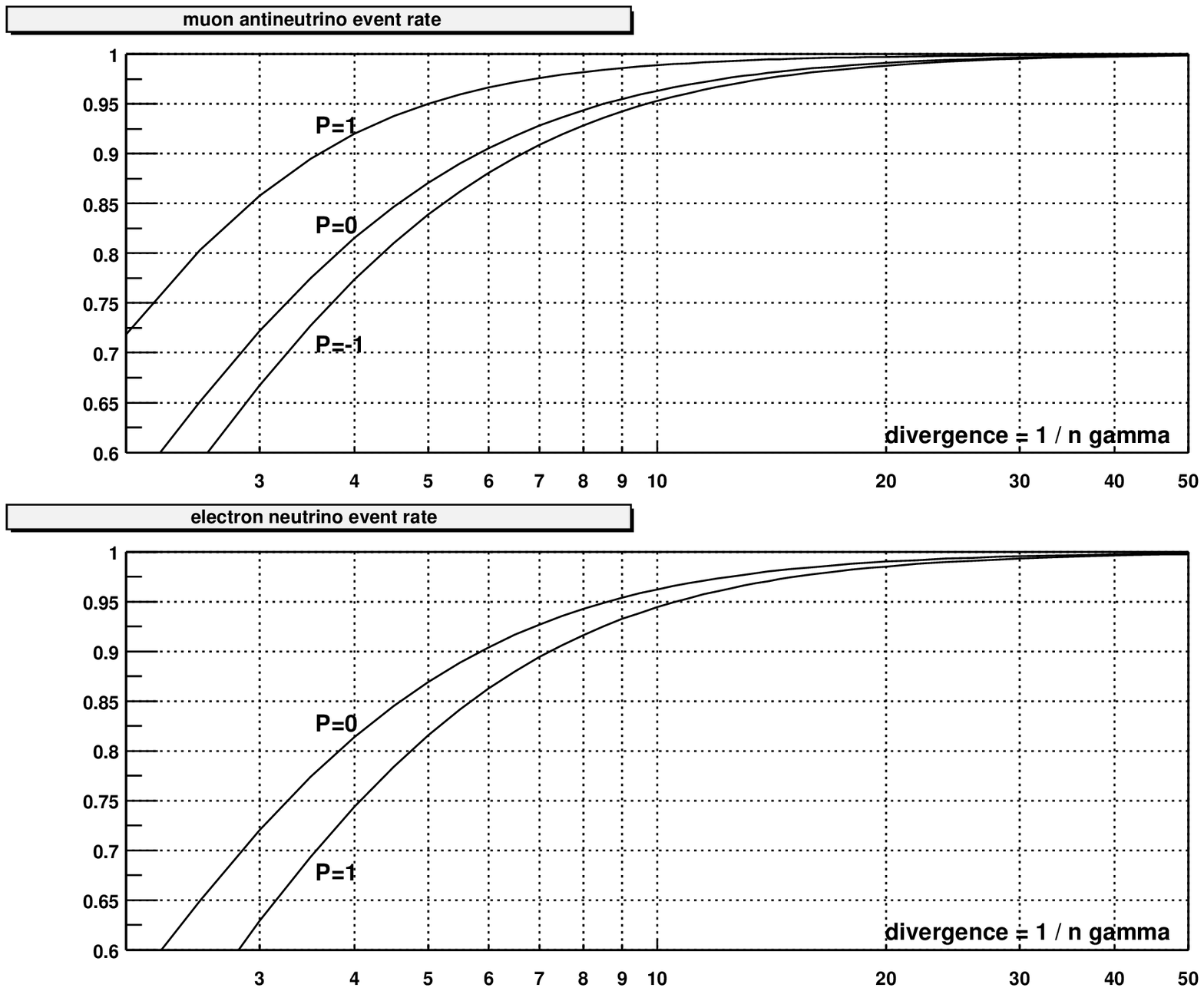}
\caption{ \sl The relative event rates for muon 
anti-neutrinos (top) and electron
neutrinos (bottom), for various polarisation values as a function of the 
beam divergence, parametrised as $1/n \gamma$.}
\label{fig:divergence}
\end{center}
\end{figure}

This effect has been studied more precisely in~\cite{Ioannis},
where event numbers are calculated for various polarisations and 
divergences. The impact of the muon beam divergence on the neutrino event rate 
can be seen in Fig.~\ref{fig:divergence}.
The first conclusion one can draw from these plots is that, for a given
number of muons, the highest flux is obtained for small muon beam
divergence. In order to keep the event rate loss due to the muon beam
divergence below 5\%, the divergence should be close to $0.1 / \gamma_{\mu}$.

\begin{figure}[ht]
\begin{center}
\includegraphics[width=0.9\textwidth]{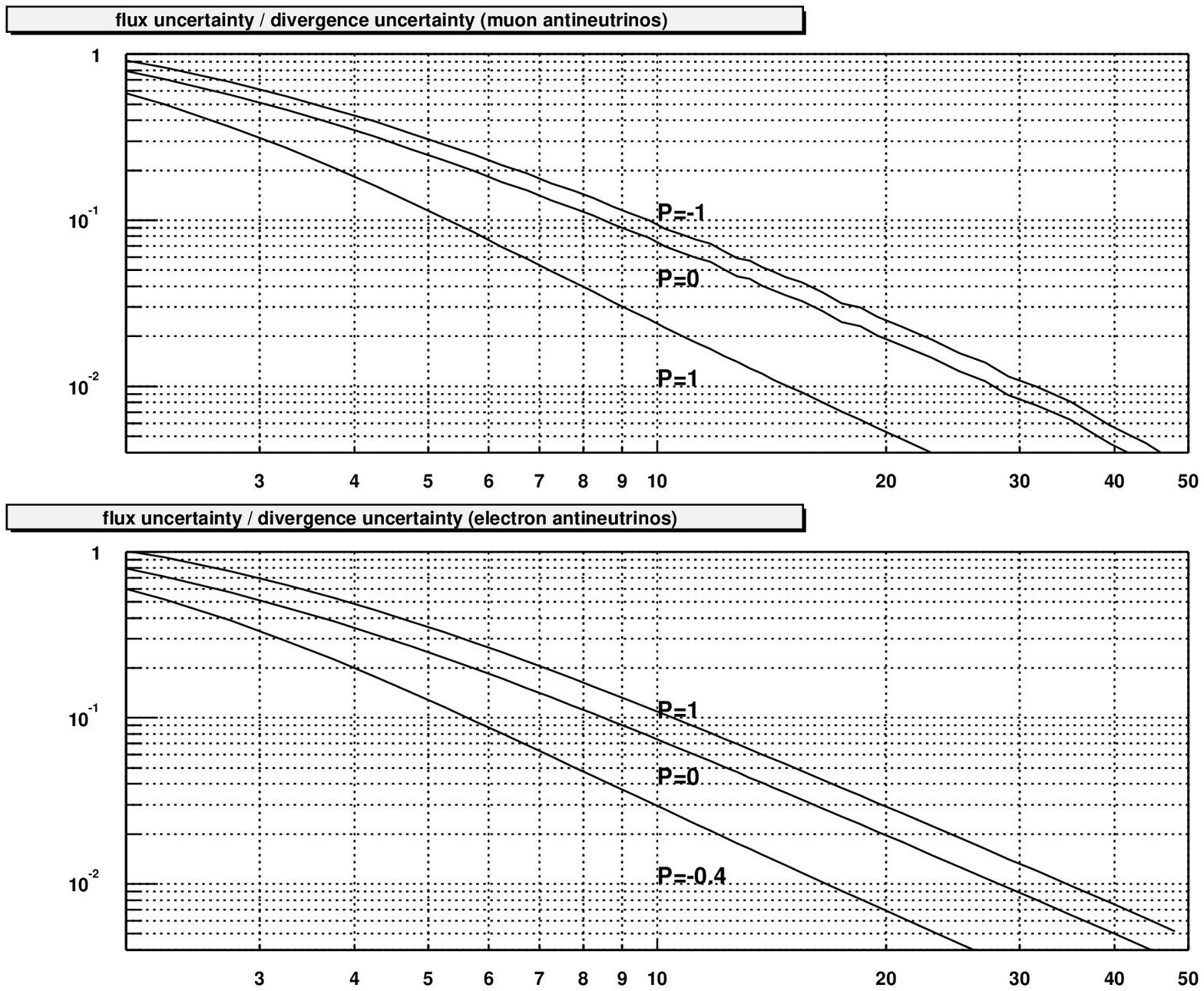}
\caption{\it The ratio of the uncertainty in the event rate over the 
uncertainty in the muon beam divergence as a function of the beam 
divergence, parametrised as $1/n \gamma$. The top
(bottom) plot corresponds to muon anti-neutrinos (electron neutrinos).}
\label{div:error}
\end{center}
\end{figure}

From the curves in Fig.~\ref{div:error}, one can determine the relative
error of the predicted event rate, given the uncertainty in the knowledge
of the beam divergence itself. For example, if the beam divergence is 
$0.1 / \gamma$ and is known with a relative precision of 10\%, the
$\overline{\nu}_{\mu}$ and $\nu_{e}$ event rates can both be predicted
with an accuracy of about 0.75\%.  For a divergence of $0.2 / \gamma$, the
uncertainty on the flux would be 2.5 \%. As we will see, however, the
knowledge of the beam divergence is unlikely to be a constant relative
fraction.

\begin{figure}[ht]
\begin{center}
\includegraphics[width=0.7\textwidth]{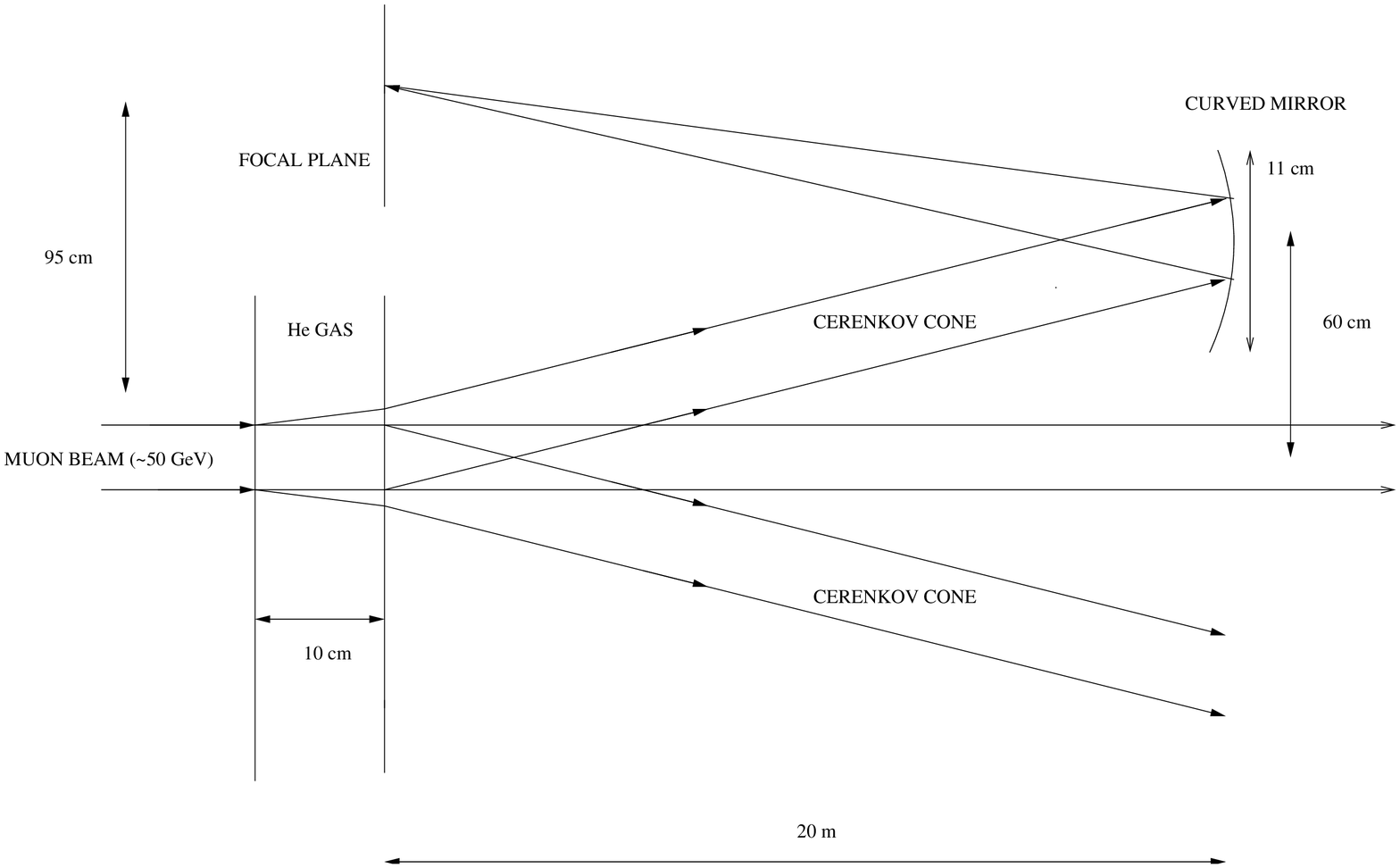}
\caption{\sl Schematic of a muon beam divergence measurement device. 
A low-pressure He gas volume is contained by windows 
(one of which must be transparent) within a straight section 
of the  the muon decay ring. The Cherenkov light is collected by a 
parallel 
to point optics in the direction of interest, so as to provide an image of the
angular distribution of particles in the focal plane. }
\label{fig:beamcerenkov}
\end{center}
\end{figure}
 
One can turn the argument around, and request that the beam divergence be
$0.1 / \gamma$ and known to a relative precision of 1.5\%, so that the
corresponding uncertainty on flux is only $10^{-3}$.  It is clear that in
this case the muon beam divergence will need to be measured.  For a beam
of 50 GeV, the beam divergence is 200 micro-radians and the requirement is
that it should be known to 3 micro-radians.

As a measurement device, one could imagine a gas Cherenkov detector
focusing the Cherenkov radiation in such a way as to make an image of 
the
muon beam direction, as sketched in Fig.~\ref{fig:beamcerenkov}. This has
been studied in~\cite{Piteira}, with the conclusion that for 200
micro-radians divergence, a precision of a few \% can be achieved.  
The additional multiple scattering introduced by the device leads to a
growth of emittance during the muon fill, by a few tens of micro-radians,
which is small and will be measured.  Since the resolution is dominated by
optical imperfections, diffraction effects and heating effects in the gas
of the Cherenkov detector, they act as an additional experimental
smearing $\sigma_{\mathrm{exp}}$ added in quadrature to the true beam
divergence $\sigma_{\mathrm{beam}}$.  In the scheme of
Fig.~\ref{fig:beamcerenkov}, the largest effect is optical diffraction,
which amounts to 30 micro-radians. It is easy to show that the correction
for experimental resolution is
$
\frac{\Delta \sigma_{\mathrm{beam}}}{\sigma_{\mathrm{beam}}}
=
\frac{\Delta \sigma_{\mathrm{exp}}}{\sigma_{\mathrm{exp}}}
\left( \frac{\sigma_{\mathrm{exp}}}{\sigma_{\mathrm{beam}}} \right)^2.
$

This makes the beam divergence progressively harder to measure as it
becomes smaller. Assuming that the experimental error is 30~micro-radians
and is known with a precision of 30\% of its value, the above gives a flux
uncertainty of $5 \times 10^{-4}$, more or less independent of the beam
divergence in the range of 0.05 to 0.2.

In conclusion, the requirement that the beam divergence be no greater than
$0.1 / \gamma$ ensures that the corrections and uncertainties to the
neutrino fluxes remain small (below 1\%), even if one should rely on the
accelerator properties themselves.  In order to achieve a higher precision
a direct measurement of the beam divergence will be necessary -- and is
probably feasible. If relaxing this condition would allow a larger muon
flux, a divergence measurement device becomes mandatory, and would ensure
that the uncertainty on the neutrino flux remains well below $10^{-3}$.

\subsubsection{Summary of uncertainties in the neutrino flux}

A first look has been given to the sources of systematic uncertainties in
the neutrino fluxes and their possible cures.

\begin{itemize}
\item 
The monitoring of the total number of muons circulating 
in the ring can be inferred from a Beam Current Transformer 
with a precision of the order of $10^{-3}$ or better. 
The decay electrons vanish quickly and are not a problem. 
\item
The theoretical knowledge of the neutrino fluxes from muon decay
is not an issue. Radiative effects have been calculated: they
amount to around $4 \times 10^{-3}$, with an error that is much 
smaller \cite{Broncano-Mena}.
\item
The muon beam polarisation determines the flux directly, 
both in shape and magnitude. It seems delicate to determine its
value  with a precision much better than a few \%. In a ring geometry, 
however, polarisation precesses and averages out with high precision 
(a few $\times 10^{-4}$). This is  a strong argument in favour of a ring 
geometry against a bow-tie geometry. 
\item 
The event rate varies like the muon beam energy to the third power, but 
the muon
beam energy
can be inferred very precisely from the muon spin precession. A 
polarimeter idea has been outlined, and the measurement should cause
no difficulty. 
Beam polarisation can be preserved if an RF system is installed in the
decay ring. 
The energy spread can be derived from the depolarisation pattern, 
in special runs
with no RF if necessary. 
\item 
The muon beam angle and angular divergence have an important effect on the 
neutrino flux. For a given number of muons, the smaller the beam divergence, 
the higher the flux. 
Thus the beam divergence in the straight section of the muon decay ring 
should be made as small as possible, but should not constitute a limit 
on the number of stored muons. 
\item 
Measurement devices for the beam divergence
will be necessary, but they can probably be designed and built to ensure 
a flux uncertainty below $10^{-3}$.
 \end{itemize}

In addition, the near detector stations should allow measurements of 
cross sections with high precision. The inverse muon decay reaction 
$\nu_{\mu} + e^- \rightarrow \mu^- + \nu_e$ offers the possibility of an 
absolute normalisation of the flux. 

We conclude that, provided the necessary instrumentation is foreseen, 
 the Neutrino Factory flux should  be known with a precision 
of the order of $10^{-3}$.

\subsection{Flux control for Beta Beams and Super Beams}

The International Scoping Study did not explicitly look at flux control 
and beam instrumentation for Beta beam and Super Beams. However, some
of the concepts developed in the context of a Neutrino Factory are
applicable to a Beta Beam.  
Similar requirements for flux control are needed at a Beta beam facility 
as for a Neutrino Factory. Polarization of the beam is not an issue
in a Beta Beam, but the number of radioactive ions in the storage 
decay ring can be determined with a Beam Current Transformer. The
divergence of the beam would need to be measured as well. A Cherenkov
detector as proposed above for a Neutrino factory would be able to measure
the divergence of a Beta Beam, provided that it did not affect the 
stability of the beam. In addition to these beam monitoring devices, a near
detector would also be needed (see section~\ref{Near_detectors}).

There is extensive experience in the design of conventional beams of neutrinos 
from pion decay, so understanding the flux control requirements for these beams 
will determine the parameters needed for beam monitoring at a Super Beam. Recent 
examples include the MINOS beam line \cite{NUMI_beam}, the CERN to Gran Sasso (CNGS) 
beam \cite{CNGS_beam} and the beam line for the T2K experiment \cite{T2K}.
                                                                                                                  
The NUMI beam at Fermilab \cite{NUMI_beam} that supplies neutrinos for the MINOS 
experiment \cite{MINOS_TDR} contains a system for flux monitoring 
of the neutrino beam. The monitoring system presently consists of ionization 
chambers \cite{MINOS_ionization} placed at the end of the decay pipe, to measure 
muons, undecayed mesons, and protons that did not react in the target, and in three 
alcoves dug into the dolomite rock to measure fluxes of muons that are produced along 
with the neutrinos. These chambers provide information to determine the neutrino beam 
alignment and as a beam monitor, to ensure target integrity and horn focusing.
                                                                                                                  
The CERN to Gran Sasso (CNGS) neutrino beam, with a $\nu_\mu$ average energy of 17.4 GeV 
\cite{Ferrari2005}, is well matched to the $\nu_\tau$ appearance experiments at the 
Laboratori Nazionali del Gran Sasso (LNGS), OPERA \cite{opera,unknown:2006ki} and 
Icarus \cite{T600}. A misalignment of the horn by 6 mm or the reflector by 30 mm, or if 
the proton beam is off-target by 1 mm, or if the CNGS beam is misaligned by 0.5 mrad, 
may cause a drop in neutrino flux of 3\%. Monitoring of these parameters is achieved by 
the Target Beam Instrumentation Downstream (TBID) and the muon ionization chambers 
installed in the muon pits downstream of the beam stop. The TBID contains secondary 
emission monitors, consisting of 12 $\mu$m thick titanium foils, and check the efficiency 
of the target conversion (by comparison with an upstream station) and the alignment of the 
beam. The muon ionization chambers measure the muon intensity, the muon profile and the centre 
of the beam. There are 17 fixed monitors in a cross, and one moveable chamber for relative 
calibration. Since OPERA and Icarus plan to perform an appearance search for tau neutrinos, 
it is not as important to measure the $\nu_\mu$ flux with a similar precision to a 
disappearance measurement. Hence, a near detector at the CNGS was not deemed to be an 
essential component of the beamline and was not built given the cost of a near cavern.
                                                                                                                  
The T2K experiment \cite{T2K} exploits an off-axis beam at angles between 2$^{\circ}$ and 
3$^{\circ}$. It monitors the muon flux on-axis, downstream from the beam dump, and serves 
as a real-time status monitor sensitive to the proton intensity, proton beam position on 
target and the performance of the horn. The detectors will be a combination of He gas ion 
chambers and semi-conductor detectors. In addition, there will be an on-line neutrino 
flux monitor, in the form of an array of iron-scintillator stacks, to determine the centre 
and profile of the on-axis neutrino beam. From the on-axis muon and off-axis flux monitors, 
one can deduce the off-axis flux, which will be compared with the ND280 (Near Detector at 
280 m from the target) \cite{ND280,T2K_2}. A similar strategy would probably have to be 
adopted for any other off-axis super beam scenario.

\section{Near detectors }
\label{Near_detectors}

\subsection{Aims}

In order to perform measurements of neutrino oscillations at a neutrino facility, 
it is necessary to establish the ratio of neutrino interactions in a near detector 
with respect to the far detector. Hence, the careful design of a near detector and of the beam instrumentation is crucial to measure the flux, energy and cross-sections of the incident neutrinos 
\cite{perez} to be able to reduce the long baseline neutrino oscillation systematic errors. 

The present generation of near detectors (e.g. for K2K and MINOS) have been concentrating on disappearance measurements,  which require the near-to-far detector comparison of the main \numu component  of the beam.  Life appears to be somewhat easier when searching for the appearance measurement, at least at first, when the statistics in the appearance channel are limited. However, the physics of the golden channel is to measure precisely the appearance probability and to compare it between neutrinos and anti-neutrinos, or neutrinos of different energies or baselines, to establish CP violation and/or matter effects. All of a sudden the ratio to worry about is not only near-to-far, but electron-to-muon neutrino cross-sections. Indeed, when measuring the CP asymmetry

\begin{equation}
A_{CP}=\frac{P(\numu \rightarrow \nue) - P(\numubar \rightarrow \nuebar )}{ P(\numu \rightarrow \nue) + P(\numubar \rightarrow \nuebar )} ,
\end{equation}

a troublesome quantity will appear, the double ratio: 
\begin{equation}
DR =\frac{  \sigma_{\numu} /  \sigma_{\nue} } { \sigma_{\numubar} / \sigma_{\nuebar}} ,
\end{equation} 

where $\sigma_{\numu} $ really means $\sigma_{\numu}  \times \epsilon - B$ including correction for efficiency $\epsilon$ and background $B$. Although it would seem that many systematic errors would cancel in this ratio, this is only partially true. The effects that ensure a deviation of this quantity from unity are quite difficult to master: 
\begin{itemize}
\item
the muon mass effect;
\item
Fermi motion and binding energy;
\item
the non-isoscalarity of the target (this is particularly relevant for water where anti-neutrinos and neutrinos interact very differently on the free protons);
\item 
the different neutrino and antineutrino $y$ distributions; and
\item 
the different appearance of the final state lepton in the detector. 
\end {itemize}

These effects are particularly relevant for the low energy neutrinos, as discussed in Appendix \ref{Double-ratio}. Experimental certification will require a dedicated design of the beam line and near detectors, and probably measurement of cross-sections for all channels quoted above, either at the absolute level or in relation with one of the four channels.   

The shape and technology of a near detector depends on the type of facility to be 
considered (whether Super Beam, Beta Beam or Neutrino Factory). The main requirements of 
near detectors are that they should measure and control the neutrino flux, the beam 
angle and direction, the neutrino energy, all the relevant cross-sections and the 
background to the far detector. Backgrounds differ depending on the far detector 
technology and the energy of the neutrino beam, so the requirements of a near detector 
for each of the facilities will be different in each case. In the following sections, 
we will look at the requirements for a near detector at a Beta Beam, a Super beam 
and a Neutrino Factory.

\subsection{Near detector at a Beta beam and Super Beam}

The near detector at a Beta Beam or a Super Beam was not considered in detail by the 
International Scoping Study. However, the average energy of the neutrino beam in these 
two scenarios will demand a detector that is capable of observing low energy neutrino
interactions, as discussed in appendix \ref{Low_energy}. 

For Super Beams, the detector will need to have a magnetic field to be able to distinguish
neutrinos from anti-neutrinos as in the Near Detector currently being designed for 
T2K \cite{ND280}. The average energy of the neutrinos will be typically from 500~MeV to a 
few GeV, so the dominant interactions will be charged current quasi-elastic and neutral current elastic
interactions, neutral and charged current single and multi-pion production, and coherent
pion production. At these energies, it is extremely important to have a detector target
with the same nuclear mass (A) as the far detector, or, at least, to understand the 
dependency of the cross-section with the nuclear mass. Other nuclear effects at low energy,
such as Pauli blocking or Fermi Motion are very important to be taken into account so, typically, one 
would aim to measure these in light nuclei. These data will be better known from the 
Miner$\nu$a experiment \cite{Minerva}, but the near detector at a high intensity Super Beam
or Beta Beam should be able to carry out these measurements with improved accuracy.

For Beta Beams, there is only one species of neutrino, so a magnetic field is not essential in the
near detector. All other considerations of cross-section measurements at low energy remain 
the same as in the Super Beam case.

\subsection{Near detector at a Neutrino Factory}

For a neutrino factory, we have discussed the beam instrumentation that will 
measure the beam angle, the divergence and the polarization of the muons in 
the storage ring. In addition, a near detector will need to be able to measure 
the neutrino flux, the neutrino beam angle and its divergence, the neutrino 
energy, the neutrino cross-sections and a measurement of the background to the 
oscillation signal at the far detector, which includes a high statistics 
measurement of the charm background from neutrino interactions. 

There is also a rich physics programme that can be carried out at a near detector 
\cite{blondel,macfarland}. Deep inelastic, quasi-elastic and resonance scattering 
reactions can be studied with unprecedented accuracy. Other measurements include 
the determination of the weak mixing angle $\sin^2 \theta_W$ from the ratio of 
neutral to charged current interactions, measurements of the parton distribution 
functions (both polarized and unpolarized) in a region of phase space that is 
complementary to those determined by HERA, a measurement of the strong coupling 
constant and other effects such as nuclear reinteractions and nuclear shadowing. 
The large sample of charm events reconstructed for the neutrino oscillation 
background studies can be used to measure the charm background to the oscillation 
signal but can also be used to measure the CKM matrix element $V_{cd}$, and  to search 
for CP violation in $D_0-\overline{D}_0$ mixing. More accurate measurements of $\Lambda$ 
polarization might shed more light on the spin content of nucleons. 

This varied physics programme requires a near detector (or detectors) with high 
granularity in the inner region that subtends to the far detector. The active target 
mass of the detector does not need to be very large. With a mass of 50 kg, one would 
obtain $10^9$ charged current neutrino interactions per year in a detector at a distance 
of 30 m from the muon storage ring, with the straight decay sections being 100 m long.

There are a number of technological choices for a near detector at a neutrino factory, 
to achieve the general aims stated above. Due to the nature of neutrino beams, one may 
choose to build a multi-purpose detector that will carry out the physics programme, or 
instead have a number of different more specialised detectors for individual topics. 
However, some of the features needed in a near detector include high granularity, to 
compare the subtended angle between near and far, a magnetic field for charge separation, 
and muon and electron identification for flavour determination. More specific needs 
also include excellent spatial resolution to be able to carry out measurements of charm 
events, the possibility of including different targets for nuclear cross-section determination 
and maybe the possibility to polarize the target for measurements of polarized parton 
distribution functions. 

\subsubsection{Flux normalization and control}

Neutrino fluxes from muon decay are given by Eqs.~\ref{eq:nufact_fluxes_1} and 
\ref{eq:nufact_fluxes_2}. These fluxes 
are readily calculable, with small theoretical uncertainties (an accuracy of better than 
10$^{-3}$), as was shown in section~\ref{Beam_Instrumentation}. 

\begin{figure}[tbhp]
\begin{center}
\includegraphics[width=0.40\textwidth]{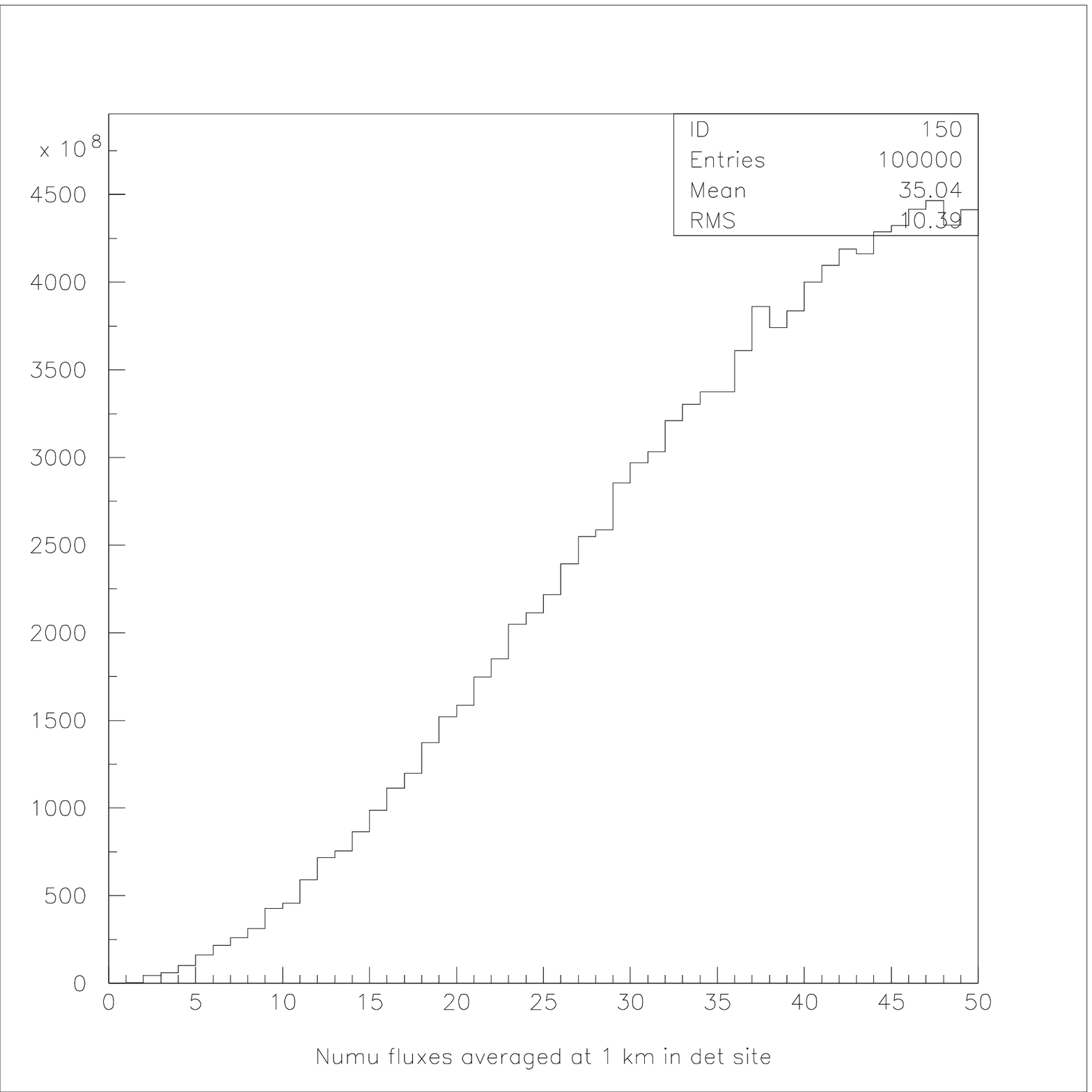}
\includegraphics[width=0.40\textwidth]{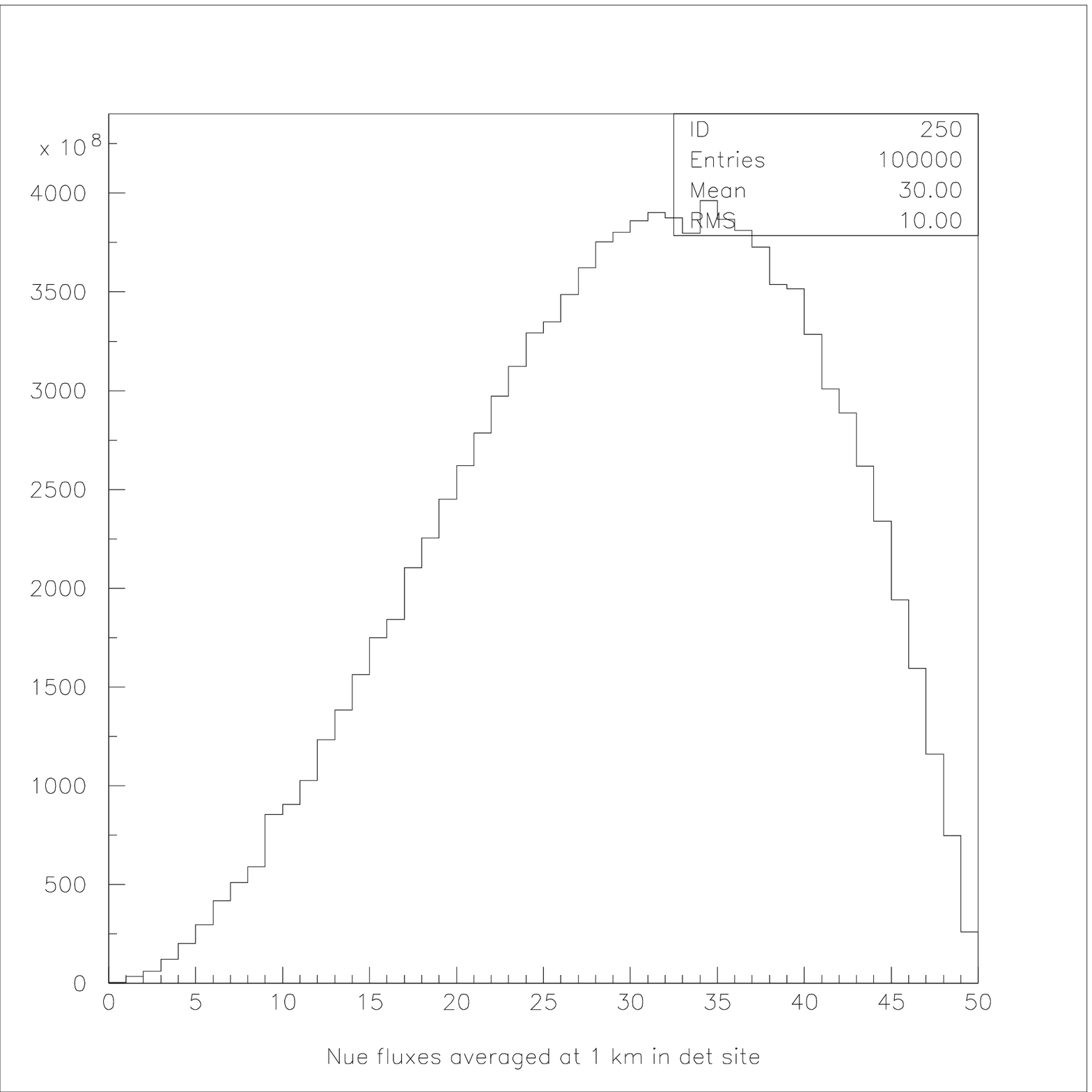}  \\
\caption
{{\sl
Flux of $\overline{\nu}_\mu$ (left panel) and $\nu_e$ (right panel) at a detector 
2500 km from a neutrino factory with a 20 m radius, subtending an angle of 8 $\mu$rad., 
from the decay of 50~GeV $\mu^+$.}}
\label{flux2500km}
\end{center}
\end{figure}

A neutrino factory offers the possibility of having an unprecedented number of neutrino 
interactions in a near detector. The position of the near detector at the end of the straight 
decay section of the muon storage ring is a crucial parameter to determine the rate and 
spectrum of the neutrino interactions. The systematic errors in the ratio of fluxes between 
the near and far detector are reduced when the spectrum in the near detector is similar to the
spectrum at the far detector. For example, a far detector at 2500~km, with a radius of 20~m 
subtends an angle of less than 8 $\mu$rad. The flux of $\overline{\nu}_\mu$ (left panel) and $\nu_e$ 
(right panel) from the decay of 50~GeV $\mu^+$ for this configuration, with average energies 
of 35.0~GeV and 30.0~GeV is shown in figure~\ref{flux2500km}. 

\begin{figure}[tbhp]
\begin{center}
\includegraphics[width=0.40\textwidth]{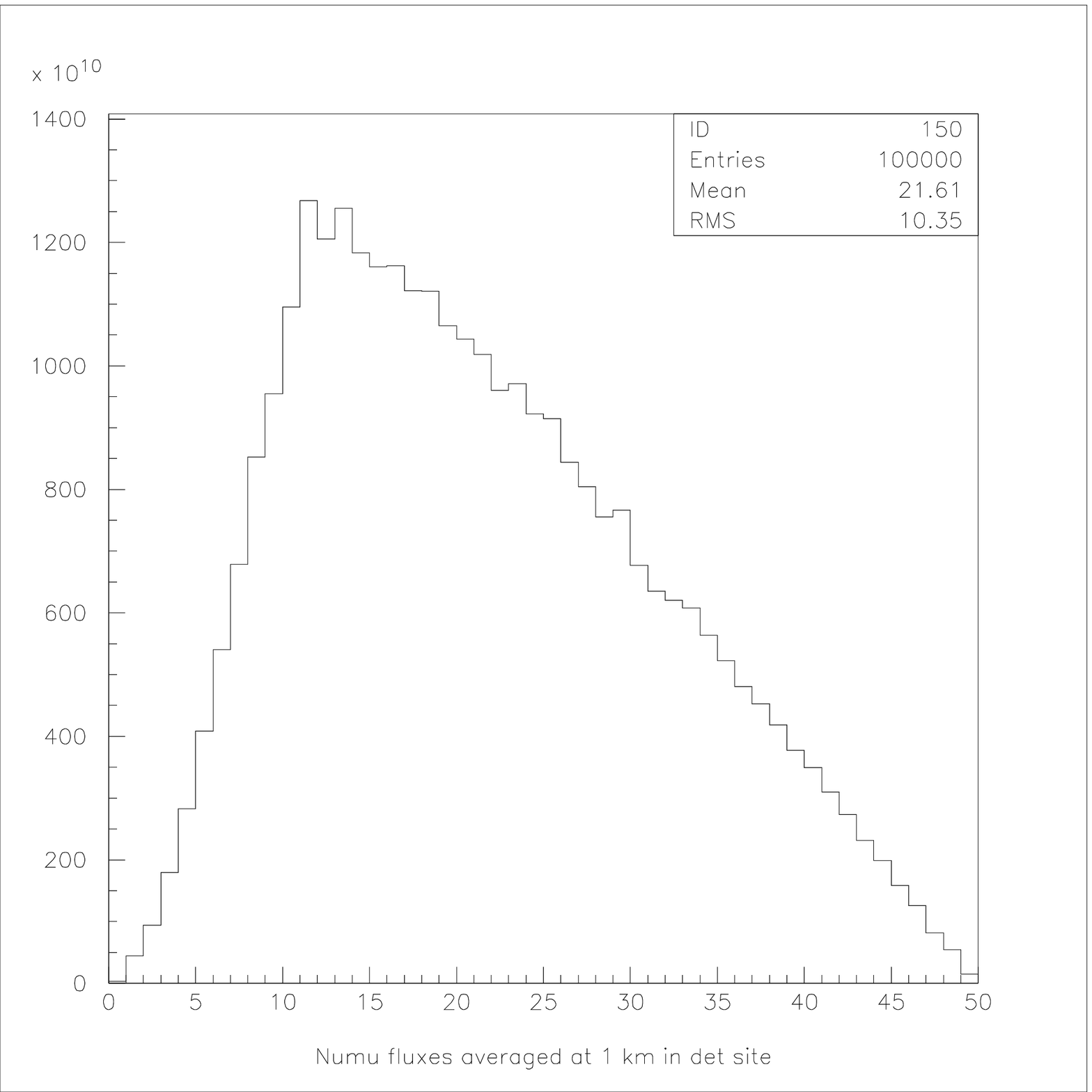}
\includegraphics[width=0.40\textwidth]{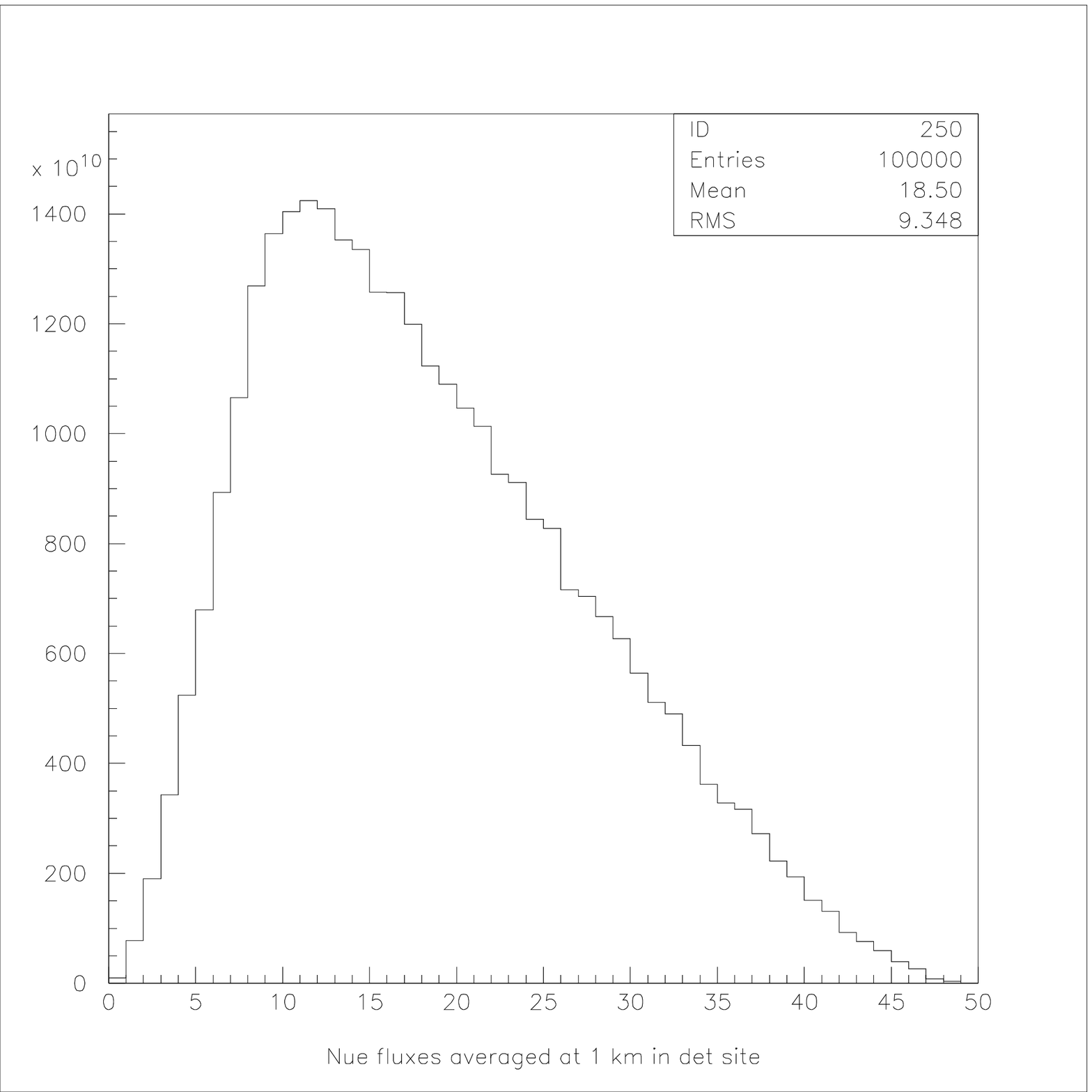}  \\
\includegraphics[width=0.40\textwidth]{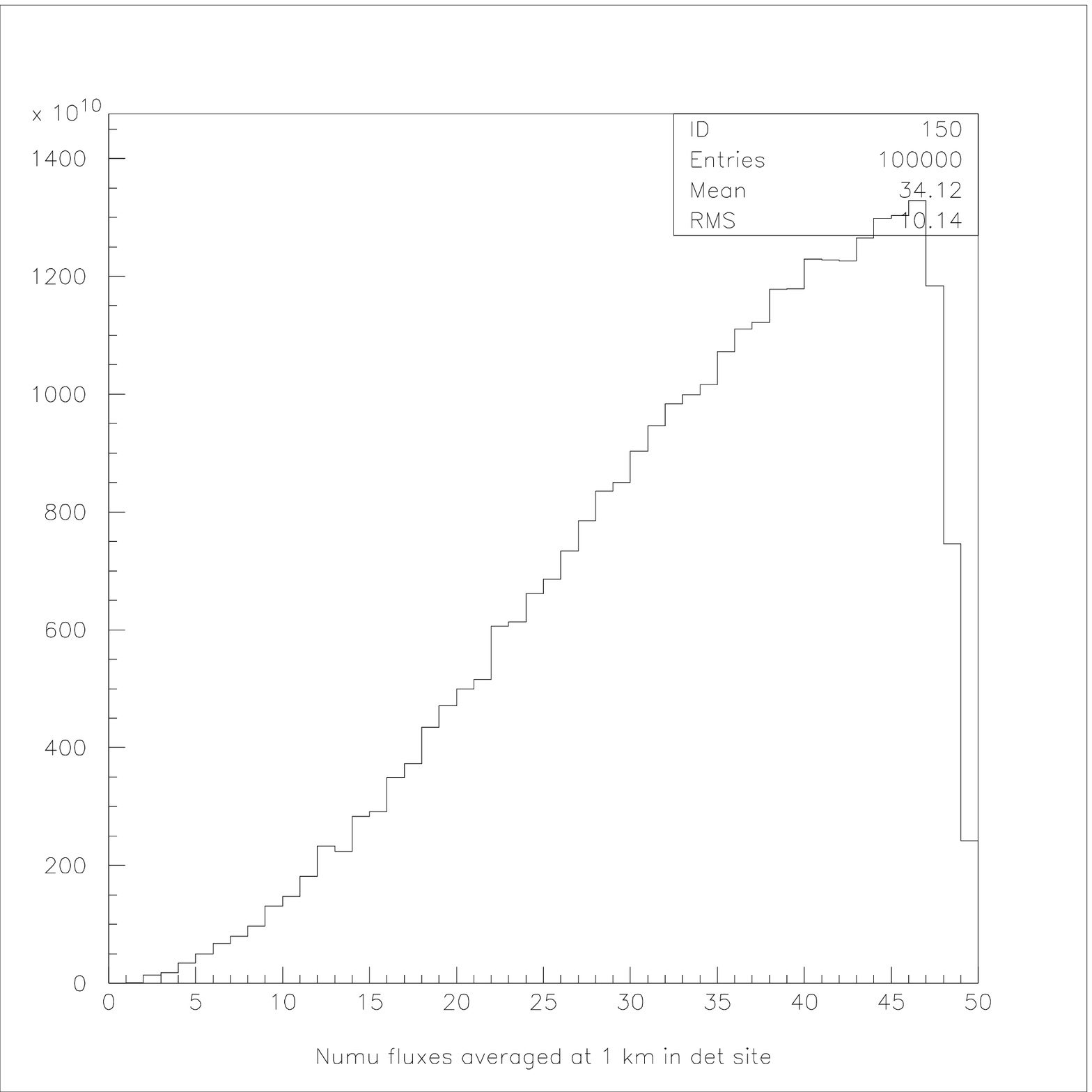}
\includegraphics[width=0.40\textwidth]{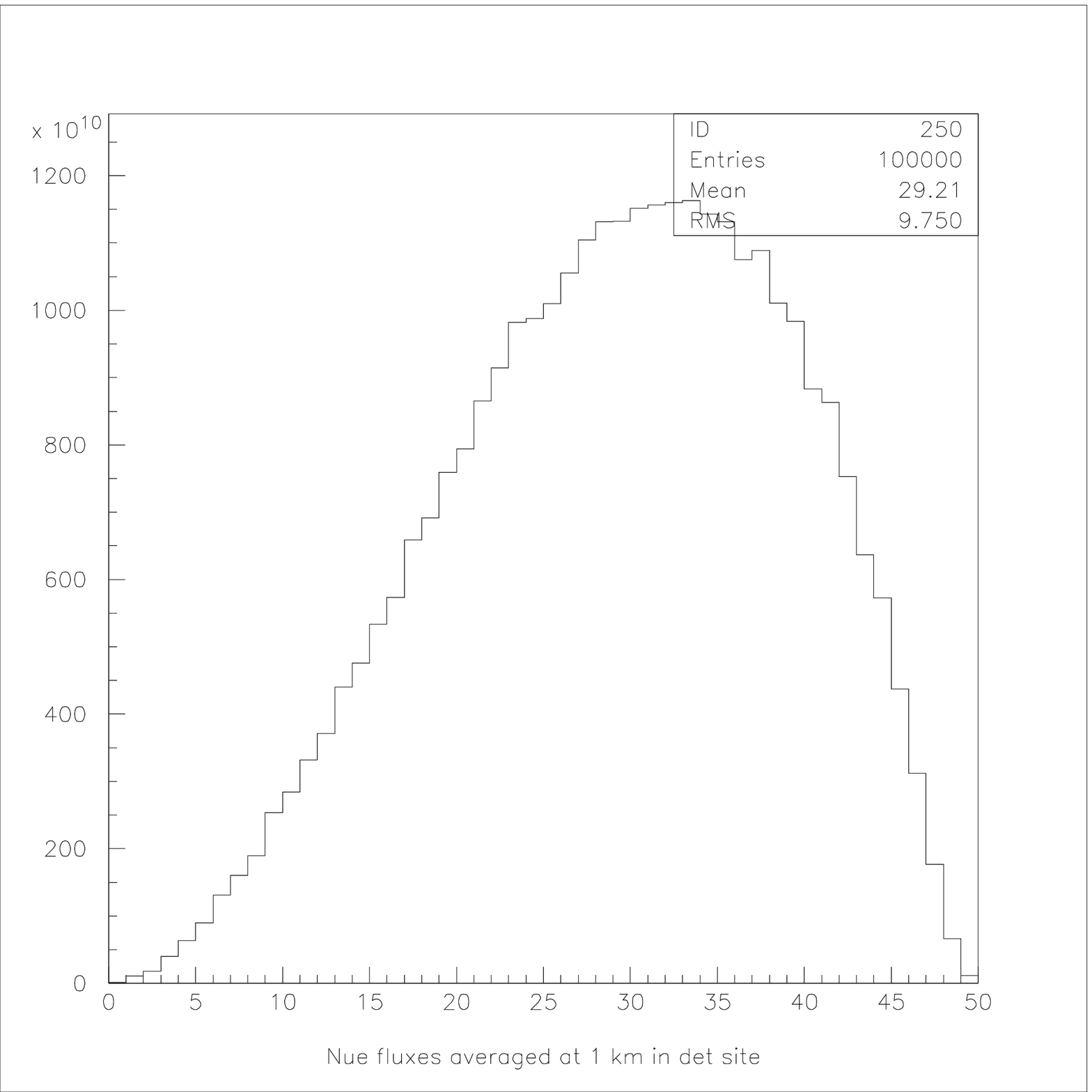}  \\
\caption
{{\sl
Flux of $\overline{\nu}_\mu$ (left panel) and $\nu_e$ (right panel) at a near 
detector with a 0.5~m radius, 130~m from the decay of a $\mu^+$ (top). Flux of 
$\overline{\nu}_\mu$ (left panel) and $\nu_e$ (right panel) at a near detector 
with a 0.5~m radius, 1~km from the decay of a $\mu^+$ (bottom).}}
\label{flux130m-1km}
\end{center}
\end{figure}

At the near detector, one needs to be able to subtend a similarly small angle, and this 
can be achieved by varying the distance to the source or by improving the spatial 
resolution of the detector. For example, as shown in figure~\ref{flux130m-1km}, at a 
distance of 130 m from the decay point of the 50~GeV $\mu^+$, one obtains distributions 
that are quite different to the far detector (average energies for $\overline{\nu}_\mu$ 
and $\nu_e$ of 21.6~GeV and 18.5~GeV), while at a distance of 1~km from the decay point 
of the $\mu^+$, the distributions now look quite similar to those of the far detector 
(average energies for $\overline{\nu}_\mu$ and $\nu_e$ of 34.1~GeV and 29.2~GeV). The 
difference in the spectra between near and far detector can be a source of systematic 
error in predicting the far detector flux from the migration of the near detector flux. 
If the near and far detector fluxes are similar, then the systematic error in the 
extrapolation from near to far can be reduced.

\begin{figure}[tbhp]
\begin{center}
\includegraphics[width=0.49\textwidth]{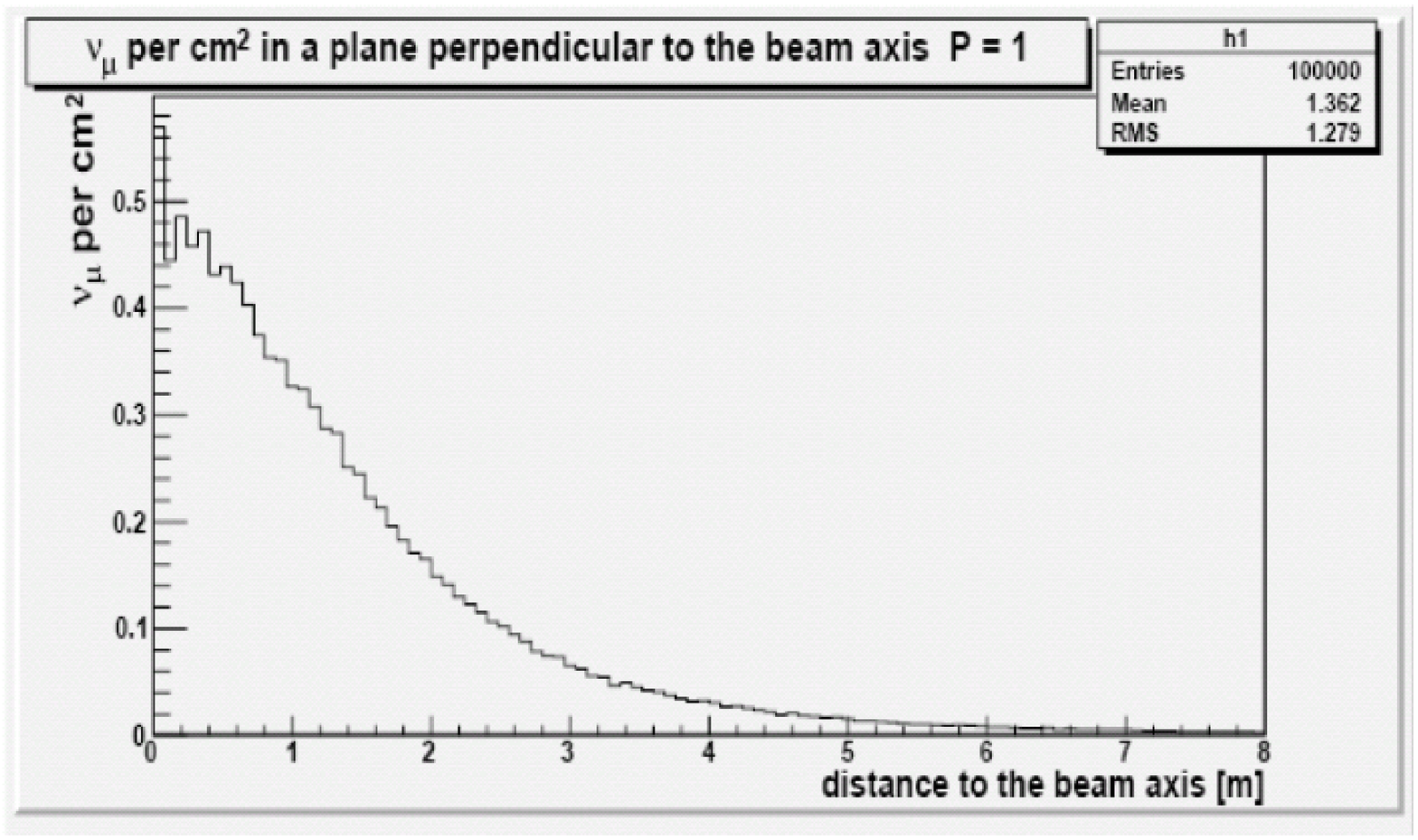}
\includegraphics[width=0.49\textwidth]{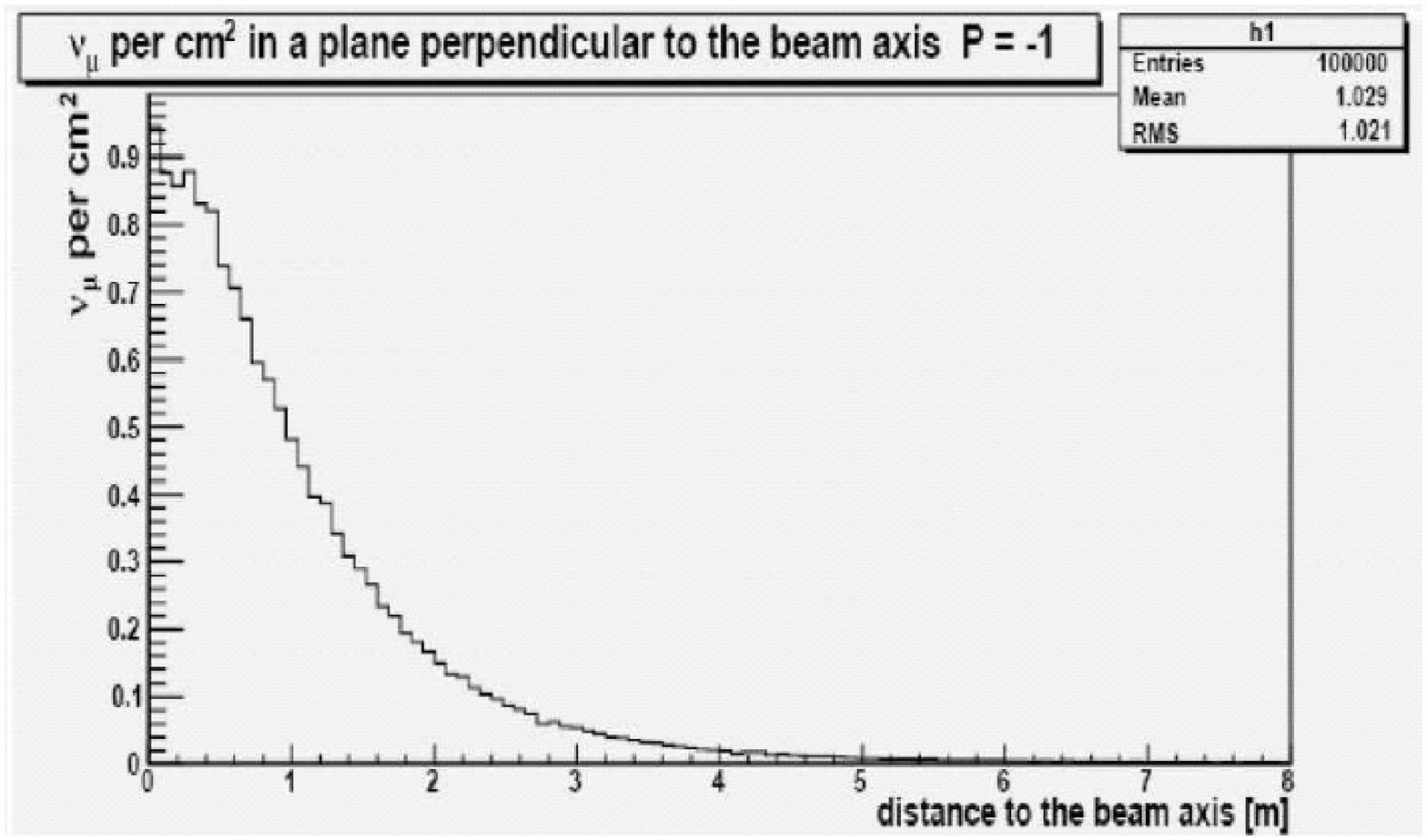}  \\
\includegraphics[width=0.49\textwidth]{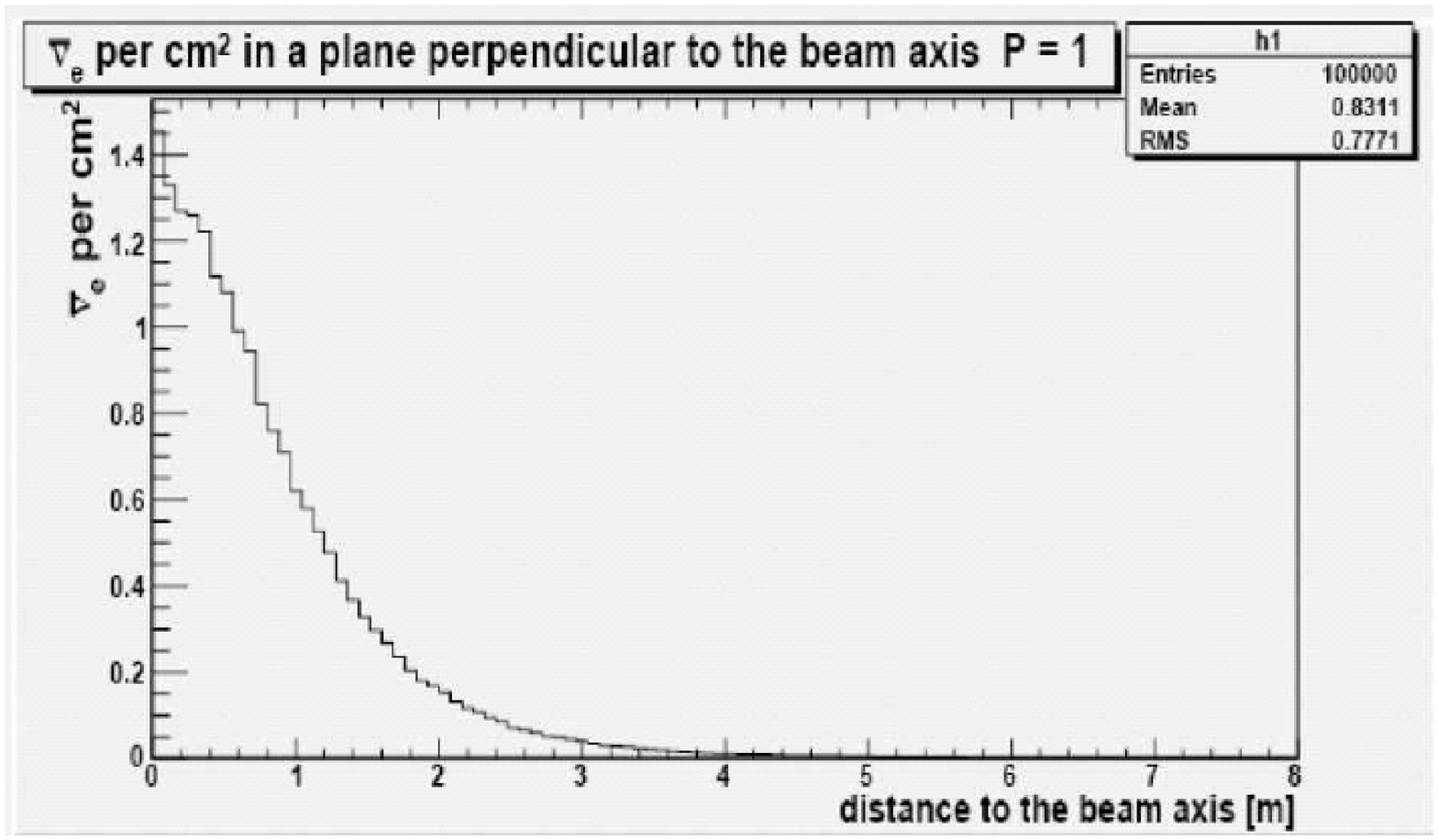}
\includegraphics[width=0.49\textwidth]{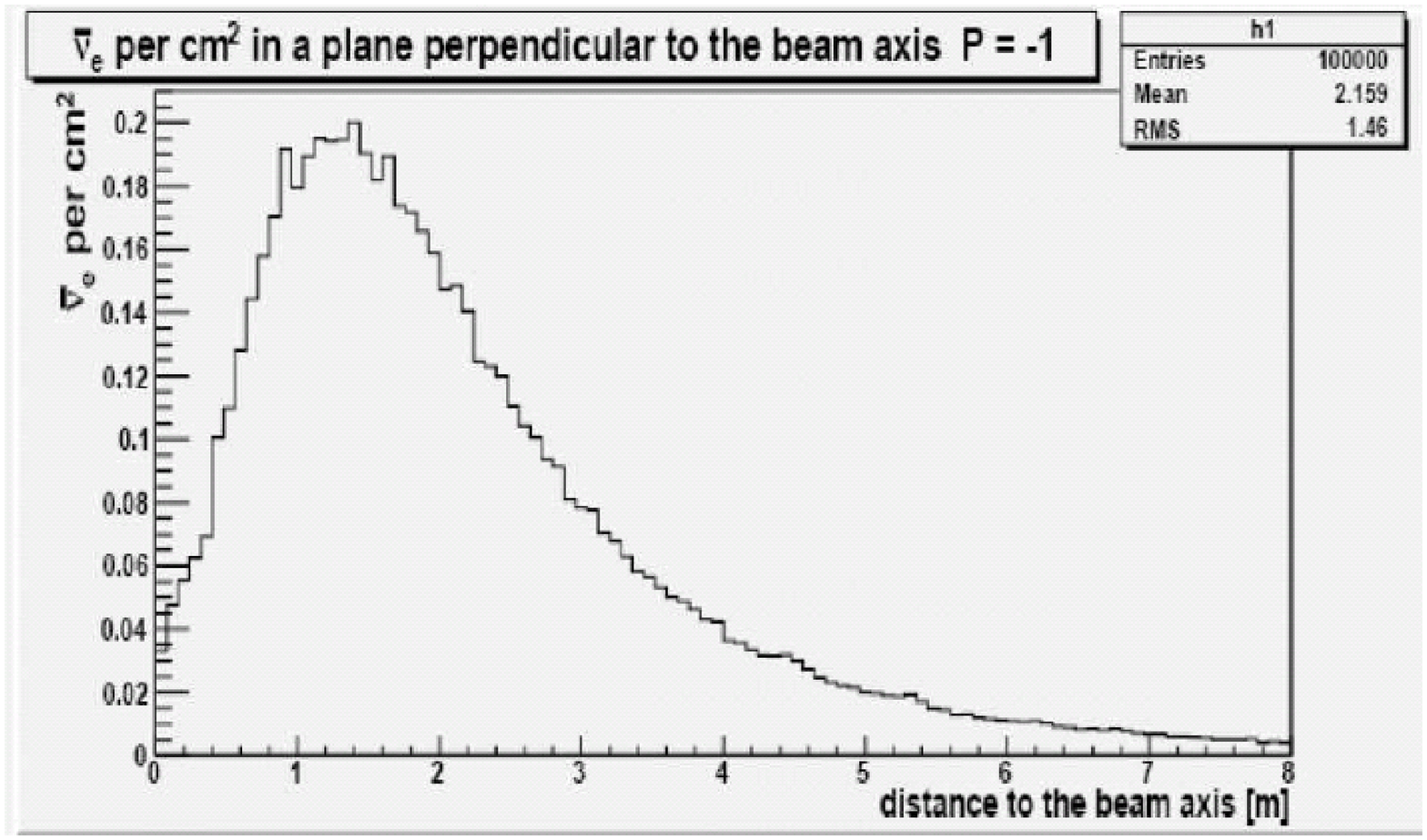}  \\
\caption
{{\sl
Number of neutrinos per cm$^2$ per 10$^5$ muon decays, at 500~m from the end of a decay 
straight of 500~m at a neutrino factory. Top left panel: $\nu_\mu$ and $P=+1$; top right 
panel: $\nu_\mu$ and $P=-1$; bottom left panel: $\overline{\nu}_e$ and $P=+1$; bottom 
right panel: $\overline{\nu}_e$ and $P=-1$.}}
\label{nu_events_500m}
\end{center}
\end{figure}

Another source of difference between the far and near detectors is that 
the far detector effectively sees a point neutrino source, while the near detector 
sees a line source, from the decay of the muons along the decay straight in the muon 
storage ring. For example, let us assume we have a straight section of length 500 m, and 
we place the near detector at a distance of 500 m from the end of the straight section. 
We assume that the muons decay uniformly along the decay section, that the angular 
distribution is Gaussian with a $\sigma_\theta = 0.5\times 10^{-3}$, and that the 
energy of the muons is 40~GeV with $\sigma_E = 80$~MeV. If negative muons $\mu^-$ 
decay, we obtain the flux distributions shown in figure~\ref{nu_events_500m}, 
for 10$^5$ muon decays simulated. We will assume 10$^{21}$ muon decays in one year 
of operation of the neutrino factory.

One of the main issues to minimise systematic errors in the near and far detector is 
to determine the flux and cross-sections separately, since normally one obtains the 
product $\Phi(E_\nu)\times \sigma(E_\nu)$.
In order to separate the latter, one can use the inverse muon decay reaction: 
$\nu_\mu + e^- \rightarrow  \nu_e + \mu^-$, with total cross-section:

\begin{equation}
\sigma(\nu_\mu e^-)=\frac{G_F^2}{\pi}\frac{\left(s-m^2_\mu\right)^2}{s}  ,
\end{equation}

and muon production through annihilation: $\overline{\nu}_e + e^- \rightarrow  \overline{\nu}_\mu + \mu^-$, 
with the following cross-section in the Standard Model \cite{Okun:1982}:

\begin{equation}
\sigma(\overline{\nu}_\mu e^-)=
\frac{2G_F^2}{\pi}\frac{\left(s-m^2_\mu\right)^2\left(E_e E_\mu + 1/3 E_{\nu 1} E_{\nu 2} \right)}{s^2}  ,
\end{equation} 

where $s= 2 m_e E_\nu$. 

The production threshold for these reactions is $E_\nu>\frac{m^2_\mu}{2 m_e}=10.9$~GeV. 
The signature is a single outgoing muon without any visible recoil energy at the interaction point, 
and with a transverse momentum kinematically constrained to be $p_T\leq 2 m_e E_\mu$, as 
was demonstrated by the measurements performed by CHARM-II \cite{CHARMII-IMD}.

Alternatively, one can also use the elastic scattering interactions: 
$\nu_\mu + e^- \rightarrow  \nu_\mu + e^-$ and 
$\nu_e + e^-\rightarrow\nu_e + e^-$ that also have calculable rates:
\begin{equation}
\frac{d\sigma(\nu_\mu e^-)}{dy}=
\frac{2G_F^2 m_e E_\nu}{\pi}\left[ \left( -\frac{1}{2}+\sin^2 \theta_W \right)^2 + 
\sin^4 \theta_W (1-y)^2 \right]
\end{equation}
and
\begin{equation}
\frac{d\sigma(\nu_e e^-)}{dy}=
\frac{2G_F^2 m_e E_\nu}{\pi}\left[ \left( \frac{1}{2} + \sin^2 \theta_W \right)^2 + 
\sin^4 \theta_W (1-y)^2 \right]  .
\end{equation}
The signature for these neutrino-electron events is a low angle forward going lepton 
with no nuclear recoil. A similar signature was used by the CHARM-II \cite{CHARMII-nue} 
detector to measure $\sin^2 \theta_W$ from neutrino-electron elastic scattering. 
An excess of events of neutrino-electron scattering can be observed  for low values of 
the $E\theta^2$ variable over the predominant background from neutral current 
$\pi^0$ production and $\nu_e$ quasi-elastic scattering. 

\begin{figure}[tbh]
\begin{center}
\includegraphics[width=0.45\textwidth]{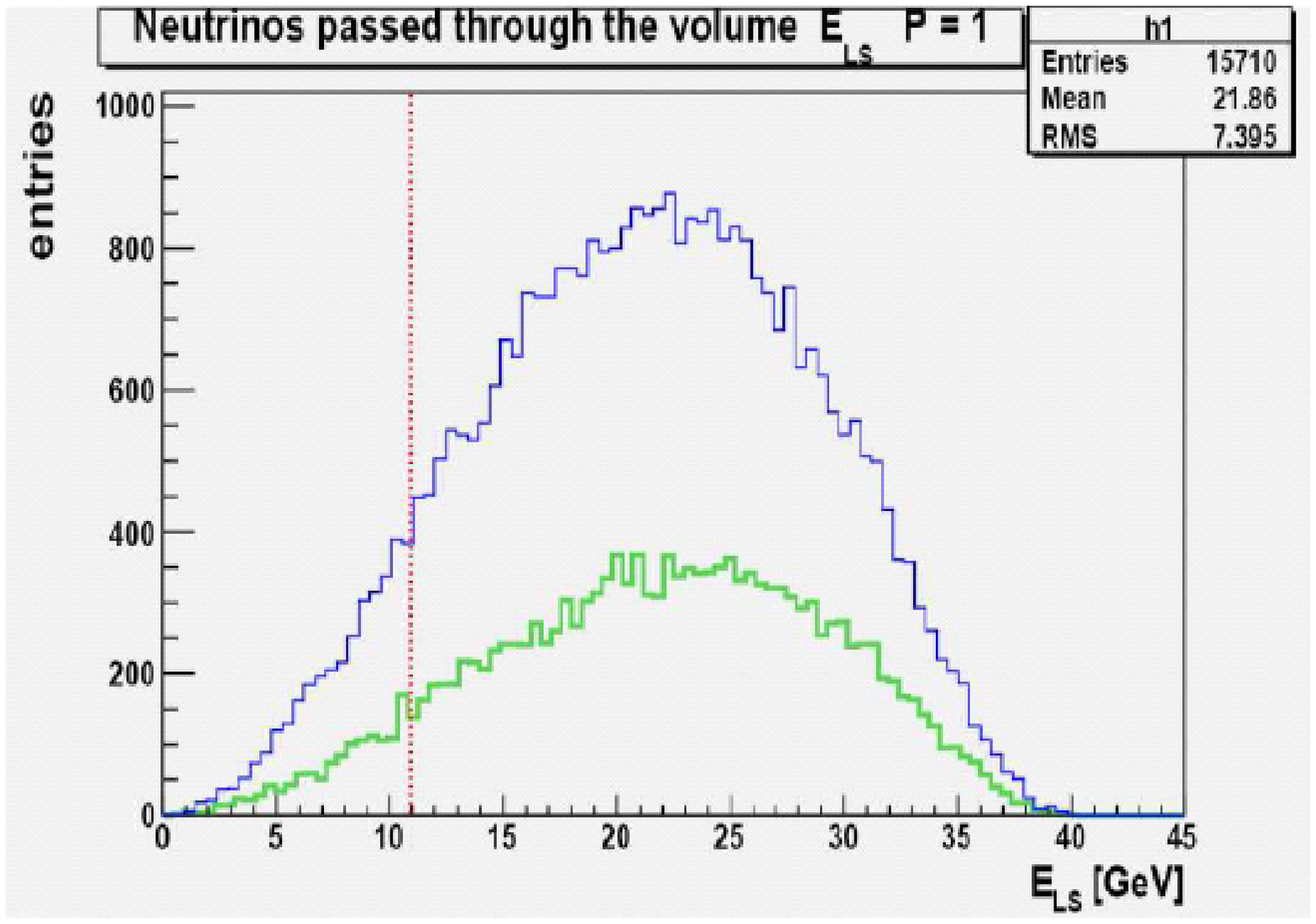}
\includegraphics[width=0.45\textwidth]{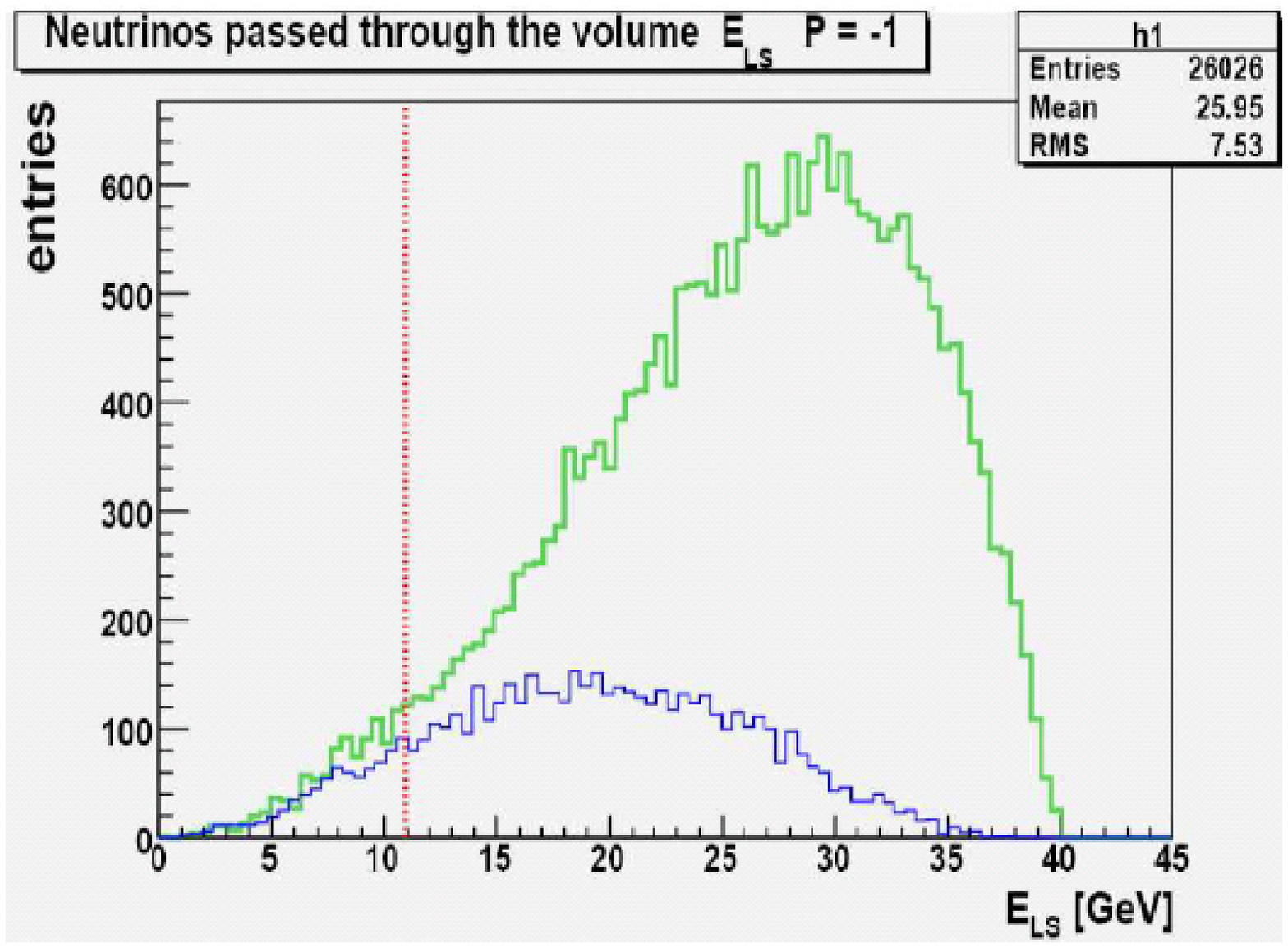}
\\
\includegraphics[width=0.45\textwidth]{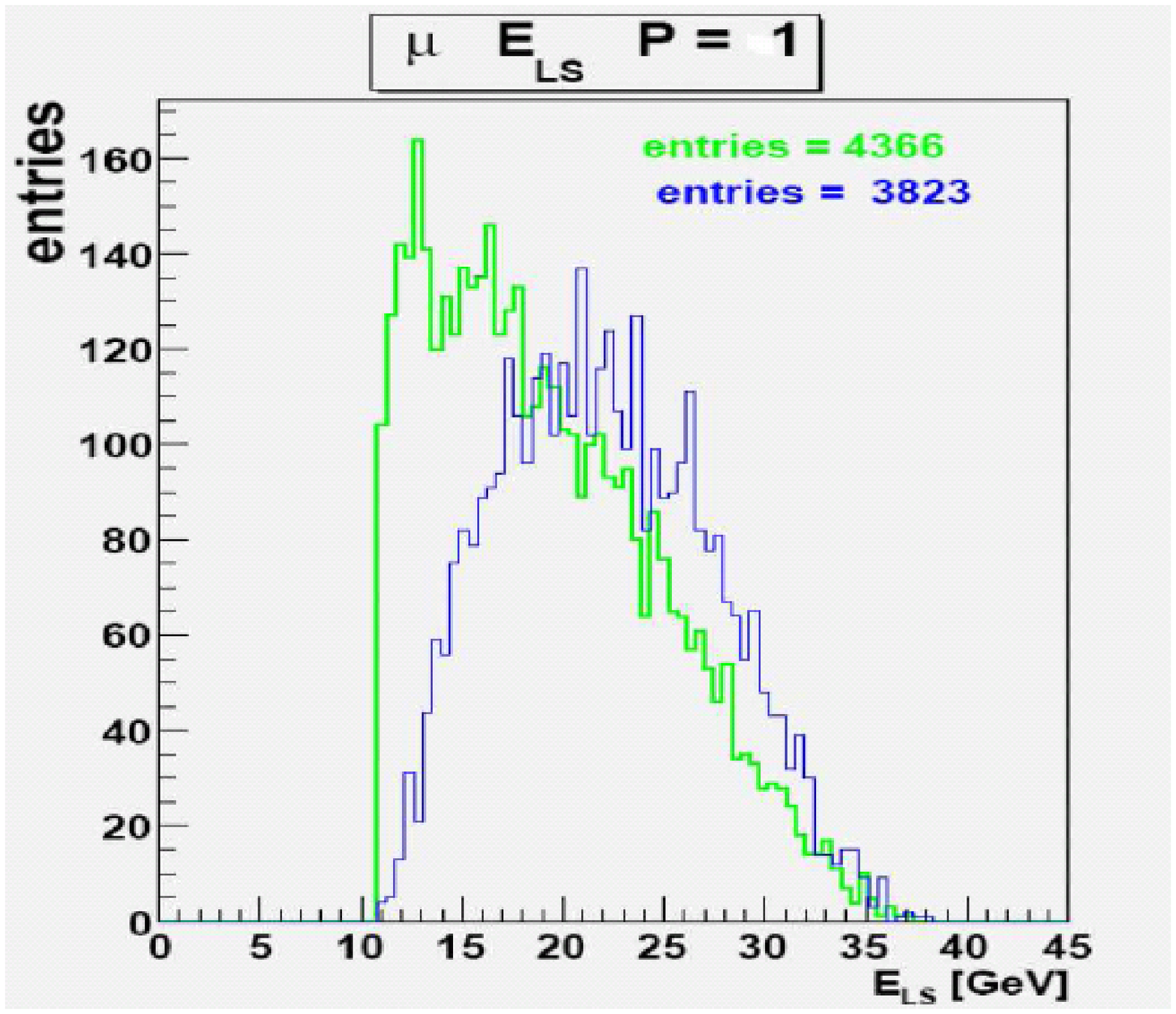}
\includegraphics[width=0.45\textwidth]{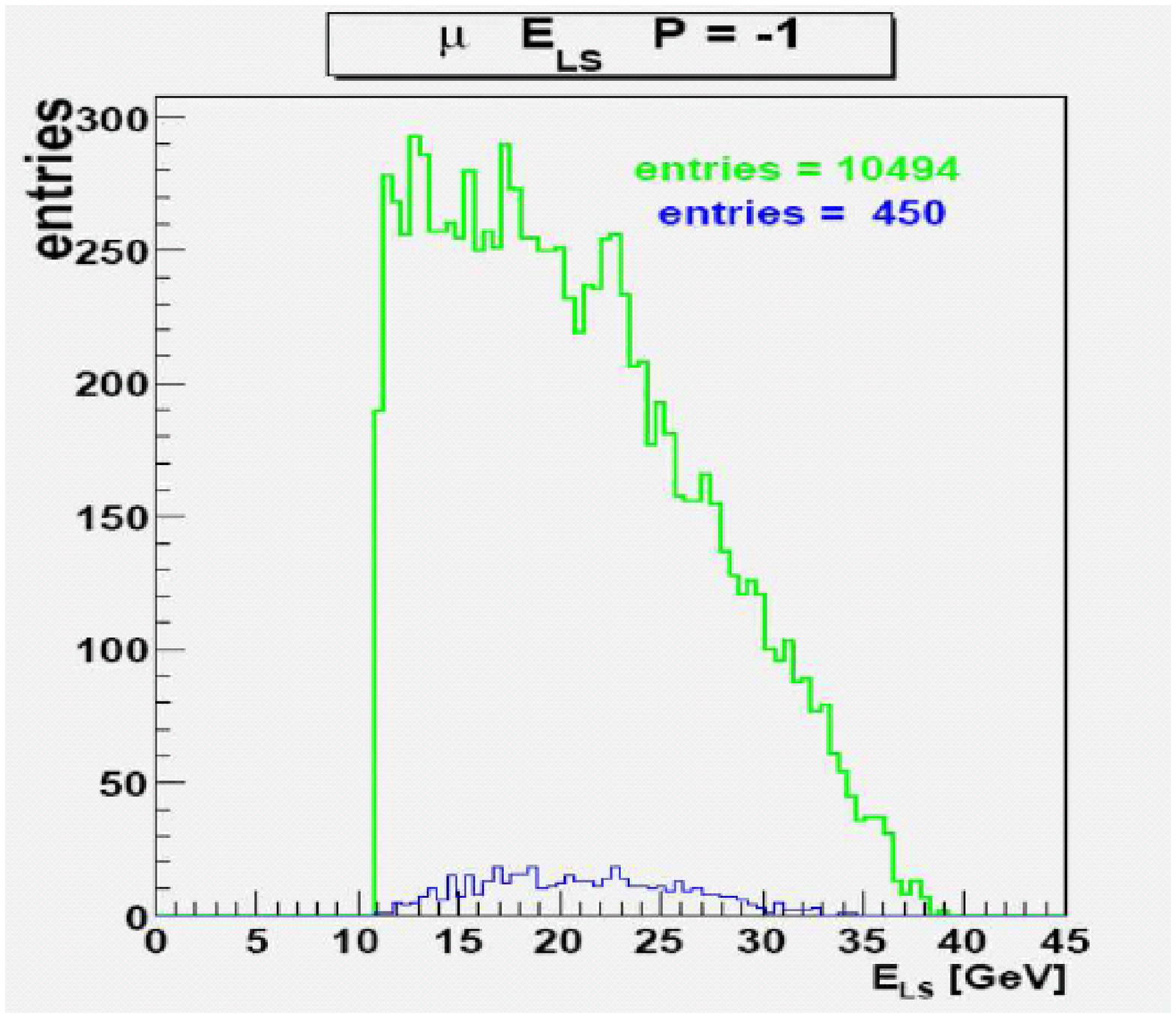}  \\
\caption
{{\sl
Energy spectrum of $\nu_\mu$ (green)  and  $\overline{\nu}_e$ (blue) passing 
through a cylinder with radius 1 m and thickness 30 cm, and at 500 m distance  
from the end of the straight section for polarization $P=+1$ (top left) 
and $P=-1$ (top right). Red line indicates the energy threshold.  
Inverse muon decay $\nu_\mu + e^- \rightarrow  \nu_e + \mu^-$ (green) and
$\overline{\nu}_e + e^- \rightarrow  \overline{\nu}_\mu + \mu^-$ (blue) events in a detector 
of radius 1 m, thickness 30 cm and density 1.032 g cm$^2$ at a distance of 
500 m from the end of the straight section of the decay ring for polarization 
$P=+1$ (bottom left) and $P=-1$ (bottom right).}}
\label{IMD}
\end{center}
\end{figure}

The reconstructed spectra of $\nu_\mu + e^- \rightarrow  \nu_e + \mu^-$ and 
$\overline{\nu}_e + e^- \rightarrow  \overline{\nu}_\mu + \mu^-$ can be seen in figure~\ref{IMD} 
in a detector of radius 1~m, thickness 30~cm filled with scintillator 
($\rho = 1.032$~g~cm$^{-2}$), for a total mass of ~1~tonne. The neutrinos originate 
from the decay of 40~GeV muons in the 500~m straight section of the decay ring at 
a neutrino factory and the detector is 500~m from the end of the straight section.

Table~\ref{tb:IMD} shows the event rate expected from the inverse muon decay 
reactions. It is clear that the event rate is strongly dependent on the polarization 
and can be used as an independent verification of the polarization of the decay 
muons. Since the two reactions ($\nu_\mu e^-$ and $\overline{\nu}_e  e^-$) are 
practically indistinguishable, the statistical error in the flux will come from 
the sum of the two, an accuracy of better than 10$^{-3}$ in the flux using these 
reactions can only be achieved for a muon energy of more than 40~GeV within one 
year of data taking. However, the efficiency and the background for these reactions 
have not been determined yet, so the statistical significance will be diminished.

\begin{table*}[htb]
\centering
\begin{tabular}{|c|c|c|c|c|}
\hline
$E_\mu$ (GeV) & Polarization & $\nu_\mu e^- \rightarrow  \nu_e  \mu^-$  & 
$\overline{\nu}_e  e^- \rightarrow  \overline{\nu}_\mu  \mu^-$  & $\nu_\mu N$ \\ \hline
40 & +1 & $6.87\times 10^5$ & $5.81\times 10^5$   & $1.92\times 10^9$  \\ \hline
40 & -1 & $1.67\times 10^6$ & $6.97\times 10^4$   & $2.81\times 10^9$  \\ \hline
30 & +1 & $2.02\times 10^5$ & $1.97\times 10^5$   & $1.32\times 10^9$  \\ \hline
30 & -1 & $5.89\times 10^5$ & $1.60\times 10^4$   & $1.91\times 10^9$  \\ \hline
20 & +1 & $1.83\times 10^4$ & $1.14\times 10^4$   & $8.07\times 10^8$  \\ \hline
20 & -1 & $7.83\times 10^4$ & $7.76\times 10^2$   & $1.14\times 10^9$  \\ \hline

\hline 
\end{tabular}
\caption{
Total number of muons per year from inverse muon decay reactions produced in a 
cylindrical detector with radius 1 m, thickness 30 cm  and density 1.032 g/cm$^3$
(scintillator, total mass ~1 ton), 500 m distant  from the end of the straight 
section of muon storage ring ($10^{21}$ muon decays per year).
The last column shows the total number of muons per year produced in the same 
cylinder from inclusive CC reactions. 
}
\label{tb:IMD}
\end{table*}    

\subsubsection{Cross-sections and parton distribution functions}

The near detector will carry out a programme of cross-section measurements, 
necessary for the far detector \cite{nuint}. Due to the experimental control of 
the flux, it will be possible to extract the cross-section of the different 
interactions to be studied, such as deep inelastic, quasi-elastic and elastic
interactions, $\Delta^{+}$ and $\Delta^{++}$ resonance and single and multi-pion 
production (see appendix~\ref{Low_energy}). The 
aim will be to cover all the available energy range, with particular emphasis 
at low energies (where quasi-elastic events dominate), since this might be 
needed to observe the second oscillation maximum at a far detector. At these 
lower energies, nuclear reinteractions and shadowing as well as the role of 
Fermi motion play a role, and these effects need to be determined. Very low 
energy interaction measurements might be achievable using a liquid argon TPC, 
or other very light tracking detector. We should envisage also the possibility 
of using different nuclear targets, as well as the direct access to nucleon 
scattering from hydrogen and deuterium targets.

\subsubsection{Charm measurements}

The wrong-sign muon signature of the neutrino oscillation ``golden channel'' 
can be identified, for example, in a magnetised iron calorimeter, by distinguishing 
between muons, hadrons and electrons, and measuring the charge of the lepton. 
The main backgrounds for this signal are due to wrong charge identification and 
to the production of wrong sign muons from the decay of a charm particle (for 
example, from a $D^-$), produced either in neutral current interactions or in 
charged current interactions where the primary muon has not been identified. 
The charm background is the most dangerous at high energies, but a combined 
cut in the momentum of the muon ($P_\mu$) and its isolation with respect to 
the hadronic jet using the variable $\qt = \pmu \, \sin^2 \theta_{\mu h}$, 
where $\theta_{\mu h}$ is the angle between the muon and the hadronic shower 
(see section~\ref{sec:mind}) can reduce the background to the $8\times 10^{-6}$ 
level for an efficiency of 45\% \cite{golden}. However, this background reduction relies 
on an accurate knowledge of the $\qt$ distribution of charm particles that 
should be measured at a near detector.  

\begin{figure}[tbhp]
\begin{center}
\includegraphics[width=0.95\textwidth]{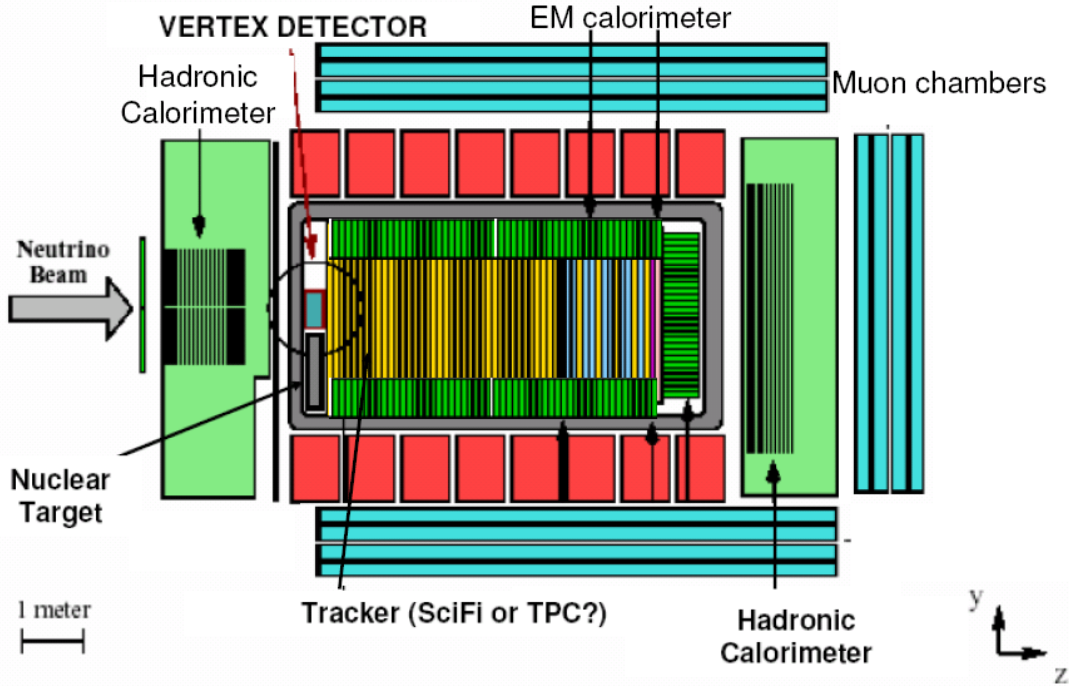} \\
\caption{
{\sl Possible geometry for a near detector at a neutrino factory.}}

\label{fig:near-detector}
\end{center}
\end{figure}

A near detector should be able to operate at a high rate and have very good 
spatial resolution, to be able to distinguish primary and secondary vertices 
needed to identify charm events. It should also have a small radiation length 
so that it may distinguish electrons from muons in a magnetic field. This can 
be achieved by a vertex detector of low $Z$ (either a solid state detector, such 
as silicon, or a fibre tracker) followed by tracking in a magnetic field and 
calorimetry, with electron and muon identification capabilities \cite{ND_Vertex05}. 
A possible near detector geometry could be fit into the NOMAD dipole magnet \cite{nomad}, 
currently being used for the T2K 280 m detector \cite{ND280} (figure~\ref{fig:near-detector}).

\begin{figure}[tbhp]
\begin{center}
\includegraphics[width=0.5\textwidth]{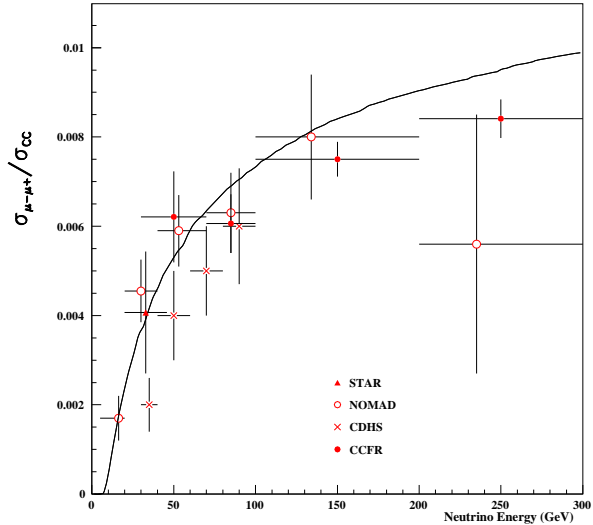} \\
\caption
{\sl Opposite sign dimuon rate of NOMAD-STAR and other experiments.
           Overlayed is a charm mass fit of 1.3 GeV/c$^{2}$ \cite{nomad_dimu}.}

\label{fig:dimuon}
\end{center}
\end{figure}

A prototype silicon detector, consisting of four passive layers of boron carbide (45 kg) 
and five layers of silicon microstrip detectors (NOMAD-STAR) was implemented within the 
NOMAD neutrino oscillation experiment \cite{STAR_1,STAR_2}. Impact parameter and vertex 
resolutions were measured to be 33~$\mu$m and 19~$\mu$m respectively for this detector. 
A sample of 45 charm candidates (background of 22 events) was identified \cite{STAR_charm}. 
The total charm meson production rate found was $7.2\pm 2.4$\% of the $\nu_\mu$ charged 
current rate, at an average energy of 33~GeV, which compares well with other experiments 
assuming the semi-leptonic branching ratio of charm particles \cite{nomad_dimu} 
(see figure \ref{fig:dimuon}). An efficiency of 3.5\% for $D^0$ and $D^+$, and an 
efficiency of 12.5\% for $D_s^+$ were achieved. Even with these low efficiencies, one 
could obtain more than $3\times 10^6$ charm events per year. However, using a fully 
active silicon pixel detector with more layers can provide further improvements. 
For example, 18 layers of 500$\mu$m thick silicon of dimensions $50\times 50$~cm$^2$ 
(total silicon area of 4.5~m$^2$) corresponds to 52~kg of silicon. Efficiencies for 
reconstructing charm events should vastly improve with this geometry. Monolithic 
Active Pixel (MAPS) \cite{MAPS_1,MAPS_2} or DEPFET \cite{DEPFET} detectors would 
be good candidates for this type of silicon technology.

Another possibility for a near detector dedicated to the study of charm is an 
emulsion cloud chamber followed by a tracking detector such as a scintillating 
fibre tracker (similar to OPERA \cite{opera} or CHORUS \cite{CHORUS}). Emulsion 
technology has already demonstrated that it is a superb medium for the study of 
charm \cite{CHORUS_1,CHORUS_2,CHORUS_3,CHORUS_4,CHORUS_5,CHORUS_6} due to its 
unrivalled spatial resolution. The main issue, however, is whether it can 
cope with the high rate that will be observed at a neutrino factory.
 
In addition to the important measurement of the oscillation background, this 
sample of charm events can be used to determine the strange quark content of the sea, 
the CKM parameter $V_{cd}$ to unprecedented accuracy and search for CP violation 
in $D_0$-$\overline{D}_0$ mixing. The sign of the lepton 
produced at the primary vertex can be used to tag the initial charm particle, 
with the decay products determining whether there was any change in the flavour 
of the charm meson \cite{macfarland}.

\subsubsection{Outlook}

The near detector at a neutrino factory is an essential ingredient in the 
overall neutrino factory complex, necessary to reduce the systematic errors 
for the neutrino oscillation signal. There are many choices for a detector 
technology that could be implemented. Liquid argon TPCs in a magnetic field 
would be able to carry out most of the near detector programme. Also, more 
conventional scintillator technology (similar to Minerva \cite{Minerva}), a scintillating 
fibre tracker or a gas TPC (like in the T2K near detector \cite{ND280}) would also be able 
to perform cross-section and flux control measurements. However, it seems 
likely that only silicon or emulsion detectors can achieve the necessary 
spatial resolution to perform the charm measurements needed to determine 
the background for the oscillation search. These options shall be further 
studied within the context of the International Design Study.

\section{Far detectors }





\subsection{Tracking calorimeters}
\label{sec:track_cal}


In a Neutrino Factory the  $\nu_e \rightarrow \nu_\mu$ oscillation channel, the 
so-called golden channel,  
provides the cleanest experimental signature since it only requires the detection 
of ``wrong-sing muons''   
(ws-muon) -- muons with the opposite charge to those circulating in the storage ring -- 
in a detector with charge measurement capabilities. Muon reconstruction is well understood and can 
be performed with high efficiency keeping a negligible background level. 
Assuming stored positive muons, the main backgrounds for the ws-muon search are \cite{lmd,golden}:

\begin{itemize}
\item right-charge muons whose charge has been misidentified, in \numubar CC events. 
\item ws-muons from hadron decays and ws-hadrons misidentified as muons 
      in \numubar or \nue NC events, 
\item ws-muons from hadron decays and ws-hadrons misidentified as muons 
      in \numubar or \nue ~charge CC 
      when the main lepton is not identified. 
\end{itemize}

A detector aiming to study the golden channel should be able to 
identify muons and measure their momenta and charge with high efficiency 
and purity. Magnetized iron calorimeters have been considered in the past 
\cite{lmd}-\cite{ino}. The ws-muon detection efficiency can be kept above 
$50 \%$ for a background level of the order of $10^{-5}$. This kind of 
detector is extremely powerful for the measurement of very small 
$\theta_{13}$, reaching values of $sin^2(2\theta_{13})$ below $10^{-4}$. 
However, they may have trouble in studying CP violation  
because the high density  of the detector prevents the detection of low 
energy neutrinos (below few \GeV), which could provide very valuable information 
for the simultaneous measurement of $\dcp$ and $\theta_{13}$.  

An alternative to iron calorimeters, which follows the \nova experiment~\cite{nova} guidelines, 
has been recently considered. 
A magnetised version of Totally Active Scintillator Detectors 
(TASD), could be very efficient for the ws-muon search, even 
for neutrino energies below 1~\GeV. 
The non-magnetised TASD detector (as \nova) would be a good candidate for lower energy 
beams in the few \GeV range, as WBB or Beta-Beams. The physics performance of such a 
detector in those scenarios has been discussed elsewhere~\cite{iss_physics_report}. 
In this section the magnetised fully active and iron calorimeters are described.

\subsubsection{Magnetised iron calorimeters}
\label{sec:mind}

The wrong-sign muon search at a neutrino
factory requires a very massive detector with good muon and muon charge 
identification capabilities.  
Magnetic iron calorimeters can fulfil these requirements using well 
known technologies. Indeed, they are conceptually similar to the existing 
MINOS detector~\cite{minos_proposal}, but with a mass one order of magnitude larger.
Several complementary studies have being conducted so far: the Magnetic Iron Neutrino Detector
(MIND) \cite{lmd,golden,mind_review} (called LMD in the past) 
and Monolith \cite{monolith,mind_review}. Recently, a new option, the Indian Neutrino Observatory (INO)~\cite{ino}, 
similar to Monolith, has been proposed to study the golden channel at 7000~\Km. 

In this section the results of the MIND study are presented. 
The conceptual design of the MIND detector consists of a sandwich of 
4~\cm thick iron plates and 1~\cm thick detection layers, with transverse 
dimensions 14$\times$14~\m$^3$. The detector has a length of 40~\m 
and a total mass of 60~\Kton. The fiducial mass is of the order of 50~\Kton. 

The nature of the detection layers is not yet specified. A possible  
choice could be either solid (as MINOS) or liquid (as \nova) scintillator bars. 
The radiation length of plastic scintillator is assumed for the moment. 
A transverse resolution, $\varepsilon$, of 1~\cm in 
both coordinates is considered. The measurement of the charge of the muon 
forces the detector to be magnetised. A realistic detector would use a toroidal 
field produced by a superconducting coil traversing the detector longitudinally (as MINOS). 
This implies however unnecessary complications from the point of view of the 
reconstruction program, at this stage of the analysis. 
In this conceptual design an average dipole field of 1~\Tesla 
(1.3~\Tesla in the iron plates) in the Y 
direction is used. From the performance point of view both are similar except by  
the small radial decrease of the toroidal field (see Fig.~\ref{fig:minos_field}), 
which can be ignored for the moment.  

To study the performance of the MIND detector a Monte Carlo simulation 
based on the GEANT 3 package~\cite{geant3} has been performed. Deep inelastic (DIS) 
neutrino interactions have been generated using the LEPTO package~\cite{lepto}.  
From the point of view of computing time a full simulation   
is not practical because background rejection has to be 
studied to the level of $10^{-6}$, which requires more than $10^6$ 
events for each kind of background. Thus the MIND study is based 
on a fast simulation in which the electronic response of the detector 
is not simulated and a smearing of the relevant physics quantities is 
used instead. The physical quantities used in the analysis are the muon momentum 
($\pmu$), the muon angle ($\theta_\mu$), the hadronic energy ($\ehad$) and the hadronic 
angle ($\theta_{h}$). In previous analyses~\cite{lmd,golden,mind_review} 
all of them were smeared as in the MINOS proposal~\cite{minos_proposal}. 
In this analysis a better hadronic angular resolution, as reported by 
Monolith~\cite{monolith_testbeam}, is used.

\vspace{0.5cm}
\noindent
{\bf \it Muon identification}
\vspace{0.5cm}

\begin{figure}[htbp]
\begin{center}
\epsfig{figure=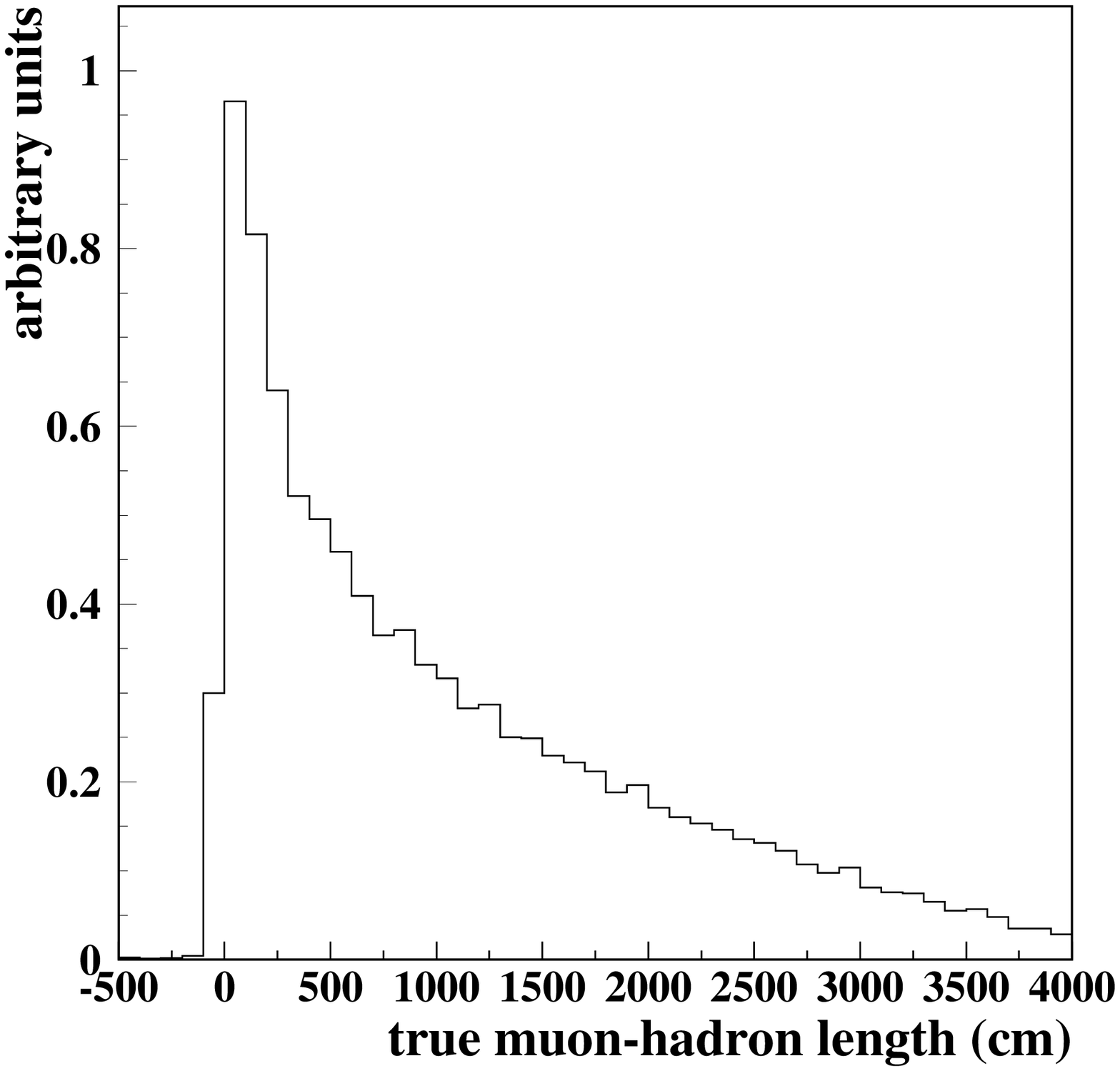, width=7cm,height=5cm}
\epsfig{figure=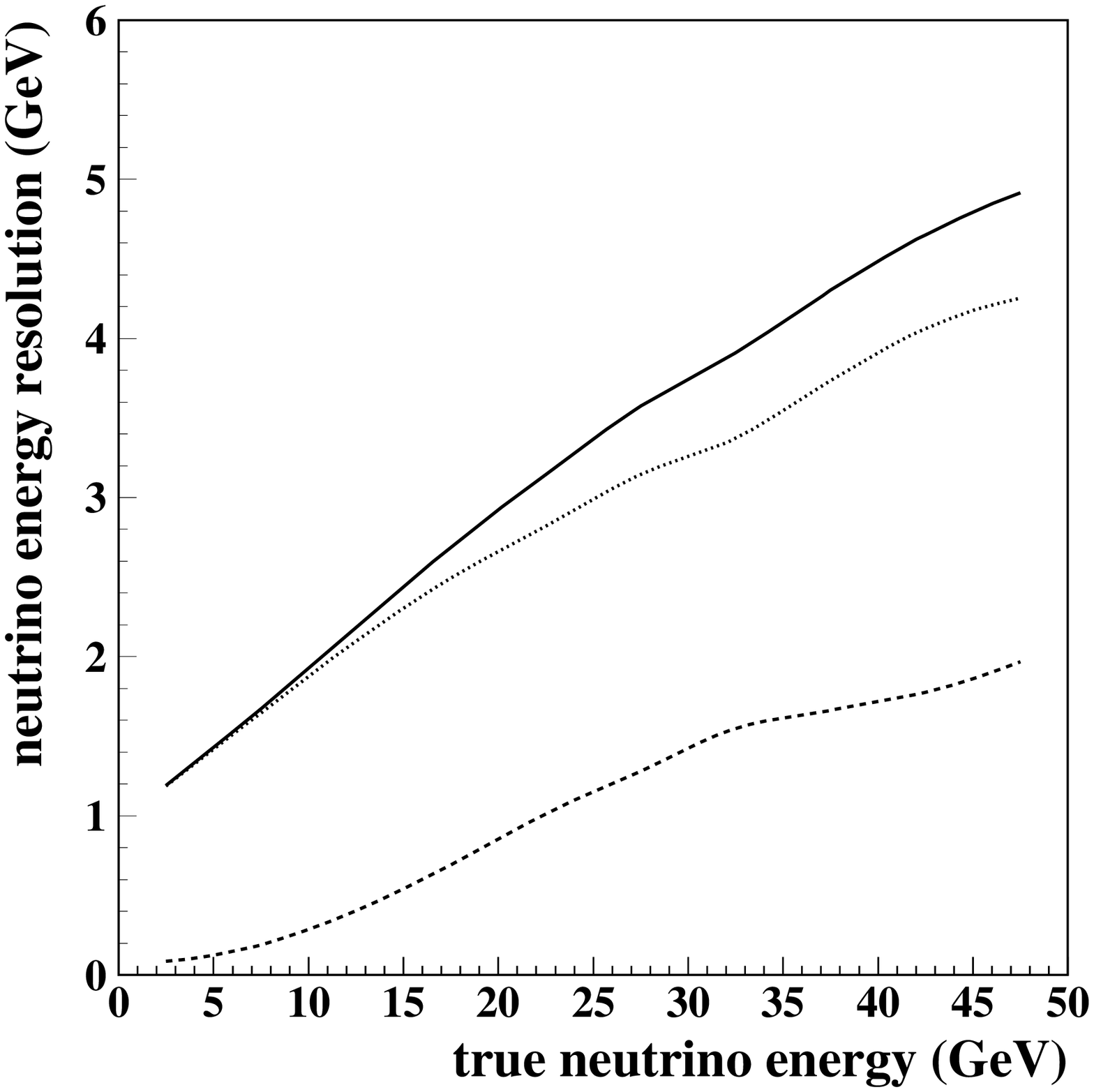, width=7cm,height=5cm}
\caption{On the left panel, distribution of the true $\Delta L$. 
On the right panel neutrino energy resolution as 
a function of the true neutrino energy (solid line). 
The dashed line corresponds to contribution of the muon momentum measurement, while the dotted 
line is the hadronic energy resolution. 
\label{fig:muonid_enu_res}}
\end{center}
\end{figure}

Neutrino interactions in such a detector have a clear signature. 
$\numu$CC or $\numubar$CC events are characterized by a muon, 
easily seen as a penetrating track of typically several metres length, 
and a shower resulting from the interactions of the final-state hadrons, 
which extinguishes at short distances. Thus, the identification of muons 
can be easily done by range. Fig. \ref{fig:muonid_enu_res}-left shows the distribution of  
$\Delta L = L_\mu - L_{h}$, 
where $L_\mu$ and $L_h$ are respectively the lengths travelled respectively 
by the longest muon and hadron in $\numubar$ CC events. 
The muon identification criterion is set as follows:
a particle will be identified as a muon if it goes a given length 
$\Delta L$ -- to be optimised -- beyond any other particle in the event. 
Notice that this is a very conservative 
approach since it assumes that the muon and the hadronic shower 
have the same direction. 

\vspace{0.5cm}
\noindent
{\bf \it Energy resolution}
\vspace{0.5cm}

An estimator of the neutrino energy, $\enu$, is the total visible energy in 
the event, $\evis$, which is the sum of the muon and hadron shower energies. 
The first can be estimated either by range or by curvature for fully contained muons 
and only by curvature when the muon escapes the detector. 
A momentum resolution of $(3.5\pmu+0.022\pmu^2)\%$, 
as an approximation to the one quoted in the MINOS proposal, is used for the range measurement, 
while the resolution obtained  
by curvature is computed using the Gluckstern formula \cite{gluckstern}. 
On the other hand, the hadron shower energy is 
computed by calorimetry, using the resolution quoted in the MINOS proposal: 
$\delta \ehad /\ehad = 0.03+0.76/\sqrt{\ehad}$. Fig.~\ref{fig:muonid_enu_res} shows the average 
$\enu$ resolution as a function of $\enu$ for $\numubar$ CC events. The contributions 
of the hadronic shower and the muon are indicated. The former clearly dominates the $\enu$ resolution. 

\vspace{0.5cm}
\noindent
{\bf \it Charge identification}
\vspace{0.5cm}

\begin{figure}[htbp]
\begin{center}
\epsfig{figure=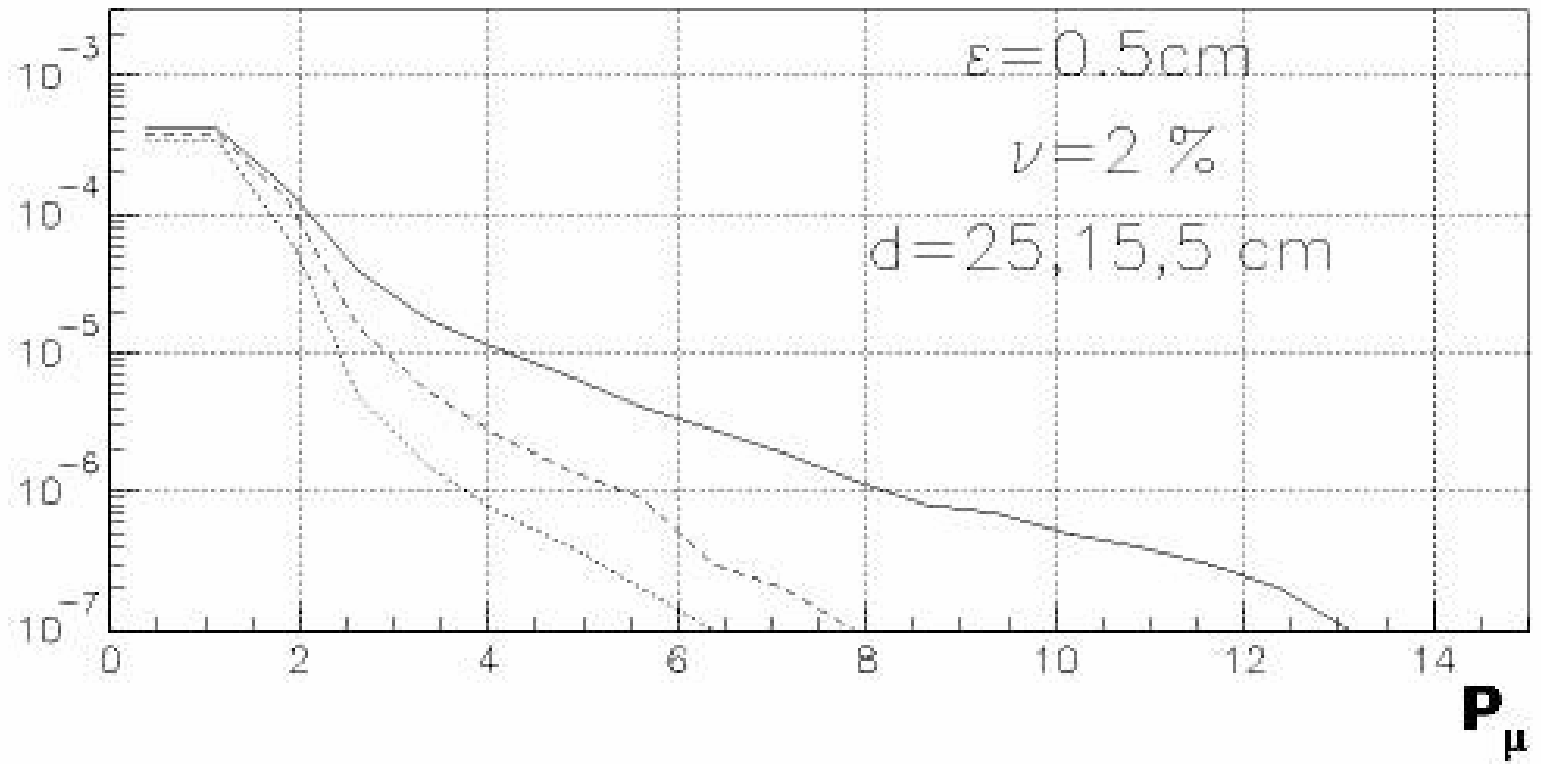,width=5.2cm,height=5cm} 
\hspace*{-0.5cm}
\epsfig{figure=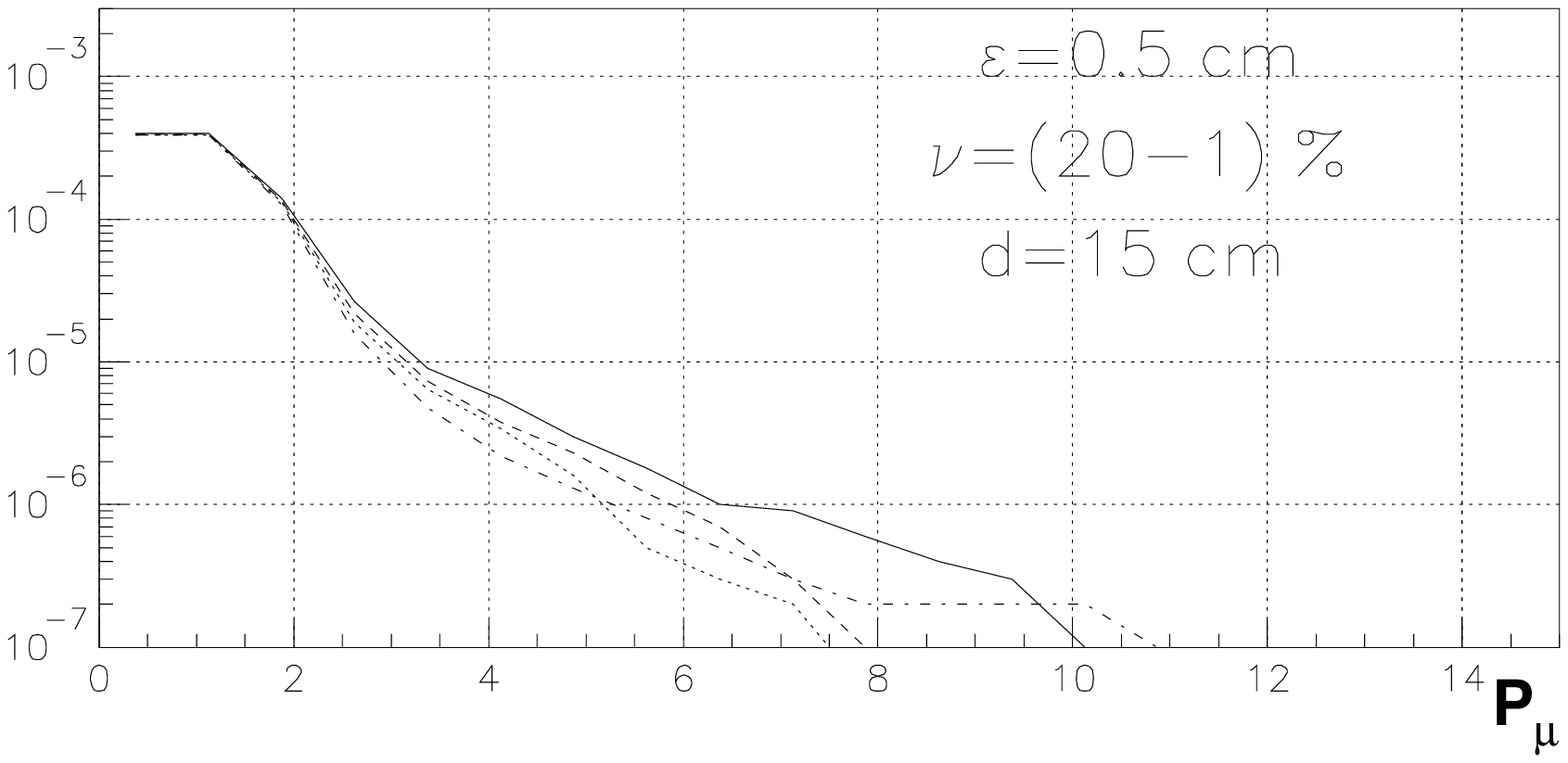,width=5.2cm,height=5cm}
\hspace*{-0.5cm}
\epsfig{figure=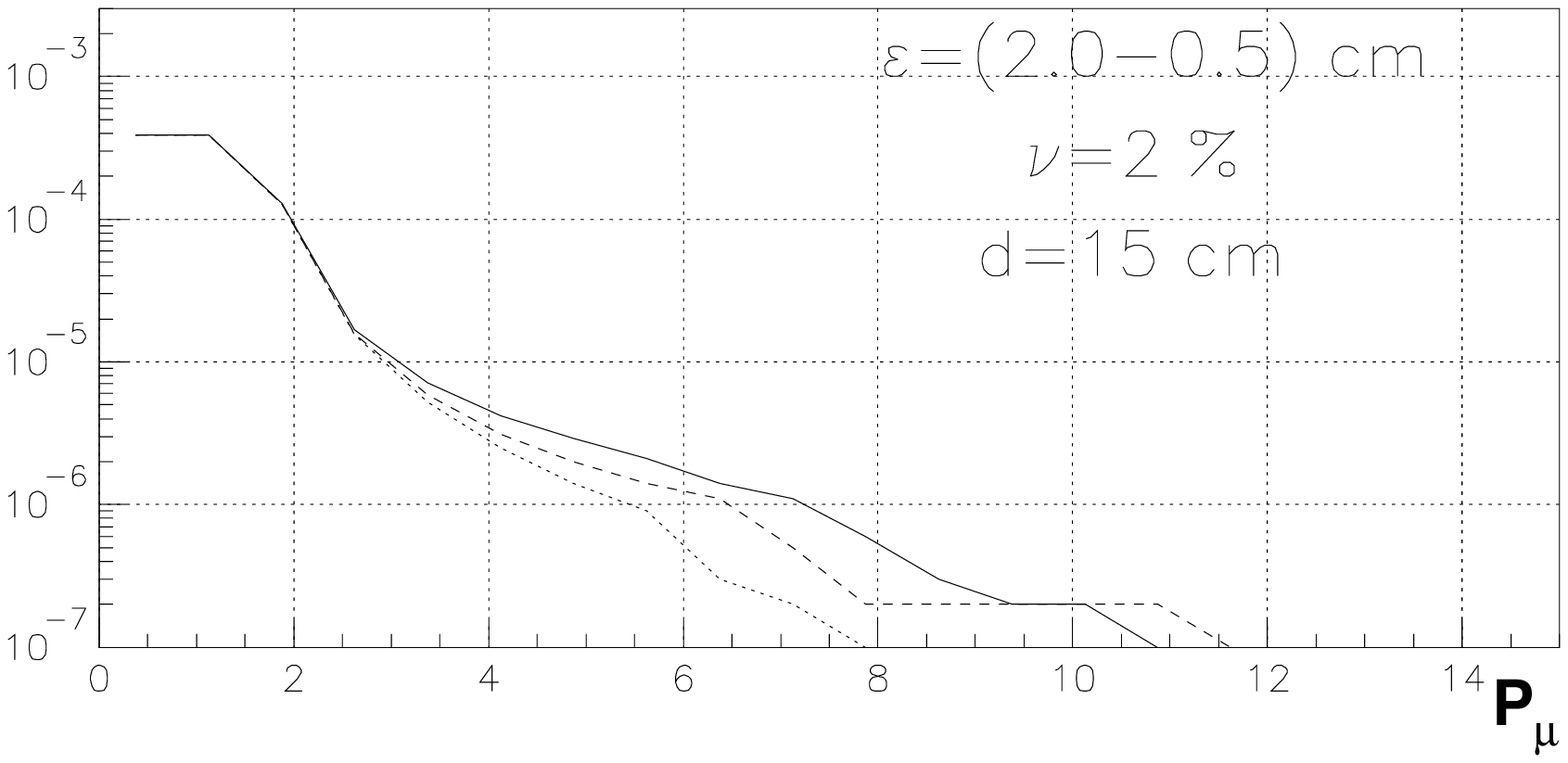,width=5.2cm,height=5cm}
\caption{Charge misidentification background as a function of momentum 
         for different configurations of MIND, assuming a constant average field of 
         1~\Tesla. 
         $\varepsilon$ is the transverse resolution, 
         $\nu$ is the hit finding inefficiency  
         and $d$ the distance betweeen measurement planes. 
\label{fig:charge}}
\end{center}
\end{figure}

As mentioned above, charge misidentification of primary muons in $\numubar$ CC 
interactions constitutes an important background to the ws-muon signal. 
Fig.~\ref{fig:charge} (from Ref.~\cite{mind_review}) shows the charge misidentification rate 
for different configurations of the MIND detector assuming a constant average magnetic field of 1~\Tesla 
(independently of the iron distribution).
The muon hits have been fitted to a cubic model taking into account multiple scattering and energy loss.  
High angle scatters have been removed by a local $\chi^2$ criteria. The charge misidentification rate is 
of the order of $10^{-6}$ for 5~\GeVc muons and close to $10^{-4}$ for 2~\GeVc muons. 
The distance between measurement planes seems to be the crucial parameter to be optimised. 
This analysis has however two main limitations: 
i) the average magnetic field is independent of the distance between measurement planes, 
which is unrealistic below some distance ($\sim$ 5\cm) since the magnetic field is only present in the iron; 
ii) all interactions were generated in the center of the detector such that there were no border effects. 

\begin{figure}[htbp]
\begin{center}
\epsfig{figure=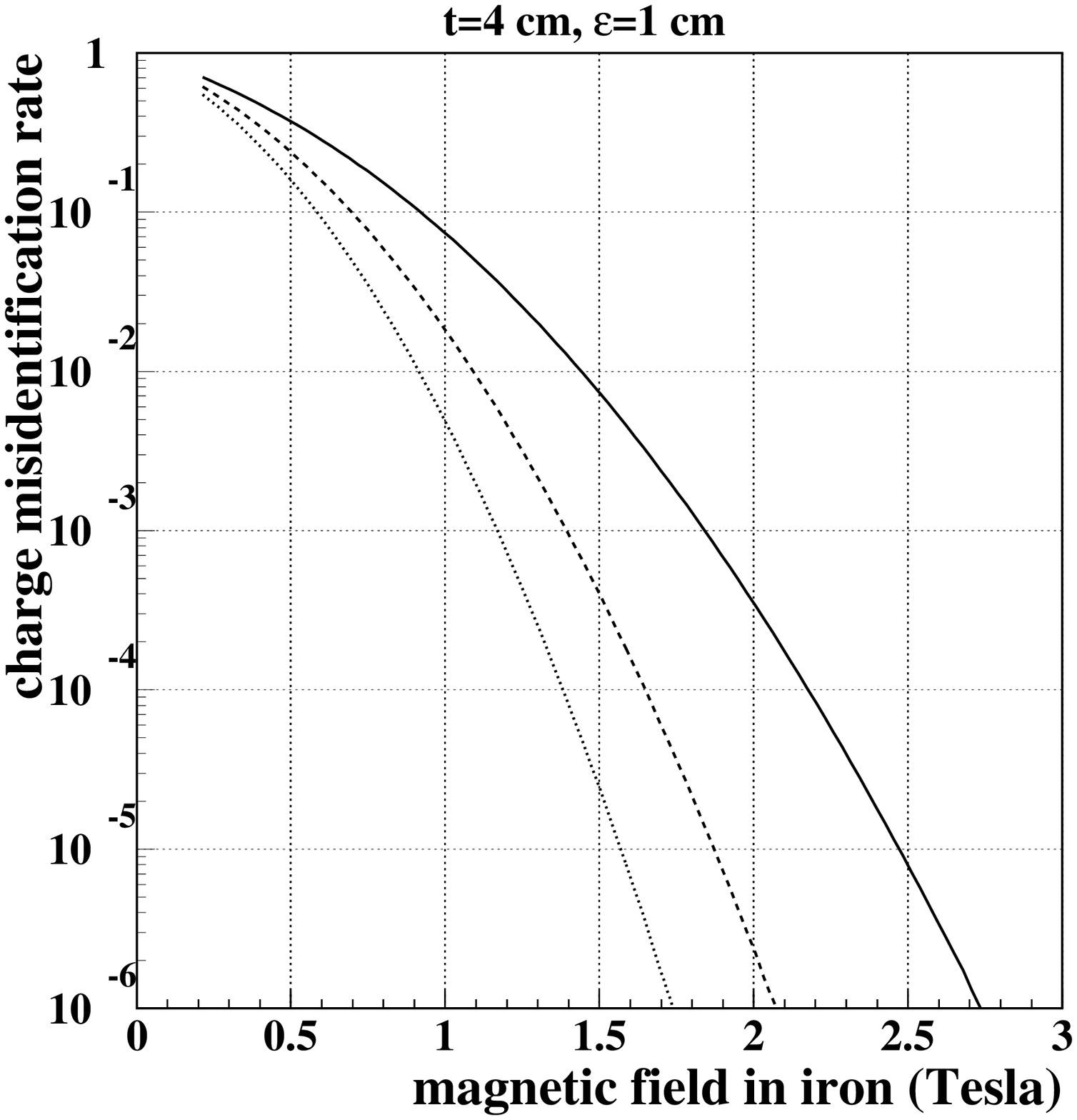, width=5cm,height=5cm}
\epsfig{figure=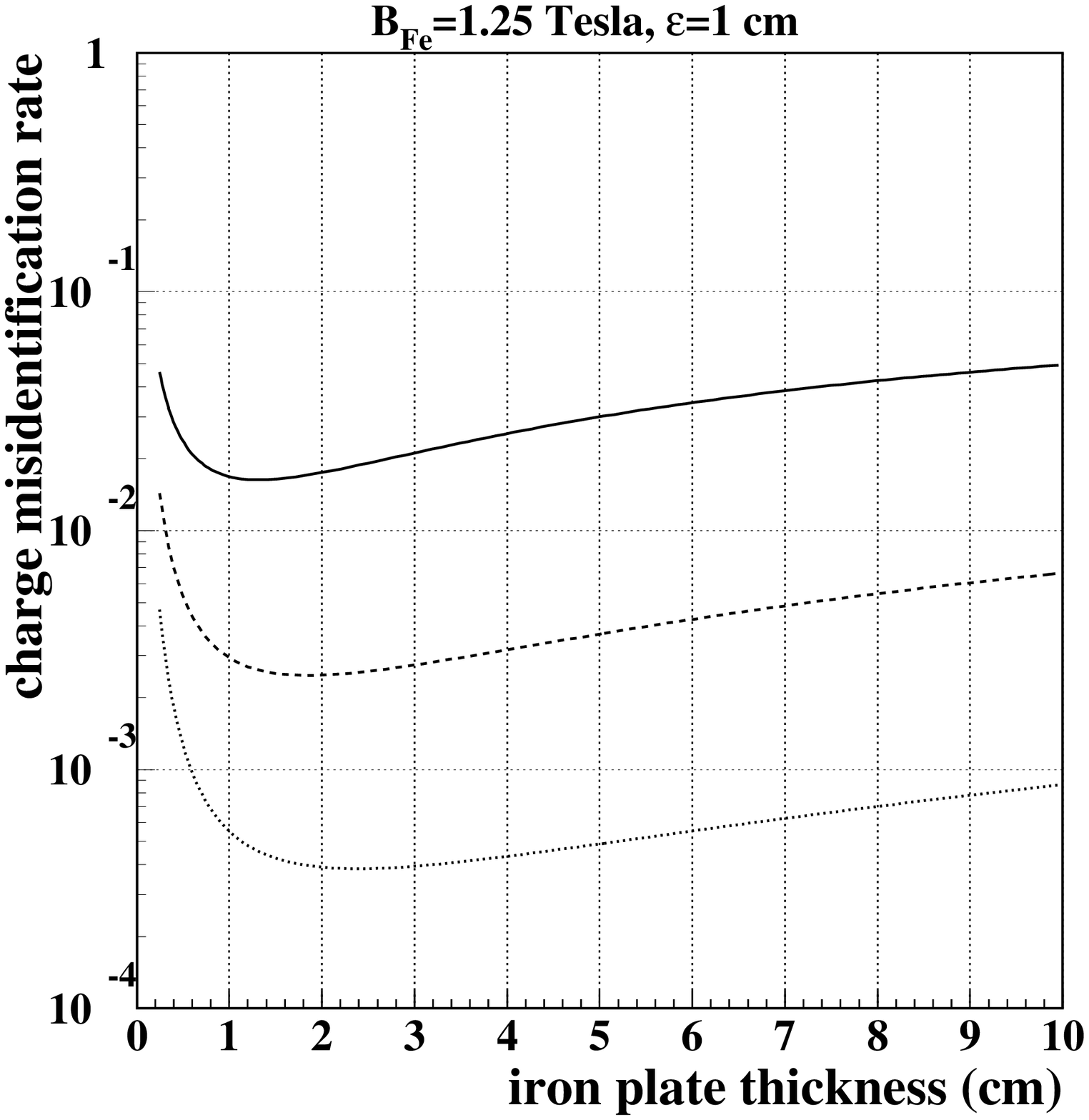, width=5cm,height=5cm}
\epsfig{figure=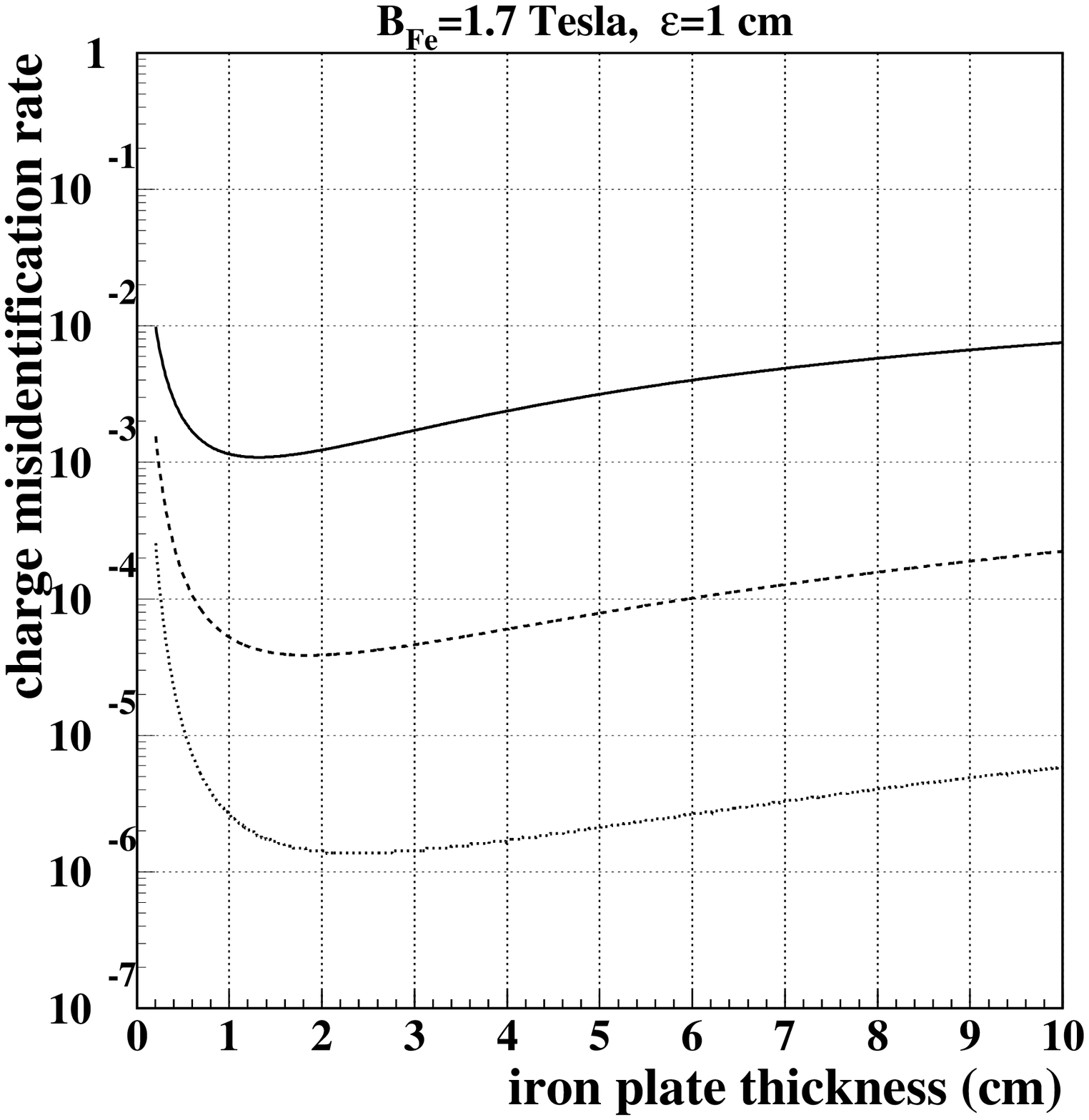, width=5cm,height=5cm}
\caption{Charge misidentification rate for different detector configurations and for different muon momenta:
1~\GeVc (solid line),   1.5~\GeVc (dashed line) and   2~\GeVc (dotted line), assuming a Gaussian multiple scattering.   
$t$ is the thickness of the iron plates (4~\cm for the default setup) and $\varepsilon$ the transverse resolution.
\label{fig:charge_misid}}
\end{center}
\end{figure}

In principle all high angle scatters can be removed by requiring the local and global $\chi^2$ 
of the track fit to be within certain limits. In this case the charge misidentification rate can be 
computed using simple equations that assume Gaussian multiple scattering and no border effects. 
Fig.~\ref{fig:charge_misid} shows the charge misidentification rate for muons of 1, 1.5 and 2~\GeVc and 
different detector configurations. For the default magnetic field 
(1.25 \Tesla in iron, corresponding to 1 \Tesla average), 
any iron plate thickness between 1 and 5 \cm seems to work, being this parameter more important at 
lower momenta. The crucial parameter is the magnetic field. At 1\GeVc the default performance is $0.3\%$. 
An order of magnitud less is obtained when the field in iron is increased from 1.25 to 1.7~\Tesla 
and another order of magnitude for 2~\Tesla.

\vspace{1cm}
\noindent
{\bf \it Signal and background efficiencies for very small $\theta_{13}$}
\vspace{0.5cm}

\begin{figure}[htbp]
\begin{center}
\numubar\ CC events  \hspace*{4cm} (\numubar +\nue) NC events  \\
\epsfig{file=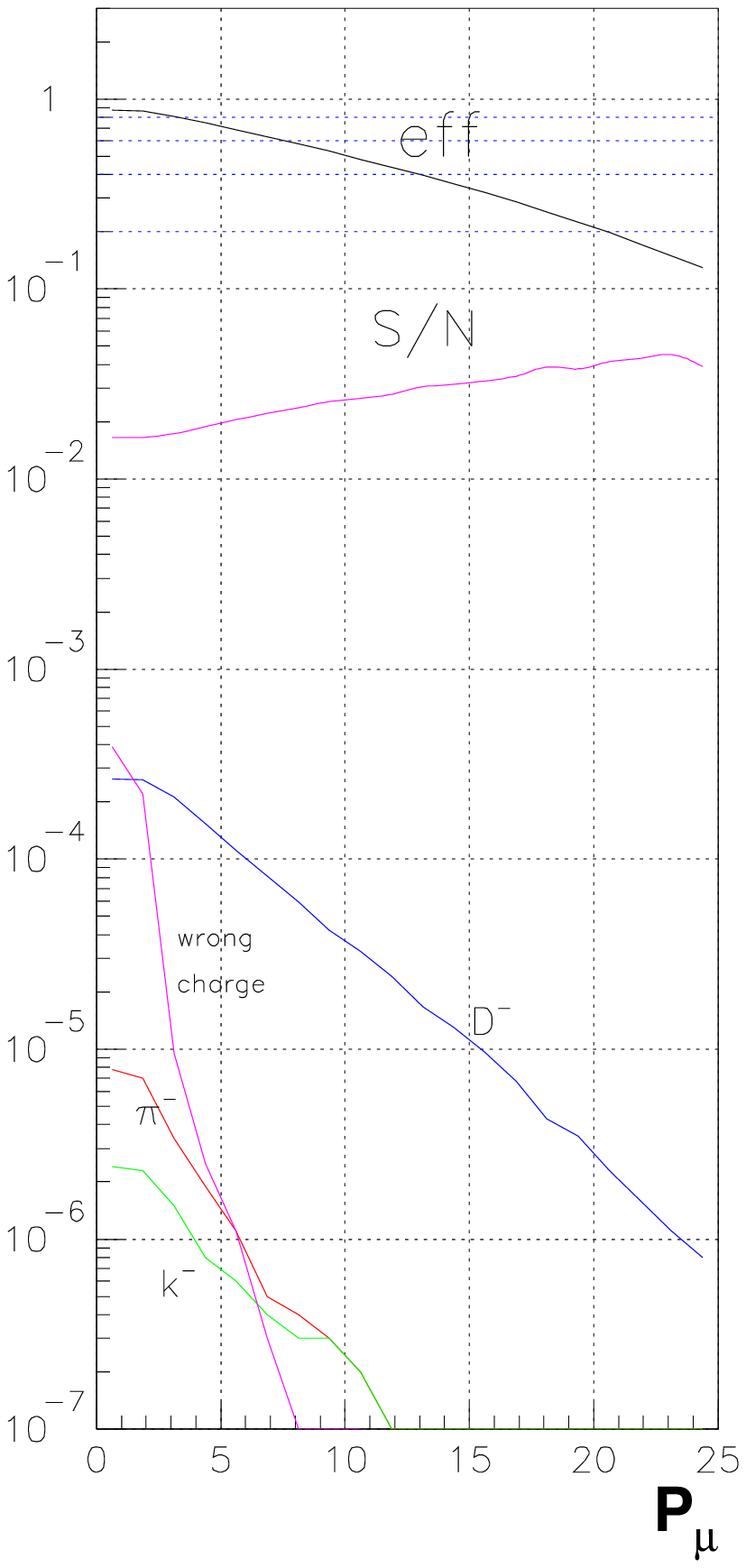,height=7cm,width=4.4cm}
\hspace*{-1cm}
\epsfig{file=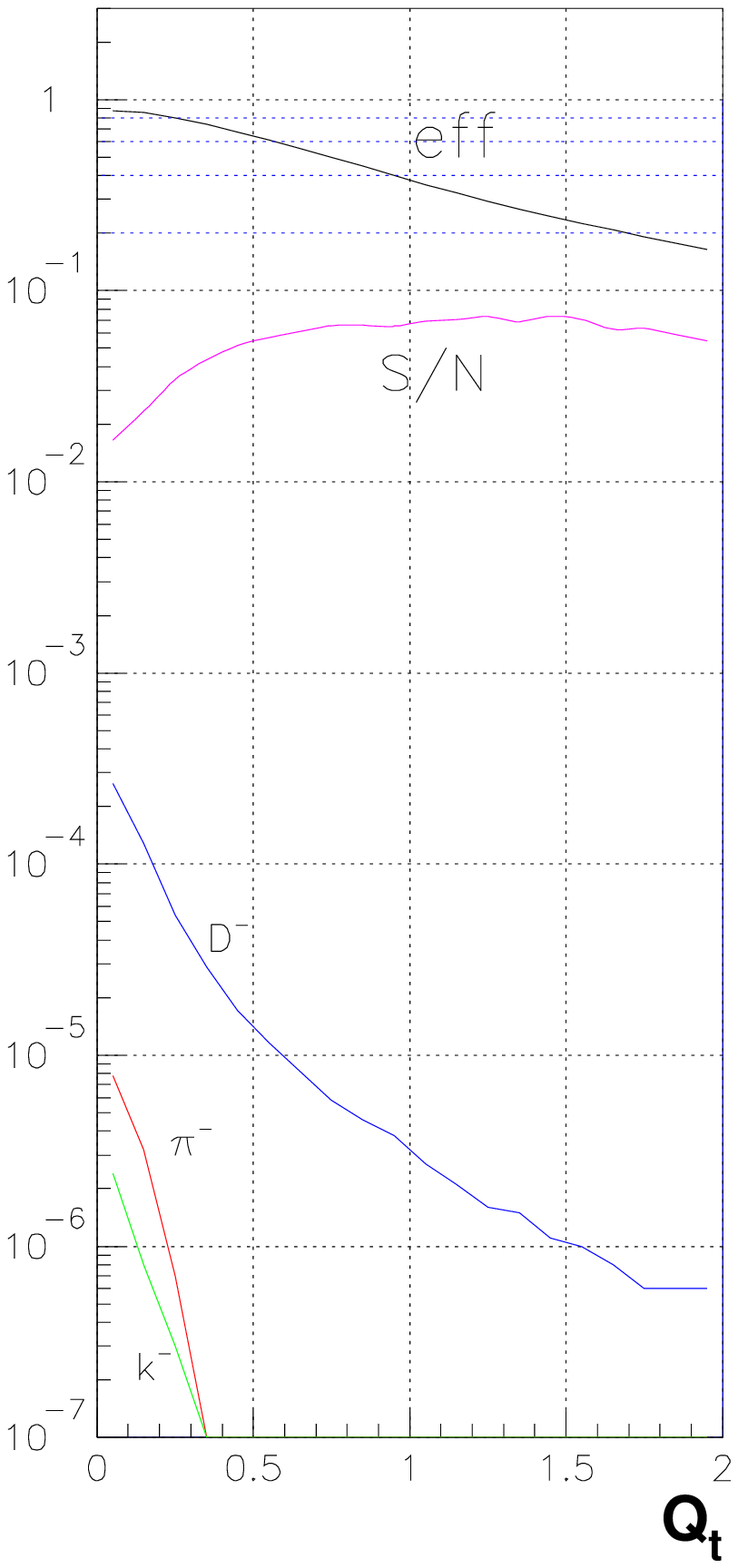,height=7cm,width=4.4cm}
\hspace*{-1cm}
\epsfig{file=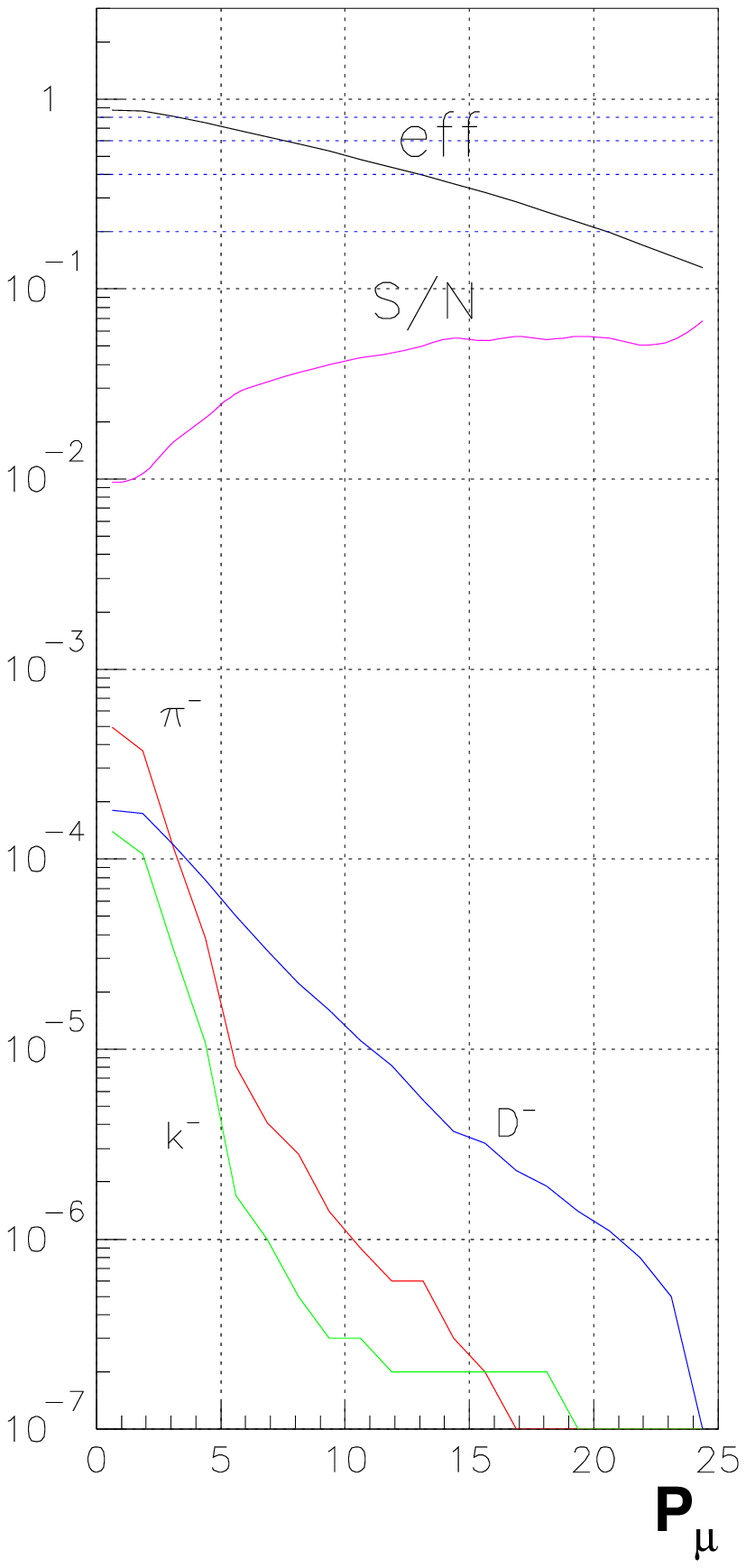,height=7cm,width=4.4cm} 
\hspace*{-1cm}
\epsfig{file=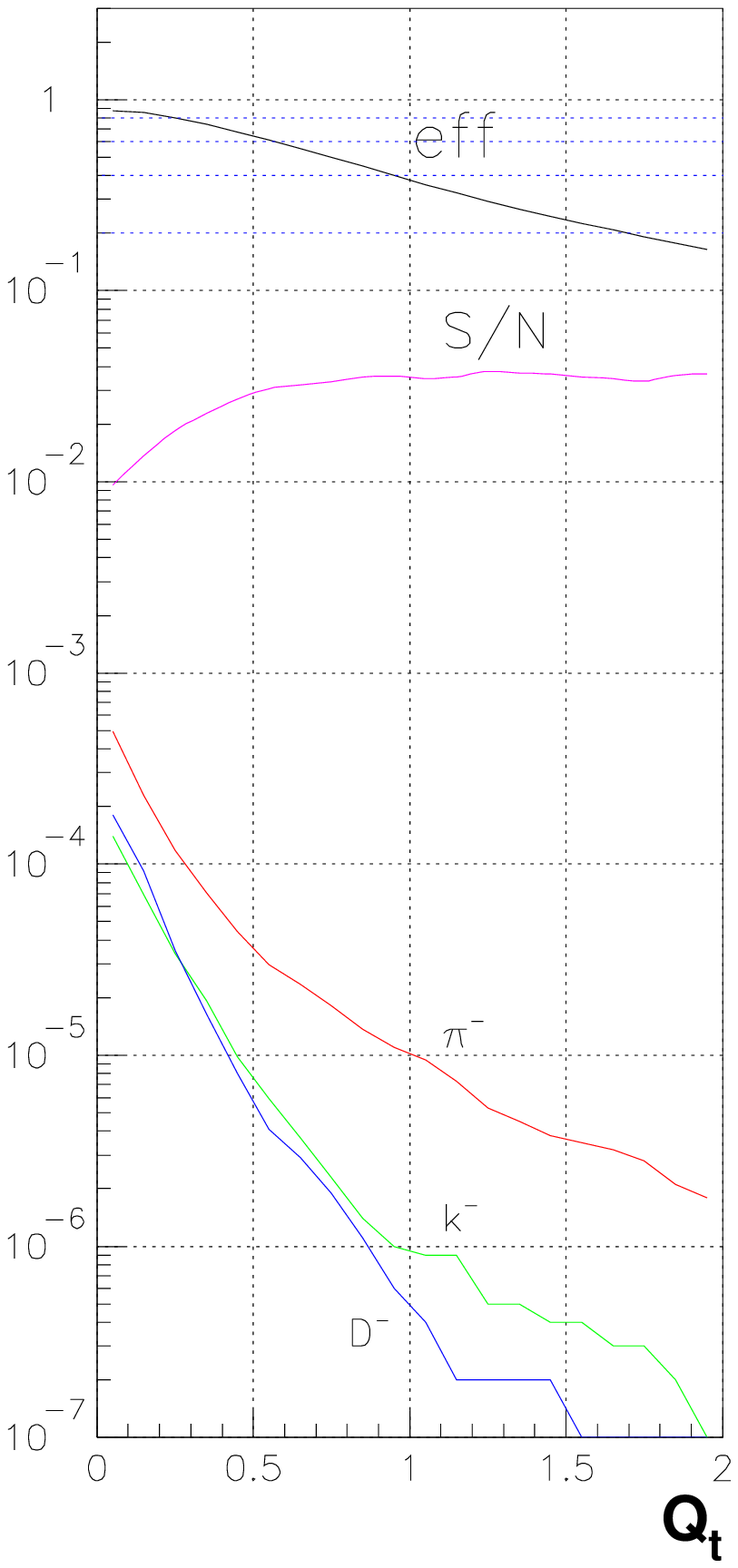,height=7cm,width=4.4cm}  
\end{center}
\caption{Fractional backgrounds from  hadron decays as a function of the cuts in 
         $\pmu$  and $\qt$ for 
         \numubar CC and (\numubar+\nue) NC interactions (for stored $\mu^+$'s). 
         The charge misidentification rate is also shown on the left for a 
         conservative configuration: d=15~\cm, $\nu$=2$\%$ and $\varepsilon$=0.5~\cm. 
\label{fig:hadron_bkg1}}
\vspace{4mm}
\end{figure}

As it was shown in Ref.~\cite{golden}, muons from the decay of hadrons (mainly 
charmed particles) in $\numubar$ CC interactions constitute the leading background 
at high neutrino energies.  
Fortunately, ``real'' wrong-sign muons ( from oscillated \nue's) 
will be in general more energetic and more isolated from the hadronic jet. 
Thus, this background can be controlled to a reasonable level by a 
a combined cut in the momentum of the muon 
($\pmu$) and its isolation with respect to the hadronic jet, which is represented 
by the variable $\qt = \pmu \, sin^2 \theta_{\mu h}$, where $\theta_{\mu h}$ is the angle 
between the muon and the hadronic shower. Fig.~\ref{fig:hadron_bkg1} shows the 
fractional bakgrounds in $\numubar$ CC events as a function of the cuts in both 
$\pmu$ and $\qt$. The optimal cuts depend on the baseline 
since signal and backgroud evolve differently with the distance (see Ref.~\cite{golden}). 
For a baseline of $3500~\Km$ the optimal cuts are $\pmu >$ 5~\GeVc and $\qt>$0.7~\GeVc 
(from Ref.~\cite{mind_review}, $\pmu >$ 7.5~\GeVc and $\qt>$1~\GeVc were used in~\cite{golden}), 
which give a total background rate of $8 \times 10^{-6}$ for an efficiency of $45\%$.

\vspace{0.5cm}
\noindent
{\bf \it Improving the signal efficiency at low neutrino energy  }
\vspace{0.5cm}

The analysis presented in Refs.\cite{golden,mind_review} and described above 
was optimised for the measurement of very small $\theta_{13}$. Values of the mixing angle below 
$0.2^o$ (corresponding to $sin^2(2\theta_{13})<5 \cdot 10^{-5}$) were accessible.  
Being the signal essentially proportional to $sin^2(2\theta_{13})$, 
a very small background level was required, motivating the strong cut on the muon momentum.  
However, this cut led to essentially no efficiency 
below 10~\GeV neutrino energy. This is not a problem for the measurement 
of $\theta_{13}$, since this parameter enters in the oscillation probability 
as a normalization factor, which can be obtained at much higher energies, 
where the neutrino flux and cross section are larger. 
However, the detection of low energy neutrinos is crucial for the simultaneous 
measurement of $\theta_{13}$ and $\dcp$. Indeed, the measurement of $\dcp$ 
is based on the experimental capabilities to distinguish the oscillation pattern of 
neutrinos from that of anti-neutrinos \cite{golden}. This CP asymmetry is maximum  for neutrino 
energies in the region of the oscillation peak ($\sim$ 7~\GeV at 3500~\Km) and below. 
Refer to the Physics Report~\cite{iss_physics_report} for more details.

\begin{figure}[htb]
\begin{center}
\epsfig{figure=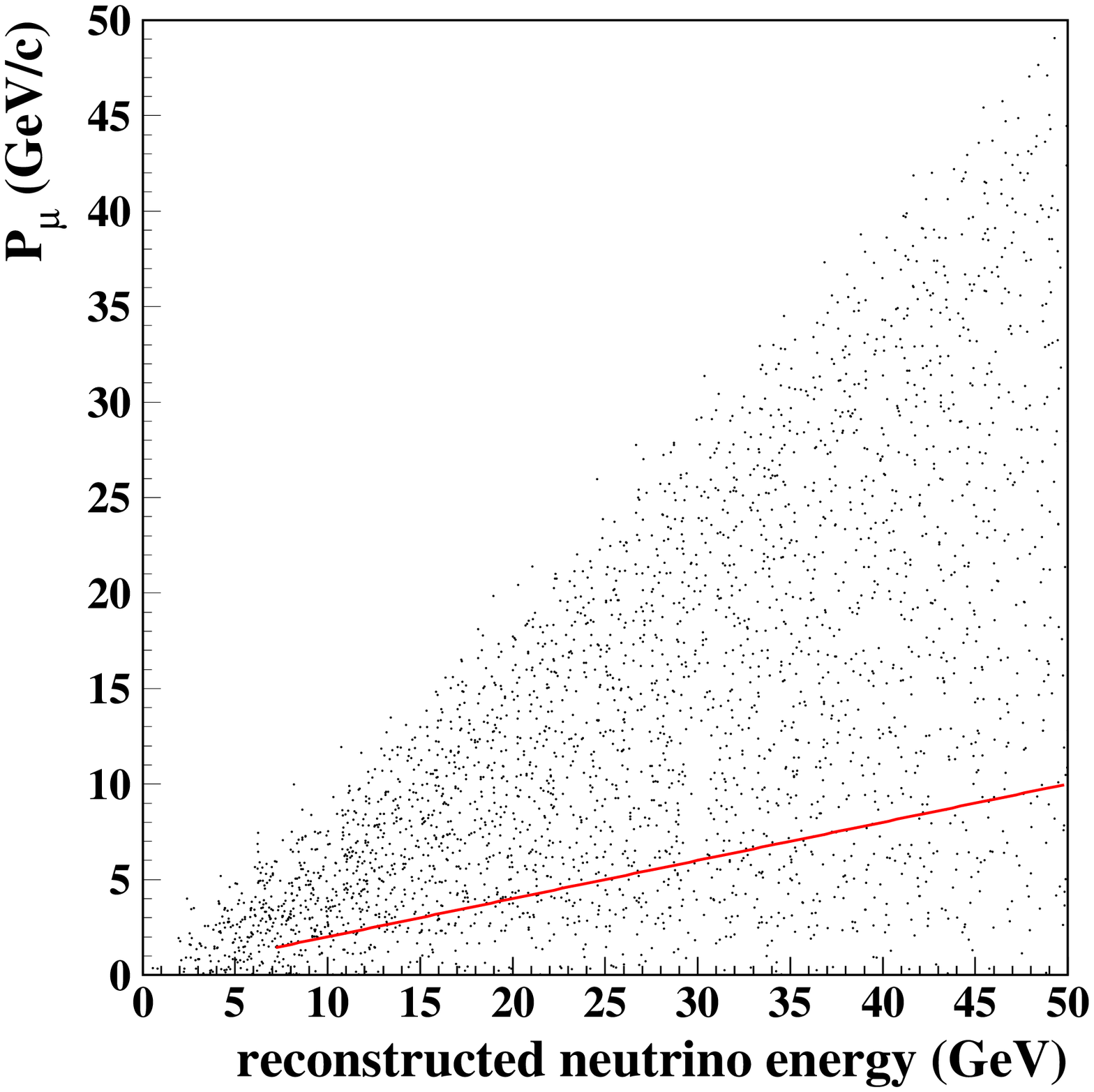, width=0.4\textwidth} 
\epsfig{figure=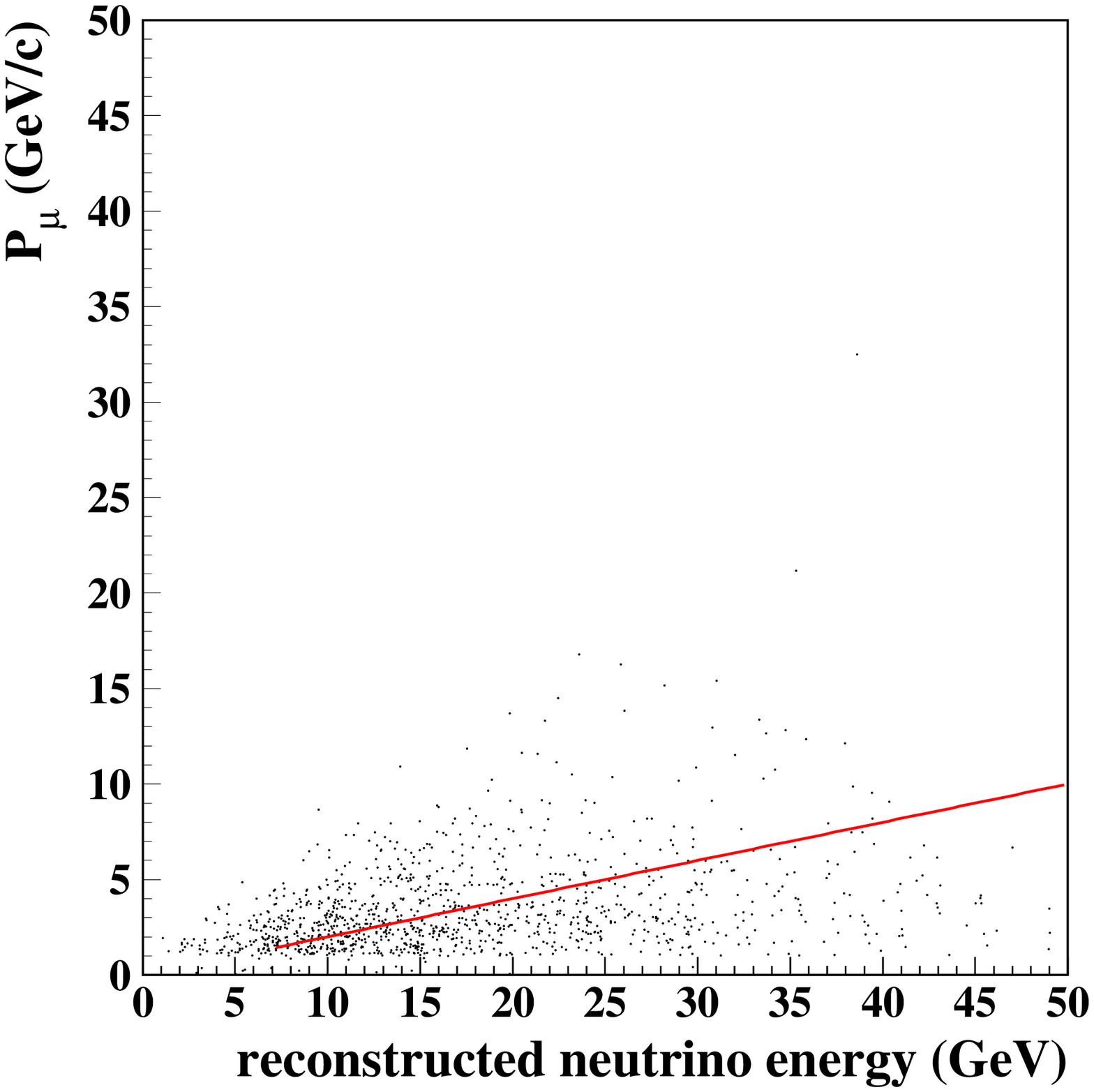, width=0.4\textwidth} \\
\epsfig{figure=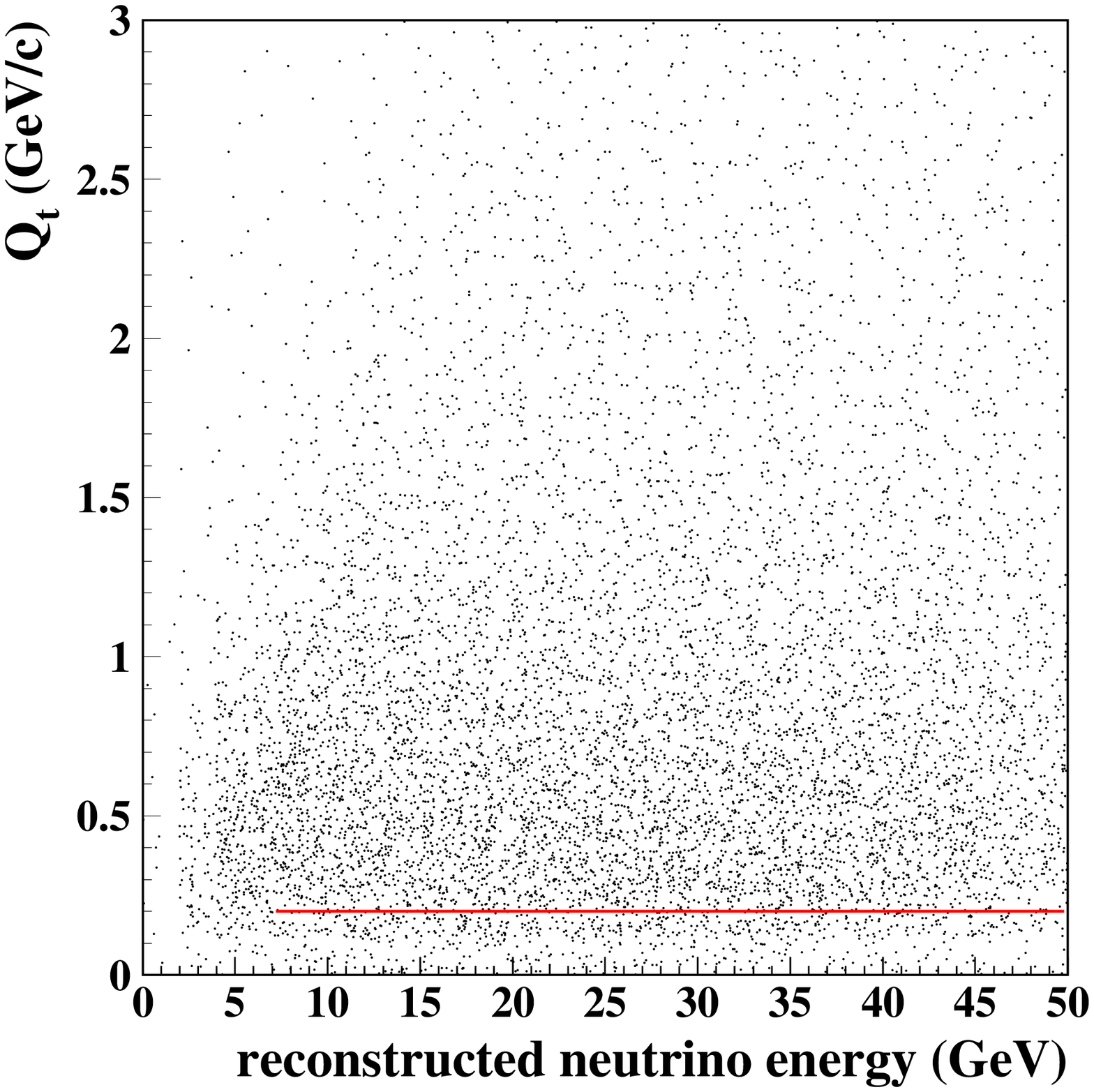, width=0.4\textwidth} 
\epsfig{figure=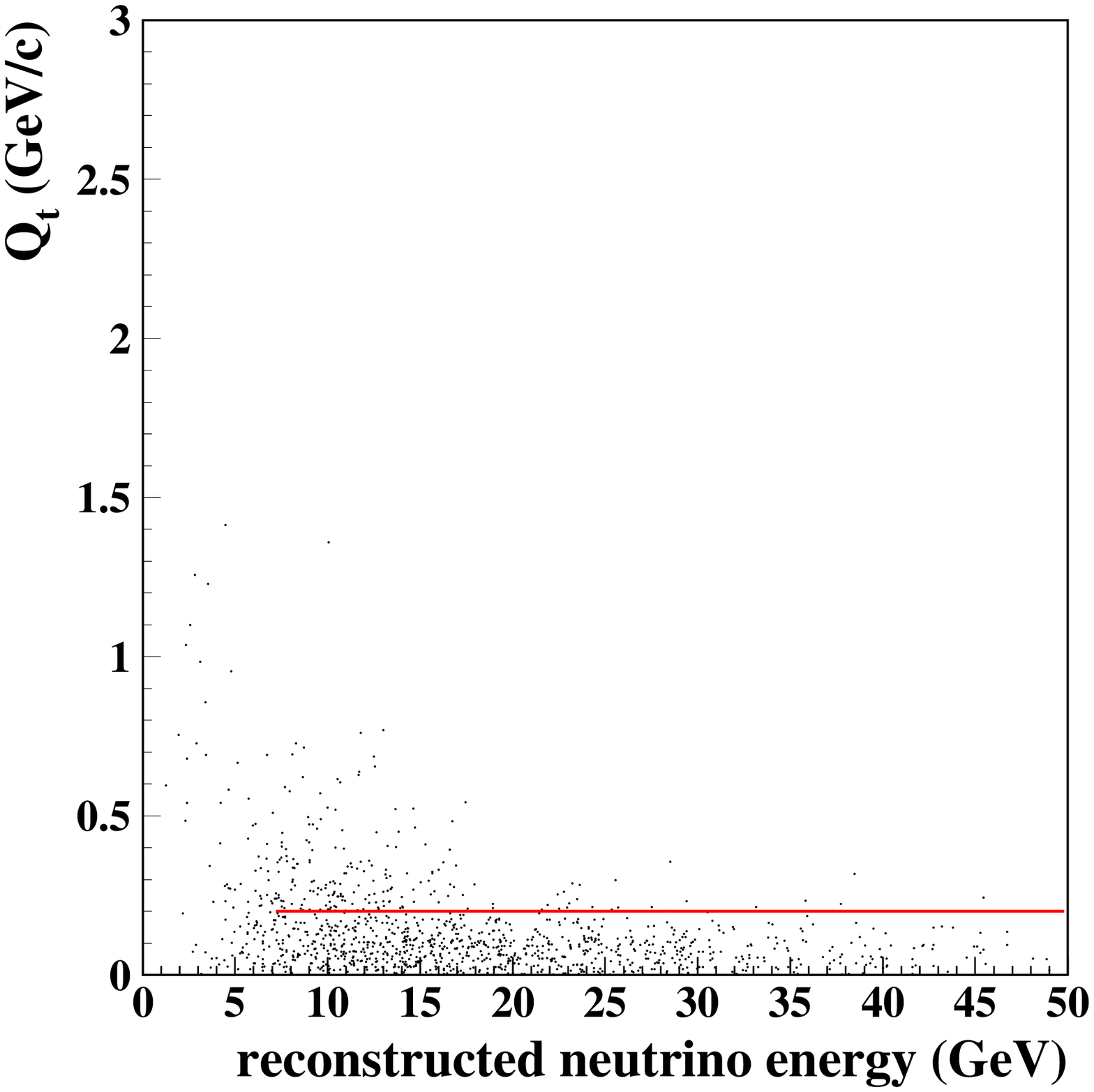, width=0.4\textwidth} \\
\caption{$\pmu$ (top panels) and $\qt$  (bottom panels) distributions as a function of $\evis$ for signal (left panels) 
and \numubar CC background (right panels) events. The kinematical cuts are also shown. 
\label{fig:pmu_qt_vs_enu}}
\end{center}
\end{figure}

\begin{figure}[htb]
\begin{center}
\epsfig{figure=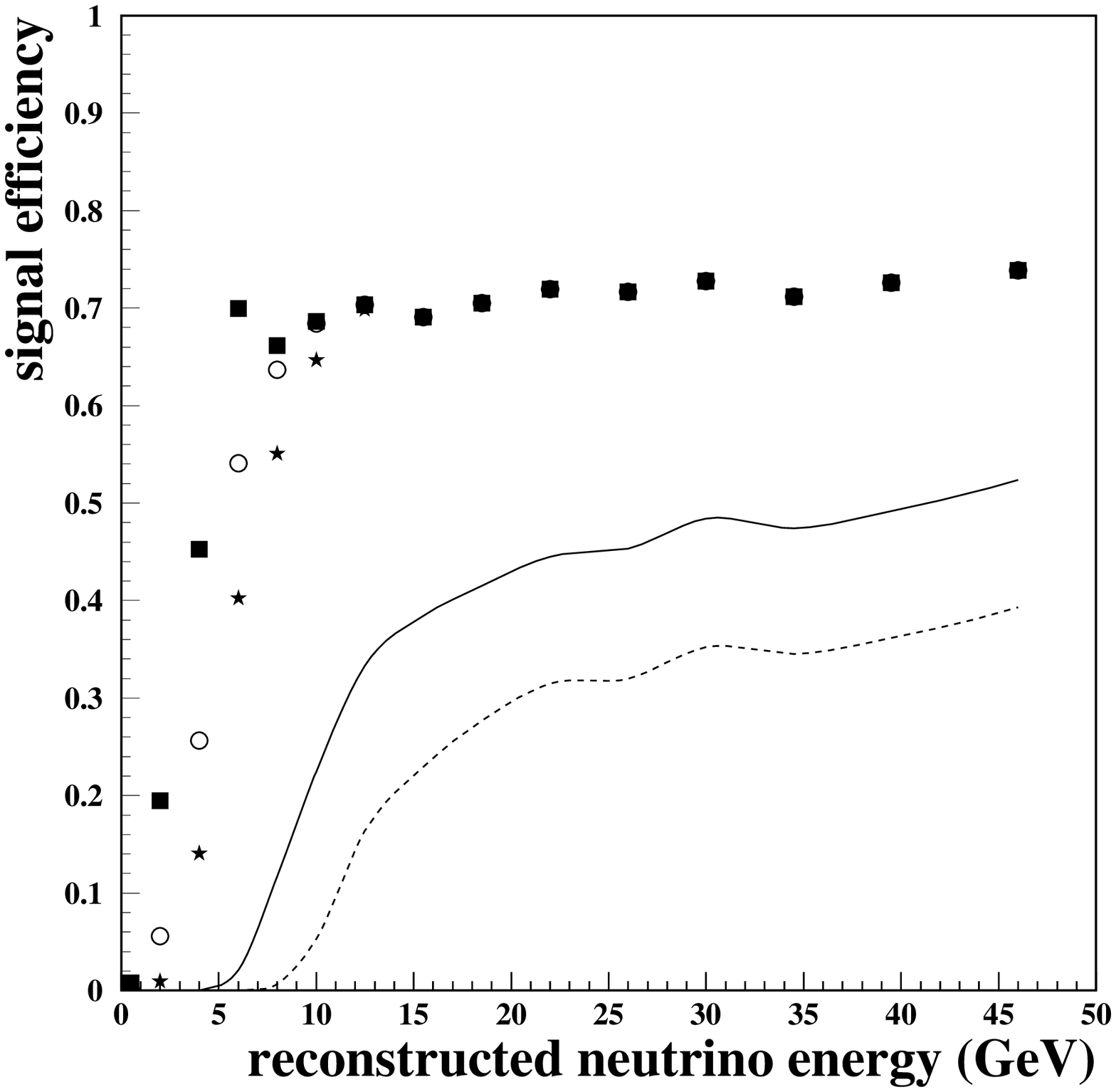, width=7cm,height=5cm}  
\epsfig{figure=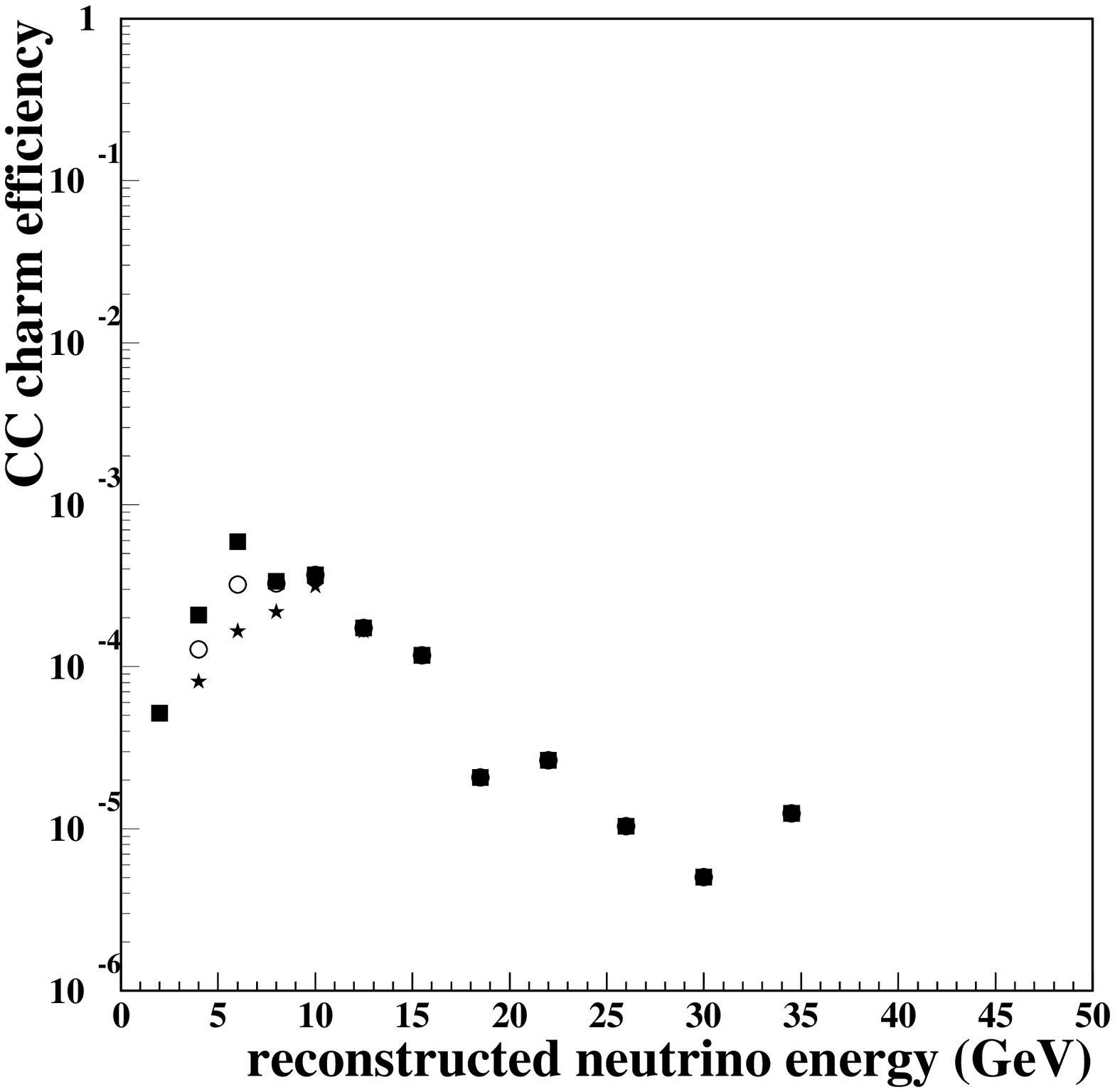, width=7cm,height=5cm}  \\
\epsfig{figure=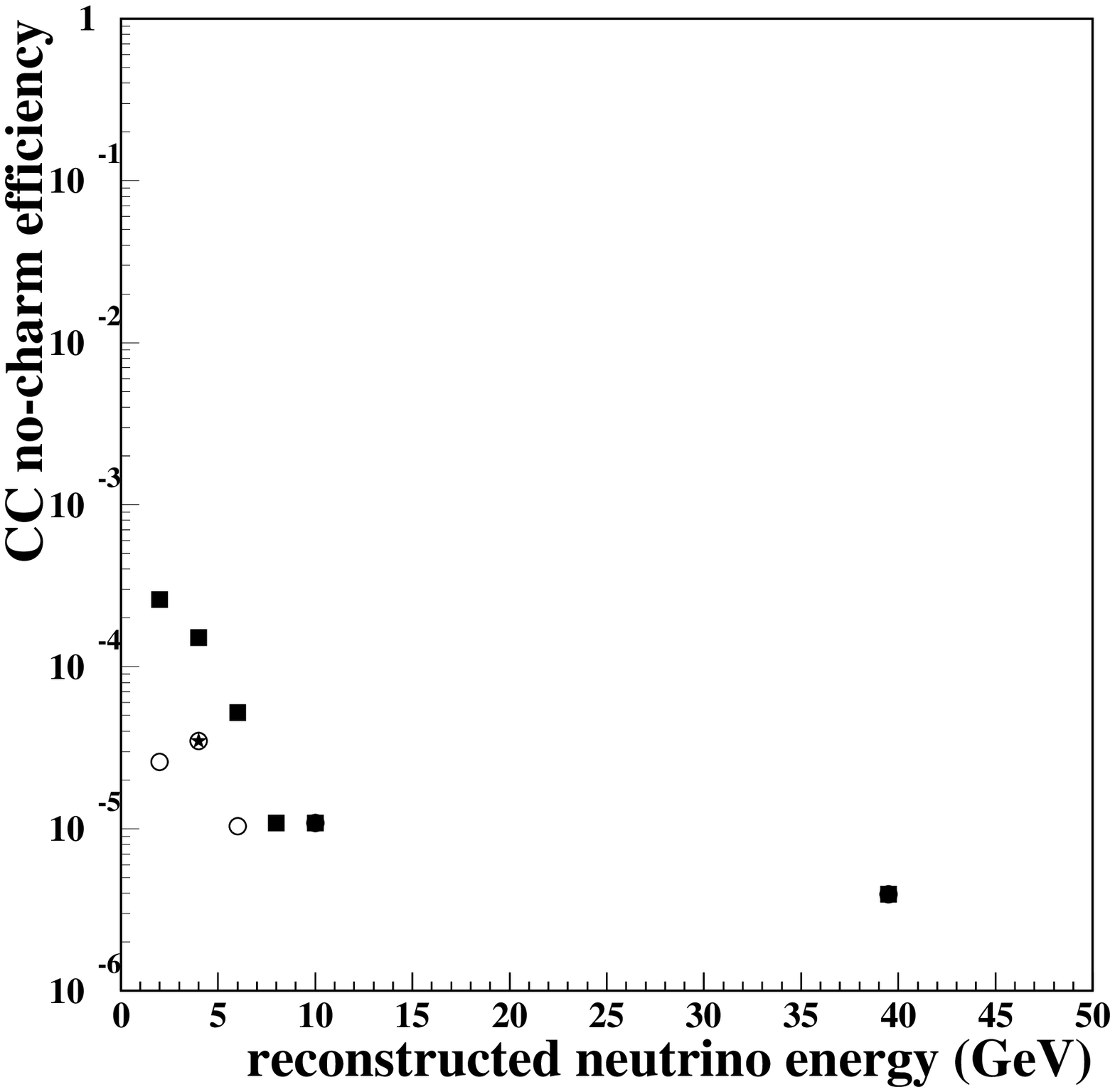, width=7cm,height=5cm} 
\epsfig{figure=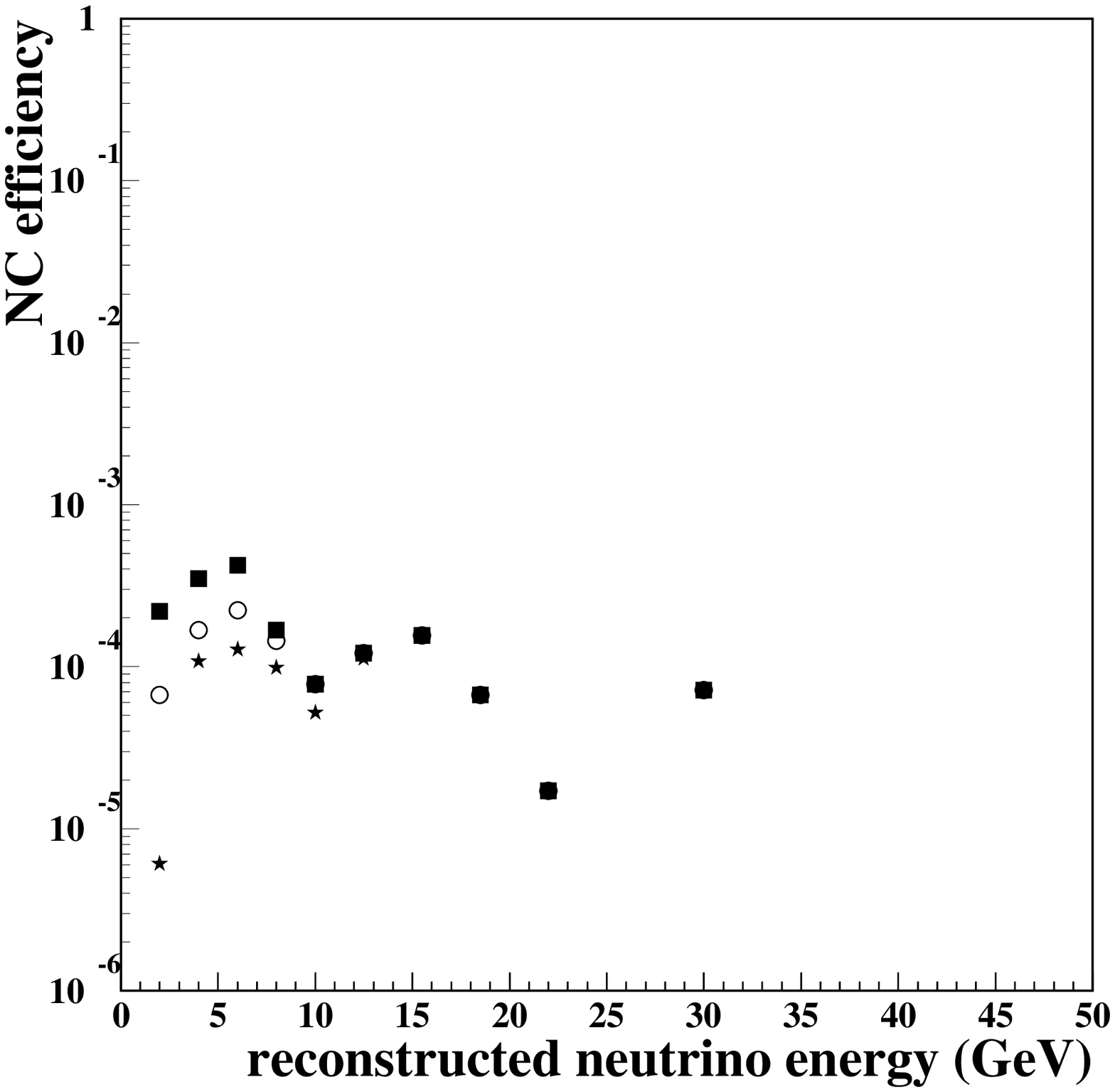, width=7cm,height=5cm}
\caption{Signal and hadronic background efficiencies as a function 
of the reconstructed neutrino energy for different cuts on the muon length:
75~\cm (black boxes), 150~\cm (empty circles) and 200~\cm (stars). top-left: $\numu$ CC; this 
plot also shows the signal efficiency obtained in previous analyses: $\pmu>$5~\GeVc and $\qt>$0.7~\GeVc 
from Ref.~\cite{mind_review} (solid line) and $\pmu>$7.5~GeVc and $\qt>$1~\GeVc 
from Ref.~\cite{golden} (dashed line).   
Top-right panel: $\numubar$ CC with charm decays. Bottom-left panel: $\numubar$ CC other than 
charm decays (mainly pion and kaon decay). Bottom-right panel: $\numubar$ NC. $5\times10^6$ events have 
been used both for $\numubar$ CC and NC interactions. The bin size has been chosen taking into account the 
$\enu$ resolution ($\sim 2 \delta \enu$). 
\label{fig:eff_sig_bkg}}
\end{center}
\end{figure}

Taking advantage of the correlation between the momentum of the muon 
and the total visible energy, the cuts can be optimised for both 
$\theta_{13}$ and $\dcp$. 
Fig.~\ref{fig:pmu_qt_vs_enu} shows the $\pmu$ (top panels) and $\qt$  (bottom panels) 
distributions as a function of $\evis$ for signal (left panels) 
and \numubar CC background (right panels) events. This Fig. also shows 
the variable cuts: $\pmu>(0.2/c)\cdot \evis$ and $\qt>0.2$ \GeVc for $\evis > 7$~GeV and no cuts below this energy. 
The resulting efficiency for the signal and 
the hadronic backgrounds is shown in Fig.~\ref{fig:eff_sig_bkg}.  
\begin{table}[htbp]
\caption{\label{tab:cuts}
  The list of the relevant cuts used in the analysis. Kinematical cuts are only applied for $\evis>7$~\GeVc.
}
\begin{center}
\begin{tabular}{cccc}
\hline
{\bf Fiducial}     & {\bf Quality}             & {\bf Muon id}              & {\bf Kinematical}  \\ 
\hline 
$z<1700$~\cm  &  $5^o<\theta_{rec}<90^o$    & $\Delta L >$ 75,150,200~\cm   &  $\qt>$0.2~\GeVc\\
$|x|,|y|<600$~\cm &                        &                               &  $\pmu>(0.2/c) \cdot \evis$\\ 
\hline
\end{tabular}
\end{center}
\end{table}


\vspace{0.5cm}
\noindent
{\bf \it Experience from MINOS and Monolith}
\vspace{0.5cm}

\begin{figure}[htbp]
\begin{center}
\epsfig{figure=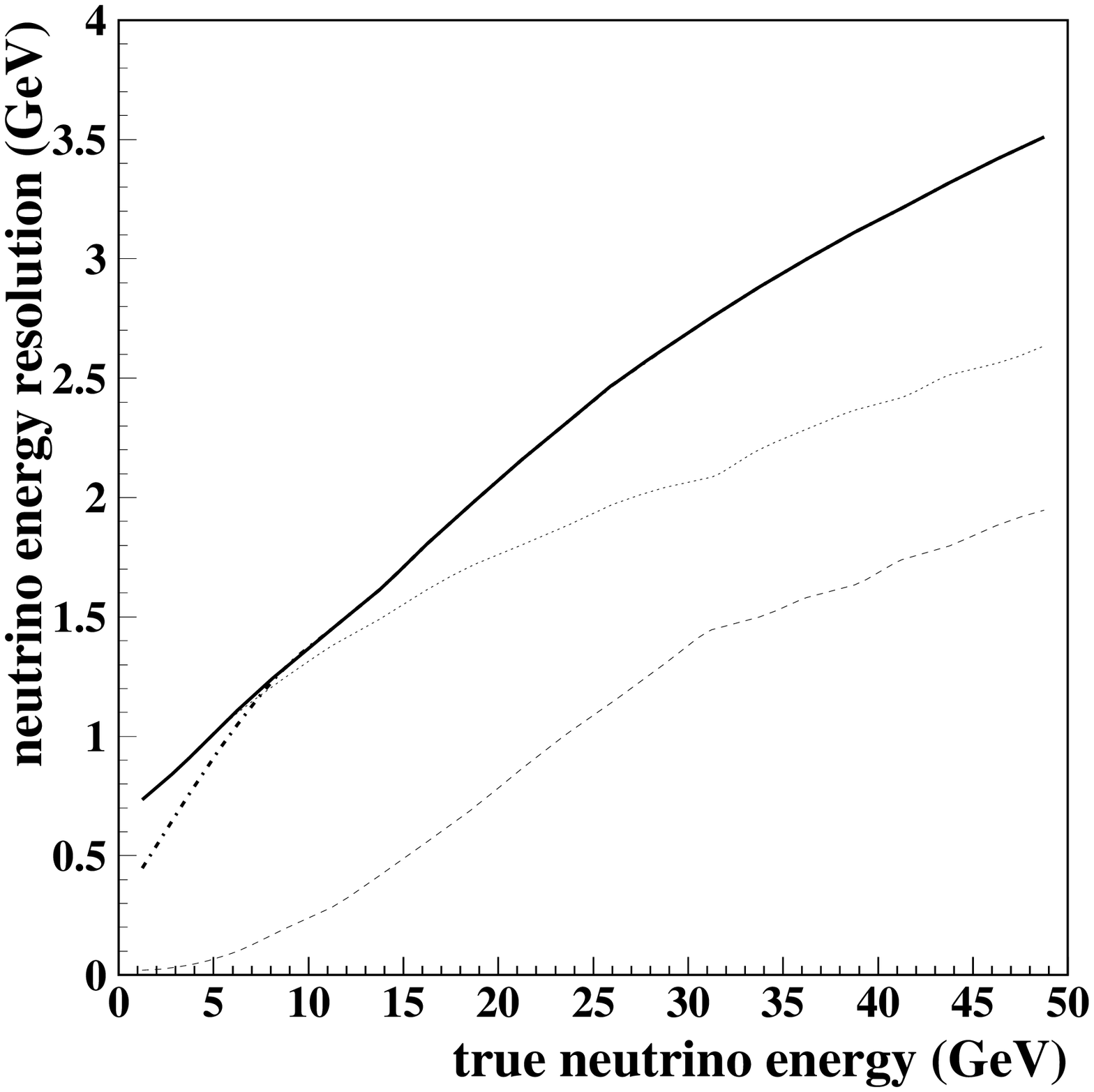, width=10cm,height=7cm}
\caption{Neutrino energy resolution as a function of the true neutrino energy (solid line). 
The dashed line corresponds to contribution of the muon momentum measurement, while the dotted 
line is the hadronic energy resolution. The dotted-dashed line shows the improvement in the resolution 
when QE events are included. 
\label{fig:enu_resol2}}
\end{center}
\end{figure}

The hadronic energy resolution obtained experimentaly 
by MINOS~\cite{minos_cal}, $\delta \ehad /\ehad = 0.55/\sqrt{\ehad}$, where $\ehad$ is in GeV,  is significantly 
better than the one quoted in the proposal and mentioned above. This should improve the current  
$\enu$ resolution, as shown in Fig.~\ref{fig:enu_resol2}.

\begin{figure}[htbp]
\begin{center}
\epsfig{figure=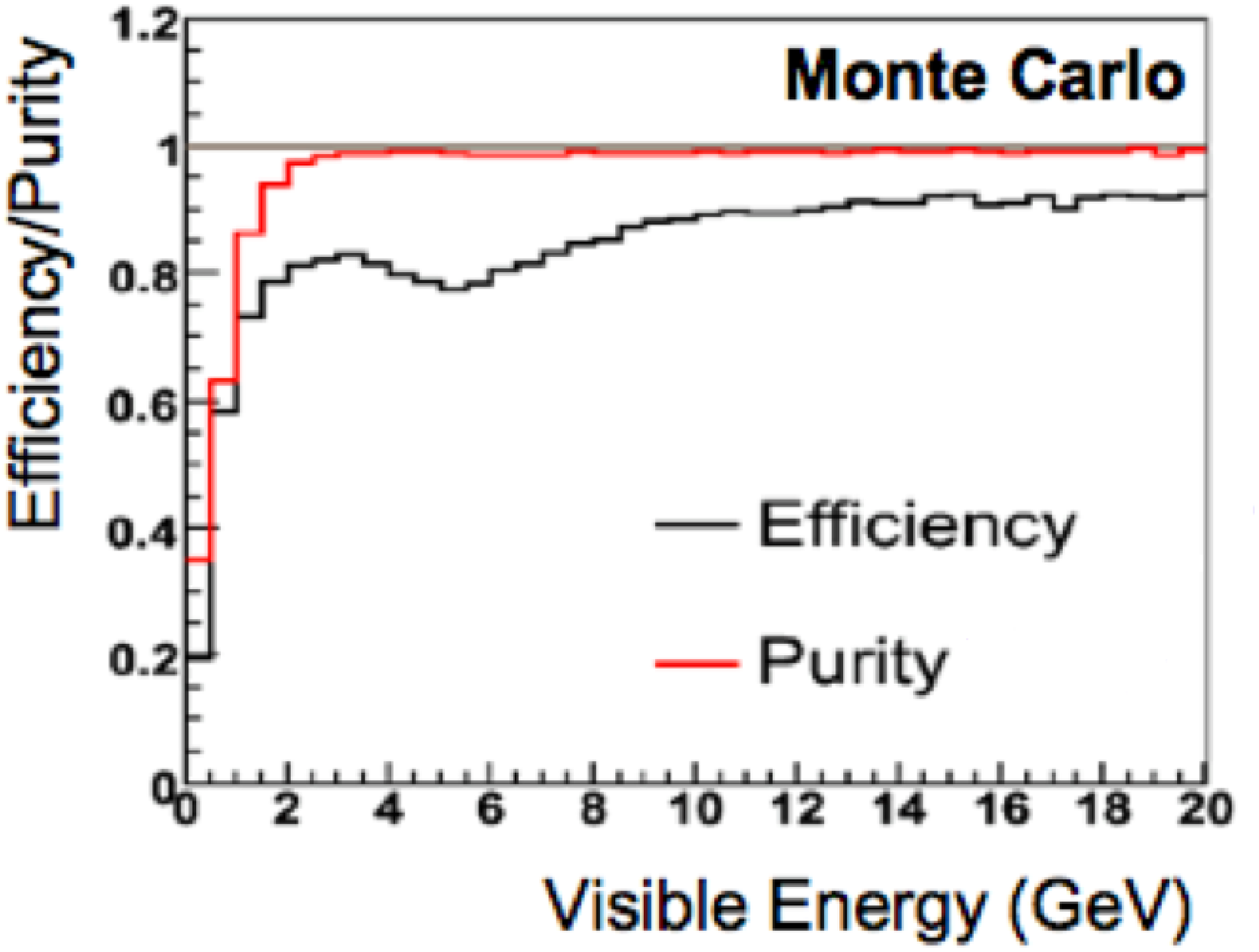, width=7cm,height=5cm}
\epsfig{figure=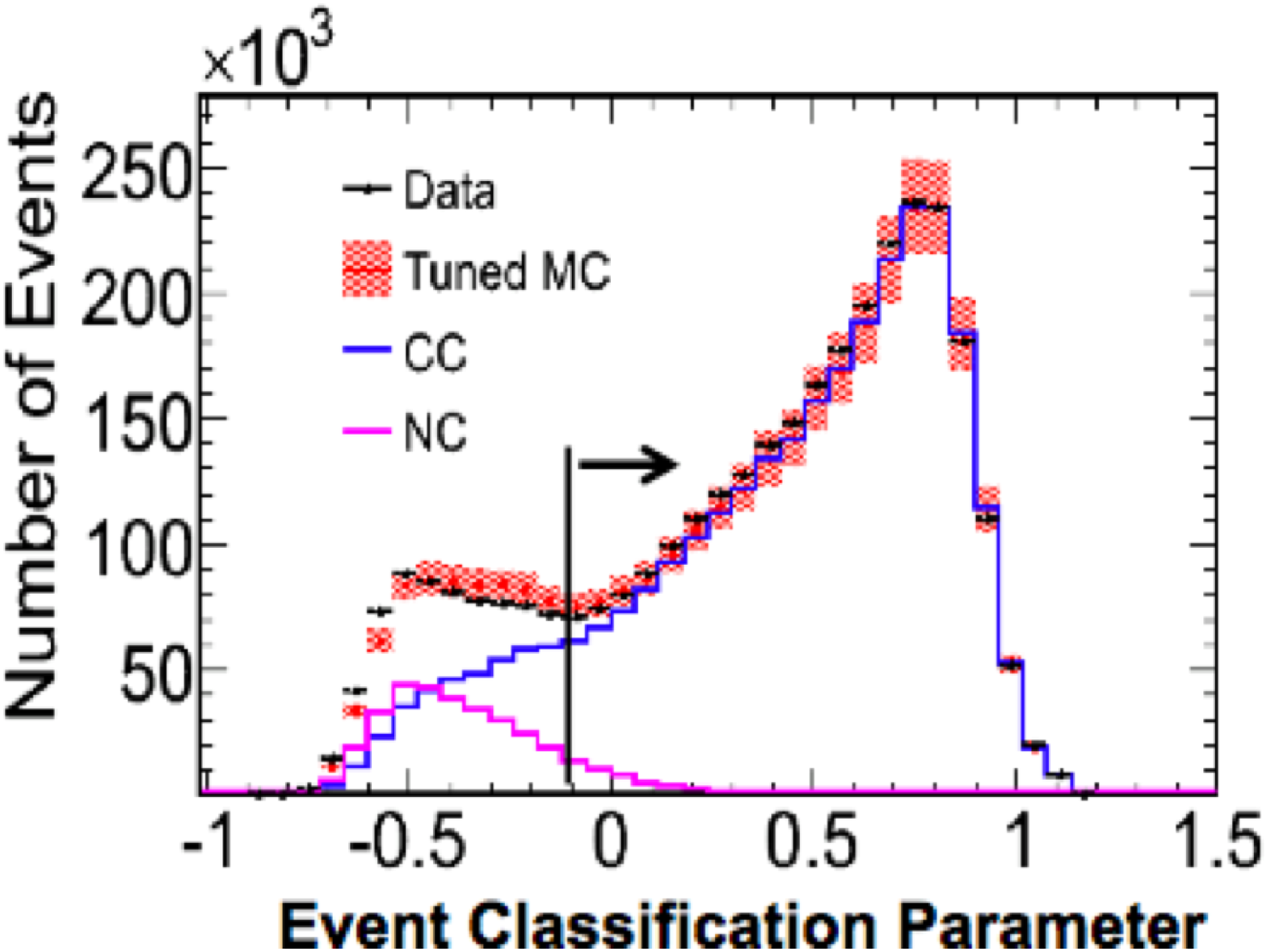, width=7cm,height=5cm}
\caption{On the left panel \numu CC selection efficiency and purity as a function of the 
reconstructed neutrino energy ($\evis$) obtained for MC data. On the right panel comparison of 
the event classification parameter (likelihood function) for real data and MC. 
\label{fig:minos_eff}}
\end{center}
\end{figure}

\begin{figure}[htbp]
\begin{center}
\epsfig{figure=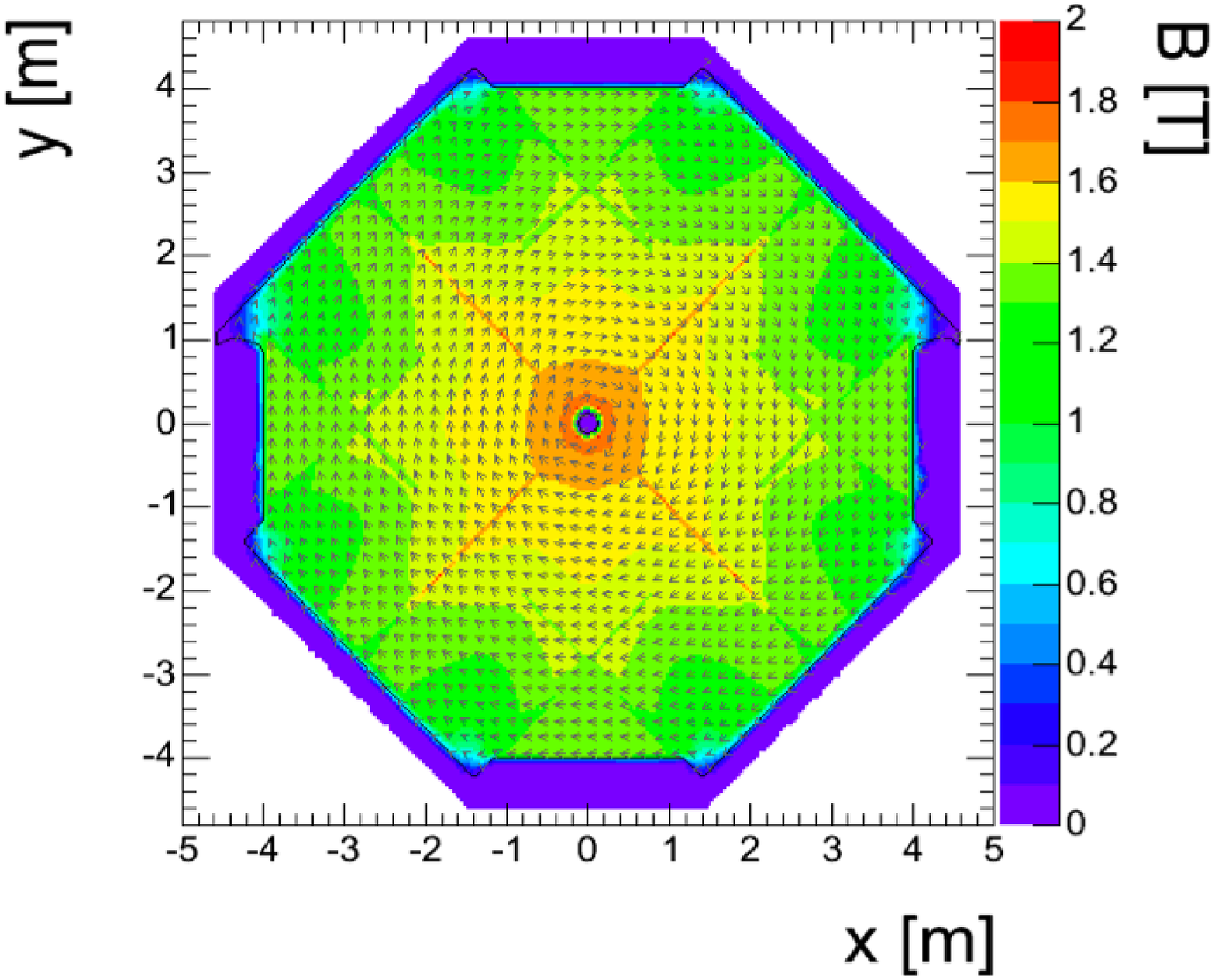,width=0.45\textwidth}
\epsfig{figure=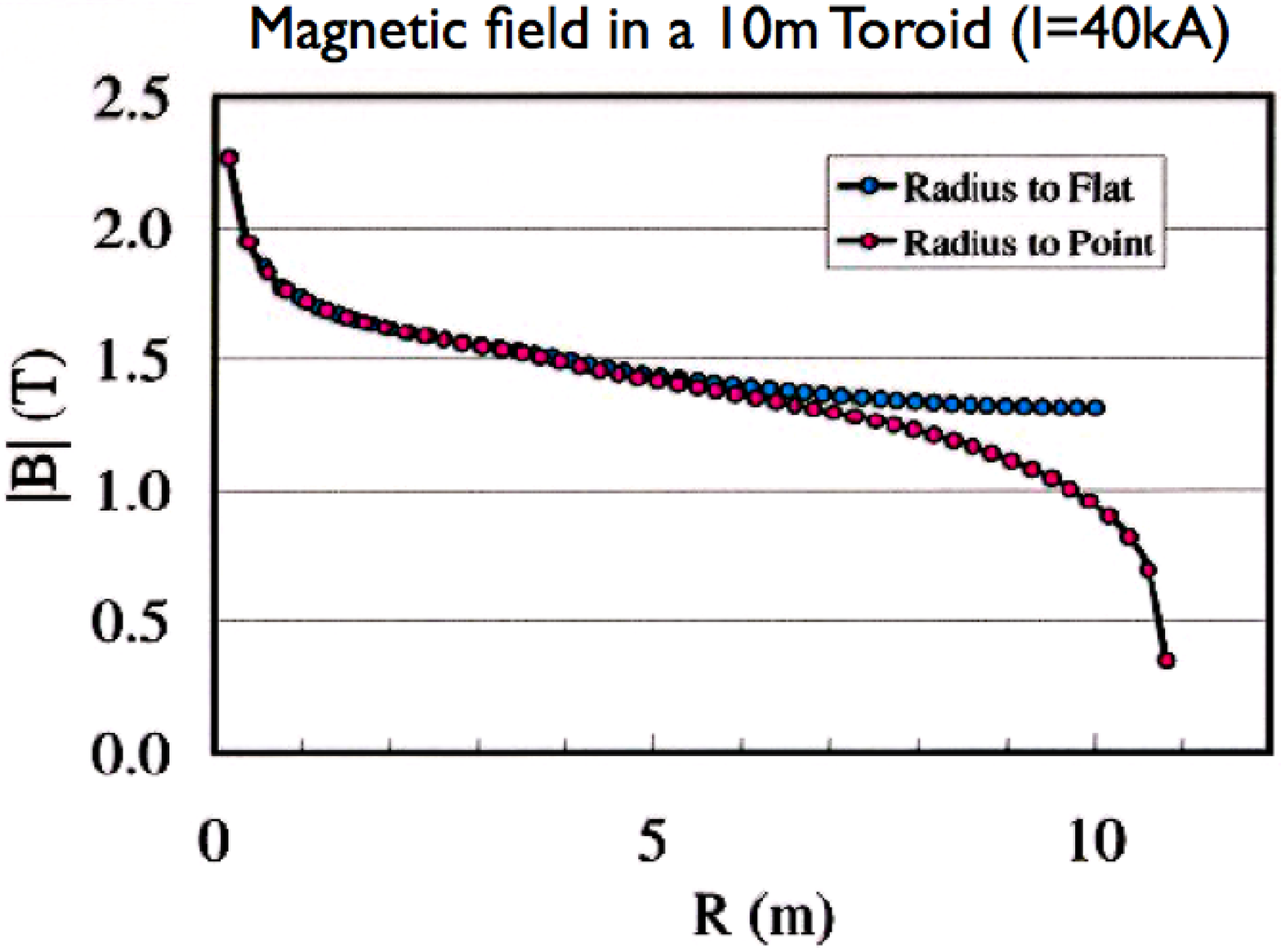,width=0.45\textwidth}
\caption{On the left panel, magnetic field as a function of the transverse coordinates in the MINOS far detector. 
On the right, extrapolation of the MINOS field to a bigger torus.  
\label{fig:minos_field}}
\end{center}
\end{figure}

The MINOS experiment has also demonstrated that \numubar+\numu CC identification 
(based mainly on muon identification) can be performed with high efficiency 
and purity down to 1~\GeV neutrino energy~\cite{harris_win07}, as shown in Fig.~\ref{fig:minos_eff}. 
The MINOS analysis uses a full simulation, with QE and RES interactions, and a full reconstruction, 
in which the effect of the pattern recognition is included. The event classification 
parameter shown in Fig.~\ref{fig:minos_eff}(right panel) combines information from 
track length and pulse height in each measurement plane. For neutrinos above 1~\GeV 
the signal efficiency is better than 70\% while the purity approaches 98\% above 2~\GeV. 
The main problem at such low neutrino energies would be the identification of the muon charge. 

Fig.~\ref{fig:minos_field} shows the magnetic field strength in the MINOS detector and the extrapolation to a 
bigger toroid of 10 \m radius. A 7~\m radius toroid, as the one proposed here, seems feasible. 

The hadronic angular resolution ($\delta \theta_{h}$) used in the current analysis was obtained 
by the Monolith group in a test beam \cite{monolith_testbeam}. For a spacing between measurement planes 
of 7~\cm they found $\delta \theta_{h} = 10.4^\circ /\sqrt{\ehad} + 10.1^\circ /\ehad$, 
which is significantly better than the resolution quoted in the MINOS proposal for a spacing of 4.4~\cm,  
$\delta \theta_{h} = 16.67^\circ /\sqrt{\ehad} + 12.15^\circ /\ehad$. This affects the $\qt$ resolution, 
which was important for the analyses presented in Refs~\cite{lmd,golden,mind_review}, since the 
$\qt$ cut was in the tail of the distribution, but it is not an issue when the cut is relaxed, as 
it is the case in the current analysis.

\vspace{0.5cm}
\noindent
{\bf \it Discussion}
\vspace{0.5cm}

Although a detailed study with a full simulation is still missing, the 
muon charge misidentification seems to be the leading background at low neutrino 
energies (below 10~\GeV). The charge misidentification rate depends primarily on the 
magnitude of the magnetic field (the curvature resolution is inversely proportional to the 
magnetic field), which must be as high as possible. 
A minimum average magnetic field of 1~\Tesla should be considered. The MINOS experience suggests 
that fields of the order of 1.5~\Tesla could be achieved. As discussed previously, a small change 
in the field of 20\% reduces the charge misidentification background by one order of magnitude. 
Thus, the magnetic field issue should be studied very carefully. 


One of the main issues in the MIND analysis is how well the signal efficiency 
can be determined at low neutrino energies. Given the high derivative of the efficiency curve 
below 10~\GeV (see Fig.~\ref{fig:eff_sig_bkg}), the accuracy on the efficiency measurement would be 
highly affected by the resolution on the neutrino energy. As discussed above, the resolution 
assumed in this analysis is worst than the one obtained by MINOS (see Figs.~\ref{fig:muonid_enu_res} 
and \ref{fig:enu_resol2}).

The current simulation does not consider  
quasi-elastic (QE) interactions and resonance production (RES), 
which should dominate below 2~\GeV neutrino energy.  
QE interactions would have a possitive impact on the $\enu$ resolution since the neutrino 
energy can be directly computed from the muon momentum and angle. For these events the $\enu$ 
resolution would approach the $\pmu$ resolution by range, which is of the order of $4\%$ at these 
energies. The average resolution can be computed using the DIS and QE cross sections and the 
corresponding $\enu$ resolutions. This is shown in Fig.~\ref{fig:enu_resol2}. Another possibility 
is to use only QE events, below a certain energy.

In the current analysis the impact of a realistic pattern recognition has been ommited. 
The cut in the muon length ensures that a sufficient number of muon hits are isolated from 
the hadronic shower. This is a reasonable approximation at high neutrino energies, 
since the primary muon generally escapes the hadronic shower  (true for muons 
above $\sim$2~\GeVc). Low energy muons, which are lost in the current analysis, 
could be recovered with an improved pattern recognition. 
The clean topologies of QE and RES events would help in this aspect. 
Pattern recognition should not be an issue for these kind of events, 
although the wrong charge assignments would be frequent for muons below 1.5~GeVc 
($\sim 2\cdot 10^{-3}$ for 1.5~\GeVc muons). 

A satisfactory charge measurement is obtained for iron plate thickness in the range 2-5 \cm. 
Thus, the longitudinal segmentation is mainly 
driven by the hadronic energy resolution and the pattern recognition efficiency. 
The former should improve if the number of samples increases, although 
the current MINOS resolution seems to be sufficient. 
On the other hand, an improved pattern recognition efficiency at low momentum could 
be very important since the cut in the muon length could be relaxed.

Tranverse resolution might be important for the charge measurement at low momentum, 
for the $\qt$ resolution and for pattern recognition. 
Anything better than 1~\cm would be sufficient for the charge and the $\qt$ measurements. 
Again, pattern recognition seems to be the main issue. 

The \numubar+\numu CC identification efficiency obtained by MINOS suggests 
that the signal efficiency in MIND could be much flatter in the energy range 
from  1 to 10~\GeV. This is the result of using a powerful pattern recognition 
and event classification algorithms.

An optimised MIND detector could reach the required performance down to neutrino energies of 
1-2~\GeV. A few questions remain open:

\begin{itemize}
   \item How well can the efficiency be measured at low neutrino energies?  
   \item What would be the effect of pattern recognition? This is partially answered by 
         MINOS, although this effect should be included in the MIND reconstruction.  
   \item What is the QE selection efficiency and purity? 
   \item What is the effect of non Gaussian effects in the charge measurement? This is one of the main 
         issues and should be answered with a prototype.  
   \item What is the maximum magnetic field that can be afforded?   
\end{itemize}

\subsubsection{Totally Active Scintillating Detectors}
\label{sec:tasd}

\begin{figure}[htbp]
\begin{center}
\epsfig{figure=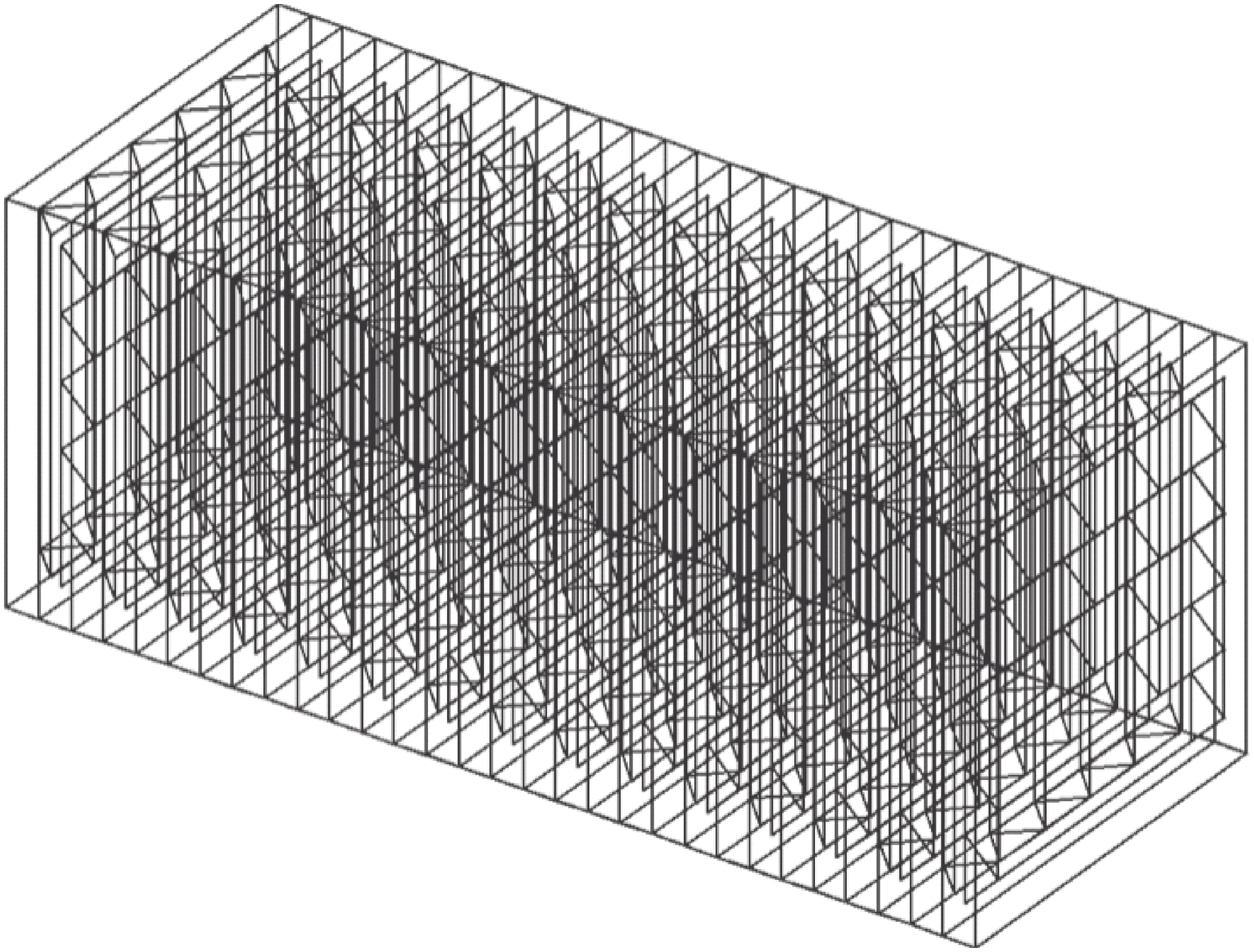, width=0.4\textwidth} \hspace*{1cm}
\epsfig{figure=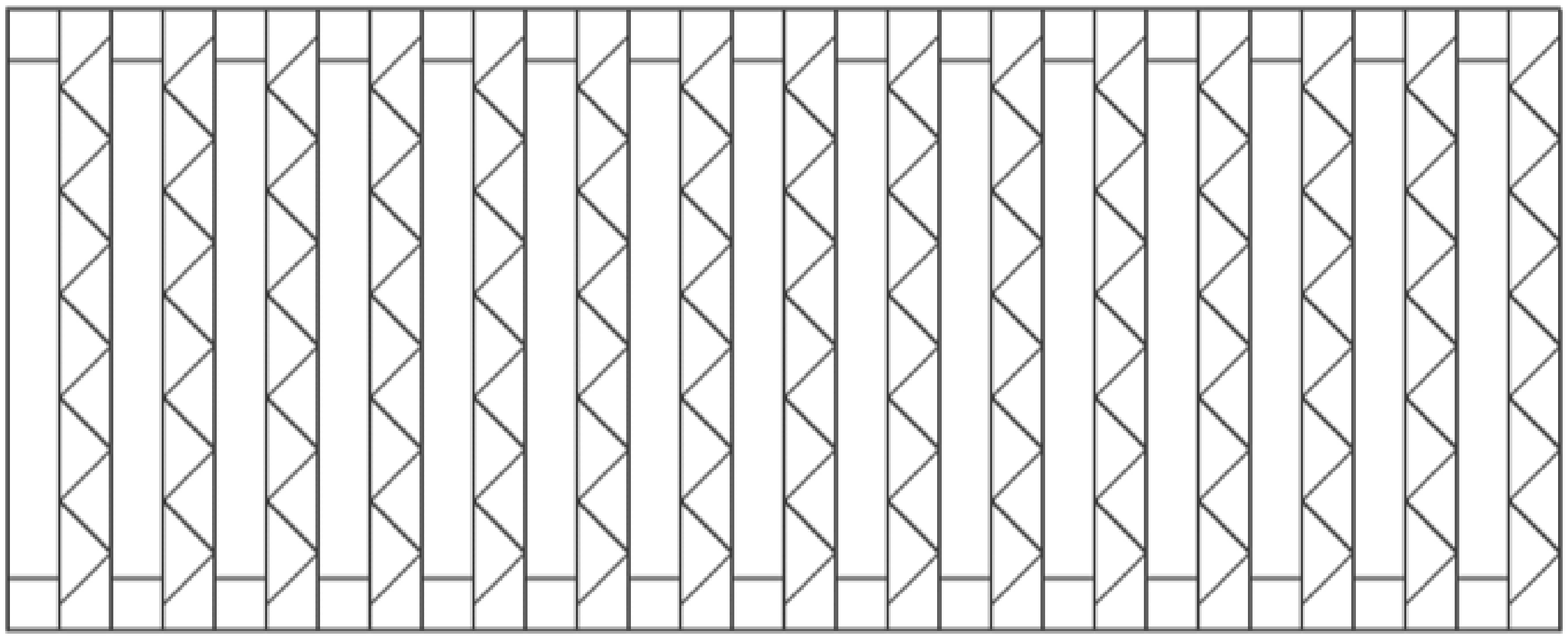, width=0.4\textwidth}
\caption{GEANT4 view of the simulated TASD detector.
\label{fig:tasd}}
\end{center}
\end{figure}

The possibility of using totally active calorimeters in a Neutrino Factory
was first considered at NuFact05~\cite{tasd}. 
A first study of the performance of this design was presented
at the ISS meeting in August 2006~\cite{tasd2}. 

The detector would consist of long scintillator bars with a triangular
cross-section arranged in planes which
make x and y measurements in a 0.5~\Tesla magnetic field. The scintillator bars
considered have a length of 15~\m and
the triangular end has a base of 3~\cm and a height of 1.5~\cm. This design is
an extrapolation of the MINER$\nu$A
experiment~\cite{minerva_www} to produce a detector with
dimensions $15 \times 15 \times 100$~\m and a mass
of approximately 22.5~\Kton.

This detector was simulated with GEANT4 version 8.1 (see Fig.~\ref{fig:tasd}) and the digitisation
took into account the dE/dx in the
scintillator slabs and a light yield extrapolated from MINER$\nu$A 
tests. The magnetic field was simulated
as a uniform 0.5~\Tesla field perpendicular to the beam axis. The performance of
the detector was studied by simulating
the passage of single muons and positrons with a momentum ranging from 100~\MeVc to 
15~\GeVc. Future studies of
this design will include a more realistic field map based on recent design
work to achieve the large magnetic
volume and the simulation of neutrino interactions.

The simulated hits were digitised with an assumed energy resolution of 2
photo electrons and the reconstruction
of clusters imposed a threshold of 0.5 photo electrons before building space
points and performing a track fit
using the Kalman Fitting package RecPack~\cite{recpack}.

\begin{figure}[htb]
\begin{center}
\epsfig{figure=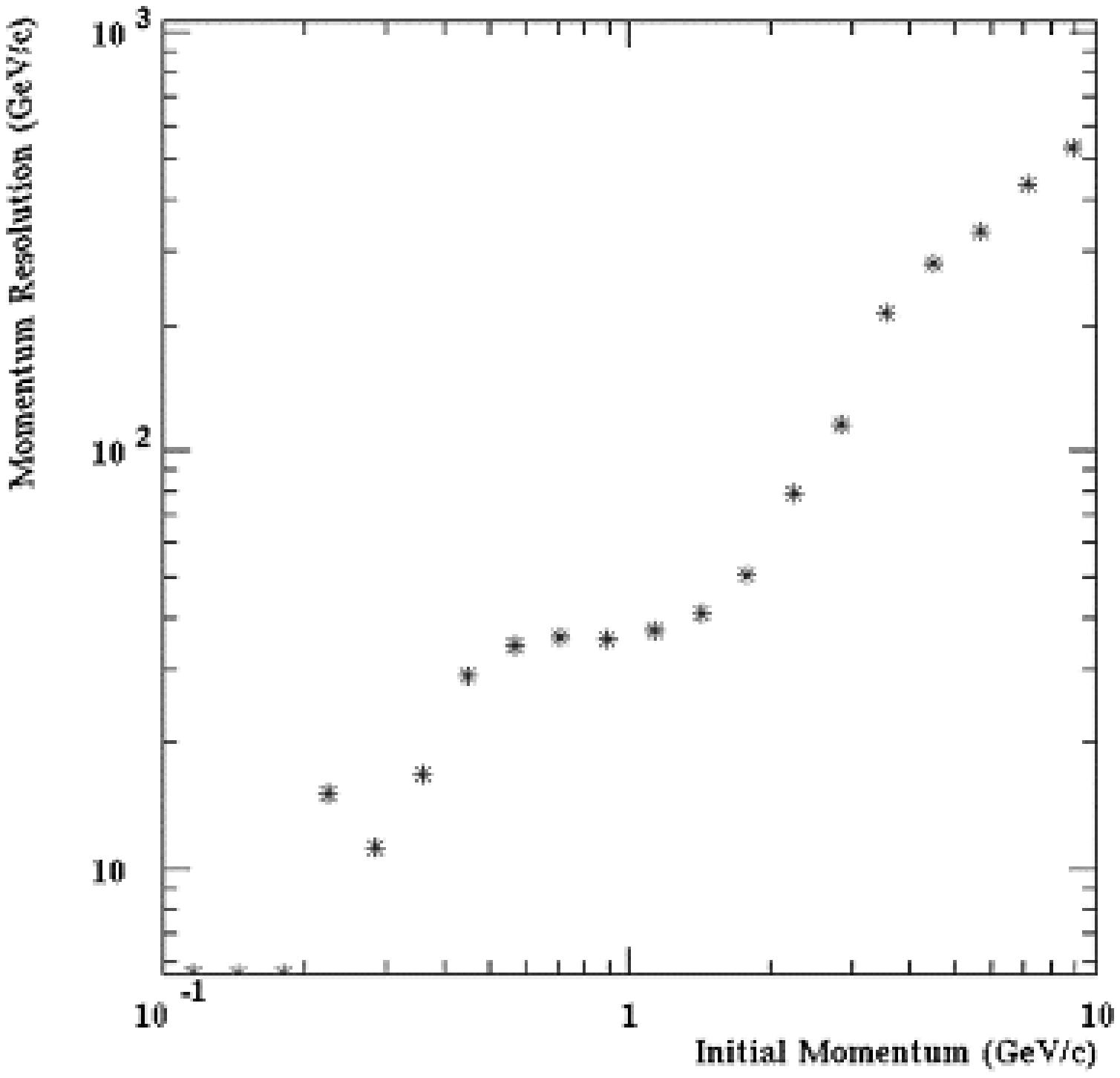, width=7cm,height=5cm}
\epsfig{figure=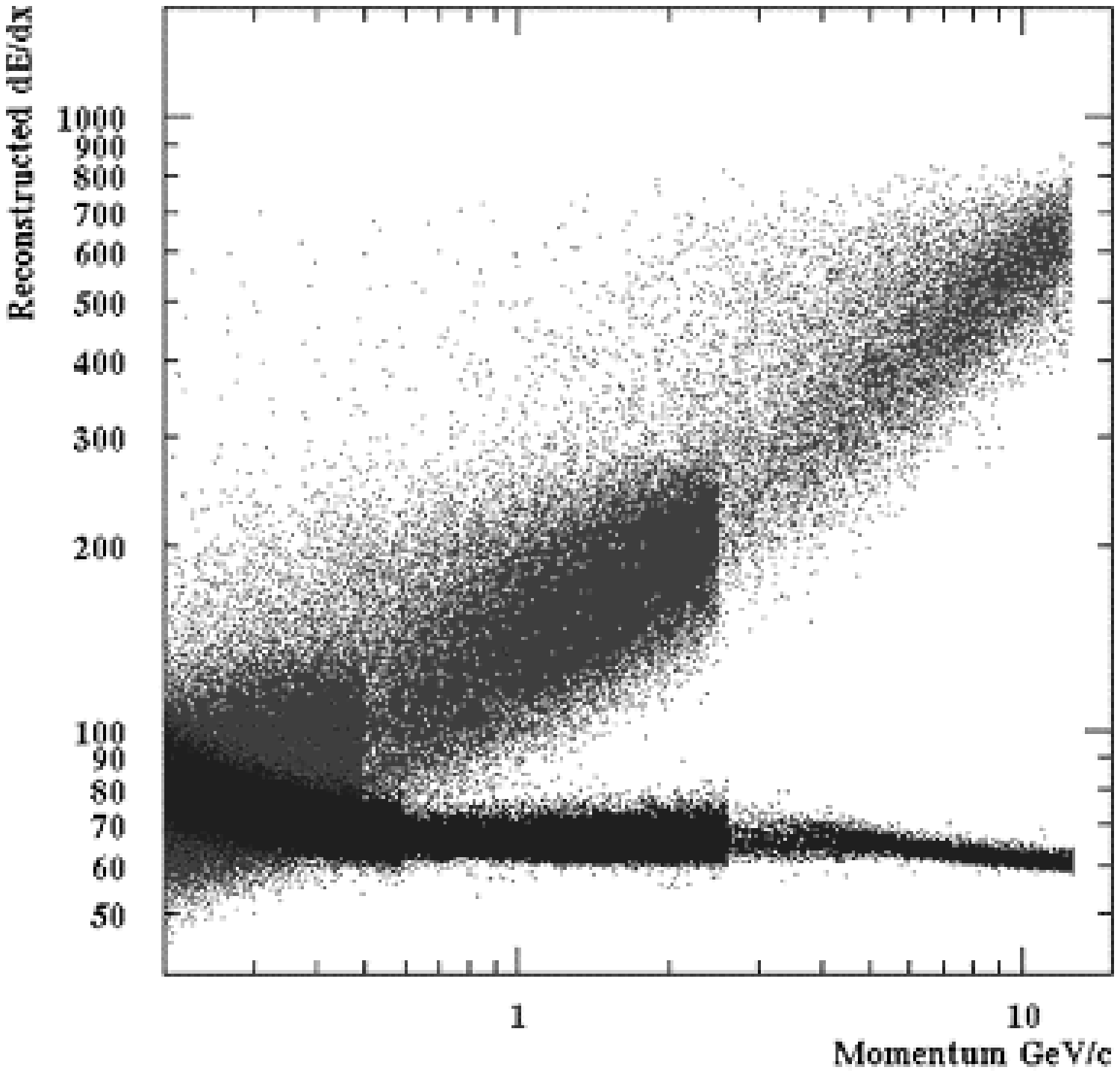, width=7cm,height=5cm} \\
\epsfig{figure=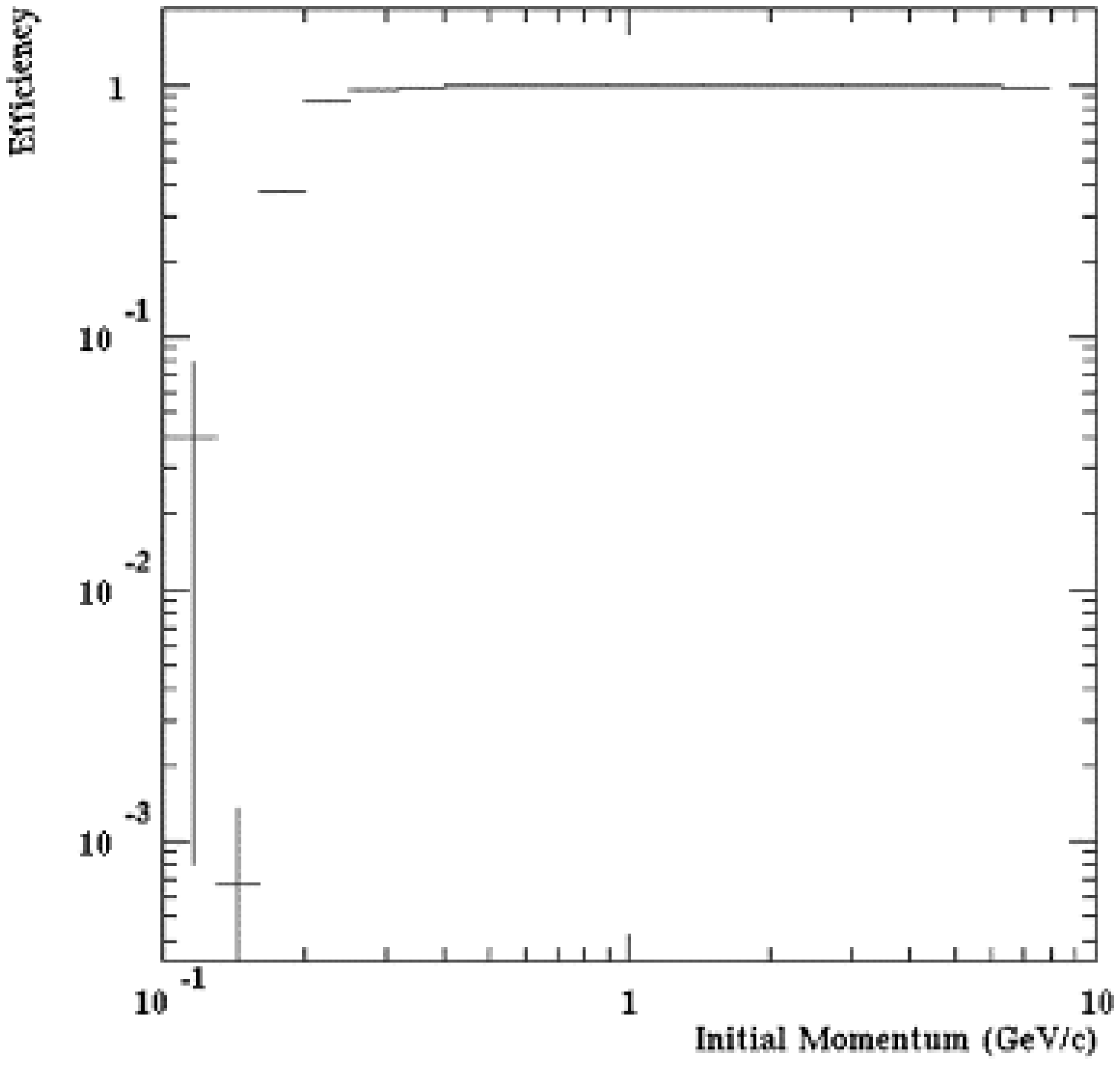, width=7cm,height=5cm} 
\epsfig{figure=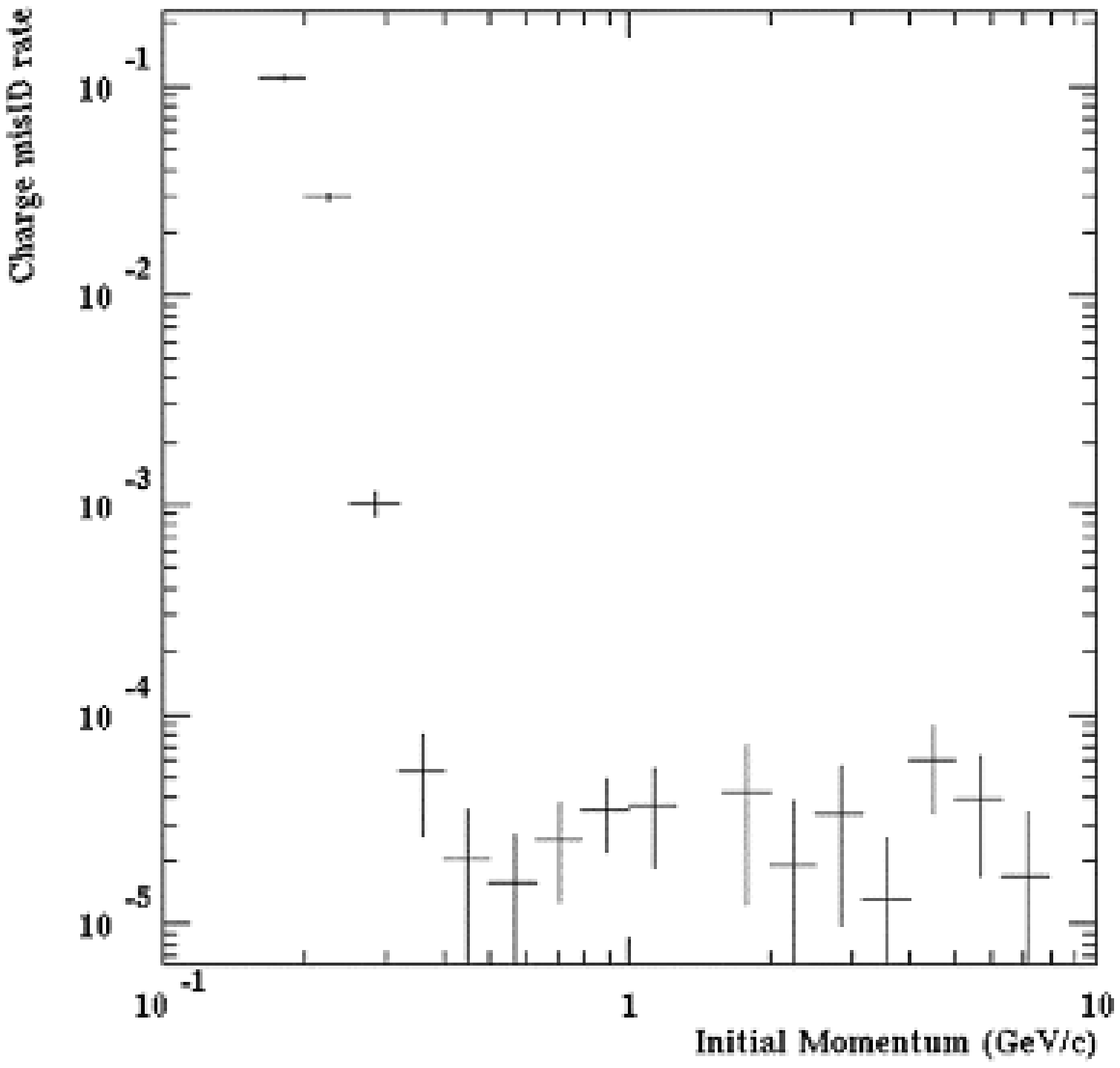, width=7cm,height=5cm} 
\caption{Performance of the Magnetised TASD detector. 
Top-Left panel: muon momentum resolution as a function of the muon momentum.
Top-Right panel: reconstructed $dE/dx$ as a function of momentum for muons (blue/clear) and 
positrons (red/dark). 
Bottom-Left panel: muon identification efficiency as a function of the muon momentum.
Bottom-Right panel: muon charge mis-identification rate as a function of the muon momentum.
\label{fig:tasd_figures}}
\end{center}
\end{figure}

In order to study the momentum resolution and the rate at which the charge
of a muon is mis-identified, 2.3 million
muons were simulated of which 1.8 million, divided equally in two flat
momentum ranges (0.1- 1~\GeVc and
1- 10~\GeVc), were analysed.
The position resolution was found to be approximately 4.5~\mm RMS with a
central Gaussian with width of 2.5~\mm.
The momentum resolution as a function of the muon momentum is shown in
Fig.~\ref{fig:tasd_figures}(top-left). The tracker achieves a resolution
of better than 10\% over the complete momentum range studied.

A first attempt to establish the particle ID performance of the detector is
summarised in Fig.~\ref{fig:tasd_figures}(top-right). This figure
shows the reconstructed $dE/dx$ versus the reconstructed momentum for muons
(blue/clear) and positrons (red/dark). It can
be seen that above approximately 600~\MeVc it should be possible to separate
muons and positrons on the basis
of the reconstructed energy.

Due to the low 
density 
of the Totally Active Scintillating Detector (TASD), 
it is possible to reconstruct muons down to
a few hundred \MeVc. Fig.~\ref{fig:tasd_figures}(bottom-left)  shows
the efficiency for reconstructing positive muons as a function of the
initial momentum of the muon. The detector
becomes fully efficient above 400~\MeVc.

The charge of the muon was determined by performing two separate Kalman
track fits, one with a positive charge
and the other with a negative charge. The charge mis-identification rate was
determined by counting the rate at
which the track fit with the incorrect charge resulted in a better
$\chi^{2}$ per degree of freedom than that with
the correct charge. Fig.~\ref{fig:tasd_figures}(bottom-right)  shows the charge mis-identification rate as a
function of the initial muon momentum.

This first investigation of the TASD 
concept has shown it to be worthy of a more
detailed study. In particular, it has led to interest in the concept of a
lower energy Neutrino Factory~\cite{iss_physics_report} 
(due to the lower threshold than the baseline magnetised iron detector) but more
work is required in order to bring the
understanding of this device to a comparable level to the baseline.

\subsection{Large Water Cerenkov detectors}


Since the pioneering age of the Kamiokande and IMB detectors, and after the success of the Super-Kamiokande detector (an extension by a factor 20 with respect to previous detectors), the physics community involved in this area is continuously growing in the three geographical regions, namely Japan, USA and Europe. 

To strengthen the know-how and R\&D exchanges, a series of International Workshops have been set up since 1999, the so-called NNN Workshop standing for ``Next Nucleon Decay and Neutrino Detectors". The last meetings were organized at Aussois (France) in 2005, Seattle (USA 2006) and Hamamatsu (Japan 2007). As it is clearly stated in the title of this Workshop, detection techniques other than Water Cerenkov are also considered, as for instance Liquid Scintillator, Liquid Argon as well as Iron detectors. 

Also, if the pioneering Water Cerenkov detectors were built to look for Nucleon Decay, a prediction of Grand Unified Theories, Neutrino physics has been the bread and butter since the beginning. Just to remind the glorious past: first detection of a Super Novae neutrino burst, Solar and Atmospheric anomaly discoveries that were explained as mass and mixing of  neutrinos, the latter being confirmed by the first long base line neutrino beams and by reactor experiments. 

Nucleon decay and neutrino physics are closely linked theoretically (ie. most if not all of the GUT theories predict nucleons to decay and neutrinos to have non zero masses and mixings). Hence, these are areas of equally strong interest to motivate the R\&D program extension of the next generation Water
Cerenkov mass to the megaton scale (about a factor 20 more than SuperKamiokande). 
One should keep in mind that, in addition to the physics addressed by the ISS, the physics potential of such a detector includes: nucleon decay,
supernovae neutrinos from bursts, relic neutrinos, solar and atmospheric neutrinos, long baseline low energy neutrinos (beta beam, super beam and combined with atmospheric neutrinos) and other astrophysical topics.

The physics performance~\cite{iss_physics_report}, scalability and robustness of Water Cerenkov detectors are well established and the R\&D efforts are concentrated now in two engineering aspects: the excavation of large cavities and the cost reduction of the photodetectors. The addition of Gadolinium salt, once it is demonstrated that it can be safely used in a 1~\Kton prototype and also in SuperKamiokande, could be a decisive ingredient for the new detectors, especially for neutrinos from Supernovae.

\subsubsection{The present detector design}

Up to now the three geographical regions have proposed three detector designs with a fiducial mass 
around 500~\Kton. Some characteristics are presented in table~\ref{WC:tab-1}. 
\begin{figure}[p]
\centering
\includegraphics[width=0.45\textwidth]{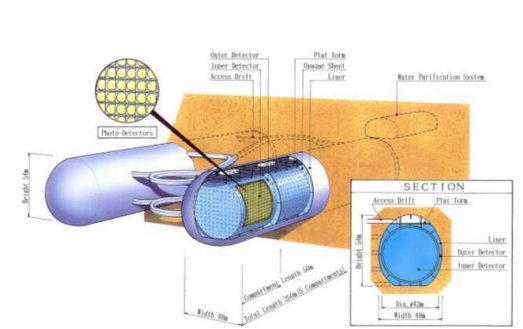}
\caption{\label{fig:HK}Sketch of the Hyper-K detector (Japan).}	
\end{figure}
\begin{figure}
\centering
\includegraphics[width=0.45\textwidth]{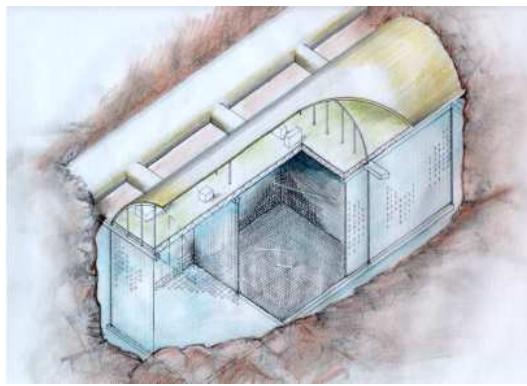}
\caption{\label{fig:UNO}Sketch of the UNO detector (USA).}	
\end{figure}
\begin{figure}
\centering
\includegraphics[width=0.45\textwidth]{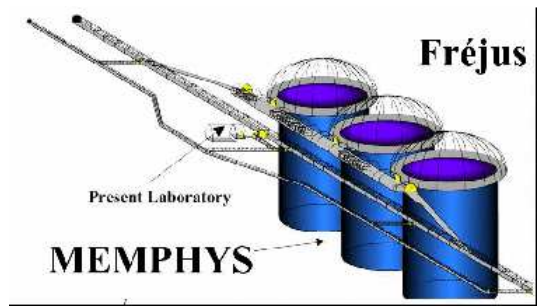}
\caption{\label{fig:MEMPHYS}Sketch of the MEMPHYS detector under the Fr\'{e}jus mountain (Europe).}	
\end{figure}

The Japanese design (Fig.\ref{fig:HK}) 
 is based on two twin tunnels with 5 optically independent cylindrical compartments, each 43~m in diameter and 50~m long, each covered by about 20,000 photodetectors to realize a 40\% surface coverage. 
The US design (Fig.\ref{fig:UNO}) 
 is composed of 3 cubic optically independent compartments ($60\times60\times60~\mathrm{m}^3$). The inner detector regions are viewed  by about 57,000 20" PMTs, with a photocathode coverage of 40\% for the central compartment  and 10\% for the two side compartments. An outer detector serves as a veto shield  of $2.5$~m depth and is instrumented with about 15,000 outward-facing 8" PMTs. The European design (Fig.\ref{fig:MEMPHYS})
 is based on up to 5 shafts (3 are enough for 500~\Kton fiducial mass), each 65~m in diameter and 65~m height for the total water container dimensions. The PMT surface defined as 2~m inside the water container is covered by about 81,000 12" PMTs to reach a 30\% surface coverage equivalent to a 40\% coverage with 20" PMTs (see sec.~\ref{sec:photodetector}). The fiducial volume is defined by an additional conservative guard of 2~m. The outer volume  between the PMT surface and the water vessel is instrumented with 8" PMTs.  
%
\subsubsection{Large underground cavities}

All the detector projects are located in underground laboratories. The water equivalent depth of the different detectors sites are: $\approx 1500$~m.w.e for the Tochibora mine in Japan, and around $4200$~m.w.e for the Homestake or Henderson mines (the two remaining sites after the NSF decision for DUSEL possible site candidates) in the USA, and $\approx 4800$~m.w.e for the Fr\'{e}jus road tunnel in Europe. 
A deeper site, so fewer cosmic ray induced background, is especially important in the case of relic supernovae and solar neutrinos, but in case of nucleon decay the detector segmentation may help also. 

The main difficulty is the non existence yet of man-made large cavities (see Tab.~\ref{WC:tab-1}) at the depth envisaged. But on an other hand, there are no a priory indications that 
one could not built such large cavities and engineering studies are undertaken in the three geographical regions. 
In Japan, a preliminary survey of the candidate place for Hyper-K is already done, and the rock properties at the Tochibora mine have been checked. The cavity model has been analyzed in a real environment. The egg transversal shape and the twin tunnel scenario is envisaged as baseline for Hyper-K. 
In the US, various engineering models have been used by different consultants. It turns out that, with the present knowledge, the UNO cavity seems feasible, although more refined work, with experimental inputs from rock quality measurements and geological fault knowledge in situ is needed to go further in the project design. 
In Europe, a pre-study has been performed  by the Italian and French companies involved in the building of the existing road tunnel. These companies have taken advantage of the numerous measurements made during the excavation of the present road tunnel and (relatively small) LSM Laboratory to establish a valid estimation of the rock quality as input for simulations. The main outcome of this pre-study is that very large cavities with a ``shaft" shape are feasible, while a ``tunnel" shape looks disfavored. The next step that can be undertaken in an European Founding framework, is to validate the rock quality at the exact detector location and to finalize the detailed shape of the cavities and access tunnels in close conjunction with the detector design optimization. 

Beyond the cavity shape and excavation scenario optimization, there is the need of an extensive R\&D on water containers (vessels versus multi-liners). This is an important aspect for radioactivity background suppression and also in detector mechanical design with its associate impacts on detector cost.   
%
\subsubsection{Photodetector R\&D}   
\label{sec:photodetector}

The surface coverage by photodetector is not yet optimized as more feedback is needed from the analysis from the SuperKamiokande I-II and III phases and from Monte Carlo studies of the foreseen detectors.  
Nevertheless, 
one may already state that the very low energy neutrino events (Super Novae neutrinos, ${}^8\mathrm{B}$ Solar Neutrinos) as well as the search of $\pi^0$ in Nucleon Decay or the $\pi^0/e$ separation in $\nu_e$ appearance experiment, all demand good coverage.

In all the detector design there are at least one order of magnitude more photodetectors than SuperKamiokande I (or III). The R\&D is largely shared among the three regions and in very close contact with the two manufacturers, namely Hamamatsu in Japan and Photonis in Europe and USA (since July 05, Photonis has acquired the DEP and Burle companies). 

The research axis on large HPDs in Japan has been mainly driven by the need to get a lower price for a new photodetector than the presently available Hamamatsu 20" PMTs, especially to get rid of the dynode amplifier system which is introduced manually in such a tube. Their measured characteristics are encouraging: single photo-electron sensitivity, wide dynamic range limited only by the readout, good timing and good uniformity over the large photo-cathode. But these HPD need to be operated at 20kV High Voltage and a low noise fast electronics. So, the cost per channel is a real challenge. 

In Europe, Photonis is very competitive on 12" PMTs and argue that the main parameter to optimize is the $cost/(cm^2 \times QE \times CE)$, including electronics. Some French laboratories are involved with Photonis in a joint R\&D programme concerning the characteristics of the 12" measurements and improvements and also concerning the integrated electronics front-end. The main idea is to adopt smart-photodetectors which provide directly digitized data. The front-end requirements are: a high speed discriminator for autotrigger on single photo-electron, a coincidence logic to reduce dark current counting rate (to be defined by MC studies),
a digitization of charge over 12 bits with a dynamical range up to 200~p.e, a digitization of time of arrival over 12 bits to provide nano-second accuracy, and a variable gain to equalize photomultiplier response and operate with a common high voltage (cost reduction). This electronics R\&D takes advantage of the R\&D from previous years and concrete realizations for OPERA, LHCb and WSi calorimeter for ILC among others.

Another R\&D line which is pursued at CERN in collaboration with Photonis is on the
so-called X-HPD, an almost spherical phototube with a cylindrical crystal
scintillator anode mounted in the centre of the sphere and read out by a small
conventional PMT (1"). The concept which is a modern implementation of Philips'
SMART tube and the QUASAR tube (Lake Baikal experiment), has been demonstrated with
a 208~mm prototype tube \cite{Braem1,Braem2} and promises excellent performance in terms of viewing
angle ($\approx 3\pi$), quantum efficiency ($\ge$ 40\% peak), collection efficiency
and timing. The radial field geometry makes the X-HPD immune to the earth magnetic
field. The X-HPD is operated around 20~kV. Due to the pre-gain of the scintillator
stage of about 30-40, gains in excess of $10^7$ are easily reached. A design for a
15" tube exists.

\begin{sidewaystable}
\centering
\begin{tabular}{rccc}
\hline\noalign{\smallskip}
 Parameters                  &        \textbf{UNO} (USA)            &    \textbf{HyperK} (Japan)          &      \textbf{MEMPHYS} (Europe)\\
\noalign{\smallskip}\hline\noalign{\smallskip}
\multicolumn{4}{l}{\textbf{Underground laboratory}}  \\ 
       location   &   Henderson / Homestake      &   Tochibora               &        Fr\'{e}jus \\
		depth (m.e.w$\pm 5\%$)	&    4500/4800                 &     1500                  &        4800  \\
Long Base Line (km)   & $1480\div2760$ / $1280\div2530$ & 290                 &        130 \\
                       & FermiLab$\div$BNL       & JAERI                     &         CERN \\
\noalign{\smallskip}\hline\noalign{\smallskip}
\multicolumn{4}{l}{\textbf{Detector dimensions}}          \\
type              & 3 cubic compartments   & 2 twin tunnels  & $3\div5$ shafts\\
                  &                        & 5 compartments  &                 \\
dimensions				& $3\times (60\times60\times60)\mathrm{m}^3$ 
									& $2\times 5 \times (\phi=43\mathrm{m} \times L=50\mathrm{m})$
									& $(3\div5)\times(\phi=65\mathrm{m} \times H=65\mathrm{m}) $ \\	  
fiducial mass (\Kton)& 440                          &       550                   & $440\div730$\\
\noalign{\smallskip}\hline\noalign{\smallskip}
\multicolumn{4}{l}{\textbf{Photodetectors$^\dag$}}          \\
           type   & 20" PMT              & 20" H(A)PD           & 12" PMT \\
 	       number	  & 38,000 (central) \& $2\times 9500$ (sides)
 	       					& 20,000 per compartment
 	       					& 81,000 per shaft \\
 surface coverage & 40\% (central) \& 10\% (sides) 
 									& 40\%
 									& 30\%   												                       \\
\noalign{\smallskip}\hline\noalign{\smallskip}
\multicolumn{4}{l}{\textbf{Cost \& Schedule}}          \\
estimated cost              &  500M\$           &  500 Oku Yen?$^*$  & 161M\euro{} per shaft (50\% cavity) \\
                  &                   &               & $+$ 100M\euro{}-infrastructure \\
tentative schedule          & $\sim 10$ yrs construction   & $\sim 10$ yrs construction        & \multicolumn{1}{l}{$t_0^{**}+8$ yrs cavities digging}  \\
                  &                   &               & \multicolumn{1}{l}{$t_0+9$ yrs PMTs production}  \\
                  &                    &               & \multicolumn{1}{l}{$t_0+10$ yrs detectors installation} \\
                  &                   &                & \multicolumn{1}{l}{Start of Non Accelerator Prog.} \\
                  &                   &                & \multicolumn{1}{l}{as soon as a shaft is commissioned} \\
\noalign{\smallskip}\hline
\end{tabular}
\caption{\label{WC:tab-1}Some basic parameters of the three Water Cerenkov detector baseline designs. ${}^\dag$: Only inner detector photodetectors are mentioned in this table.
*:Target cost, no realistic estimate yet.**: The $t_0$ date envisaged is 2010.}
\end{sidewaystable}
%

\subsection{Liquid Argon TPCs}
\label{sec:la_tpcs}

The liquid Argon Time Projection Chamber (LArTPC)~\cite{Rubbia:1977, Aprile:1985xz} is
a powerful detector for uniform and high accuracy imaging of massive active volumes.
It is based on the fact that in highly pure Argon, ionization tracks can be drifted
over distances of the order of meters.
Imaging is provided by position-segmented electrodes at the end of the drift path, continuously recording the
signals induced. The absolute timing of the event is given by the prompt scintillation light, providing the $T_0$ 
reference signal for the TPC. Such a device allows real-time imaging of events with bubble chamber quality, 
with a longitudinal granularity of the order of a percent of a radiation length. 
An example of a simulated neutral-current event in a LArTPC detector can be seen in  Fig.~\ref{fig2}.

The use of the LArTPC in high energy
physics was pioneered by the ICARUS collaboration~\cite{3tons, Cennini:ha, Arneodo:2006ug}.
The successful operation of the ICARUS T600 half-module ($\sim$300
tons) demonstrated the feasibility of the technique on this mass
scale~\cite{T600}. Building very large mass LArTPCs necessary for
long-baseline neutrino physics will require new techniques. 

Two different R\&D efforts are described in the next two sections.
The GLACIER project investigates a scalable concept based on an
industrial Liquified Natural Gas (LNG) tank to build very large
LArTPCs with masses up to 100~\Kton. It includes feasibility
studies to magnetize a LArTPC of a few 10~\Kton, allowing for
charge discrimination -- a necessary requirement at a Neutrino
Factory. The North American LArTPC effort, again based on the industrial LNG tank concept,
is towards the design of an unmagnetized detector for use in experiments involving a
``standard'' neutrino super-beam.

\subsubsection{The GLACIER project}
\label{sec:glacier}




A very large LArTPC with a mass ranging from $\sim 10$ to 100~\Kton would deliver extraordinary physics output
owing to the excellent event reconstruction capabilities.
Coupled to future Super Beams~\cite{Ferrari:2002yj,Meregaglia:2006du}, Beta Beams or Neutrino Factories
it could greatly improve our understanding of the mixing matrix in the lepton sector with
the goal of measuring the CP-phase. At the same time, it would
allow to conduct astroparticle experiments of unprecedented sensitivity~\cite{Rubbia:2004yq}.
Preliminary simulations show that a ``shallow depth'' operation at about 200~m rock
overburden would not significantly affect the physics performance, including the
astrophysical observations~\cite{Bueno:2007um}.



The possibility to complement the features of the LArTPC
with those provided by a magnetic field
would open new possibilities~\cite{Rubbia:2001pk,Bueno:2001jd}: charge discrimination,
momentum measurement of particles escaping the detector ($e.g.$ high energy muons),
and precise kinematics.
The magnetic field is required in the context of the Neutrino Factory~\cite{Rubbia:2001pk}:
(1) a low field, $e.g.$ B=0.1~T, for the measurement of  the muon charge (CP-violation);
(2) a strong field, $e.g.$ B=1~T for the measurement of the muon/electron charges (T-violation).

\begin{figure}[ht]
 \begin{center}
   \includegraphics[width=12cm]{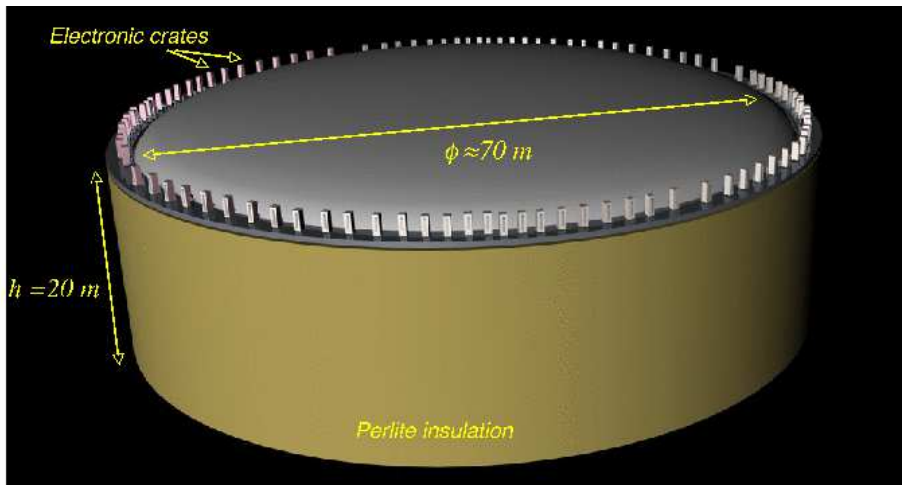}
 \caption{\label{glacier} Tanker for a 100~\Kton LArTPC based on industrial LNG technology}
 \end{center}
 \end{figure}
A concept for a LArTPC, scalable up to 100~\Kton (see
Fig~\ref{glacier}), has been proposed~\cite{Rubbia:2004tz}. It
relies on (a) industrial tanks developed by the petrochemical
industry (no R\&D required, readily available, safe) and their
extrapolation to underground or shallow depth LAr storage, (b) novel
readout method for very long drift paths with e.g. LEM readout, (c)
new solutions for very high drift voltage, (d) a modularity at the
level of 100~\Kton (limited by cavern size) and (e) the possibility
to embed the LAr in a magnetic field.

Such a scalable, single LAr tank design is the most attractive solution from the point of view
of physics, detector construction, operation and cryogenics, and finally cost.
The first experimental prototype of a magnetized
LArTPC has been operated~\cite{Badertscher:2004py,Badertscher:2005te}.
These encouraging results allow to envision a
large detector with magnetic field~\cite{Ereditato:2005yx}.
Beyond the basic proof of principle, the main challenge to be addressed is
the possibility to magnetize a very large mass of Argon, in a range of 10~\Kton or more.
The most practical design
is that of a vertically standing solenoidal coil producing vertical field lines, parallel
to the drift direction, by immersing a superconducting
solenoid directly into the LAr tank.

A rich R\&D program is underway with the aim of optimizing the design of future large mass LArTPC
detectors~\cite{Ereditato:2005ru} and is briefly summarized below.


The development of suitable charge extraction, amplification and collection devices is a crucial issue
and related R\&D is in progress. A LEM-readout is being considered and has been shown to
yield gains up to 10000 with a double stage LEM in gaseous Ar at cryogenic temperature.
Experimental tests are presently ongoing on charge extraction from the LAr phase, coupled with a LEM-based
amplification and collection in gaseous argon.

The understanding of charge collection under high pressure for events occurring at the bottom
of the large cryogenic tank is also being addressed.
For this purpose, a small chamber will be pressurized to 3-4~bar to simulate
the hydrostatic pressure at the bottom of a future 100~\Kton tank,
to check the drift properties of electrons.

Another important subject is the problem of delivering very high voltage to the
inner detectors trying to avoid the use of (delicate) HV feedthroughs.
A series of device prototypes were realized based on the Greinacher or Cockroft-Walton circuit allowing the
feeding into the vessel of a relatively low voltage and operation of the required amplification directly
inside the cryogenic liquid.
Tests reaching 120~kV in cold have been successfully performed.

The realization of a 5 m long detector column will allow to
experimentally prove the feasibility of detectors with long drift
path and will represent a very important milestone. The vessel for
this detector has been designed by a collaboration of the
University of Bern, ETH Zurich and University of Granada and will be mounted in Bern in
2007.
The device will be operated with a reduced electric field value in
order to simulate very long drift distances of up to 20 m. Charge
readout will be studied in detail together with the adoption of
possible novel technological solutions. A high voltage system
based on the previously described Greinacher approach will be
implemented.

For the immersed magnetic coil solenoid, the use of
high-temperature superconductors (HTS) at the LAr
temperature would be an attractive solution, but is at the moment
hardly technically achievable with the 1st generation of HTS
ribbons. We have started a R\&D program to investigate the    
conceptual feasibility of this idea~\cite{strauss} with BSCCO HTS
wires from American Superconductor~\cite{amsuper} and are now
investigating the performance of second generation YBCO wires from
American Superconductors and from SuperPower,
Inc.~\cite{superpower}.

Technodyne International Limited, UK~\cite{Technodyne}, which has unique
expertise in the design of LNG tanks, has produced
a feasibility study in order to understand and clarify all the issues related to the operation of a large
underground LAr detector. The study led to a first engineering design, addressing the
mechanical structure, temperature homogeneity and heat losses, LAr process, safety, and preliminary
cost estimate.
Concerning the provision of LAr, a dedicated, likely not underground but nearby, air-liquefaction
plant was foreseen.

The further development of the industrial design of a large volume tank able to operate underground
should be pursued.
The study initiated with Technodyne should be considered as a first ``feasibility'' step meant to
select the main issues that will need to be further understood and to promptly identify possible ``show-stoppers''.
This work should proceed by more elaborate and detailed industrial design of the
large underground (deep or shallow depth) tank also including the details of the detector instrumentation.
Finally,  the study of logistics, infrastructure and safety issues related to underground sites should also progress,
possibly in view of the  two typical geographical configurations: a tunnel-access underground
laboratory and a vertical mine-type-access underground laboratory.

In parallel, a program to study the technical feasibility of a large scale purification system
needed for the optimal operation of the TPC is being planned in collaboration
with the cryogenic department at  Southampton University (UK) and the
Institut f\"ur Luft und K\"altetechnik (ILK, Dresden, Germany).

The strategy to eventually reach the 100~\Kton scale
foresees an R\&D program leading to the detailed design study for
a tentative 100~\Kton non-magnetized and 25~\Kton magnetized detector,
including cost estimates. A 1~\Kton engineering module
could be foreseen to investigate the
tank concept, large scale purification, shallow depth operation, etc.
A 10~\Kton detector would have complementary physics reach
to the Superkamiokande detector currently in operation.

In addition to a successful completion of the technological R$\&$D, in the medium term a
measurement campaign on charged particle beams is envisaged with the goal to demonstrate
$e{^\pm}/\pi{^0}$ separation.
Also a 100 ton LArTPC is being considered for the T2K 2km site, which will provide a
high statistics sample of neutrino interactions.

\subsubsection{Off-axis NuMI or Wide-band Superbeam Detector}
\label{sec:la_us}

The purpose of future long-baseline neutrino experiments is to
observe $\nu_\mu \to \nu_e$ transitions. While this doesn't give a direct measurement of
$\sin{(2\theta_{13})}$ or the mass hierarchy, a combination of results from experiments with different
baselines and results from reactor neutrino experiments could allow for the extraction of the neutrino
parameters. In the United States there is the NuMI facility~\cite{NuMI} at Fermilab which provides a
$\nu_\mu$ beam for the MINOS experiment located 732 km away in a mine in the state of Minnesota. The beam
has been operating since January 2004. 

\begin{figure}[tbhp]
\begin{center}
\includegraphics[width=0.95\textwidth]{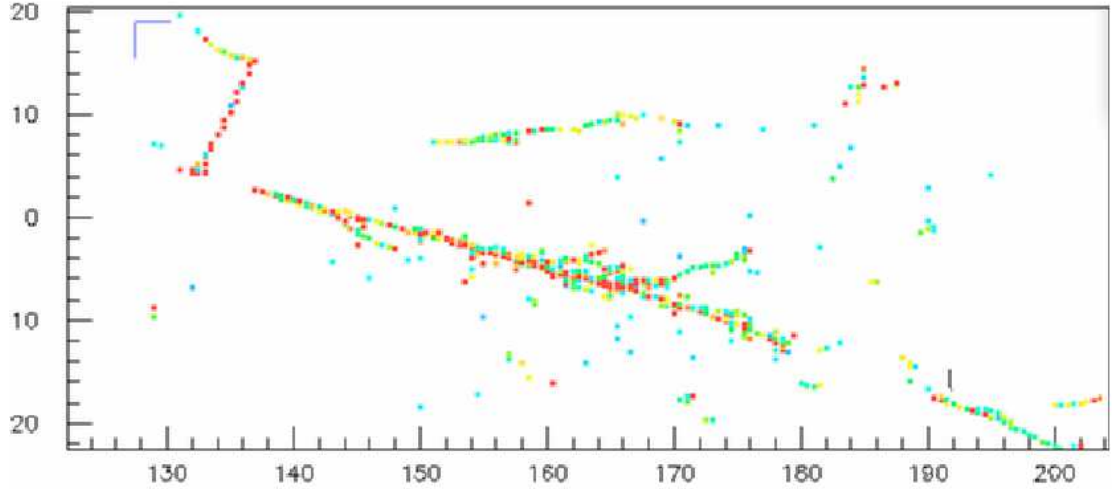}
\caption{\label{fig2}A simulated neutral current event with a 1 GeV $\pi^0$
  ($\nu_{\mu} + n \rightarrow \nu_{\mu} + \pi^+ + \pi^- + \pi^0 +
  n$).  Sampling rate is every 3.5\% of a radiation length in all
  three views.}
\end{center}
\end{figure}

The ultimate background to a $\nu_e$ appearance experiment is the
inherent $\nu_e$ content of the $\nu_\mu$ beam. The other serious background
to the $\nu_e$ appearance signal (i.e., electron
appearance from charged-current $\nu_e$ interactions) is $\pi^0$'s produced in neutral-current
events. Reducing this puts a premium on detectors that can differentiate electrons from photons.
The image of a simulated neutral-current event with a 1 GeV $\pi^0$
($\nu_{\mu} + n \rightarrow \nu_{\mu} + \pi^+ + \pi^- + \pi^0 + n$)
in a LArTPC detector, as
simulated by a GEANT3-based Monte Carlo, is shown in  Fig.~\ref{fig2}.
The lower photon shower is clearly identifiable in LAr based on the displacement
from the vertex and the high pulse height at the shower start. The efficiency for detecting
$\nu_e$s in a LArTPC is $\sim$80--90\% with a negligible neutral-current
$\pi^0$ event background.

A group of physicists from some 
North American universities and Fermilab have collaborated over the past several
years in an effort to design a large (15 to 50~\Kton) LArTPC as the detector for a long-baseline 
$\nu_\mu \to \nu_e$ appearance experiment~\cite{canoe}. 
In the baseline 15~\Kton detector, the LAr argon is stored in
a large, cylindrical, industrial Liquified Natural Gas (LNG) tank. The tank is 29.1 m in diameter
and 25.6 m high. The design employs 8 distinct drift regions with 3 metres between cathode planes
and signal wires. The drift field is 500 V/cm giving a drift velocity of 1.5 m/ms and a maximum
drift time of 2 ms. Following ICARUS, each signal ``plane'' contains three wire planes -- a vertical
collection plane and two induction planes strung at $\pm 30^\circ$ to the vertical. 
The wire pitch is 5 mm. There are also a number of new ideas, including utilizing wire-wrapped ``panels''
instead of wire planes, which are described in Ref.~\cite{dave}.  

\begin{figure}
\begin{center}
\epsfig{figure=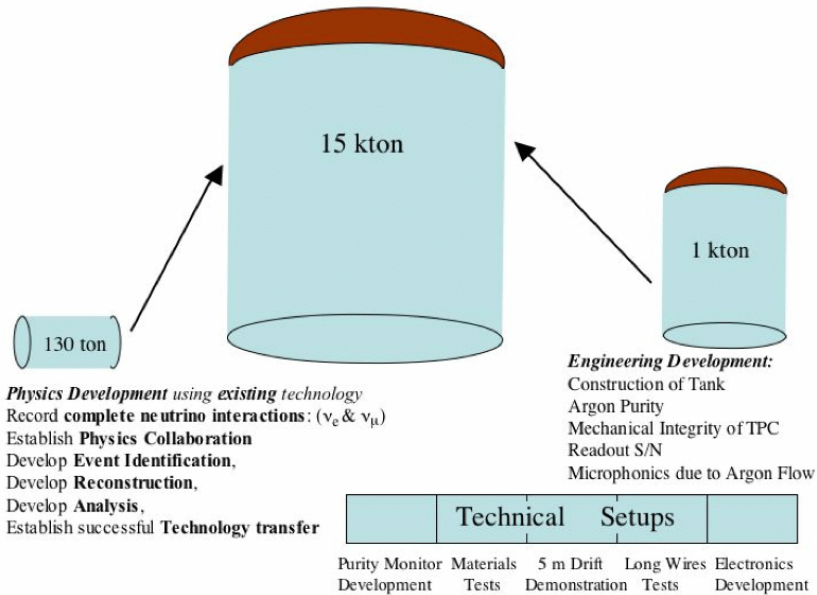, width=0.7\textwidth}
  \caption{\label{fig3}Proposed R\&D programme towards realization of a large
LArTPC. }
\end{center}
\end{figure}

A schematic of the R\&D programme that was proposed in the fall of 2005 is shown in  Fig.~\ref{fig3}. 
The programme included:

\begin{enumerate}
\item A series of technical test setups directed to answering specific
questions pertaining to a massive LArTPC (e.g., long drift, argon purity, wire tensioning, 
etc.). A number of these have been accomplished, as described in Ref.~\cite{dave}.
\item The construction of a 30--50 ton fiducial mass ($\sim$100--130 ton total
  argon mass) detector in which
  electron-neutrino interactions can be fully reconstructed and a
  range of 2 GeV neutrino interactions studied.  This detector
  will operate where it can obtain a sizeable number of neutrino
  interactions from the Fermilab NuMI and/or Booster Neutrino beams. This is still in the proposal stage.
\item The construction and partial outfitting of a commercial tank of
  $\sim$1~\Kton capacity using the same techniques as proposed for the 15-50~\Kton
  tank. This will serve as the test-bed to understand the issues of
  industrial construction. 
\end{enumerate}

In conclusion, there is a vigorous programme under way in North America towards the design and testing 
of a large liquid argon TPC for use in long-baseline neutrino physics. Specifically, the LArTPC is the
ideal detector for a $\nu_e$ appearance experiment as it is very efficient for reconstructing $\nu_e$ 
events while allowing for almost complete rejection of the neutral current background.

\subsection{Emulsion Cloud Chambers }

\subsubsection{Introduction}
An ideal detector for a Neutrino Factory should be able to exploit all the oscillation channels that are available with the well defined neutrino flux composition: $\nu_e\rightarrow\nu_\mu$ (the so-called $golden$ $channel$), $\nu_e\rightarrow\nu_\tau$ (the so-called $silver$ $channel$), $\bar{\nu}_\mu\rightarrow\bar{\nu}_e$ (the so-called $platinum$ $channel$) and $\bar{\nu}_\mu\rightarrow\bar{\nu}_\tau$ when a $\mu^+$ circulates into the decay ring and their CP conjugates in the case of a $\mu^-$ circulating. Therefore, an ideal detector should perform a complete and accurate kinematical reconstruction of neutrino events and be able to:
\begin{itemize}
	\item measure the momentum and the charge of the leptons (muons and electrons);
	\item identify the decay topologies of the $\tau$ leptons.
\end{itemize}

So far, the previous tasks have been separately tackled by using different techniques. A magnetized iron calorimeter is being optimized for the study of the golden channel requiring the muon detection and the charge determination with a high efficiency and a small pion to muon misidentification probability (Sec.~\ref{sec:mind}). 
The task of identifying electrons and of measuring their charge is very tough and so far only a study based on a magnetized liquid argon detector has been presented (Sec.~\ref{sec:glacier}), although totally active scintillating detectors are potentially able to do it 
(Sec.~\ref{sec:tasd}).

A detector $\grave{a}$ la OPERA \cite{opera,unknown:2006ki}, based on the Emulsion Cloud Chamber (ECC) technique \cite{kaplon,emulsion2}, has been proposed to search for the silver channel through the direct detection of the $\tau$ muonic decay thanks to the micrometric space resolution of the nuclear emulsions \cite{Donini:2002rm,Autiero:2003fu}. 

Here, the idea of using an ECC detector placed in a magnetic field (Magnetized ECC, MECC) is discussed. This combination provides good charge reconstruction and momentum determination capabilities, while providing at the same time the micrometric space resolution and compactness of an ECC.  Such a detector has, in principle, the ambitious aim to fulfill all the requirements for an ideal detector for a Neutrino Factory.

\subsubsection{The Emulsion Cloud Chamber}

The ECC consists of a sequence of passive material plates interspersed with emulsion films. It combines the high-precision tracking capabilities of nuclear emulsions with the large mass achievable by employing passive material as a target. By assembling a large quantity of ECC modules, it is possible to realize a $\mathcal{O}$(\Kton) fine-grained vertex detector for the direct observation of the $\tau$'s produced in $\nu_{\tau}$ charged current interactions. This concept has been adopted by the OPERA Collaboration for a long-baseline search of $\nu_{\mu}\rightarrow\nu_{\tau}$ oscillations in the CNGS beam \cite{tappearance}.

The basic element of the OPERA ECC is a $cell$ made of a 1~mm thick lead plate followed by an emulsion film, which consists of $44~\mu$m thick emulsion layers on either side of a $205~\mu$m plastic base \cite{emulsion1}. The number ($15$-$20$) of grains of metallic silver produced after the chemical development in each emulsion layer ensures redundancy in the measurement of particle trajectories and allows the measurement of their energy loss that, in the non-relativistic regime, can help to distinguish different particle masses.

Thanks to the dense ECC structure and to the high granularity provided by the nuclear emulsions, the detector is also suited for electron and $\gamma$ detection, with an efficient electron/pion separation \cite{Arrabito:2007rq}. The energy resolution for an electromagnetic shower is about 20\%. By measuring the number of grains associated to each track a two-track resolution of  $\sim 1~\mu$m or even better \cite{Toshito:2004tc} can be achieved. Therefore, it is possible to disentangle single-electron tracks from electron pairs coming from $\gamma$ conversion in lead. The outstanding space resolution can also be used to measure the angle between subsequent track segments with an accuracy of about 1~mrad \cite{DeSerio:2005yd}. This allows the use of Coulomb scattering to evaluate the
particle momentum with a resolution of about 20\% \cite{Lellis:2003xt} and to reconstruct the kinematical  event variables \cite{Kodama:2002dk}.

A lead-emulsion detector has been proposed \cite{Donini:2002rm,Autiero:2003fu} to study the silver channel $\nu_e\rightarrow\nu_{\tau}$ at a Neutrino Factory, with a detector similar to OPERA but with a total mass of 4~\Kton. The main limitation factor of this detector is the possibility of measuring the charge only for muons, by an external magnetic spectrometer. The fraction of the $\tau$ decays which can be exploited is thus given by the muonic decay branching ratio, about 20\%.

\subsubsection{The Magnetized Emulsion Cloud Chamber}

\begin{figure}[htbp]
\begin{center}
\epsfig{figure=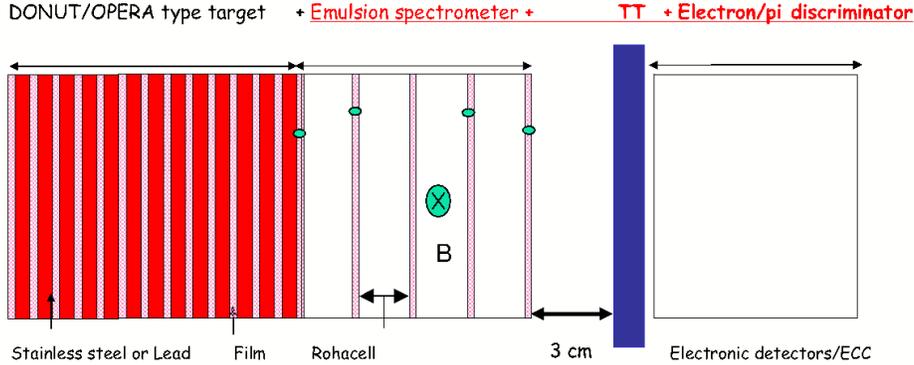, width=0.8\textwidth}
\caption{Schematic view of a Magnetized Emulsion Cloud Chamber.
\label{fi:structure}}
\end{center}
\end{figure}

The MECC here envisaged has the modular structure shown in Fig.~\ref{fi:structure}. The upstream part ({\it target}) is a sandwich of passive plates and nuclear emulsions.  
The length of the target section in terms of radiation lengths must be such 
to prevent the majority of the electrons to shower before their charge has been measured by the downstream modules. More work has to be done for the optimization of the passive material. Here the stainless steel is presented as a possible choice. 

An {\it emulsion spectrometer} is located downstream of the target. It consists of a sandwich of nuclear emulsions and a very light material called {\it spacer}, providing gaps in between emulsion films. The function of the spacer is to provide a lever arm between two consecutive emulsions films (tracking devices) with a stable mechanical structure. A few centimeter thick Rohacell plate fulfills this requirement. The trajectory measured with the emulsion films which precede and follow the spacer provides the measurement of the charge and momentum of the particle. The target and the spectrometer could mechanically form a single {\it brick} of about 10 cm length.

Downstream of the spectrometer, an electronic $target$ $tracker$ has the aim of providing the time stamp of the events. The time information is mandatory in order to match the emulsion information with the information from the electronic detectors allowing the identification of charged-current and neutral-current events.  The scanning of the emulsion films should be carried out without any track prediction. 

The most downstream element of the detector is the  {\it electron/pion discriminator}. Its aim is to provide the electron identification, having already measured the charge and momentum of the primary tracks in the spectrometer sector. A good electron identification with a low pion misidentification probability could be achieved at the same time either by a conventional electronic detector or by an emulsion calorimeter (emulsion-lead sandwiches). The choice between the two will be done according to a cost/effectiveness optimization. 


The MECC performance both for minimum ionizing particles (MIP) and electrons has been studied by considering different parameters: particle energy in the 1 to 10 GeV range, spacer thickness in the 2-5 cm range and three values of the magnetic field (0.25, 0.5 and 1 T). The same nuclear emulsion films as used by the OPERA experiment were considered. The thickness of the stainless steel plates has been taken to be 1\,mm with a total of 35 plates (about 2.5 $X_0$). The number of spacers is four.

\begin{figure}[htbp]
\begin{center}
\epsfig{figure=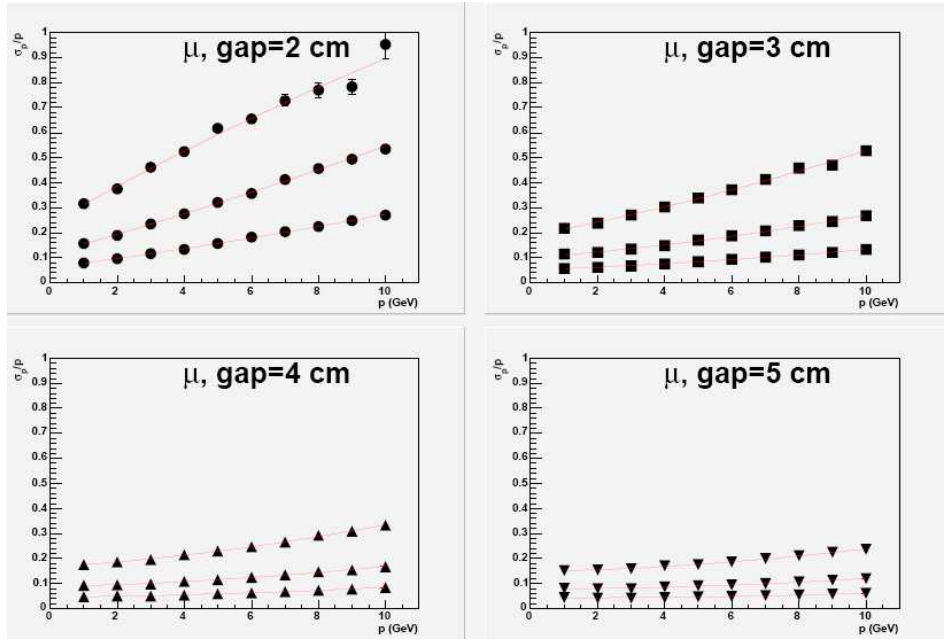, width=0.8\textwidth}
\caption{Muon momentum resolution as a function of the momentum for different spacer thicknesses and different values of the magnetic field: B=0.25 T, B=0.5 T and B=1.0 T for the upper, middle and lower curves, respectively.
\label{fig:MECCfig1}}
\end{center}
\end{figure}

\begin{figure}[htbp]
\begin{center}
\epsfig{figure=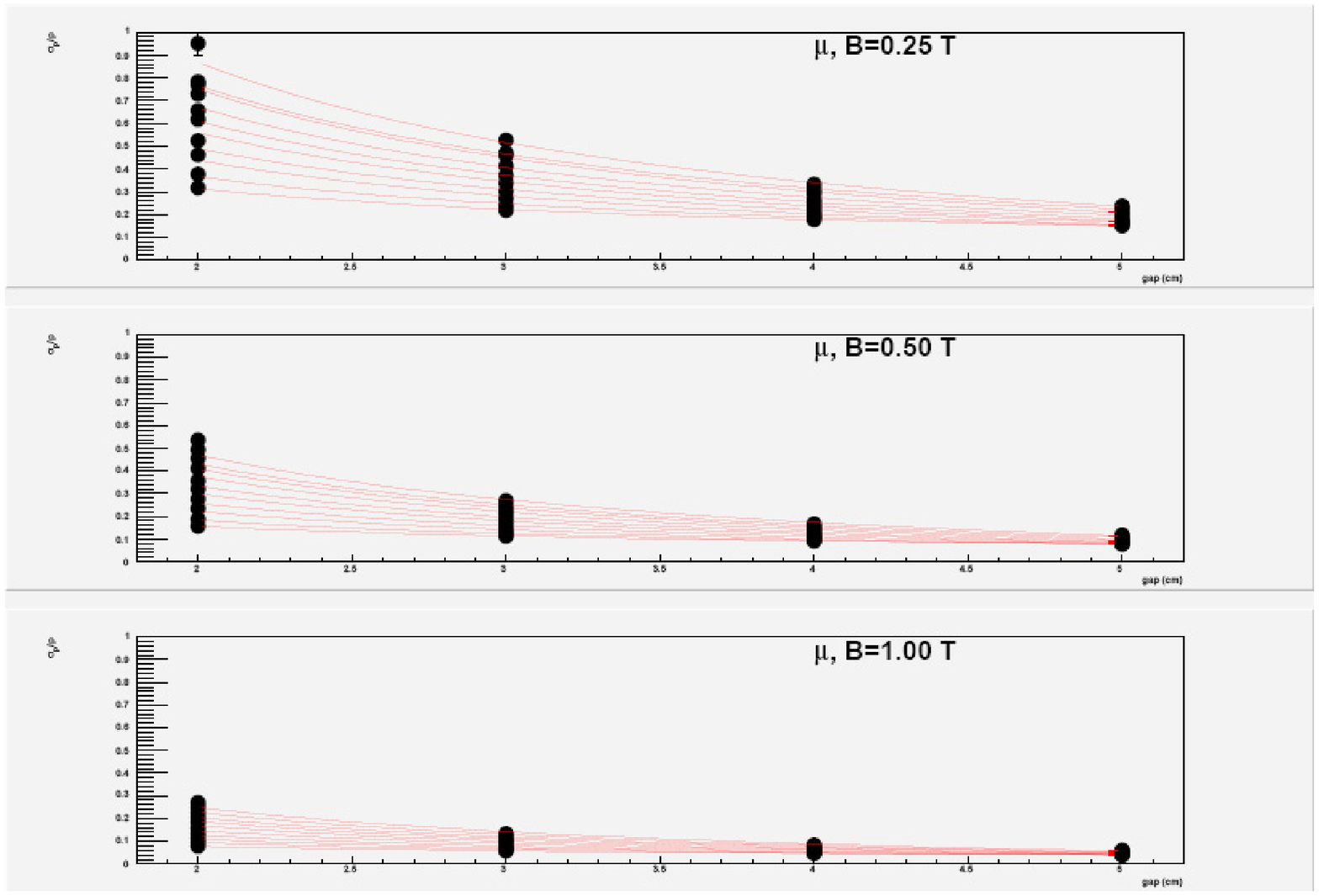, width=0.8\textwidth}
\caption{Muon momentum resolution as a function of the spacer thickness for different momenta (from 1 GeV to 10 GeV)  and different values of the magnetic field: B=0.25 T for the upper panel, B=0.5 T for the middle panel and B=1.0 T for the lower panel.
\label{fig:MECCfig2}}
\end{center}
\end{figure}

\begin{figure}[htbp]
\begin{center}
\epsfig{figure=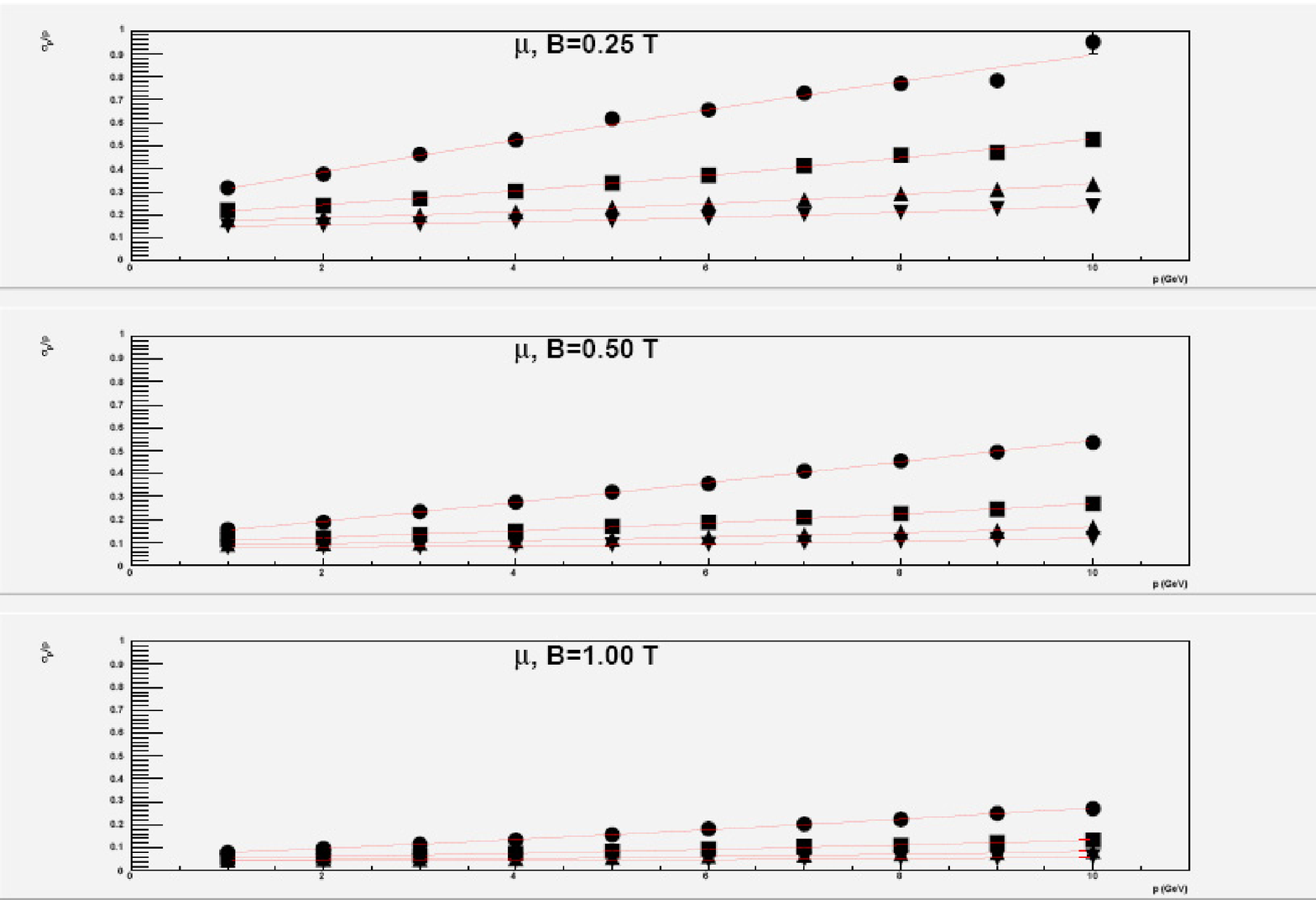, width=0.8\textwidth}
\caption{Muon momentum resolution as a function of the momentum for different spacer thickness and different values of the magnetic field: B=0.25 T for the upper panel, B=0.5 T for the middle panel and B=1.0 T for the lower panel.
\label{fig:MECCfig3}}
\vspace*{-1cm}
\end{center}
\end{figure}

\begin{figure}[htbp]
\begin{center}
\epsfig{figure=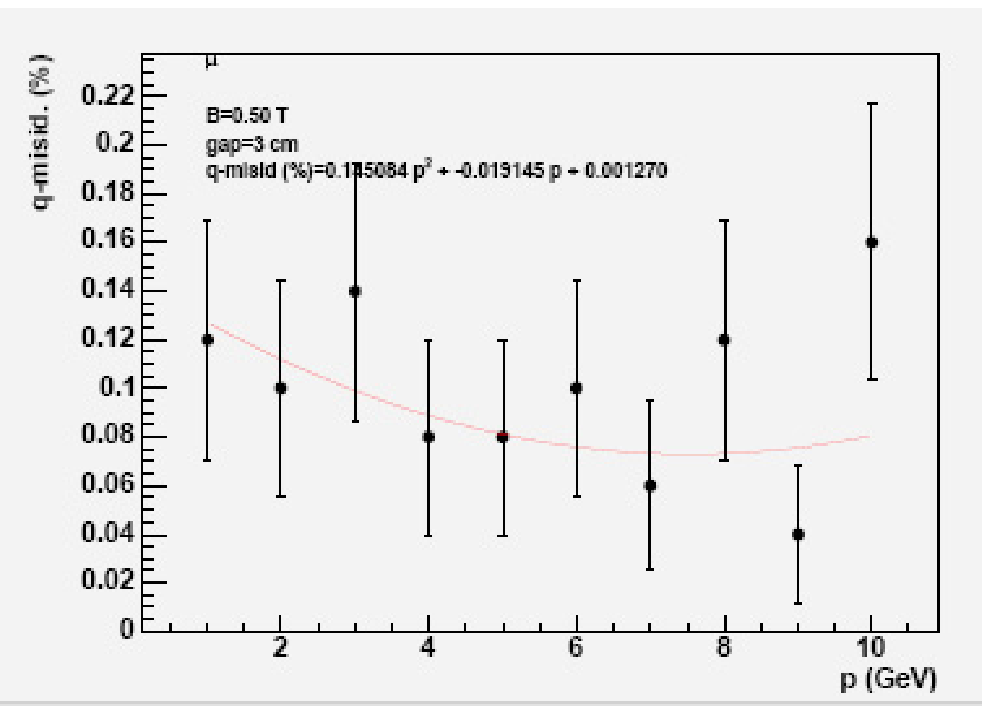, width=0.45\textwidth}
\epsfig{figure=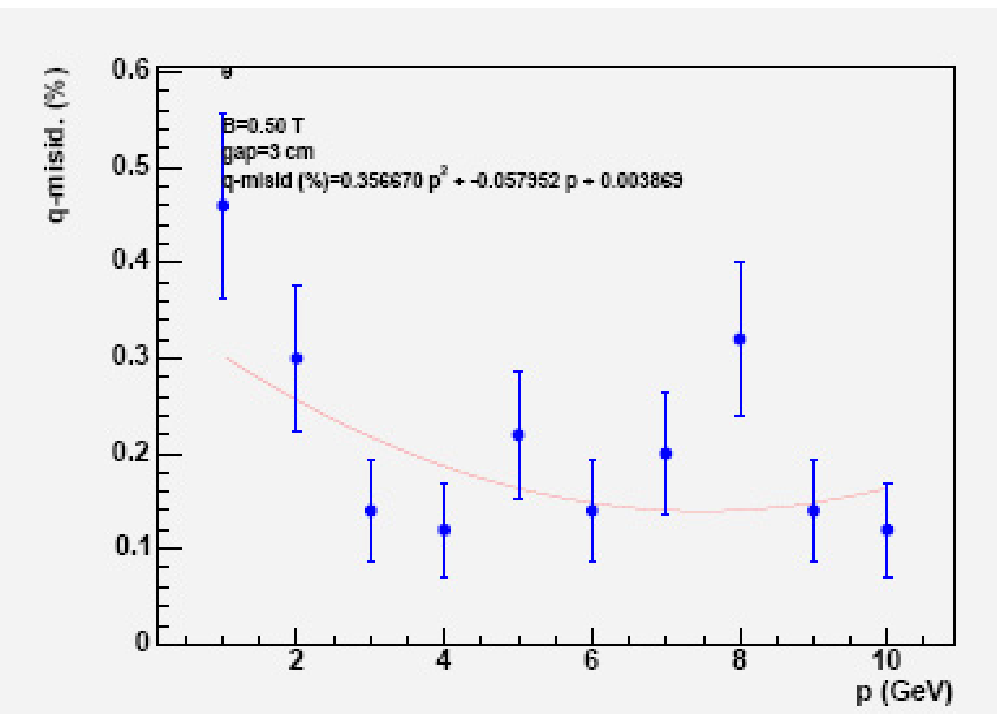, width=0.45\textwidth}
\caption{ Charge misidentification as a function of the momentum for minimum ionizing particles (left panel) and electrons (right panel), assuming a 3 cm spacer thickness and 0.5 T magnetic field. 
\label{fig:MECCchargemis}}
\end{center}
\end{figure}

Monte Carlo simulations have been performed in order to evaluate the momentum resolution and the charge identification efficiency. The momentum and the charge of the particles have been measured with four different methods, for consistency checks: slope measurement, sagitta measurement, parabolic global fit and Kalman filter. In the following only the results obtained with the Kalman filter are shown. The muon momentum resolution has been studied in the 1-10 GeV range as a function of the detector parameters that have to be optimized: the spacer thickness and the magnetic field intensity. The results are shown in Figs. \ref{fig:MECCfig1}, \ref{fig:MECCfig2}, \ref{fig:MECCfig3}. With a spacer thickness of 3 cm (more would be better but the detector would be too long) and a magnetic field of 0.5 T, a 30\% (10\%) momentum resolution at 10 (1) GeV can be achieved. The charge misidentification rate, shown in Fig. \ref{fig:MECCchargemis}(left panel), is better than 1\% below 10 GeV. 

The electron momentum and charge measurements are strongly affected by the showering. It has been shown that only 30\% of the electrons with energy in the range 1 to 10 GeV exit the target region without showering. For these events the momentum resolution and the charge identification efficiency, shown in Fig. \ref{fig:MECCchargemis}(right panel), are similar to those obtained for muons (left panel). It is worth noting that the electron reconstruction has been performed at the true hit level, i.e. without taking into account the error in the reconstruction. In this respect, it is optimistic. On the other hand, it does not take into account showering electrons for which a pattern recognition program could allow the track reconstruction, hence the charge and momentum measurement. 

Finally, the previous results have been obtained by considering a single emulsion spectrometer. Better results can be obtained, at least for MIP particles, by combining the information from consecutive emulsion spectrometers.

Another important issue is related to the number of interactions that can be stored in a brick preserving the capability of connecting unambiguously the events occurring in the emulsion target with the hits recorded by the electronic detectors. It has been shown \cite{pmKEK} that by using a tracker made of 3 cm strips up to 100 events may be stored into a single brick. This is a very conservative number that ensures the capability of the detector to stay in the beam for several years.

\begin{figure}[htbp]
\begin{center}
\epsfig{figure=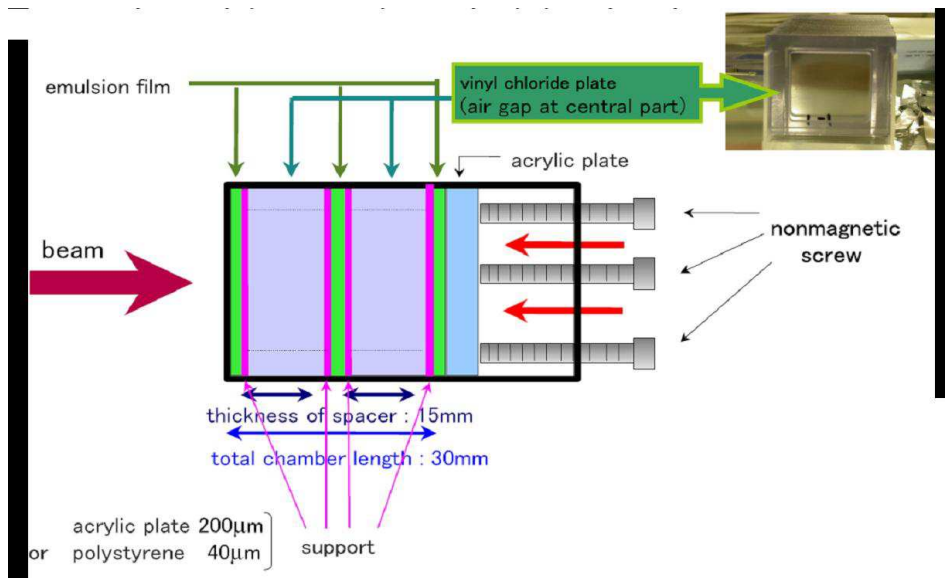, width=0.8\textwidth}
\caption{ Schematic view of the MECC exposed at the KEK-PS T1 pion beam.
\label{fig:MECCToho}}
\end{center}
\end{figure}

\begin{figure}[htbp]
\begin{center}
\epsfig{figure=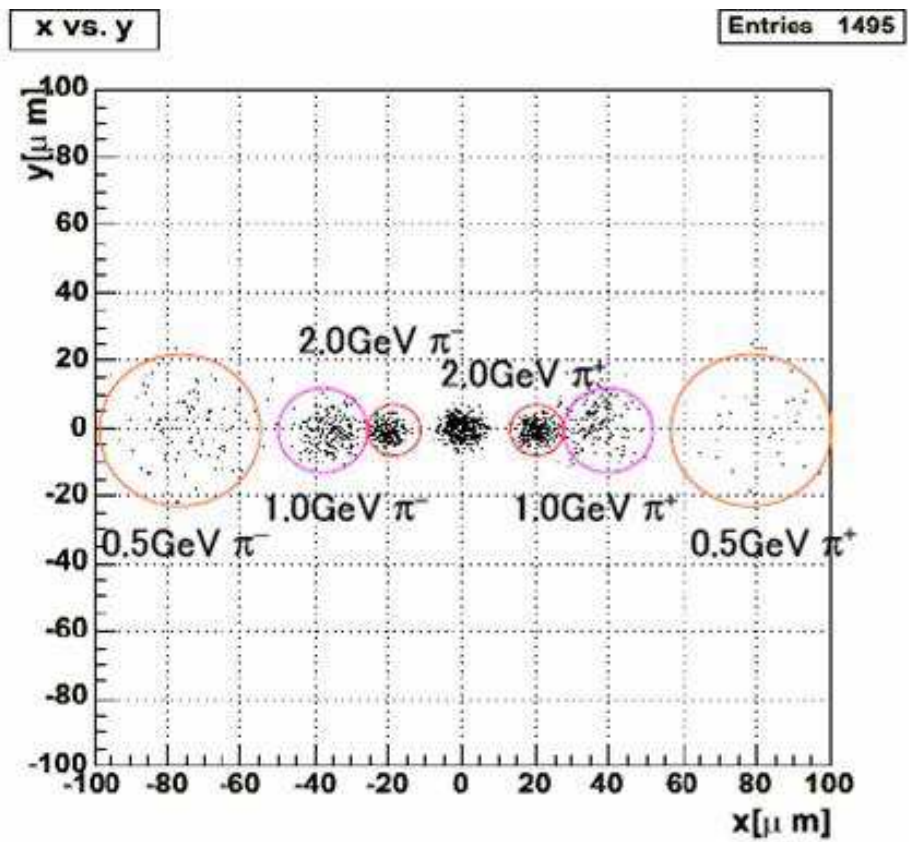, width=0.8\textwidth}
\caption{ Spatial distribution in the transverse plane of the beam spots of different energies impinging onto the emulsion spectrometer in the KEK-PS test.
\label{fig:KEKresults}}
\end{center}
\end{figure}

A first test of an emulsion spectrometer exposed to a pion beam has been performed in a KEK-PS T1 pion beam \cite{testToho}. The setup is shown in Fig. \ref{fig:MECCToho}. It consisted of 2 spacers of 1.5 cm thickness sandwiched with 3 emulsion films, for a total length of 3 cm. They were located inside a 1 T permanent magnet. The emulsion spectrometer has been exposed to pion beams with momenta 0.5, 1 and 2 GeV. The beam spots in the emulsions are shown in Fig. \ref{fig:KEKresults}. The results have been presented in \cite{luilloRAL}. The achieved momentum resolution is $\Delta p/p \sim 0.14 $, and almost constant in the studied energy range. This test shows that it is possible to study the performance of a MECC in a simple way, given the high modularity of the setup. Notice also that in the measurement performed, the alignment among the elements of the spectrometer is much more accurate than in the complete MECC structure (a few microns with respect to about ten microns). Conversely, the smaller number of spacers (2 with respect to 4 of the proposed MECC) and the thinner spacers (1.5 cm  with respect to 3 cm of the proposed MECC) determine a worsening of the resolution with respect to the standard emulsion spectrometer.

 \subsubsection{Conclusion and outlook}
The Magnetized Emulsion Cloud Chamber (MECC) should be able to detect $\tau$ decays measuring the charge of muons, electrons and hadrons. It should also be possible to study the golden channel by using an associated electronic detector. Before assessing its physics reach the maximum mass affordable in terms of scanning power and cost should be quantified. A smaller scale MECC detector would be suitable as a near detector. 

The first tests that have been carried out gave promising results. In order to have a realistic estimate of the physics reach, in the future the following studies should be performed:
\begin{itemize}
	\item define, also on the basis of the
experience with OPERA, the maximum MECC mass that can be affordable in terms of scanning power and cost, as well the minimum mass to have good sensitivity to the silver channel;
	\item carry out a realistic and cost effective design of the magnet;
	\item  study the synergy with other detectors that could act as the electron/pion discriminator. This will open the possibility to search for the golden, the silver and the platinum channels with the same detector;
	\item once  the previous points have been studied, a full simulation with neutrino events has to be performed in order to evaluate the detector sensitivity for the golden and the silver channels, and for the oscillations that produce an electron in the final state.
\end{itemize}

\subsection{Hybrid detectors}

All detectors mentioned above use different technologies and are suitable 
for different kind of measurements. However a number of interesting synergies can be found.  

In the previous section the possibility of merging an emulsion-based detector with other 
detectors (acting as pion/electron discriminator) has been mentioned. 
As described in Sec.~\ref{sec:tasd}, the TASD detector could   
efficiently discriminate between electron and muons/pions for momenta above $\sim$0.5~\GeVc. 
In addition it could also act as a spectrometer for the measurement of the lepton momentum 
and charge. Thus, an ECC-TASD hybrid would be able to measure golden (in TASD), silver (ECC-TASD) 
and platinum channels (ECC and TASD). 
An important issue concerning channels involving muons (golden and silver) 
is the background from pion to muon decay and pion/muon mis-identification due to the 
low density of liquid scintillators. 

The combination ECC-MIND would be interesting for the golden and silver channels, 
but not for platinum, since pion/electron separation in iron is very poor. 
The golden channel would be measured by MIND alone. For platinum, 
MIND would act as spectrometer for the measurement of the muon momentum and charge, 
and also as a muon identifier (by range), 
while the target and the tau vertex detection would be provided by the ECC. 
It is worth noting that MIND should be fully efficient and have very little background 
in the energy range of interest for the silver channel. 

Combinations with LArTPCs could be also considered.

An interesting combination would be the one between MIND and TASD. In this case the detector would 
consist of a sandwich between MIND and TASD modules of about 1 \m thick each.  
MIND would provide most of the target mass,  muon identification, and would act as an hadronic 
shower container. TASD would allow the measurement of the muon charge for low energy muons and 
the detection of electrons. 

MIND would help TASD in triggering hadronic showers, avoiding the potential background 
from pion to muon decay and pion/muon mis-identification. TASD would help MIND in measuring the charge 
of low momentum muons. 

The above arguments should be taken with the appropriate care since none of the 
combinations mentioned have been bench-marked with simulations yet.

\section{Baseline Detectors and Conclusion}
                                                                                                             The detector group of the International Scoping Study set out to determine the baseline detector options for each of the possible neutrino beams and to define a Research and Development (R$\&$D) plan necessary to develop those detector options (Appendix~\ref{RandD}). This programme of work will continue throughout the International Design Study in order to achieve the optimal configuration for a future neutrino facility. The baseline detectors defined by the ISS for each neutrino beam energy can be found in table~\ref{tab:baseline} and are summarised below:
                                                                                                                            
\begin{enumerate}
\item   {\bf Sub-GeV Beta Beam (BB) and Super Beam (SB)} 
A very massive (Megaton) water Cherenkov (WC) detector is the baseline option. The main R$\&$D
necessary for this detector option is the development of an inexpensive photosensor technology and the cost
and engineering for the cavern and infrastructure needed for such a detector. 
\item   {\bf 1-5 GeV (high energy) Beta Beam (BB) and Super Beam (SB)}. 
 There are a number of possibilities in this scenario, and a totally active scintillating detector (TASD)
a liquid argon TPC or a water Cherenkov detector would possibly be able to operate in this regime. 
The R$\&$D for these detector options include photosensor technology once more, and the R$\&$D for
liquid argon detectors (including long drifts and wires, Large Electron Multipliers, etc.).

\item   {\bf 20-50 GeV high energy neutrino factory from muon decay beams}.  
Magnetic detectors are necessary, so the baseline is a 100 kton magnetized
iron neutrino detector (MIND) for the wrong sign muon final states (golden channel),
or the possibility of $\sim$ 10 kton of a hybrid neutrino magnetic emulsion cloud chamber (NM-ECC) 
detector for wrong sign tau detection (silver channel). A full physics simulation
of these detectors is needed to demonstrate the efficiency as a function of energy
and to determine the charge identification at low momenta. 
\end{enumerate}

\begin{table}[htb]
\begin{center}
\begin{tabular}{llll}
\hline
Beam energy                  &  Beam type          & Far detector   & R$\&$D           \\  \hline 
Sub-GeV                      & BB and SB            & Megaton WC     & Photosensors, cavern   \\
                             &                      &                & and infrastructure   \\   
\\
1-5 GeV                      & BB and SB            & TASD           &   Photosensors and detectors.   \\
                             &                      & or LAr TPC      &   Long drifts and wires, LEMs, etc   \\ 
                             &                      & or Megaton WC  &    \\

\\
20-50 GeV                    & Nufact               & 100 kton MIND (golden)     & Simulation + physics studies  \\
                             &                      & + 10 kton NM-ECC (silver) & Charge at low momenta         \\   
\hline
\end{tabular}
\caption{\label{tab:baseline}
Baseline detectors for each beam energy range.}
\end{center}
\end{table}

Furthermore, there are more exciting possibilities of detectors that go beyond the baseline, which
could achieve improved performance to the physics parameters in question if these detectors are found to 
be feasible and affordable. These are summarised in table~\ref{tab:beyondbaseline}.
Finally, some beam instrumentation and near detector options have also been defined for each of the 
neutrino beams and energy ranges. These are summarised in table~\ref{tab:neardetectors}.

The International Scoping Study (ISS) has laid the foundations to proceed towards a full
International Design Study (IDS) for future high intensity neutrino facilities. The aim of the 
community is to have a full Conceptual Design Report of a future neutrino facility by the year 2012. The detector options covered in this ISS Detector Report and the R\&D programme identified in Appendix~\ref{RandD} will form a road map towards defining the detectors at future high
intensity neutrino facilities that will be included in the Conceptual Design Report.

\begin{table}[htb]
\begin{center}
\begin{tabular}{llll}
\hline
Beam energy                  &  Beam type          & Far detector   & R$\&$D           \\  \hline 
Sub-GeV                      & BB and SB            & 100 kton LAr TPC  & Clarify  advantage  of\\
                             &                      &                & LAr with respect to WC  \\   
\\
1-5 GeV                      & BB and SB            & TASD           &   Photosensors and detectors.   \\
                             &                      & or LAr TPC      &   Long drifts and wires, LEMs, etc   \\ 
                             &                      & or Megaton WC  &    \\

\\
20-50 GeV                    & Nufact               & Platinum detectors     &  Engineering study.  \\
                             &                      & Magnetised TASD &   Large volume magnet.        \\  
                             &                      & Magnetised LAr &  Simulations, physics. studies      \\  
                             &                      & Magnetised ECC &          \\ 
\hline
\end{tabular}
\caption{\label{tab:beyondbaseline}
Detectors beyond the baseline for each beam energy range.
}
\end{center}
\end{table}

\begin{table}[htb]
\begin{center}
\begin{tabular}{lll}
\hline
Beam energy                  &  Beam instrumentation          &  R\&D           \\  
                             &  Near Detectors                 &                 \\  \hline 
Sub-GeV                      & T2K concept             & Concept simulations, theory.  \\  
\\
1-5 GeV                      & No$\nu$a concept           &  Concept simulations, theory.  \\
                             & for precision measurement  &                                 \\ 
\\
20-50 GeV                    & Beam intensity (BCT)  & Need study.                     \\
                             & Beam energy, polarization  & Need study.                       \\
                             & Beam divergence & Need study.     \\  
                             & Shielding & Need concept.         \\ 
                             & Leptonic detector & Simulation and study.         \\ 
                             & Hadronic detector & Simulation, study and vertex detector R\&D.   \\ 
\hline
\end{tabular}
\caption{\label{tab:neardetectors}
Beam instrumentation and Near Detectors for each beam energy range.
}
\end{center}
\end{table}

\newpage
\appendix

\section{R$\&$D program}
\label{RandD}
The Research and Development (R$\&$D) programme for detectors at future neutrino facilities
will rely on a number of international initiatives aimed at delivering the optimal technology
for each of the possible neutrino beam options. The aim is to define the R$\&$D needed over the 
next four years to be able to carry out a Conceptual Design Study of the combined accelerator-
detector system. The following sub-sections will define the R$\&$D tasks that need to be carried
out in each of the detector systems to carry out the Conceptual Design Study and to be able to
perform a critical comparison of the neutrino facilities as a whole.

\subsection{Magnetized Iron Neutrino Detector (MIND) and Totally Active Scintillator Detector (TASD)}

\begin{itemize} 
\item Design, cost and engineering solutions for the magnet system for an iron calorimeter.
\item Design, cost and engineering solutions for the magnet system for a large volume 
totally active scintillation detector. 
\item R$\&$D on magnetic field resistant photon detector technology, which could include testing
of Multi-Pixel Photon Counters (MPPC), Silicon Photo-multiplier tubes (SiPM), Avalanche Photo
Diodes (APD) or other similar technologies. 
\item Feasibility and cost of long strips of extruded scintillator with optic fibre readout.
\item Building proptotype scintillator-fibre detection systems of varying lengths (5-20 m) and
measurements of the attenuation of the signal as a function of the length of scintillator, 
measurement of the number of photoelectrons collected and studying the optimal geometry
for the scintillator strips (for example, a comparison of the performance of square versus
triangular cross-section of the scintillator strips).
\item Study whether a different detector technology (such as Resistive Plate Chambers, RPC) would 
deliver the same performance at a reduced cost.
\item Build a prototype to put in a suitable test beam and test its performance inside a magnetic
field.
\end{itemize} 

\subsection{Water Cherenkov detector}

The detector R$\&$D on large water Cherenkov devices is based on the experience of running the
Super-Kamiokande detector. However, for a Megaton scale water Cherenkov device, further
R$\&$D is needed on a variety of topics:
\begin{itemize} 
\item Engineering and cost of cavern excavation for Megaton water Cherenkov detectors at different sites, 
including the optimal modularity of such a system. 
\item R$\&$D on photon detectors, such as large area Hybrid Photon Detectors (HPD), or 
standard Photo Multiplier tubes, including the reduction of the photon detection cost, reducing
the risk of implosion, electronics readout costs and reduction of energy threshold through the
selection of low activity materials for the detectors and associated mechanics. 
\item Engineering studies of the mechanics to support the photon detectors.
\item Studies of energy resolution of water Cherenkov detectors, especially at low energy (ie ~250 MeV).
\end{itemize} 

\subsection{Liquid Argon detector}

The Liquid Argon R$\&$D programme is well advanced in the USA and Europe. The main R$\&$D
issues include: 

\begin{itemize} 
\item Feasibility and cost of using industrial tankers developed by the petrochemical industry 
 and their deployment for underground liquid argon storage. 
\item Demonstration of detector performance for very long drift paths, including liquid argon purification.
\item R$\&$D on detectors for charge readout (for example, with a Large Electron Multiplier, LEM).
\item Photon detector readout options (for example, wavelength shifting coated photomultiplier tubes).
\item R$\&$D on ASICs for electronics readout and data acquisition system.
\item Development of new solutions for drift in a very high voltage (such as the Cockcroft-Walton 
style Greinacher circuit).
\item The possibility to embed the liquid argon in a B-field has been conceptually proven. However, the
magnetic field strength needs to be determined by physics requirements and the
feasibility and cost of the magnetic field design for large liquid argon volumes needs to be
established. Study of high temperature superconducting coils to operate at liquid argon temperatures
is an essential R$\&$D task to demonstrate this feasibility.
\item Dedicated test beams to study prototype detectors and to perform tracking and reconstruction
of clean electron and $\pi^0$ samples.
\end{itemize}

\subsection{Emulsion Cloud Chamber}
There has been a significant amount of R$\&$D done on the use of emulsion for particle physics
experiments, such as CHORUS, Donut and, more recently, OPERA. The main issues associated with
the emulsion cloud chamber that need to be addressed in further R$\&$D are:
\begin{itemize} 
\item Improvement to the automated scanning stations to reduce the overall scanning time and to
improve the scanning accuracy.  
\item Further R$\&$D on operating emulsion-iron sandwich systems in a magnetic field and adapting
the scanning algorithms to recognise tracks inside a magnetic field.
\end{itemize}

\subsection{Near Detectors}

\begin{itemize} 
\item Silicon vertex detector for the study of the charm background at a neutrino
factory: study a comparison of performance and cost of pixel versus strip detectors. 
Possible solutions could include standard hybrid strip or ``stripxel" detectors, 
hybrid pixel detectors, Monolithic Active Pixels (MAPS) or DEPFET pixel detectors. 
The latter are currently being studied in the context of the linear collider, so could
provide useful synergy between the two projects. Study whether layers of passive material 
(boron carbide, graphite or other low Z material) are necessary as a neutrino target.

\item Tracking device: determine the tracking medium at a near detector. A possibility 
could be to use a scintillating fibre tracker that serves both as a target and a tracking 
medium. Determine its performance, feasibility and cost. Are there any other options for 
the tracker such as drift chambers or a gas Time Projection Chamber (TPC)?

\item  Determine the performance needs for the other sub-detectors within the near
detector. For example, what is the required energy resolution for a
calorimeter? Is particle identification necessary in  the near detector? An example
of a particle identification system could be the use of a DIRC (Detection of Internally
Reflected Cherenkov Light) \cite{DIRC} such as the one used in Babar. What detector technology should
be used for the muon chambers of the near detector? 

\item Determine the accuracy of the neutrino flux measurement using the near detector 
design and determine whether it meets the specification of 0.1\% flux error. 
Perform a study of the charm background for the wrong sign muon signal. 
and measure the effect of a $Q_t$ cut to reduce the charm background.

\item Determine the accuracy of cross-section measurements as a function of 
energy. Above 5 GeV, where it is dominated by deep-inelastic scattering, the 
aim is to perform a measurement at the 0.1\% level. For less than 5 GeV, 
determine ways of measuring the different components. The near detector 
should be able to go to an energy threshold, at least as low as the far 
detector.

\end{itemize}

\section{Large magnetic volumes }


\subsection{Introduction }

All detector concepts for the Neutrino Factory (NF) require a magnetic field in 
order to determine the sign of muon (or possibly the electron) produced in the 
neutrino interaction. For the baseline detector, this is done with magnetized 
iron. Technically this is very straightforward, although the 100~kT baseline 
detector does present challenges because of its size. The cost of this magnetic 
solution is felt to be manageable. Magnetic solutions for the other NF detectors 
become much more problematic. No serious consideration has been given to 
magnetizing a MT water Cerenkov detector, but we have considered magnetizing 
volumes as large as 60,000 $m^3$ for a liquid Argon detector or a totally-active 
sampling scintillator detector (TASD). In addition the magnetic emulsion cloud chamber 
(MECC) would 
also require a relatively large magnetic volume. We have considered the 
following technologies:

\begin{itemize}
\item	Room Temperature Coils (Al or Cu)
\item   Conventional Superconducting Coils
\item	High Tc Superconducting Coils
\item	Low Temperature Non-Conventional Superconducting Coils
\end{itemize}

\begin{figure}[htbp]
\begin{center}
\epsfig{figure=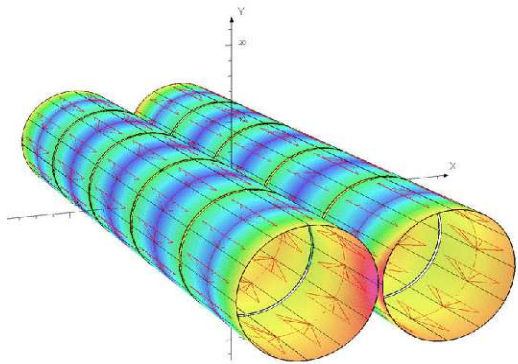, width=7cm,height=5cm}
\caption{Magnetic cavern configuration
\label{fig:mag_cavern}}
\end{center}
\end{figure}

For the cases of the TASD, the MECC, and the LAr approach currently being 
studied by a US-Canadian group providing the required magnetic volume using 
10 solenoids of roughly $15 m$ diameter $\times  15 m$ long has been considered 
with the solenoids configured into a magnetic cavern as shown in Fig.~\ref{fig:mag_cavern}.  We 
have considered a number of field strengths, but chose the baseline to be 0.5T.  
For the LAr concept being developed by the Glacier collaboration, 
field coils could be wound inside the large LAr 
tank.  In addition, we have also considered a dipole configuration for a TASD 
based on a concept that would use coils similar to those used in the Atlas 
toroids.

\subsection{Conventional Room Temperature Magnets}

In order to get adequate field strength with tolerable power dissipation, 
conventional room-temperature coils would have to be relatively thick.  We first 
considered Al conductor operating at 150K.  We then determined the amount of 
conductor necessary to produce a reference field of 0.1T.  In order to keep the 
current density at approximately 100$A/cm^2$, 10 layers of 1 $cm^2$ Al conductor 
would be required for our 15 m diameter, 15 m long reference solenoid.  Using a 
\$20/kg cost for conventional magnets~\cite{mag_1}, the estimated cost for 1 
solenoid is \$5M.  The power dissipation (assuming R=$1 \times 10^{-8}$ Ohm-m) is 
approximately 1 MW.  Ten magnets would then be \$50M and we felt that this 
number would be acceptable for a large NF detector.  However, the operating 
costs for 10 MW of power would  be \$13M/year (based on typical US power costs).  
The cost of the magnet system including 10 years of operation is thus \$180M.  
If the cost of cooling the coils to 150K is included, the costs increase 
substantially.  Studies have shown~\cite{mag_2} that there is little cost 
benefit to operating non-superconducting (Al or Cu) coils at low temperature vs. 
room temperature.  If we consider that the power dissipation at room temperature 
for Al coils triples (vs. 150K operation), then the operating cost for 
conventional room temperature magnets of this size will be unmanageable.  
Obviously trying to reach our baseline goal of 0.5T with room temperature magnets 
is totally unmanageable. 

\subsection{Conventional Superconducting Coils}

\begin{figure}[htbp]
\begin{center}
\epsfig{figure=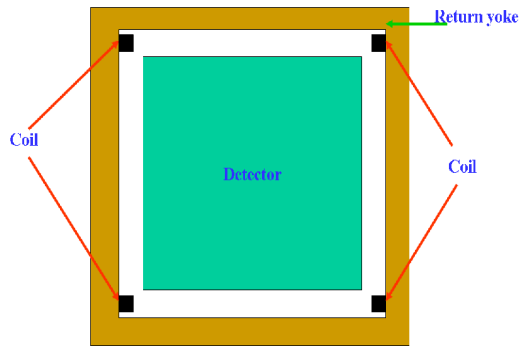, width=7cm,height=5cm}
\caption{Magnetic dipole configuration
\label{fig:mag_dipole}}
\end{center}
\end{figure}

One of the first configurations that we considered used superconducting coils 
similar 
to the coils used in the Atlas toroids to magnetize a roughly 30 kT TASD as 
shown 
in Fig.~\ref{fig:mag_dipole}.  In this
configuration, 10 coils are used along each side of the detector. 
 We estimated that the coil cost (extrapolated from the Atlas experience) 
would be on the order of \$120M and was considered acceptable. The field strength 
for this design was chosen to be 0.15T and at this field a 5 sigma determination 
of muon sign could be obtained at a muon momentum of 2 GeV/c.  However, we 
determined 
that the field quality in this configuration was not adequate.  In addition, the 
amount 
of iron required for the return flux was quite large.

Conventional superconducting solenoids are certainly an option for providing the 
large magnetic volumes that are needed.  Indeed coils of the size we are 
considering were engineered (but never built) for the proposed GEM experiment at 
the SSC.  A cylindrical geometry (solenoid) does imply that a fraction of the magnetic 
volume will not be outside the volume of the active detector which will likely be rectangular in cross section.  This is
certainly a disadvantage in terms of efficient use of the magnetic volume, but would provide 
personnel access paths to detector components inside the magnetic cavern.  It is certainly possible to consider
solenoids of rectangular cross section and thus make more efficient use of the magnetic volume, but the engineering 
and manufacturing implications of this type of design have not been evaluated.

Technically, superconducting magnets of this size could be built, but 
at what cost?  There have been a number of approaches to estimating the cost of 
a superconducting magnet and we will mention two of those there.  The first 
comes from Green and St. Lorant~\cite{mag_3}.  They looked at all the magnets 
that had been built at the time of their study (1993) and developed two formulas 
for extrapolating the cost of a superconducting magnet: one scaling by stored 
energy and one scaling by magnetic volume times field.  They are given below:
\[C = 0.5E_s^{0.662}  \]
and
\[C = 0.4(BV)^{0.635},  \]
where $E_s$ is the stored energy in MJ, B is the field in Tesla, V is the volume in 
$m^3$ and C is the cost in M\$.  The formulas given above give a cost for each 
15 m diameter, 15 m long,  0.5T magnet of approximately \$20M (based on $E_s$) and 
\$38M (based on magnetic volume). As another reference point, we used the CMS 
coil~\cite{mag_4} (B=4T, V=340 $m^3$, Stored energy = 2.7 GJ, Cost  = \$55M). 
The Green and St. Lorant formulas give costs for the CMS magnet of \$93M and 
\$41M based on stored energy and magnetic volume respectively. From these data 
we can make  ``Most Optimistic" and ``Most Pessimistic" extrapolations for our 
baseline NF solenoid.  The most optimistic cost comes from using the formula 
based on stored energy and assume that it over-estimates by a factor of 1.7 
(93/55), based on the CMS as built cost.  This gives a cost of  \$14M for each 
of our NF detector solenoids.  The most pessimistic cost extrapolation comes 
from using the formula based on magnetic volume and conclude that it under-
estimates the cost by a factor of 1.3 (55/41), based on the CMS as built cost. 
This then gives a cost of  \$60M for each of our NF detector solenoids.  There 
is obviously a large uncertainty represented here.

Another extrapolation model was used by Balbekov et. al.~\cite{mag_5} based on a 
model developed by A. Herve~\cite{herve}.  The extrapolation formulae are given below:
\[ P_0 = 0.33S^{0.8} \]
\[ P_E = 0.17E^{0.7} \]
and
\[ P = P_0 + P_E  \]
where $P_0$ is the price of the equivalent zero-energy magnet in MCHF, $P_E$ is the 
price of magnetization, and P is the total price. S is the surface area ($m^2$) 
of the cryostat and E (MJ) is the stored energy.  This model includes the cost 
of power supplies, cryogenics and vacuum plant.  From the above equations you 
can see that the model does take into account the difficulties in dealing with 
size separately from magnetic field issues.  Balbekov et. al. used three ``as-
builts" to derive the coefficients in the above equations:

\begin{itemize}
\item	ALEPH (R=2.65m, L=7m, B=1.5T, E=138MJ, P=\$14M)
\item	CMS (R-3.2m, L=14.5m, B=4T, E=3GJ, P=\$55M)
\item	GEM (R=9m, L=27m, B=0.8T, E=2GJ, P=\$98M)
\end{itemize}

The GEM magnet cost was an estimate based on a detailed design and engineering 
analysis.  
Using this estimating model we have for one of the NF detector solenoids: 
$P_0 = 0.33(707)^{0.8} = 63MCHF$,  $P_E = 0.17(265)^{0.7} = 8.5MCHF$.  The 
magnet cost is thus 
approximately \$57M (which is close to our most pessimistic extrapolation given 
above).  
One thing that stands out is that the magnetization costs are small compared to 
the total cost. 
 The mechanical costs involved with dealing with the large vacuum loading forces 
on the vacuum 
cryostat assumed to be used for this magnet are by far the dominant cost.

\subsection{High Tc magnets}

We did not explore in detail the possibilities of building a NF detector 
solenoid with high Tc superconductor,
 but we recognized the potential in this area.  Currently the cost of  high Tc 
superconductor is 100-200 times~\cite{mag_6}
 that of conventional SC for the same field and there are many engineering 
issues that would have to be investigated first 
if we are to conclude that this technology was applicable (cost + 
manufacturability) to our application. 
 However since the technological status of high Tc superconductor is moving so 
fast, we did do some zeroth-order 
estimates regarding one of these NF detector solenoids fabricated with high Tc 
superconductor.  We assumed a 
low-temperature operation of 35K. This might still allow for a non-vacuum insulated 
(foam) cryostat and thus have no 
vacuum loading to give higher current carrying capacity.  The cost of the 
superconductor for 10 NF detector 
solenoids was estimated to be \$50M.  Based on studies that have been done on 
foam-insulated vessels for GLACIER, 
we estimated the cost of the cryostats also at \$50M.  Assembly and engineering 
could not be reliably 
estimated in that they will depend on the particulars of the conductor being 
used and the currently 
existing manufacturing and assembly capabilities for high Tc superconducting 
magnets are not yet at the 
stage where reliable estimates can be made.  However the possible cost savings 
afforded by using non-vacuum 
insulated cryostats are large and high Tc superconductor cable technology is 
advancing very rapidly.

\subsection{Low Temperature Non-Conventional Superconducting Coils}

\begin{figure}[htbp]
\begin{center}
\epsfig{figure=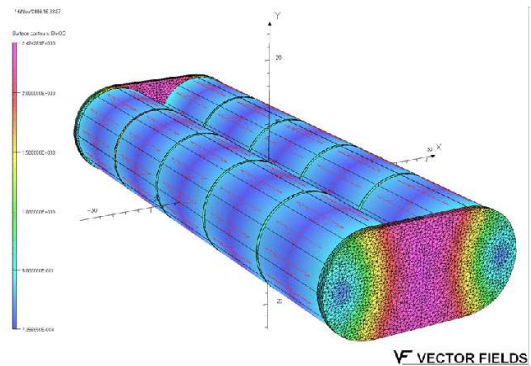, width=7cm,height=5cm}
\caption{STL Solenoid Magnetic Cavern Simulation
\label{fig:solenoid}}
\end{center}
\end{figure}

In this concept we solve the vacuum loading problem of the cryostat by using the 
superconducting transmission line (STL) that was developed for the Very Large 
Hadron Collider superferric magnets~\cite{mag_7}. The solenoid windings now 
consist of this superconducting cable which is confined in its own cryostat.  
Each solenoid consists of 150 turns and requires $\sim$7500 m of cable. There is 
no large vacuum vessel in this design. We have performed a simulation of the 
Magnetic Cavern concept using STL solenoids and the results are shown in 
Fig.~\ref{fig:solenoid}.  With the iron end-walls (1 m thick), the average 
field in the XZ plane is approximately 0.58~T at an excitation current of 50 kA.

\begin{figure}[htbp]
\begin{center}
\epsfig{figure=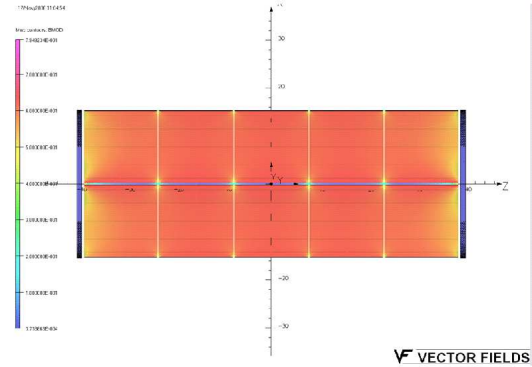, width=7cm,height=5cm}
\caption{STL Solenoid Magnetic Cavern Field Uniformity in XZ plane
\label{fig:field_uni}}
\end{center}
\end{figure}

The maximum radial force is approximately 16 kN/m and the maximum axial force 
approximately 40 kN/m.  The field uniformity is quite good with the iron end-
walls and is shown in Fig.~\ref{fig:field_uni}.

\subsection{Superconducting Transmission Line}

\begin{figure}[htbp]
\begin{center}
\epsfig{figure=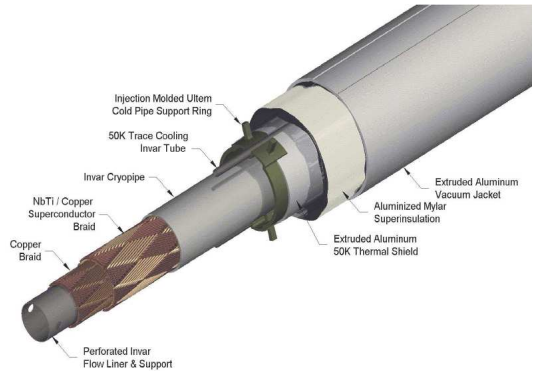, width=7cm,height=5cm}
\caption{Superconducting transmission line
\label{fig:stl}}
\end{center}
\end{figure}

The superconducting transmission line (STL) consists of a superconducting cable 
inside a cryopipe cooled by supercritical liquid helium at 4.5-6.0 K placed 
inside a co-axial cryostat. It consists of a perforated Invar tube, a copper 
stabilized superconducting cable, an Invar helium pipe, the cold pipe support 
system, a thermal shield covered by multilayer superinsulation, and the vacuum 
shell. One of the possible STL designs developed for the VLHC is shown in 
Fig.~\ref{fig:stl}. Its overall diameter is approximately 83 mm.

The STL is designed to carry a current of 100 kA at 6.5 K in a magnetic field up 
to 1~T. This provides about a 50\% current margin with respect to the required 
current in order to reach a field of 0.5T.  This operating margin can  
compensate for  temperature variations, mechanical or other perturbations in the 
system. The superconductor for the STL could be made in the form of braid or in 
the form of a two-layer spiral winding using Rutherford cable. The braid 
consists of 288 NbTi SSC-type strands 0.648 mm in diameter and arranged in a 
pattern of two sets of 24 crossing bundles with opposite pitch angle about the 
tube. A conductor made of Rutherford cables consists of 9 NbTi cables that were 
used in the SSC dipole inner layer. A copper braid is placed inside the 
superconductor to provide additional current carrying capability during a quench. 
The conductor is sandwiched between an inner perforated Invar pipe, which serves 
as a liquid helium channel, and an outer Invar pressure pipe that closes the 
helium space. Both braided and spiral-wrapped conductors and the 10 cm long 
splice between them have been successfully tested with 100 kA transport current 
within the R\&D program for the VLHC.  The STL has a 2.5-cm clear bore, which is 
sufficient for the liquid helium flow in a loop up to 10 km in length. This 
configuration allows for cooling each solenoid with continuous helium flow 
coming from a helium distribution box. 

The thermal shield is made of extruded aluminum pipe segments, which slide over 
opposite ends of each support spider. The 6.4-mm diameter Invar pipe is used for 
50 K pressurized helium. It is placed in the cavities at the top and the bottom 
of both the shield and the supports. The shield is wrapped with 40 layers of a 
dimpled super insulation. The vacuum shell is made of extruded aluminum or 
stainless steel.  Heat load estimates for the described STL are:

\begin{itemize}
\item 	Support system: 53 mW/m at 4.5 K and 670 mW/m at 40 K
\item	Super insulation: 15 mW/m at 4.5 K and 864 mW/m at 40K
\end{itemize}

The estimated cost of the described STL is approximately \$500/m. Further STL 
design optimization will be required to adjust the structure to the fabrication 
and operating conditions of the desired NF detector solenoids and to optimize 
its fabrication and operational cost.  Although what has been described here has 
been directed at the Magnetic Cavern concept, the STL could also be used in a 
very large LAr detector following the Glacier concept.  The fact that the STL 
would be operating in liquid Argon would allow for a simplified STL design since 
the heat-load environment would be very different.

\subsection{STL Solenoid Power}

The relatively low inductance of the STL solenoids (0.3 H/solenoid) allows 
powering all solenoids from a single 50 kA power supply. A power supply with a 
voltage of ±50 V will allow ramping the magnet system up or down in less than 1 
hour. A single pair of 50 kA current leads is required for powering the 
solenoids. These could either be conventional copper leads or current leads 
based on High-Temperature Superconductor.  The cryogenic wall power associated 
with the conventional 50 kA leads could be reduced by a factor of 4 with high Tc 
leads.

\subsection{Conclusions}

Magnetizing volumes on the order of 30,000 to 60,000 $m^3$ at fields up to 0.5T 
presents technical challenges, but is certainly within the current engineering 
capabilities.  The cost, however, in most scenarios is prohibitive.  The use of 
room temperature Cu or Al conductor could provide a modest field ($\le$0.1T), but 
operating costs are likely to be excessive.  Conventional superconducting magnet 
technology could provide the necessary field at acceptable operating costs, but 
the magnet construction costs using a conventional vacuum-insulated cryostat are 
not affordable.  High Tc superconducting coils using foam insulated cryostats 
show promise, especially given the rapid pace in which this technology is 
developing.  The current state-of-the-art in high Tc cable might present an 
affordable technical solution to this problem, but much more R\&D on coil 
assembly, magnet quench performance and cryostat would need to be done.  Using 
the STL concept presents some very interesting possibilities.  It eliminates the 
cost driver of large conventional superconducting coils, the vacuum-insulated 
cryostat, and has already been prototyped, tested, and costed during the R\&D 
for the VLHC.  A full engineering design would still need to be done, but this 
technique has the potential to deliver the large magnetic volume required with a 
field as high as 1T with very uniform field quality and at an acceptable cost.  
Developments with high Tc superconducting cable could also have an impact on the 
STL design concept, with potential cost savings.

\section{Matter effects}

The matter effect causes different oscillation patterns for neutrinos and antineutrinos,
depending on the mass hierarchy.  
Observing this difference is the most feasible way to determine the mass hierarchy. 
The difference may be observable with baselines longer than about 1000 km, 
depending on the quality and quantity of achievable data and oscillation parameters.

The difference is most visible at the MSW resonance, where the oscillations of one channel are enhanced
and those of the other suppressed.
For the usual neutrino parameters the resonance energy is about 10 GeV in the lithosphere, 
about 7 GeV in the mantle at depths relevant for the magic baseline, 
and about 3 GeV in the core. (The uncertainties of neutrino parameters cause 
an uncertainty of about 20 \% at $3\sigma$ for this prediction.)
For energies much higher than the resonance energy all oscillations are suppressed, 
and for energies well below the resonance energy the oscillations can be treated as in vacuum.

The detailed simulation of the propagation of neutrinos through the Earth requires a sufficiently accurate knowledge 
of the density profile along the baseline. 
The uncertainties of the density profile cause correlations in the parameter space that
complicate the analysis and reduce the accuracy of results. 
For a large $\theta_{13}$ the density uncertainty of 5 \% may cause rather large errors
while 1 \% accuracy would make the correlations ignorable. 
With smaller $\theta_{13}$ the requirements for the accuracy are milder, and with $\sin^22\theta_{13}<10^{-3}$ 
the dominant error comes from elsewhere and any reasonable density model will be sufficiently accurate. 
The correlations can be also reduced by a suitable choice of multiple baselines and channels \cite{0606111}.

Within first order, one can use the average density uncertainty of the baseline an indicator of goodness. 
Uncorrelated local variations around the average mostly smooth out for realistic density profiles, when not in resonance, 
and all small-scale density variations with a scale up to a few kilometres are completely irrelevant. However, a better
error analysis in a variable density requires numerical treatment, as different densities contribute differently, and 
particularly the resonance case should be studied with more care.

According to geophysical studies, the difference between the density of the Earth 
and the density defined by a standard spherical Earth model (e.g. PREM\cite{PREM}) does not exceed 
5~\%\footnote{The errors here and throughout this section do
not correspond to Gaussian distributions, but are rather ``maximal reasonable deviations".
For any decisions on the location of experiments we need more than $1\sigma$ certainty.}.
The uncertainties are due to both global or systematic effects for the average density distribution and unknown local variations. 
The local variations can be rather large, particularly for complicated zones like active mountains, subduction zones, 
hot spots, plumes or superplumes. Such variations may extend down to the border of the inner core.
Also it is to be noted that the inner core is in rotation relative to the mantle, even its axis deviates from the 
rotation axis of the Earth. 
The detailed models for the inner parts are not yet free from inconsistencies,
and therefore must be treated with care. 

Using the data of local and regional geophysical measurements one can construct 
local models much better than the 1-dimensional PREM model.
Specific local and regional models may reach up to 1 \% accuracy.  
With good geophysical measurements one can obtain knowledge to define the density profile even for complicated regions. 
Nevertheless, for most part of the Earth, particularly oceans, sufficiently accurate measurements cannot be done, and one has
to rely on general models. The models for ocean crust are usually very simple, but one should be careful when using
such models as the simplicity may be due to our ignorance.

\begin{figure}[tbp]
\begin{center}
\includegraphics[angle=0, width=0.9\textwidth]{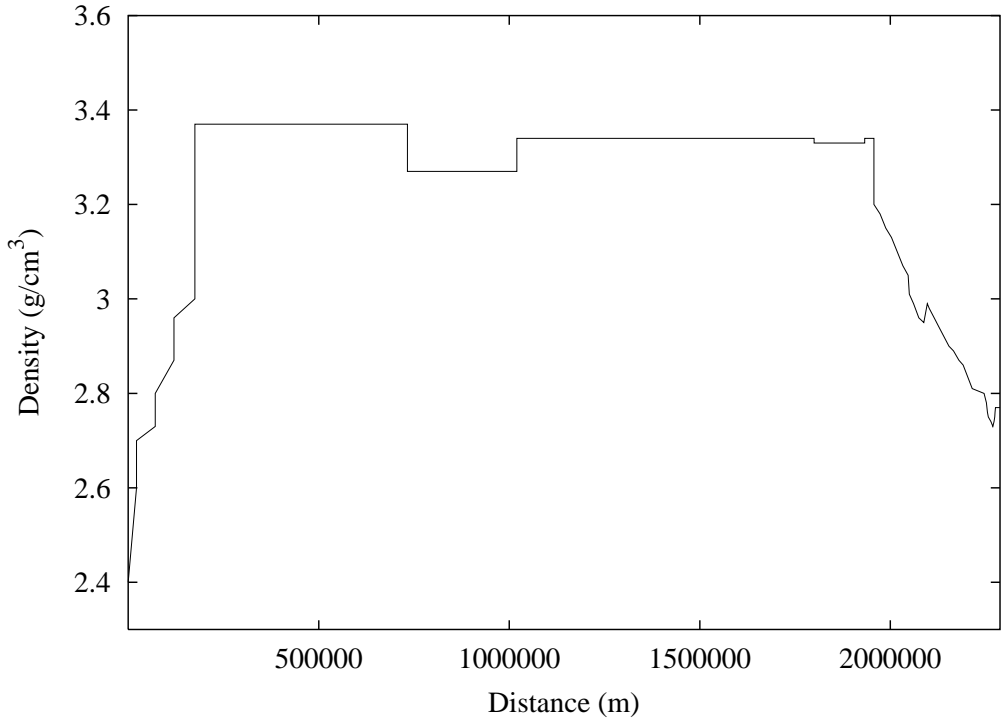}
\caption{The estimated density profile for the baseline CERN-Pyh\"asalmi (Finland).}
\label{rho.ps}
\end{center}
\end{figure}

A specific model for the baseline CERN-Pyh\"asalmi was constructed in Ref.~\cite{Kozlovskayaetal2003} (Fig.~\ref{rho.ps}).
For this specific baseline there are abundant geophysical data, and a realistic density profile
can be built up, despite some parts of the baseline being rather complicated. The most challenging part
is the upswelling asthenosphere under Germany which causes the largest uncertainty.
It was concluded that one can reach about 1\% accuracy in estimating regional density variations 
(e.g. density inhomogeneities of more than several dozens kilometers) for baselines from CERN to Pyh\"asalmi.
All later geophysical studies support the previous view, and no surprises have occurred.

It was shown explicitly in Ref.~\cite{Kozlovskayaetal2003} that the uncertainties in this model
do not cause any significant error in the interpretation of the data, with realistic experimental scenarios.

Similar studies for other baselines would be welcome (see \cite{0303112} for a study in Japan). 
While waiting for other studies, we can extrapolate
the experiences from modelling of the above baseline and from general considerations, to predict the accuracies of
other profiles. Also, opinions different from those above have been expressed \cite{Warner}.

In order to get the best accuracy for the density profile, the following general conclusions can be drawn:
\begin{enumerate}
\item It is recommendable to use well known continental areas passing tectonically stable flat regions.
\item One should avoid complicated zones like high mountains and seismically active or volcanic areas. 
\item One should avoid oceans where little data are available.
\item Similarly one had better avoid baselines passing underdeveloped or politically challenging countries where 
geophysical measurements will be too risky.
\end{enumerate}
These conditions may be rather contradictory: some of the most complicated zones are also the most studied,
like Japan. On the other hand, particularly challenging zones are the Atlantic ridge and most of the Pacific that are both 
complicated and difficult to be studied.

To reach the best accuracy for the density profile, the favoured beam directions are:
\begin{itemize}
  \item From CERN towards North-East. Baseline lengths up to 2700 km are achievable
with 1 \% accuracy for the density. On-going and planned geophysics measurements can improve the accuracy even more.
  \item Across North-America. Similar accuracies are reachable for the USA when the USArray gives data. Baseline lengths 
up to 4000 km are possible from BNL to West Coast of the USA, and baselines up to 4000--5000 km 
can be achieved through Canada to Alaska.
\end{itemize}
Geophysically disfavoured directions include beams from CERN to Canary Islands, Azores, Madeira or Iceland,
as well as any baseline around Japan.

For other long baselines the accuracy of density may not be better than 2--3 \%.
The above favoured baselines cannot be extended due to firm geographic constraints, and hence
the longer baselines necessarily must pass through complicated or worse known zones. 
Baselines 4500--6000 km may be particularly difficult when the baseline crosses the transition zone and touches tangentially 
boundary layers at the depths of 400 km and 660 km, with density jumps of 5 \% and 10 \%, respectively.
In such a case a small error in the model may cause a considerable error in the baseline density profile.
For the most difficult oceanic baselines one can hardly reach 5 \% accuracy for the average density.

\begin{figure}[tbhp]
\begin{center}
\includegraphics[angle=0,width=0.70\textwidth]{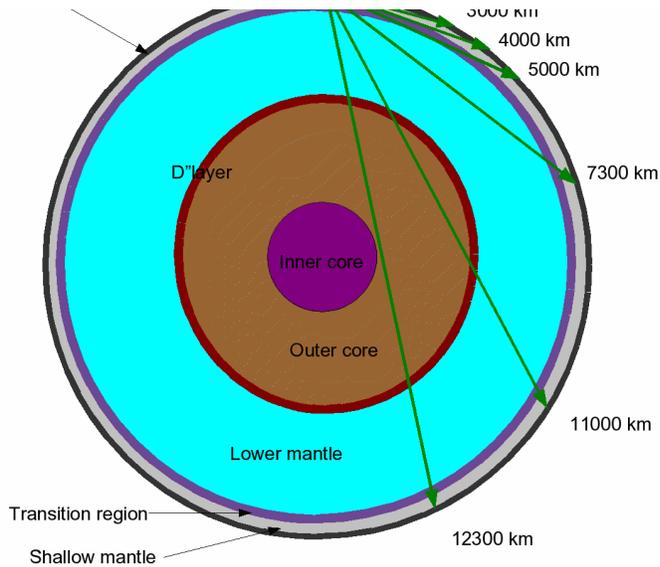}
\caption{\small Several baselines projected through the model of the Earth interior (not to scale). 
A baseline of O(5000 km) may be problematic as it largely
intersects with the transition zone, where the density changes quite abruptly from 4.0 to 4.4, 
and errors on its depth may result in large errors in the density profile. 
We see that the second and third magic baselines (11000 km and 12300 ) traverse through the
outer core that dominates their refraction lengths.}
\label{earth-baselines.gif}
\end{center}
\end{figure}

When the baseline length equals the refraction length or its multiples 
one can do a clean measurement of the $\theta_{13}$ mixing angle, 
independent on the CP-phase \cite{0112119,0301257,0610198}. 
These baselines are called magic, and can be solved analytically in constant density,
but in varying density they must be computed numerically, for example by solving the equation: 
\begin{equation}
  \int_0^{L_{\rm magic}} \exp{\left(i\int_0^x V(y) {\rm d}y \right)} {\rm d}x =0,
\label{eq:magic_baseline}
\end{equation}
where $V(y)$ is the interaction potential in matter, which is proportional to the electron density. 
Equation~\ref{eq:magic_baseline} gives a good first order approximation \cite{0506064,0610198} to the 
magic baseline. 
Integrating the above using the PREM model and two extreme cases with arbitrary 5~\% uncertainties for the density, 
and a 5 km uncertainty for the core-mantle boundary, 
the first magic length turns out to be $(7300\pm 300)$ km long, 
the second $(10 060_{-50}^{+70})$ km and the third ($12 280^{+170}_{-140} $) km.
For the first magic baseline, the dominant error comes from the deep mantle, and 
detailed knowledge of the crust in start and end points is rather irrelevant.
For the other two magic baselines the uncertainty of the length is surprisingly small, considerably smaller 
than for the first one. This may sound paradoxical, but is understandable from Fig.~\ref{earth-baselines.gif}.
These baselines pass through the dense outer core which gives the largest contribution 
to the total refraction length, and also to the error. 
For these baselines, the details of the lithosphere are completely ignorable, but 
the quoted 5 \% accuracy for the core density and particularly the 5 km accuracy
for the core-mantle boundary may be rather optimistic.

The first magic baseline is not very sensitive to such uncertainties \cite{0612158} (See also
respective sections in ISS Physics Report for analyses and references).
On the other hand, the second and particularly the third magic lengths are more
sensitive to errors, which makes them less usable for neutrino studies until better certainty on the core conditions can be reached.
Alternatively, it has been suggested to use neutrinos to measure the density of the 
mantle or core \cite{0111247,0105293,0502097,0612002,0612158}.

There is no geophysically optimal candidate for a magic baseline from the proposed sites of the accelerator.
In any case it is safest to use continental baselines, and avoid oceans and complicated zones. Most important is 
to choose the baseline so that we can maximize the accuracy in the deepest parts of the trajectory,
while the properties of the lithosphere at the end points are less relevant.
CERN to Eastern Siberia or Northern China may be closest to optimal, and from Japan the best direction is towards 
Northern Europe. 

We conclude that it is possible to obtain sufficient accuracy for the density profile to avoid correlations.
Future measurements may improve the accuracy, and if necessary, a dedicated geophysical measurement campaign 
for the selected baseline can be made, at a cost which is marginal to total cost. 
However, in practice such measurements are possible only in limited parts of
the Earth, and particularly oceanic measurements will remain unrealistic for a long time. If the mixing angle is small enough, 
density uncertainties are irrelevant and any baseline is good enough. For defining the length of the magic baseline, however, 
uncertainties of the density are relevant for all parameters, but in practice the physics is not very sensitive to them.


\begin{figure}[tbhp]
\begin{center}
\includegraphics[angle=0,width=0.7\textwidth]{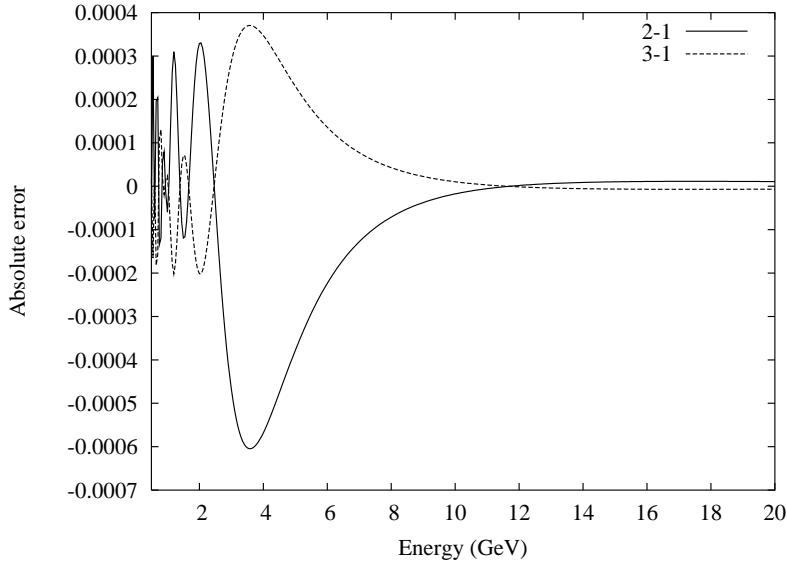}
\caption{A sample plot presenting the effect of the uncertainties in the density profile of Fig.~\ref{rho.ps} to the muon
neutrino appearance probability due to errors in density. These correspond to the absolute deviation in the probability
 with typical parameters.
}
\label{Absoluteerror.ps}
\end{center}
\end{figure}


\section{Low energy cross sections}
\label{Low_energy}


  Existing cross-sections measurements cover properly the high-energy regime, above 5~GeV, but not the low-energy where
many of the new oscillation experiments will operate. In this region, the energy is crossing several threshold of $\nu$ interactions. The knowledge of
the cross-section in this regime is very limited, see~\cite{CrossSections} for a recent compilation. In addition to the intrinsic knowledge of
the interaction, the final state particles are affected by nuclear effects like nuclear re-interactions, Pauli blocking and Fermi motion that
alters the topology and kinematics of the outgoing particles.

  The final state interactions could change the momentum and nature of nucleons and pions produced in the $\nu$ interactions. Both charged and neutral
pions contribute to the background in disappearance (charged pions faking a muon) and appearance (neutral pion faking an electron) experiments and
should be understood to a 10\% level for the next generation of superbeams~\cite{T2K_2}.

 The nuclear effects also alter the kinematics of the final state muon in charged current interactions by inhibiting the reaction (Pauli blocking) or
changing the center of mass energy where the reaction takes place (Fermi Motion). These phenomena change basic kinematic properties of the interaction
like the $q^2$ or the threshold of the reaction. The dependency of the cross-section with the nuclear mass (A) has to be considered, since most of the measurements
are done in light nuclei (deuterium, carbon, oxygen, etc.). The measurement of the dependency of cross-section with A is part of the experimental program of
the Miner$\nu$a experiment~\cite{Minerva}.

 The dominant neutrino interactions from 500~MeV to few GeV are :

\begin{itemize}

\item Charged current quasi-elastic and neutral current elastic interactions.

\item Neutral and charged current single pion production.

\item Neutral and charged current multi pion production and more inelastic interactions.

\item Neutral and charged current coherent pion production.

\end{itemize}

\begin{figure}[htbp]
\begin{center}
\epsfig{figure=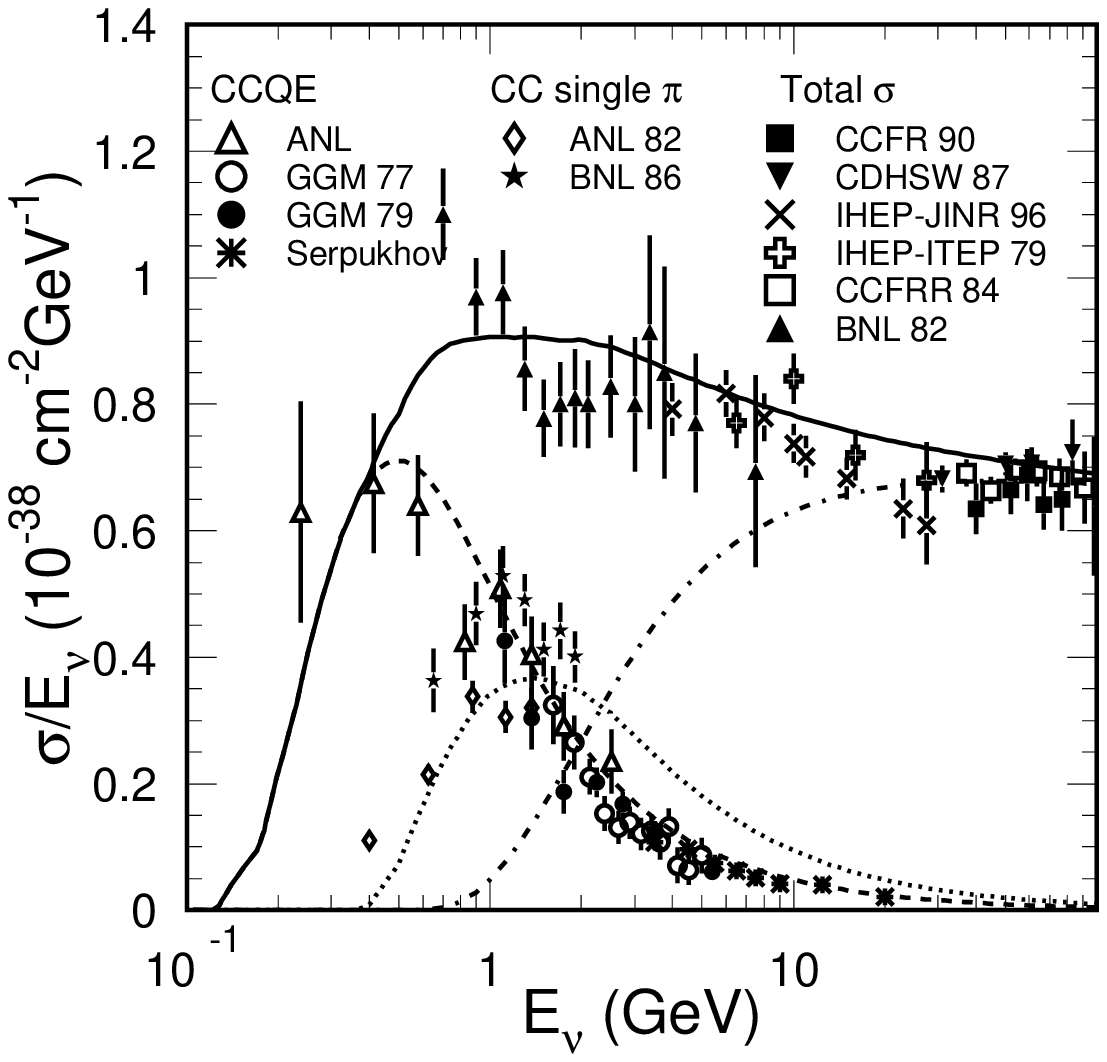, width=0.4\textwidth}
\epsfig{figure=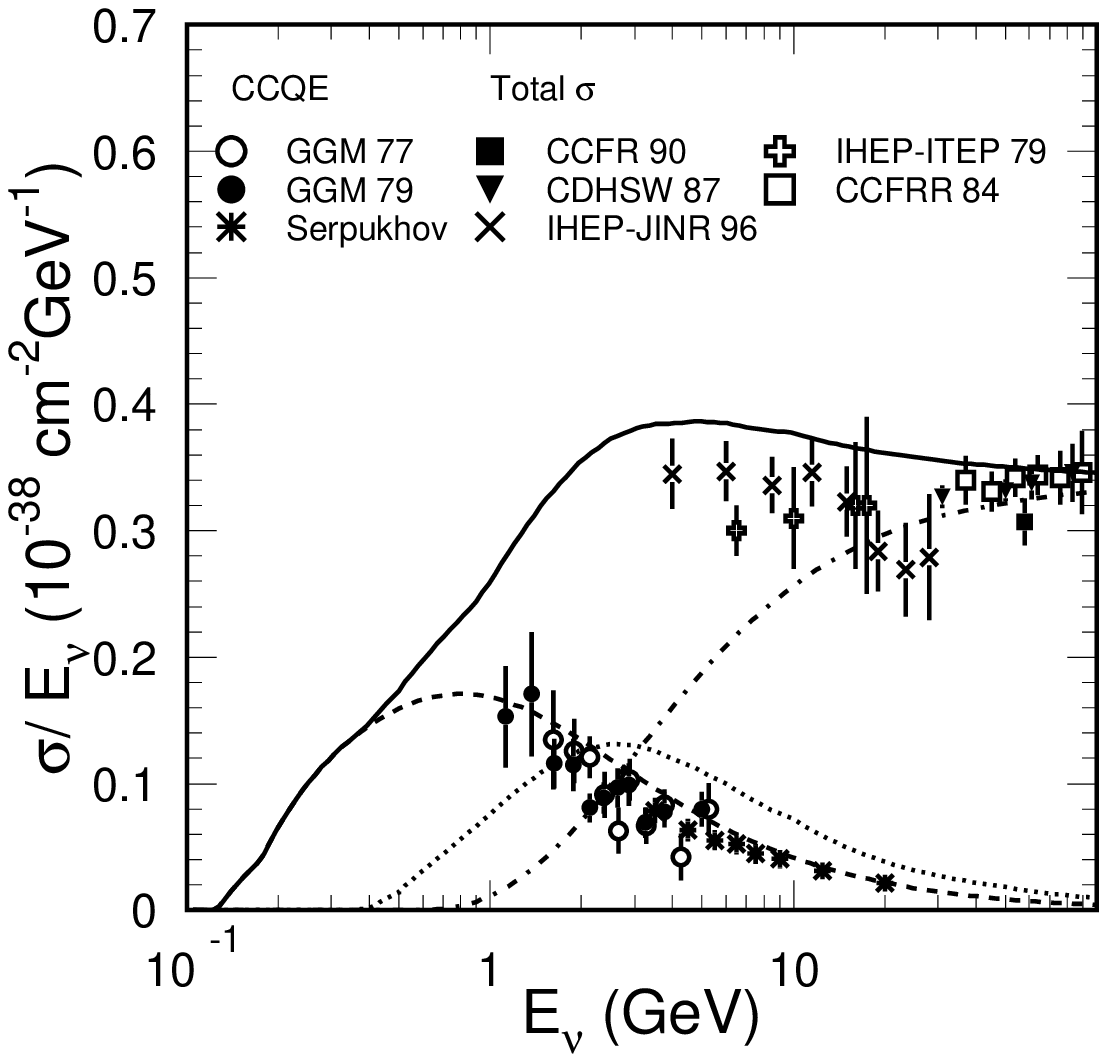, width=0.4\textwidth}
\caption{ Cross-section experimental values as a function of the neutrino energy (left). Results are compared to NEUT~\cite{Neut} Monte Carlo simulation. Points show
       the experimental data: BNL 82~\cite{BNL82}, CCFR 90~\cite{CCFR90}, CDHSW 87~\cite{CDHSW87}, IHEP-JINR 96~\cite{IHEP-JINR96},
    IHEP-ITEP 79~\cite{IHEP-ITEP79},   CCFRR 84~\cite{CCFRR84}, ANL 82~\cite{ANL82}, BNL 86~\cite{BNL86}, ANL~\cite{ANL}, GGM 77~\cite{GGM77},
        GGM 79~\cite{GGM79} and Serpukhov~\cite{Serpurkhov}.
 Cross-section experimental values as a function of the anti-neutrino energy (right). Results are compared to NEUT~\cite{Neut} Monte Carlo simulation.
       Points show the experimental data: CCFR 90~\cite{CCFR90}, CDHSW 87~\cite{CDHSW87}, IHEP-JINR 96~\cite{IHEP-JINR96},
       IHEP-ITEP 79~\cite{IHEP-ITEP79},   CCFRR 84~\cite{CCFRR84},  GGM 77~\cite{GGM77},
       GGM 79~\cite{GGM79}, and Serpukhov~\cite{Serpurkhov}. }
\label{Fig:cross-sections}
\end{center}
\end{figure}

 A compilation of actual knowledge on cross-sections is shown in Fig.\ref{Fig:cross-sections} for charged current neutrino and
anti-neutrino interactions.

 In general, the available data is old (from the 70's and 80's), normalized to charged current quasi-elastic using obsolete form factors and the beam spectrum
and flux was based on dubious hadron production models. The nuclear corrections are also not well documented or inconsistent, the data is sparse, low statistics
and some times inconsistent. The panorama is even worse when we consider production of more than one pion in the final state.

 Note that all existing cross-sections measurements above 200~MeV refer always to muon neutrinos and anti-neutrinos. The $\nu_{\tau}$ and $\nu_e$ cross sections
have not being measured due to the intrinsic difficulties to produce the appropriate neutrino beam and due to neutrino detection techniques. The cross-section can
be safely assumed to be equal to that of muon neutrinos, except when we are close to the threshold and the mass of the final state lepton together with
the nuclear effects play an important role. This is specially critical in the case of the low-$\gamma$ $\beta$ beams. The $\beta$ beams search for the transition
of $\nu_{e}$ to $\nu_{\mu}$, the low $\gamma$ version is being designed for energies from 100~MeV to 500~MeV. This is the energy region that has the largest uncertainties
in the relative cross-sections between  $\nu_{e}$ and $\nu_{\mu}$. Dedicated experiments will be needed in this case to control the systematic errors to the required
level, 0.1 \%.

\subsection{ Neutral current elastic and charge current quasi-elastic interactions}

 This interaction is of vital importance since it provides a method to reconstruct the neutrino energy. The actual knowledge of the cross-section is not
better than 20\%,  Theory is based on Conserved Vector Current
(CVC), Partially Conserved Axial Current (PCAC) and form factors
measured in electron nucleus scattering. The axial form factor is
not known and it is normally parametrized as a dipolar form factor
with the axial mass as a free parameter. It should be noticed that
this parameter changes the total cross-section and the $q^2$ of the
interactions. Both methods had been used to measure the parameter,
coming to contradictory results as it was noted in~\cite{MA}. Future
experiments~\cite{Minerva,T2K_2,ND280} will be able to measure if the 
axial form factor departs from the simplistic dipole format.

 The neutral current elastic scattering is not of relevant importance for oscillation experiments, although they can be used to determine the strange quark content
inside nucleons.

\subsection{ Charge and Neutral Current resonance: single and multi pion production }

 The production of charged and neutral pions are important backgrounds to both disappearance and appearance experiments. The knowledge of the resonance
cross-section is difficult to model. To the lack of knowledge of the standard
axial form factors we have to add the uncertainties on the amplitude
of high mass resonances in the transition region to the deep
inelastic. There are also models~\cite{Nieves} showing that the
non-resonant contributions could be relevant and affect the
cross-sections very close to threshold. The non-resonant
contribution is clearly present in $\nu_\mu n$ channels. Nieves~\cite{Nieves}
argued that it is probably necessary to depart from $C_5^A(0)\sim
1.2$, which is the PCAC dictated value of the leading axial form factor for the
$\Delta$ excitation.

 The neutral current resonant pion production should also be measured since they are background for appearance and also disappearance experiments, with the
pion being identified as a neutrino flavor tagging lepton. The
nuclear reinteractions are very relevant at this stage altering the
sign of the pion leaving the nucleus. The nuclear reinteraction
cross-sections are known to a 20 to 30\% and they are difficult to
measure in standard neutrino experiments. It is possible that T2K
will be able to address this measurement with the near detector that
has good particle identification  capabilities and momentum
resolution, see~\cite{T2K_2}.

\subsection{Neutral and charged current multi pion production and deep inelastic interactions}

  Deep inelastic cross-sections have been measured at high energies. The theoretical framework, based on structure functions, is well established and it has been
measured in different experimental conditions. But, there are still
some unclear items: nuclear effects, low $q^2$ region and the
transition region to the resonant (single and multi pion) neutrino
interactions.

As an example of the situation, the implementation of the transition
region in the NEUT Monte Carlo is done as a mixture of experimental
results and standard Monte Carlo tools. NEUT produces pions in the
final state according to FNL-7~\cite{FNL} results for a region where
1.3~GeV~$<$~W~$<$~2.0~GeV (W is the invariant mass of the hadronic
current) and according to JETSET 7.4~\cite{JETSET} above this value.

\subsection{ Charge and Neutral Current coherent pion production }

 The neutral current coherent pion production has been measured at relatively high energies (2.0~GeV) and heavy nuclei. The values for light nuclei and low
energies are not available and they might depend on the theoretical model for extrapolations. Miner$\nu$a~\cite{Minerva} and the near detector of T2K~\cite{ND280}
will be able to provide measurements for these reactions that are very important to determine the background on $\nu_e$ appearance. Anyhow, this background will be
mainly produced by interations of high energy neutrinos.

 The charged current coherent production is related to the neutral current cross-section at higher energies but the relation might be distorted at low energies
as it was suggested by a recent K2K result~\cite{CohK2K} due to the mass of the muon~\cite{ReinSehgal}.

\subsection{The cross-section double ratio}
\label{Double-ratio}

As discussed already in section \ref{Near_detectors}, the precise measurement of the CP asymmetry
\begin{equation}
A_{CP}=\frac{P(\numu \rightarrow \nue) - P(\numubar \rightarrow \nuebar )}{ P(\numu \rightarrow \nue) + P(\numubar \rightarrow \nuebar )} ,
\end{equation}
or precise measurement of any appearance probability, will require knowledge of the cross-section, efficiency and background of both the initial channel (for the near detector normalization) and of the appearance channel. The ratio to worry about is the electron-to-muon neutrino cross-sections. Indeed, the troublesome quantity is the double ratio:
\begin{equation}
DR =\frac{ \sigma_{\numu}/ \sigma_{\nue} } { \sigma_{\numubar}/ \sigma_{\nuebar}} ,
\end{equation}

where $\sigma_{\numu} $ really means $\sigma_{\numu}  \times \epsilon - B$, including a correction for efficiency $\epsilon$ and background $B$. Although it would seem that many systematic errors would cancel in this ratio, this is only partially true. The effects that ensure a deviation of this quantity from unity are quite difficult to master:
\begin{itemize}
\item
the muon mass effect;
\item
Fermi motion and binding energy;
\item
the non-isoscalarity of the target (this is particularly relevant for water where anti-neutrinos and neutrinos interact very differently on the free protons);
\item
the different neutrino and antineutrino $y$ distributions; and
\item
the different appearance of the final state lepton in the detector.
\end {itemize}

\begin{figure}[htbp]
\begin{center}
\includegraphics[width=0.49\textwidth]{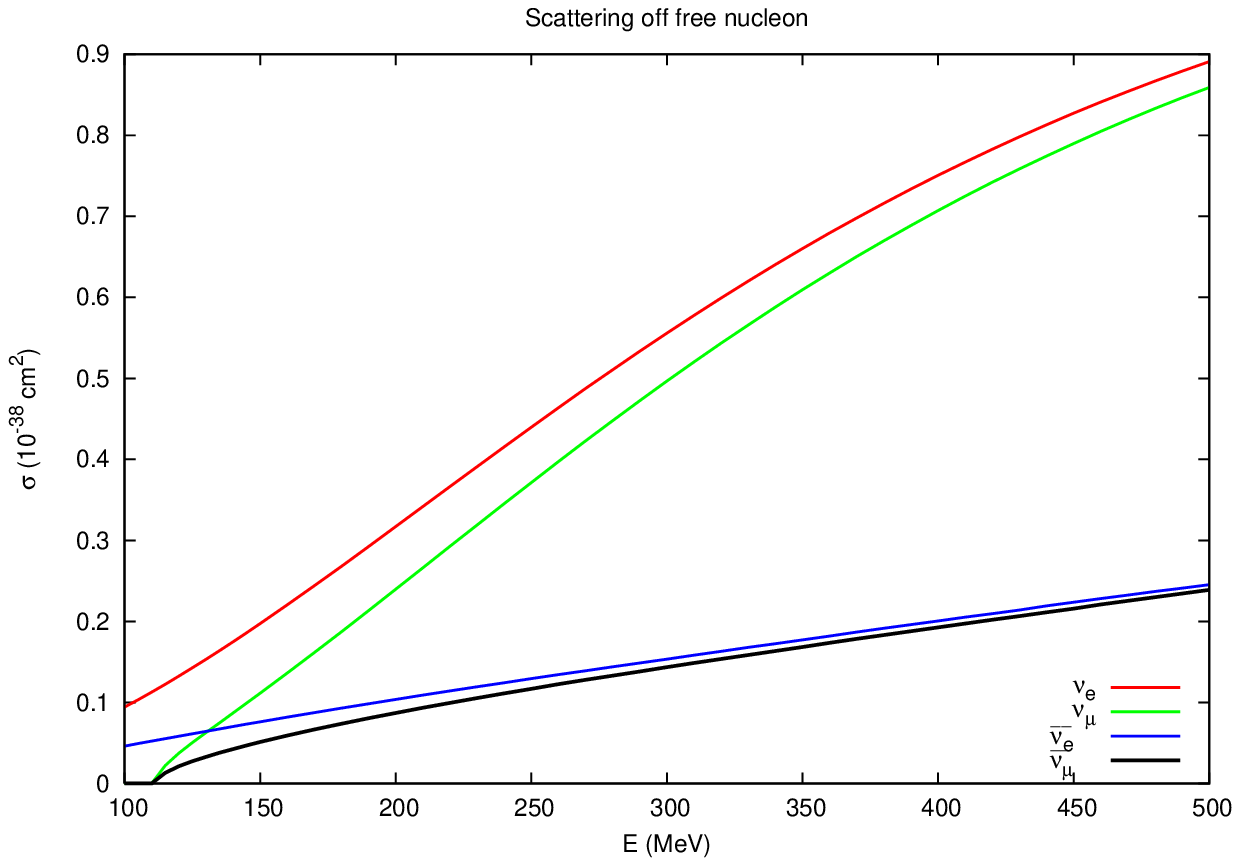}
\includegraphics[width=0.49\textwidth]{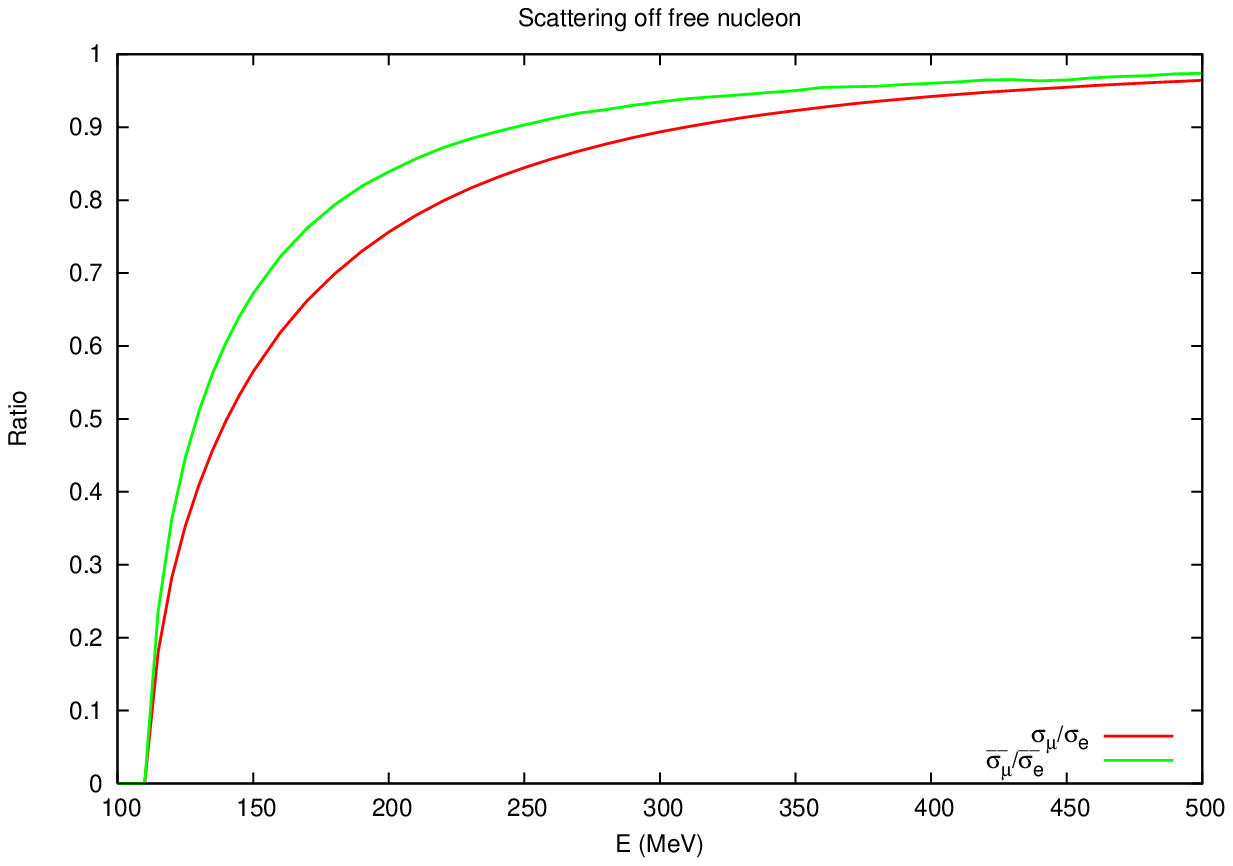}
\includegraphics[width=0.49\textwidth]{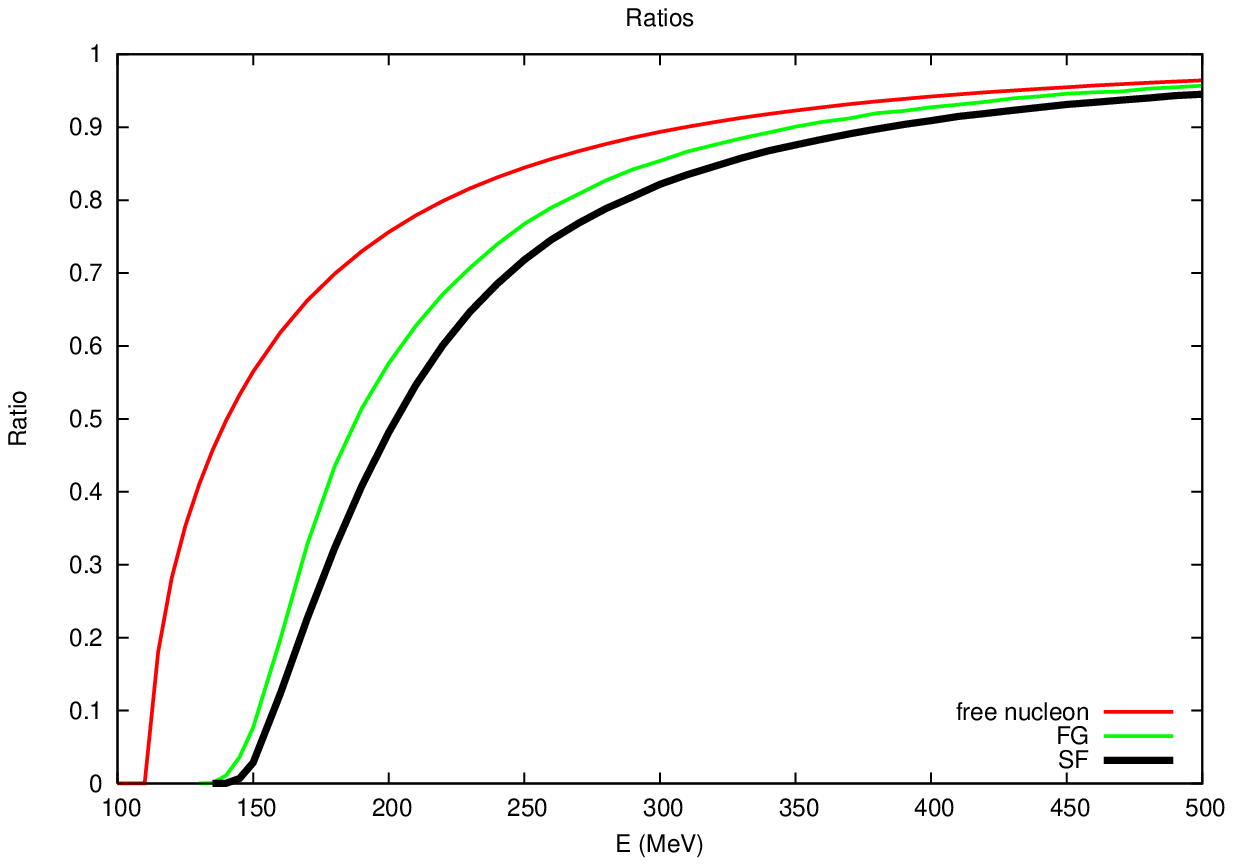}
\includegraphics[width=0.49\textwidth]{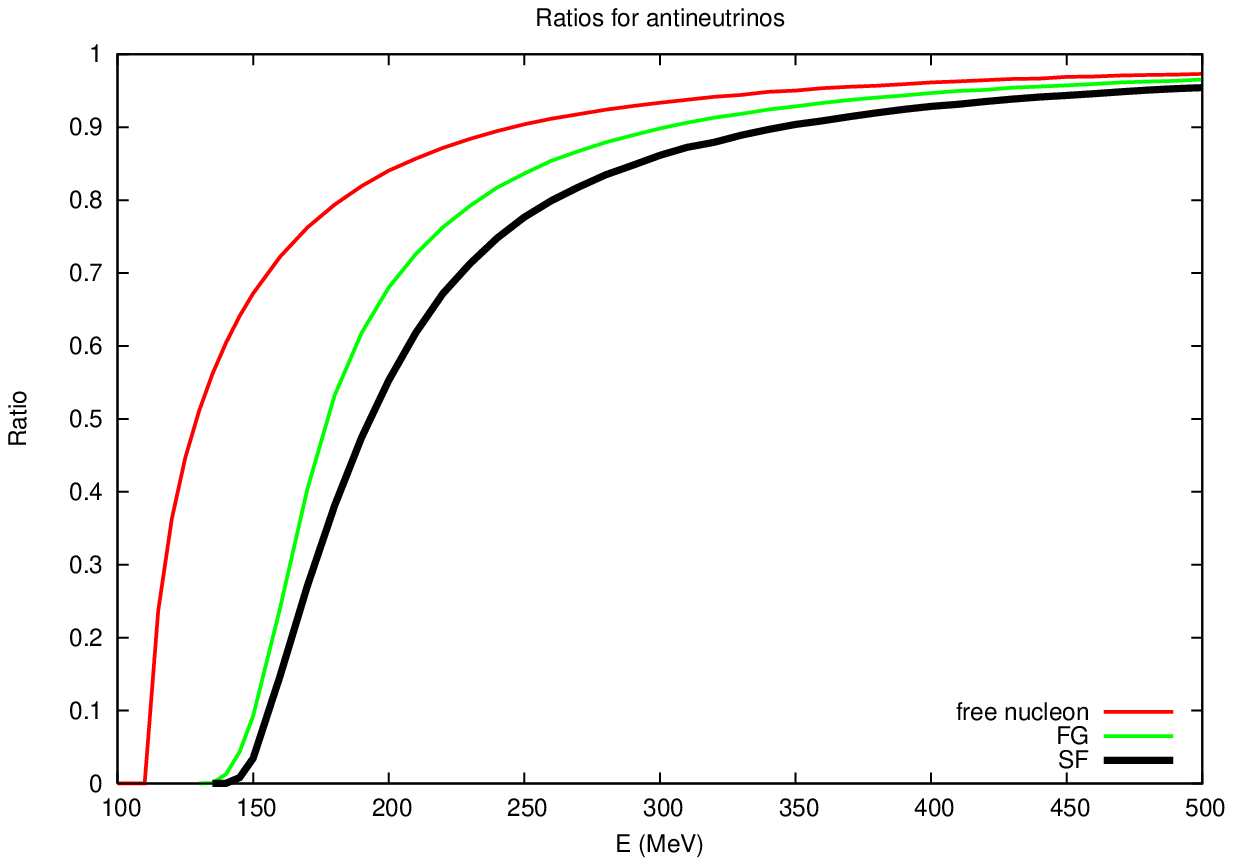}
\includegraphics[width=0.49\textwidth]{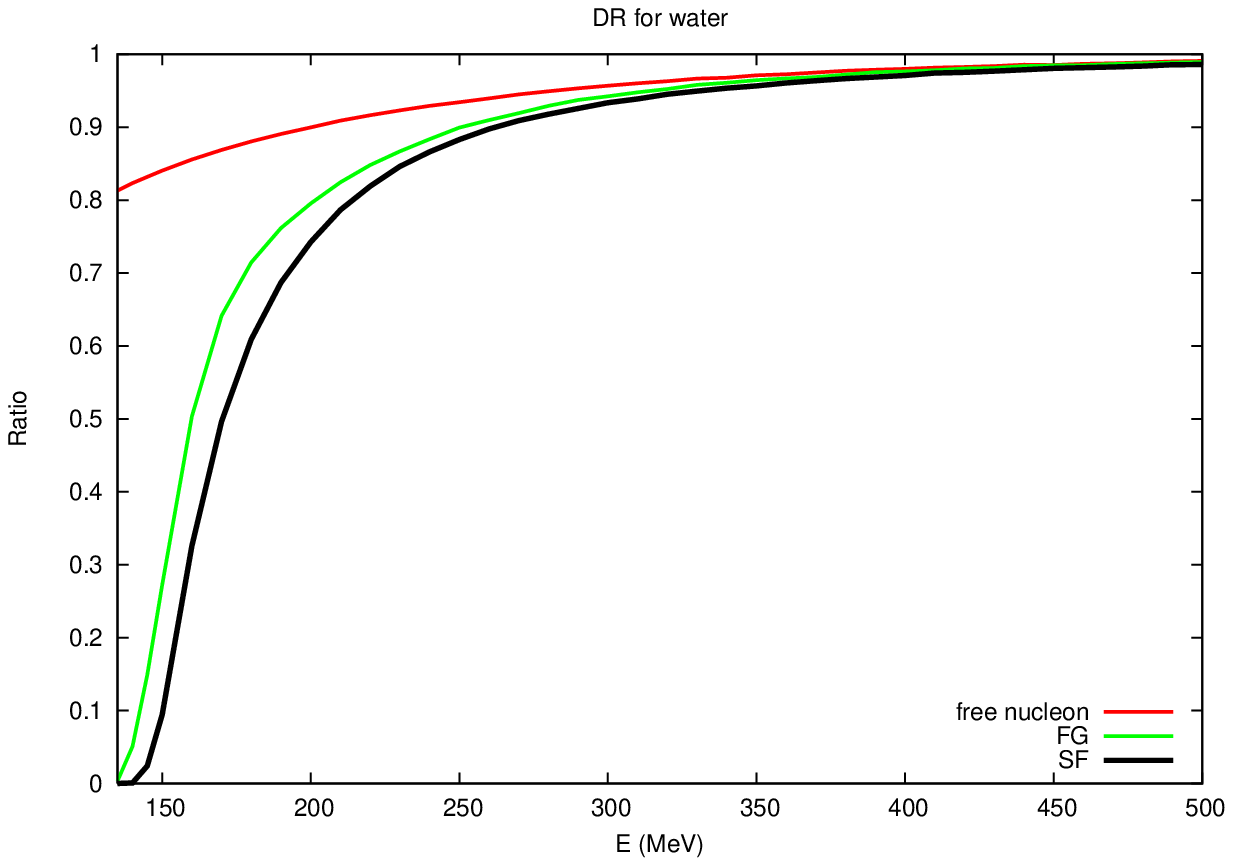}
\includegraphics[width=0.49\textwidth]{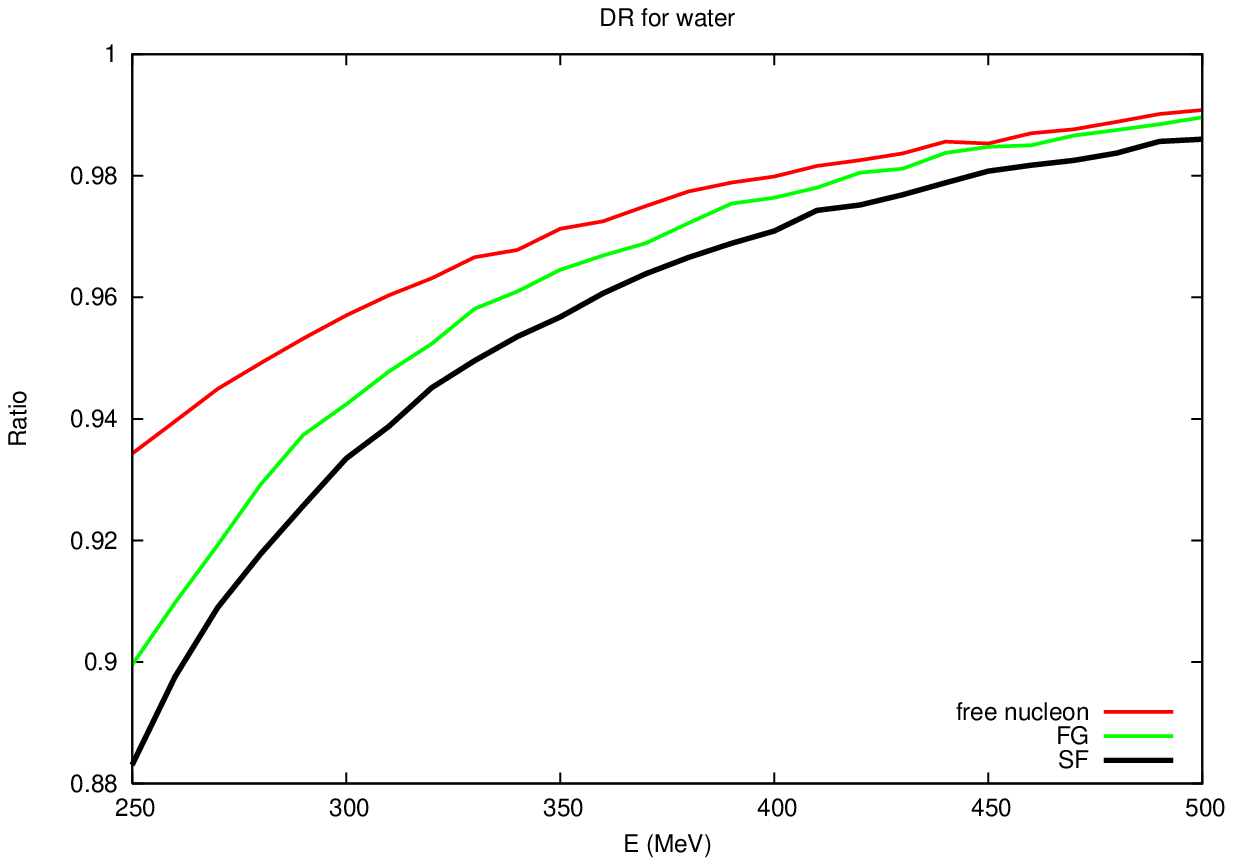}
\caption{Top Left: quasi-elastic cross-sections on free nucleon
(neutron for neutrinos and protons for anti-neutrinos) for electron
and muon neutrinos and antineutrinos.  The muon threshold is clearly
visible. Top Right: the  \numu to \nue and \numubar to \nuebar
cross-section ratios showing the effect of the different
y-distributions. Middle: the cross-section ratios between muon- and
electron neutrinos (left) and antineutrinos (right) 
taking into account
nuclear effects, compared to those on free nuclei. The binding
energy shows up as a shift in the threshold, but the exact
description of this is considered uncertain; the curves correspond
to modelling the nucleus with the Fermi Gas Model (FG) or with the
Spectral Function approach. Bottom
left: the double ratio in water from threshold to 1 GeV, and in the
'reliable' region above 250 MeV (right). } \label{cross4}
\end{center}
\end{figure}

These effects are particularly relevant for the low energy
neutrinos, as will be discussed here. One can legitimately wonder
whether everything needs to be measured or if theory cannot help by
predicting the double ratio using safe assumptions. Such an analysis
was developed by Jan Sobczyk and
collaborators \cite{Sobczyk}. If one concentrates on low energies,
the dominant cross-sections will be quasi-elastics. The
cross-sections for the four relevant species of neutrinos are shown
on the top line of Figure~\ref{cross4}.

The muon threshold effect is clearly visible. Due to the different
inelasticity (or $y$ distribution) of  neutrinos vs antineutrinos, the
muon mass correction is however not the same for neutrinos and
antineutrinos, by an amount that can be quite large (20\%).

The next thing to worry about are nuclear effects, which are nucleus dependent and particularly relevant in water where antineutrinos can interact on the free protons, while neutrinos cannot. These can be broadly separated in two classes, binding energy and Fermi motion.  The description of the effect of  binding energy is considered to be quite uncertain given that the debris of the nucleus from which the struck nucleon originates probably take away some of the binding energy in the reaction, and it cannot entirely be attributed to the struck nucleon. The resulting effect on the double ratio is extremely large at low energies, because of the existence of antineutrino interactions on the free protons. The region below 250 MeV probably cannot be trusted and the region above should be seen as having an uncertainty given by the following factors.

\begin{figure}[htbp]
\begin{center}
\includegraphics[width=10cm]{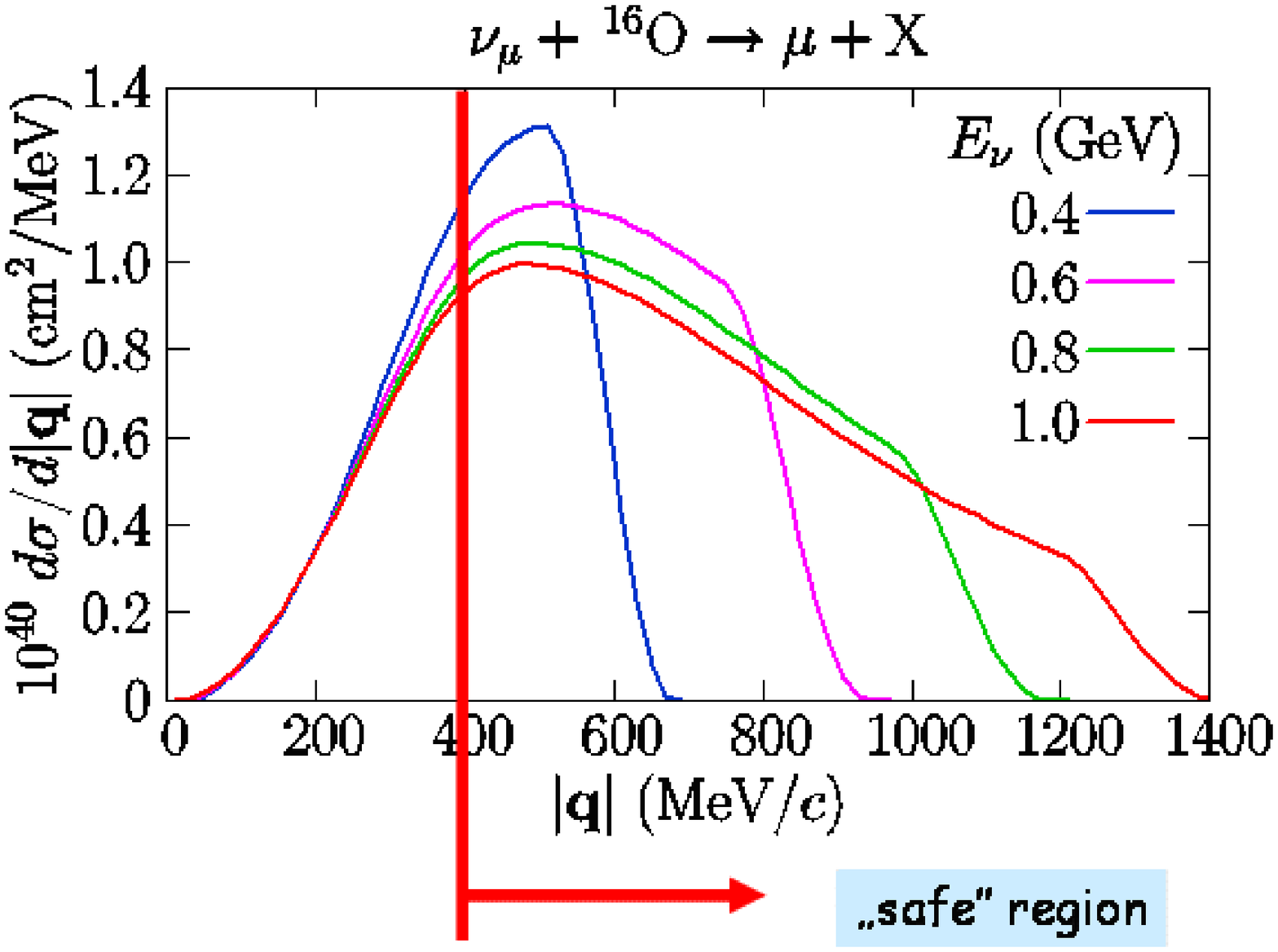}
\caption{Differential cross section of $\numu$ on $^{16}O$ as a function of momentum transfer,
at several values of neutrino energy. The Impulse Approximation is only reliable in the
region with $|\vec q| > 400$~MeV. } \label{fig:400MeV}
\end{center}
\end{figure}

\begin{itemize}
\item The uncertainty on the description of Fermi motion could be evaluated with
the guidance given by the difference between the Spectral Function approach and the  Fermi Gas model. Around 250 MeV this leads to an uncertainty of about 2\% on the double ratio.
\item The uncertainty due to the binding energy modelling. A
shift by, say, 50\% of the binding energy itself would change the double ratio by another 2\%.
\item There is also a large uncertainty related
to the Impulse Approximation (IA) used in cross section
computations. The IA assumes that the relevant degrees of freedom are
individual nucleons. The analysis of electron scattering data
clearly shows that the IA is reliable only for momentum transfers 
$|\vec q|>\sim 400$~MeV \cite{Ankowski_Sobczyk}. 
On the other hand, at a
neutrino energy $\sim 400$~MeV, about $40\%$ of the cross section
calculated within the IA corresponds to lower values of $|\vec q|$ (Fig.~\ref{fig:400MeV}). This
is a source of large uncertainty which is difficult to estimate. Of
course, one can be optimistic and believe that the ratios are not affected 
much by the use of the IA, but it is a source of additional systematic error.
\end{itemize}

Thus from considerations on total cross-sections alone, a
fundamental uncertainty of the order  of 3-4 \% can be ascertained.
The energy of 250 MeV incidentally corresponds to the oscillation
maximum for the distance between CERN and Fr\`ejus. Taking into
account the difficulties that will be associated with the different
energy spectra and detection efficiencies for muons and electrons,
it seems very unlikely that an uncertainty of less than 5\% on the
double ratio $DR$ can ever be achieved at low energies from a
combination of simulations and theory.

\newpage



\end{document}